\documentclass[twocolumn]{aastex631}
\usepackage{amsmath}
\usepackage{pbox}

\newcommand{\revone}[1]{\textcolor{black}{#1}}
\newcommand{\revtwo}[1]{\textcolor{black}{#1}}
\newcommand{\revthree}[1]{\textcolor{black}{#1}}
\newcommand{\revfour}[1]{\textcolor{black}{#1}}
\newcommand{\revfive}[1]{\textcolor{black}{#1}}

\shorttitle{Color transformation of optical/NIR wide-field surveys}
\shortauthors{Toptun et al.}

\graphicspath{{./}{figures/}}

\begin{document}

\title{Color Transformations of Photometric Measurements of Galaxies in Optical and Near-Infrared Wide-Field Imaging Surveys\footnote{An online color transformation calculator is available at: \url{https://colors.voxastro.org/}}}

\correspondingauthor{Victoria Toptun, Igor Chilingarian}
\email{victoria.toptun@voxastro.org, igor.chilingarian@cfa.harvard.edu}

\author[0000-0003-3599-3877]{Victoria A. Toptun}
\affiliation{European Southern Observatory, Karl Schwarzschildstrasse 2, 85748, Garching bei München, Germany}
\affiliation{Sternberg Astronomical Institute, M.V. Lomonosov Moscow State University, Universitetsky pr., 13, 119234, Moscow, Russia}
\affiliation{Department of Physics, M.V. Lomonosov Moscow State University, Moscow, Russia}
\author[0000-0002-7924-3253]{Igor V. Chilingarian}
\affiliation{Center for Astrophysics -- Harvard and Smithsonian, 60 Garden St. MS09, Cambridge, MA 02138, USA}
\affiliation{Sternberg Astronomical Institute, M.V. Lomonosov Moscow State University, Universitetsky pr., 13, 119234, Moscow, Russia}
\author[0000-0003-3255-7340]{Kirill A. Grishin}
\affiliation{Universit\'e Paris Cit\'e, CNRS, Astroparticule et Cosmologie, F-75013 Paris, France}
\affiliation{Sternberg Astronomical Institute, M.V. Lomonosov Moscow State University, Universitetsky pr., 13, 119234, Moscow, Russia}
\author[0000-0002-6425-6879]{Ivan Yu. Katkov}
\affiliation{New York University Abu Dhabi, Abu Dhabi, UAE}
\affiliation{Center for Astro, Particle, and Planetary Physics, NYU Abu Dhabi, Abu Dhabi, UAE}
\affiliation{Sternberg Astronomical Institute, M.V. Lomonosov Moscow State University, Universitetsky pr., 13, 119234, Moscow, Russia}

\begin{abstract}
Over the past 2 decades, wide-field photometric surveys in optical and infrared domains reached a nearly all-sky coverage thanks to numerous observational facilities operating in both hemispheres. However, subtle differences among exact realizations of Johnson and SDSS photometric systems require one to convert photometric measurements into the same system prior to analysis of composite datasets originating from multiple surveys. It turns out that the published photometric transformations lead to substantial biases when applied to integrated photometry of galaxies from the corresponding catalogs. Here we present photometric transformations based on piece-wise linear approximations of integrated photometry of galaxies in the optical surveys SDSS, DECaLS, BASS, MzLS, DES, DELVE, KiDS, VST ATLAS, and the near-infrared surveys UKIDSS, UHS, VHS, and VIKING. We validate our transformations by constructing $k$-corrected color-magnitude diagrams of non-active galaxies and measuring the position and tightness of the `red sequence.' We also provide transformations for aperture magnitudes and show how they are affected by the image quality difference among the surveys. We present the implementation of the derived transformations in {\sc python} and {\sc idl} and also a web-based color transformation calculator for galaxies. By comparing DECaLS and DES, we identified systematic issues in DECaLS photometry for extended galaxies, which we attribute to the photometric software package used by DECaLS. As an application of our method, we compiled two multi-wavelength photometric catalogs for over 200,000 low- and intermediate-redshift galaxies originating from CfA FAST and Hectospec spectral archives.
\end{abstract}

\keywords{\href{http://astrothesaurus.org/uat/269}{Color equation (269)}; \href{http://astrothesaurus.org/uat/1233}{Photometric systems (1233)}; \href{http://astrothesaurus.org/uat/611}{Galaxy photometry (611)};
\href{http://astrothesaurus.org/uat/586}{Galaxy colors (586)}}

\section{Introduction} \label{sec:intro}

Over the past decades, significant progress has been achieved in the quality and depth of wide-field imaging sky surveys in virtually all wavelength domains. However, different instruments implement even `standard' photometric systems slightly differently, and methods of data processing and analysis also vary. Therefore, for many scientific applications which require multiple sources of photometric data (e.g. to search astrophysical sources based on specific features of their spectral energy distributions or compare physical properties of objects of the same type originating from several surveys), one needs to homogenize and post-process these datasets. Homogeneous photometric and spectroscopic datasets possess a great discovery potential for finding new statistical properties of populations of astronomical objects of a given type (e.g. a universal color--color--magnitude relation of non-active galaxies found by \citealp{2012MNRAS.419.1727C}) or using sophisticated data mining techniques to identify rare sources with a frequency among the total population of 1:10,000 (e.g. compact elliptical galaxies \citealp{CZ15} or intermediate-mass black holes \citealp{Chilingarian+18}) or even 1:100,000 (e.g. diffuse extended post-starburst galaxies \citealp{2021NatAs...5.1308G}).

A few years ago our team released the Reference Catalog of Spectral Energy Distributions of galaxies (RCSED; \citealp{RCSED}), which contains the results of analysis of 800 k spectra of non-active galaxies accompanied by homogenized photometric data from GALEX \citep{2005ApJ...619L...1M}, SDSS \citep{2009ApJS..182..543A}, and UKIDSS \citep{2007MNRAS.379.1599L}. The photometric measurements in RCSED originated from multiple sources, however, measurements in every photometric band came only from one particular survey. This situation changed drastically in RCSEDv2, which contains a substantially larger sample of galaxy spectra, about 4.7~million, over the entire sky and photometric data from over a dozen of different wide-field sky surveys originating from different telescopes and instruments. However, all optical and near-infrared measurements were made in the photometric systems implementing either SDSS \citep{1996AJ....111.1748F} or Johnson \citep{1966CoLPL...4...99J}. 


Creating a photometric catalog of galaxies, which covers almost the entire sky requires a combination of several photometric surveys, hence, leading to the problem of converting all the measurements to a `standard' photometric system.
Color transformations for all the surveys used in RCSEDv2 were described in the literature, however, they were calculated using stars \citep{2015MNRAS.451.4238S, 2021ApJS..255...20A, 2019AJ....157..168D}, which cover a limited range of colors. Hence, \revfour{even though they work reasonably well for aperture and PSF magnitudes of point sources,} it is not surprising that if one applies them to galaxies having different colors and extended shapes, some serious systematic errors will emerge. \revone{We should consider all the emerging systematics for extended targets and, therefore, the calculations from ideally corrected point sources will not suit our purposes. For example, aperture corrections become very important if one starts from point sources. They are dependent on the image quality, which differs quite substantially from one area of the sky to another even within each particular ground-based survey. Moreover, aperture corrections for galaxies require the knowledge of their seeing-deconvolved surface brightness profiles. This is the main motivation for us to use total magnitudes of galaxies to re-compute the transformations. This should also significantly mitigate the problem of varying depths delivered by different surveys we included in this study.}

\begin{figure}
    \centering
    \includegraphics[width=\hsize]{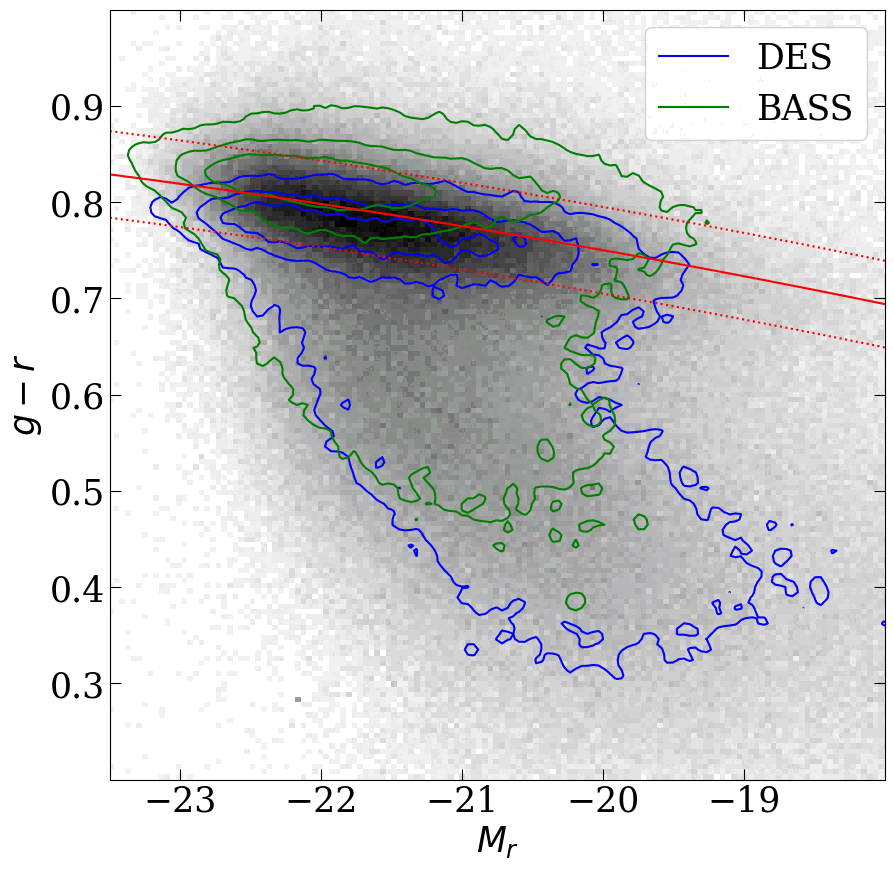}
    \caption{A color--magnitude diagram for low-redshift ($0.02<z<0.2$) galaxies constructed from SDSS (gray shaded image), DES (blue contours), and BASS (green contours). DES and BASS data were converted into the SDSS photometric system using the published transformations. The red sequence as defined by the RCSED catalog is shown by the red solid line and its r.m.s. by the red dotted lines.}
    \label{fig_rs_nocorrphot}
\end{figure}

As an ultimate quality test of the color conversions in the optical bands one can use a unique property of non-active galaxies, their color bi-modality, in particular the very tight `red sequence' formed mostly by early-type galaxies in the color--magnitude diagrams constructed from $u$ or $g$ (blue) and $r$, $i$, or $z$ (red) data \citep{1977ApJ...216..214V,1978ApJ...223..707S}. In order for this test to work properly, we need to know the redshifts and compute accurate $k$-corrections, that is a correction of a photometric measurement of an extragalactic source, which brings it to either rest-frame or to a specifically chosen redshift value. If we use galaxy flux measurements (Petrosian magnitudes, \citep{1976ApJ...209L...1P}) from  Dark Energy Survey and published color conversions between DES and SDSS from \citet{2021ApJS..255...20A} for a sample of low-redshift ($z<0.2$) galaxies from RCSEDv2, there will be a systematic offset of the red sequence in the $g-r$ color by about $-0.05$~mag, and the slope will be shallower. On the other hand, if we use BASS, the Northern part of DESI imaging surveys, the red sequence turns up redder by $+0.06$~mag, also with a somewhat different slope from that derived using SDSS data (see Figure~\ref{fig_rs_nocorrphot}). Therefore, the intrinsically thin (0.04~mag) red sequence in the combined dataset will puff up to $>0.1$~mag hampering many scientific applications, which require SED fitting, e.g. photometric redshift estimates, search for members of specific galaxy types, inferring star formation histories.

This brings our main motivation to re-compute color transformations between several variants of the optical SDSS ($ugriz$) and near-infrared Johnson ($YJHK$) photometric systems using a carefully selected sample of low-to-intermediate redshift galaxies with spectroscopic redshifts, all measured in a similar fashion in terms of fluxes (Petrosian magnitudes). We can then test the derived transformations by reconstructing the color--magnitude diagrams of galaxies in different bands and measuring the position and thickness of the red sequence.


\section{Data}
\label{sec:data}

\begin{figure*}
\includegraphics[width=0.47\hsize]{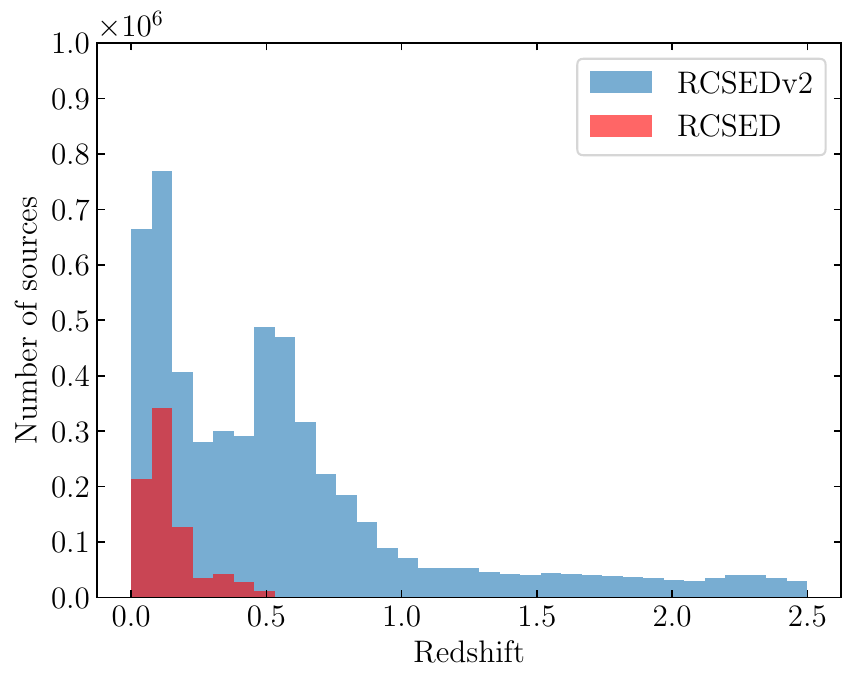}
\includegraphics[width=0.47\hsize]{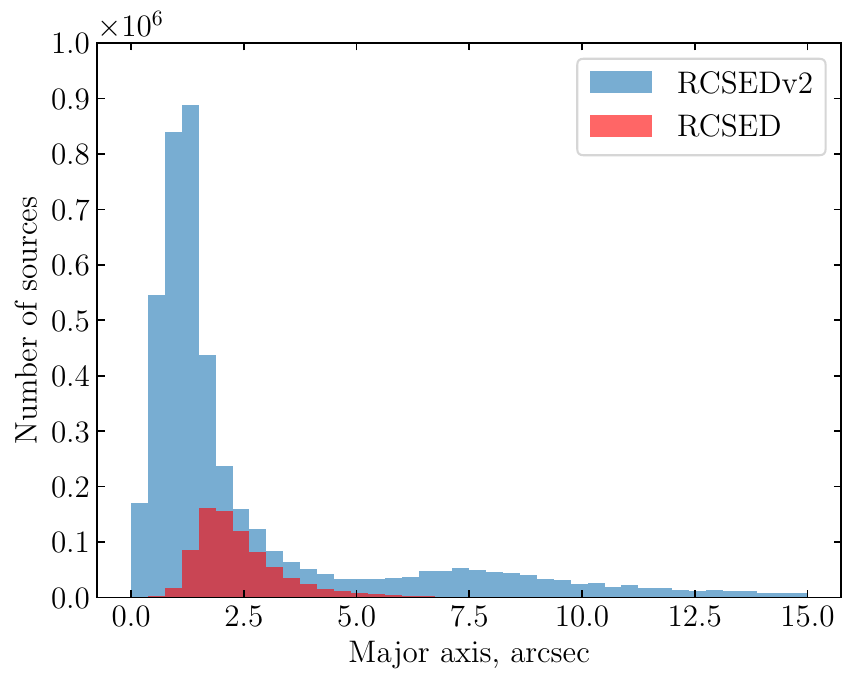}\\
\vspace{-0.4cm}
\caption{Distributions of redshift (left) and major axis (right) for RCSEDv2 (blue) and RCSED (red). 
A large amount of sources on z=0.6 originates from eBOSS.\label{hist}}
\end{figure*}

\begin{figure*}
\centering
\includegraphics[width=1\hsize]{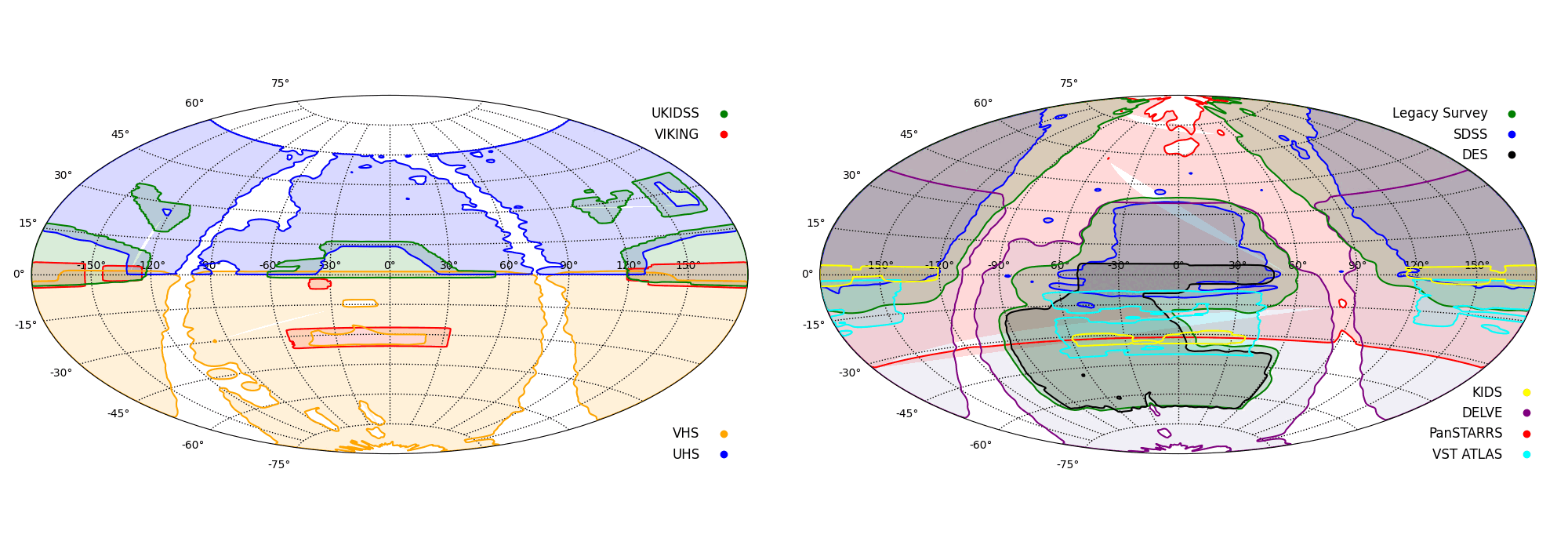}
\caption{Coverage footprints of optical (right) and infrared (left) \revone{photometry for the parent sample of 4.7M galaxies.}\label{cover}}
\end{figure*}

The input photometric catalog is based on 4.7M galaxies with redshifts in publicly available wide-field surveys covering almost completely the northern and southern hemispheres of the sky. Redshifts are needed for further validation of photometry through building a red sequence (see \ref{subsection:rs}), as well as for quickly separating galaxies from stars. To exclude the stars, we selected only objects with redshifts greater than 0.0016 (corresponding to radial velocity $\sim$ 480 km/s) or those classified as galaxies in the Hyperleda database \citep{2014A&A...570A..13M}. The sample comprises galaxies originating from the following spectroscopic surveys and catalogs: SDSS DR16 \citep{2020MNRAS.498.2354R}, 2dFGRS \citep{2001MNRAS.328.1039C}, 6dFGS \citep{2004MNRAS.355..747J}, FAST \citep{2021AJ....161....3M}, Hectospec \citep{2005PASP..117.1411F}, LAMOST \citep{2019ApJS..240....6Y}, LEGA-C \citep{2021ApJS..256...44V}, DEEP2/DEEP3 \citep{2013ApJS..208....5N}, WiggleZ \citep{2012PhRvD..86j3518P}, GAMA DR4 \citep{2020MNRAS.496.3235B}, 2dFLenS \citep{2016MNRAS.462.4240B}.

To investigate the differences in photometric systems and calculate the conversions, we used publicly available wide-field surveys in the optical and infrared spectral \revtwo{domains} with photometric systems similar to or implementing those of SDSS and UKIDSS. The final photometric catalog was assembled from the following sky surveys:  SDSS, DESI Legacy Surveys \citep{2019AJ....157..168D}, Dark Energy Survey \citep{2021ApJS..255...20A}, DELVE \citep{2022ApJS..261...38D}, PanSTARRS PS1 \citep{2020ApJS..251....7F}, VST ATLAS \citep{2015MNRAS.451.4238S}, KiDS \citep{2017A&A...604A.134D} (optical), UKIDSS, UHS \citep{2018MNRAS.473.5113D}, VHS \citep{2012A&A...548A.119C}, VIKING \citep{2013Msngr.154...32E} (NIR) and unWISE \citep{2019ApJS..240...30S} (NIR/mid-IR). The datasets were retrieved using Virtual Observatory access services. We list all the data sources in Table~\ref{tab_sample}.

The unWISE data are not used to calculate conversions, but to remove quasars and strong type-I AGN from the sample. We exclude quasars, which often show significant broadband variability in optical and NIR domains exceeding 0.2~mag and therefore are not suitable for calibration purposes by using an empirical filter based on unWISE magnitudes: $w1-w2<0.2$ \citep{2013ApJ...772...26A}. In addition, to exclude sources with poor quality of photometry, we did not use measurements with statistical uncertainties exceeding 0.3~mag.

\begin{table*}
\caption{List of surveys used in this study\label{tab_sample}}
\begin{center}
\begin{tabular}{lllllll} 
 \hline
 Name of survey& Instrument & Coverage &  Bands & Type & Apertures & Source \\
  &  & (deg$^2$) &   &  & Diameter ($''$) &  \\
 \hline 
  SDSS DR17 & SDSS & 14,555  & \textit{ugriz} & Petrosian & 2, 3 & SDSS CasJobs\footnote{\url{https://skyserver.sdss.org/CasJobs/}}\\
  DES DR2 & DECam & 5000  & \textit{griz} (DECam) & Petrosian & \pbox{2.5cm}{0.49, 0.97, 1.46,\\ 1.95, 2.92, 3.90,\\ 4.87, 5.84, 6.82,\\ 7.79, 11.69, 17.53} & Astro Data Lab\footnote{\url{https://datalab.noirlab.edu/}}\\
  DELVE DR2 & DECam &  17,000  & \textit{griz} (DECam) & Petrosian & none & Astro Data Lab\\
  \pbox{2.5cm}{Legacy\,Survey\,\revone{DR9}\\(DeCALS, MzLS+BASS)} & \pbox{2.5cm}{DECam\\90Prime\\Mosaic-3} & 19,000  & \textit{grz}  & \revone{Model} & \pbox{2.5cm}{1, 1.5, 2, 3, 4,\\7, 10, 14} & Astro Data Lab\\
  PanSTARRS \revone{DR1} & PanSTARRS & 30,000 & \textit{griz} (PS) & Kron, PSF & none & Vizier TAP\footnote{\url{https://tapvizier.u-strasbg.fr/adql/}} \\
  VST ATLAS \revone{DR3} & VST OmegaCAM & 4500 & \textit{ugriz} (VST) & Petrosian & 2, 2.8, 5.7 & Vizier TAP \\
  KiDS \revone{DR3} & VST OmegaCAM & 1350 & \textit{ugriz} (VST) & Kron & none & Vizier TAP \\
  UKIDSS \revone{DR11} & WFCAM &  3700  & \textit{YJHK} (UKIRT) & Petrosian & 2, 2.8, 5.7 & Astro Data Lab \\
  UHS \revone{DR1} & WFCAM &  12,700  & \textit{YJHK} (UKIRT) & Petrosian & 2, 2.8, 5.7 & WFCAM Science Archive\footnote{\url{http://wsa.roe.ac.uk/}} \\
  VHS \revone{DR6} & VIRCAM & \revtwo{16,730} & \textit{YJHK} (VISTA) & Petrosian & 2, 2.8, 5.7 & Astro Data Lab \\
  VIKING \revone{DR5} & VIRCAM & \revtwo{1500} & \textit{YJHK} (VISTA) & Petrosian & 2, 2.8, 5.7 & VISTA Science Archive\footnote{\url{http://horus.roe.ac.uk/vsa/}} \\
\hline 
 \end{tabular} 
 \end{center} 
 \end{table*}

\section{Method description}
\subsection{Data preprocessing}

Before starting data homogenization, it is necessary to reduce all the magnitudes to common zero-points and aperture radii, as well as clear the sample from obviously incorrect values that will clog up the statistics. In the beginning, magnitudes with NaN values such as --~9999, 999, etc. (the exact value depends on the survey) were removed from the sample. Surveys that are not originally in the AB system are converted to AB using published zero-point differences: \revone{\citet{2006MNRAS.367..454H} for the UKIRT surveys UKIDSS and UHS  (+0.634, +0.938, +1.379, +1.9 for \textit{YJHK} respectively) and \citet{2018MNRAS.474.5459G} for the VISTA surveys VHS and VIKING (+0.6, +0.916, +1.366, +1.827 for \textit{YJHK} respectively)}. Also, to fit the data we use measurements only for galaxies with small angular sizes having radii containing 50\% of Petrosian flux smaller than 2.$''$5 to minimize the effects of potentially poor sky subtraction for very extended targets.


\begin{figure}
\includegraphics[width=0.98\hsize]{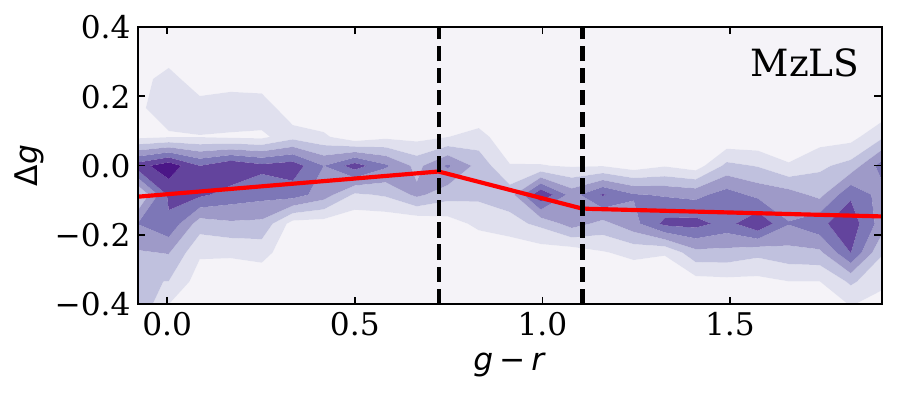}\\
\includegraphics[width=0.98\hsize]{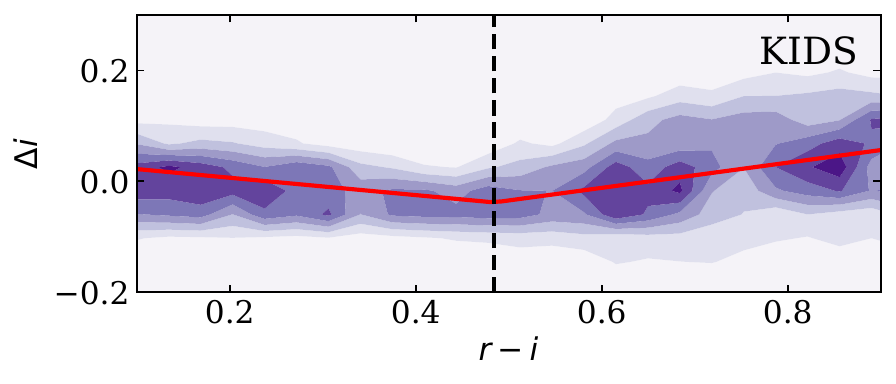}\\
\vspace{-0.4cm}
\caption{Example of a piecewise linear regression in the \revtwo{Legacy-MzLS $g$ band (left) and KiDS $i$ band} (right). Black dashed lines show the breakdown into linear regression segments. The color indicates the weights. The red line is the \revtwo{fitting} result. The Y-axis shows the difference between the correcting band and the corresponding SDSS band, the X-axis shows the color in the correcting survey.\label{fit}}
\end{figure}

\subsection{Calculation of color transformations}

After preprocessing we convert all magnitudes to some ``standard'' photometric system: this is a crucial step when one needs to use photometric measurements originating from different sources. Despite the same name, e.g. ``SDSS g,'' the actual implementations of a given photometric systems by different survey have significantly different filter throughput curves. Therefore, one needs to adopt one reference photometric system and convert all the available measurements into it. We adopted the filter sets from SDSS in the optical and UKIDSS in the near-infrared as our internal reference. One has to pay special attention to the fact that ``integrated'' photometry from different surveys may have different magnitude types, e.g. Petrosian in SDSS vs Kron \citep{1980ApJS...43..305K} in PanSTARRS. Also, Petrosian radii depend on the survey depth, therefore they are not the same for, e.g. SDSS and DES, leading to inconsistent integrated flux measurements. Nevertheless, for the computation of color transformations we used total magnitudes because the main goal of our work was to be able to build integrated spectral energy distributions of galaxies using data from different surveys, which would be consistent among them. We also discuss the applicability of our transformations to aperture magnitudes, which do not depend on the survey depth, \revtwo{but do depend on the image quality.}

We approximate a relation between the magnitude difference of the survey being converted and the reference survey and an observed color of a galaxy. A simple $\chi^2$ fitting of a 2D scattered dataset with the weights of each point corresponding to the uncertainties of original measurements will not produce satisfactory results because densely populated regions of the color--magnitude space will dominate the statistics and the solution will ` `ignore'' sparsely populated ``edges.'' \citet{2012MNRAS.419.1727C} proposed a weighted approach to fit 2D and 3D scattered datasets from SDSS and GALEX to infer the parameterisation of the color--color--magnitude relation for non-active galaxies: a weight assigned to each measurement depends on the local density of data points in the parameter space inferred from 2D or 3D histograms. Here we use a variant of this approach, which includes 2 steps.

The first step is to compute a 2D color--magnitude histogram in the survey we are converting, e.g. $g$ vs $(g-r)$ if we correct DES $g$ into SDSS $g$ using DES $g-r$ color. We keep 98\%\ of data along each dimension by first computing 1D distributions of all data points along color and magnitude and then rejecting 1\%\ of the strongest outliers on each side of the distribution. Then the remaining color--magnitude space is split into 50$\times$50 bins to compute a 2D histogram. Then the 1st-step weight is computed as the inverse of a count in the 2D histogram to account for the non-uniform distribution of galaxies in the observed color--magnitude space. For the subsequent analysis we reject the measurements, which fall into the bins of a 2D histogram containing fewer than 15 data points. 

On the second step we compute a set of 1D distributions of magnitude difference between the survey being converted (e.g. DES $g$) and the reference (e.g. SDSS $g$) in 17 bins on the observed color (e.g. DES $g-r$) assigned as follows: (i) the 80\%-ile of the data is split into 11 equally sized bins; (ii) the 80\%-98\%-ile is split into 3 bins on each side of the 90\%-ile distribution; (iii) the remaining 1\%\ of outliers by color on either side are rejected. Then in each of the 17 bins we compute a 1D histogram on the magnitude difference (e.g. $g_{\mathrm{DES}}-g_{\mathrm{SDSS}}$) in 150 bins rejecting 10\%\ of the strongest outliers on each side, i.e. keeping 80\%-ile of the data. This apparently high number is needed to properly handle the central very dense part of the distribution. Then, each of these 1D histograms is normalized by its maximum and the 2nd-step weights for each photometric measurement is computed as the inverse of the counts in the 1D distributions. 


At the end, we multiply 1st- and 2nd-step weights to assign a final weight to each measurement and use a linear regression to compute the color transformation in the form:
\begin{equation*}
    mag_{\mathrm{target}} = mag_{\mathrm{ref}} + a + b \cdot color_{\mathrm{ref}}
\end{equation*}

\begin{figure*}
\includegraphics[width=1\hsize]{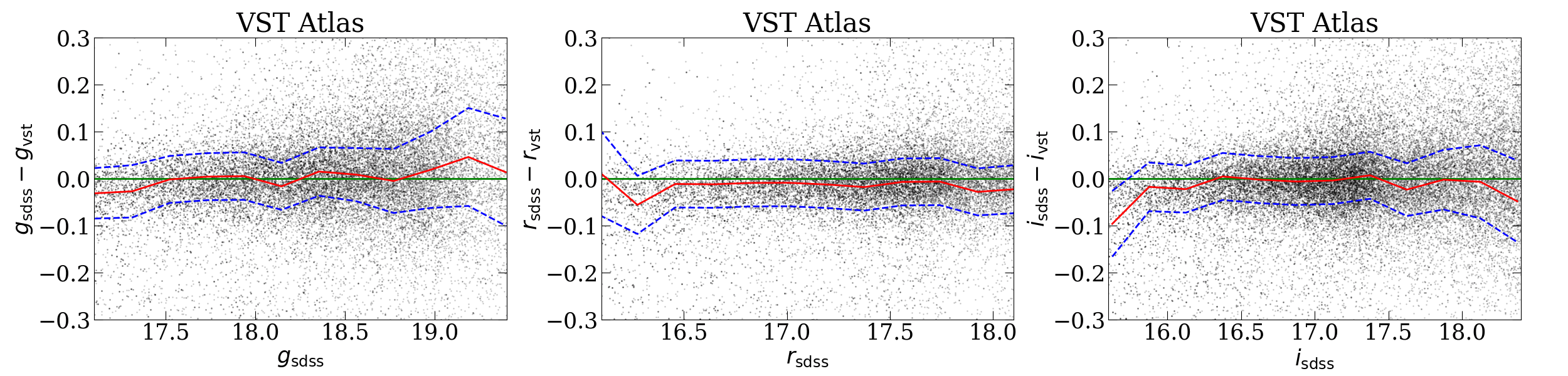}\\
\vspace{-0.4cm}
\caption{\revone{Magnitude residuals for corrected VST Atlas and SDSS as a function of the magnitude used in the fitting. The result does not show any magnitude-dependent effects.} \label{vst_mag_depedence}}
\end{figure*}

\revtwo{In most cases}, a linear regression cannot properly describe the photometric transformation in the entire range of observed colors. In such cases, we use a piece-wise linear regression with up-to three color sub-ranges (see Figure~\ref{fit}). We preferred a piece-wise linear fit to a nonlinear (e.g. 2-nd or 3-rd degree polynomial) solution because the latter case could lead to very high residuals in the tails of the color distribution.
\revone{We do not \revtwo{impose any constraints} on  magnitude range and \revtwo{compute the regression} for the color dependence only, because the result does not show any magnitude dependent effects (see Figure~\ref{vst_mag_depedence}).}

\begin{figure}
\centering
\includegraphics[width=1\hsize]{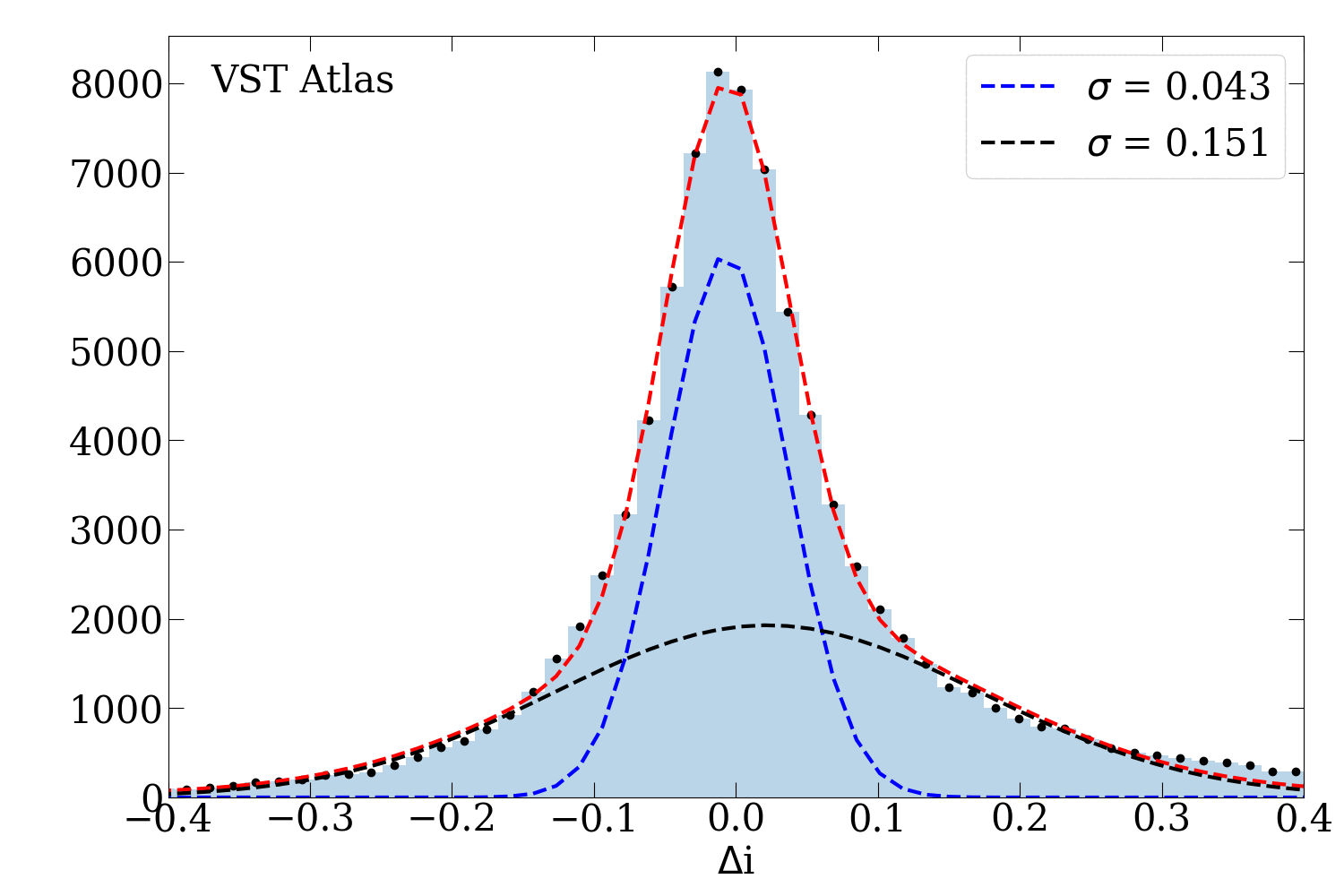}\\
\caption{A histogram of the fitting residuals for the VST Atlas $i$-band approximated by two Gaussians, one for a `background' level and another one for a sharp peak.\label{errors}}
\end{figure}

To calculate the quality of the derived transformation, we could not use a standard r.m.s. because some of the photometric measurements used in the fitting have rather large uncertainties of up-to 0.3~mag (see the previous section). Therefore, when we compute a histogram of the fitting residuals, we see a `background' level looking like a large Gaussian with a sharp peak on top of it: if we compute an r.m.s. value without taking this into account, we will severely overestimate the r.m.s. of the residuals. Therefore, the values of residuals presented in the next section of the paper correspond to the moments of the peak on top of the background distribution shown in Figure~\ref{errors}.

\section{Results}\label{sec:results}

\begin{figure*}
\includegraphics[width=0.33\hsize]{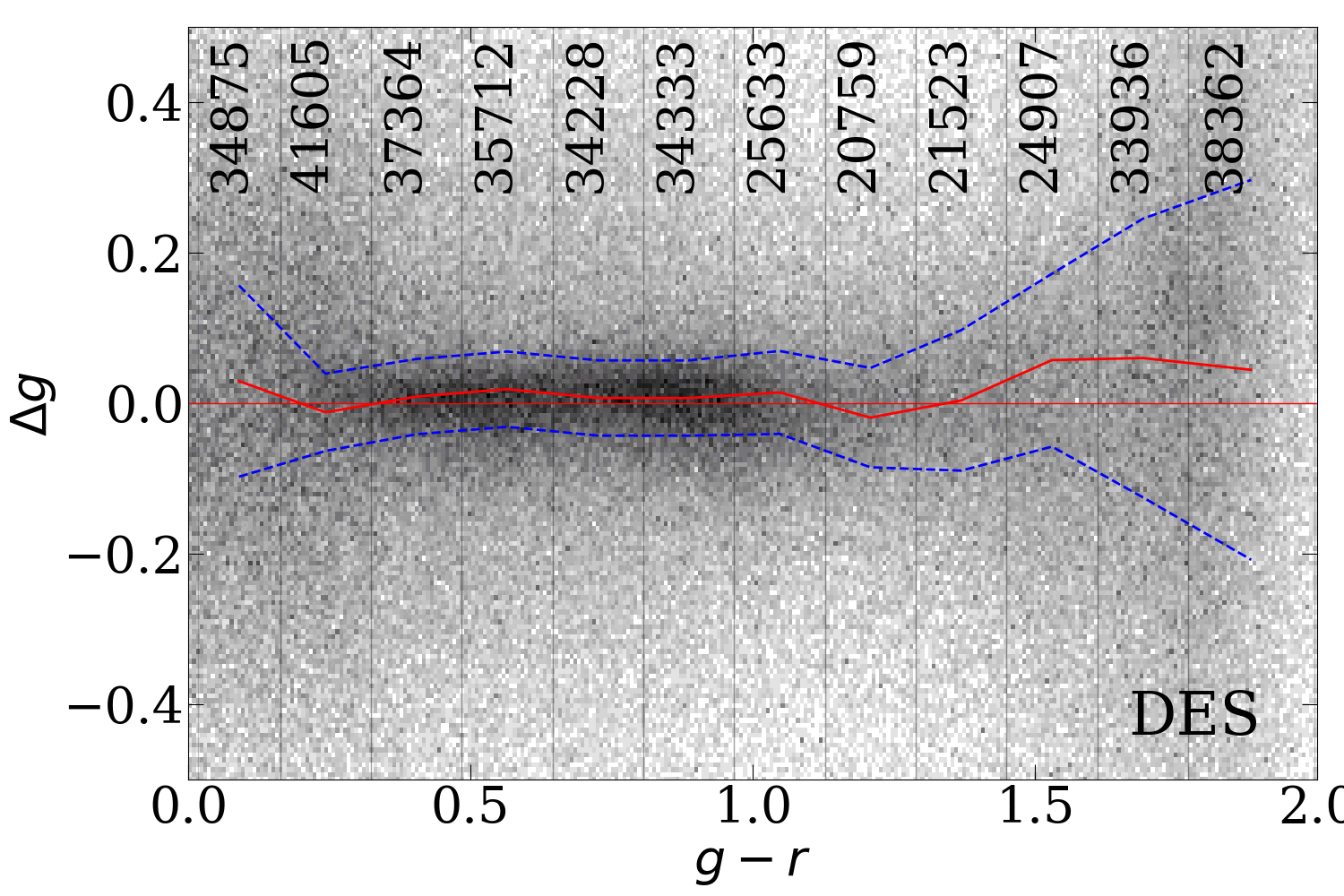}
\includegraphics[width=0.33\hsize]{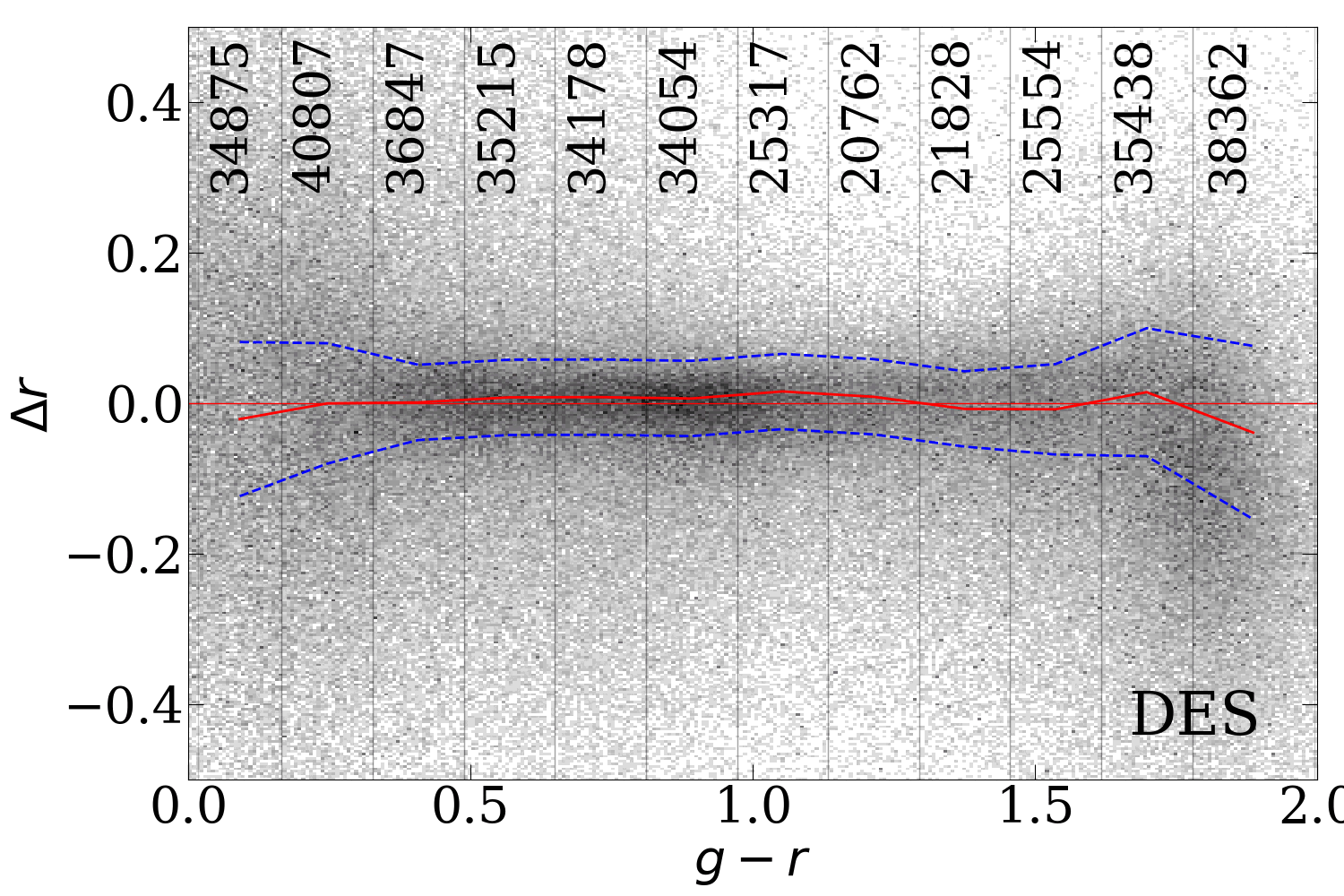}
\includegraphics[width=0.33\hsize]{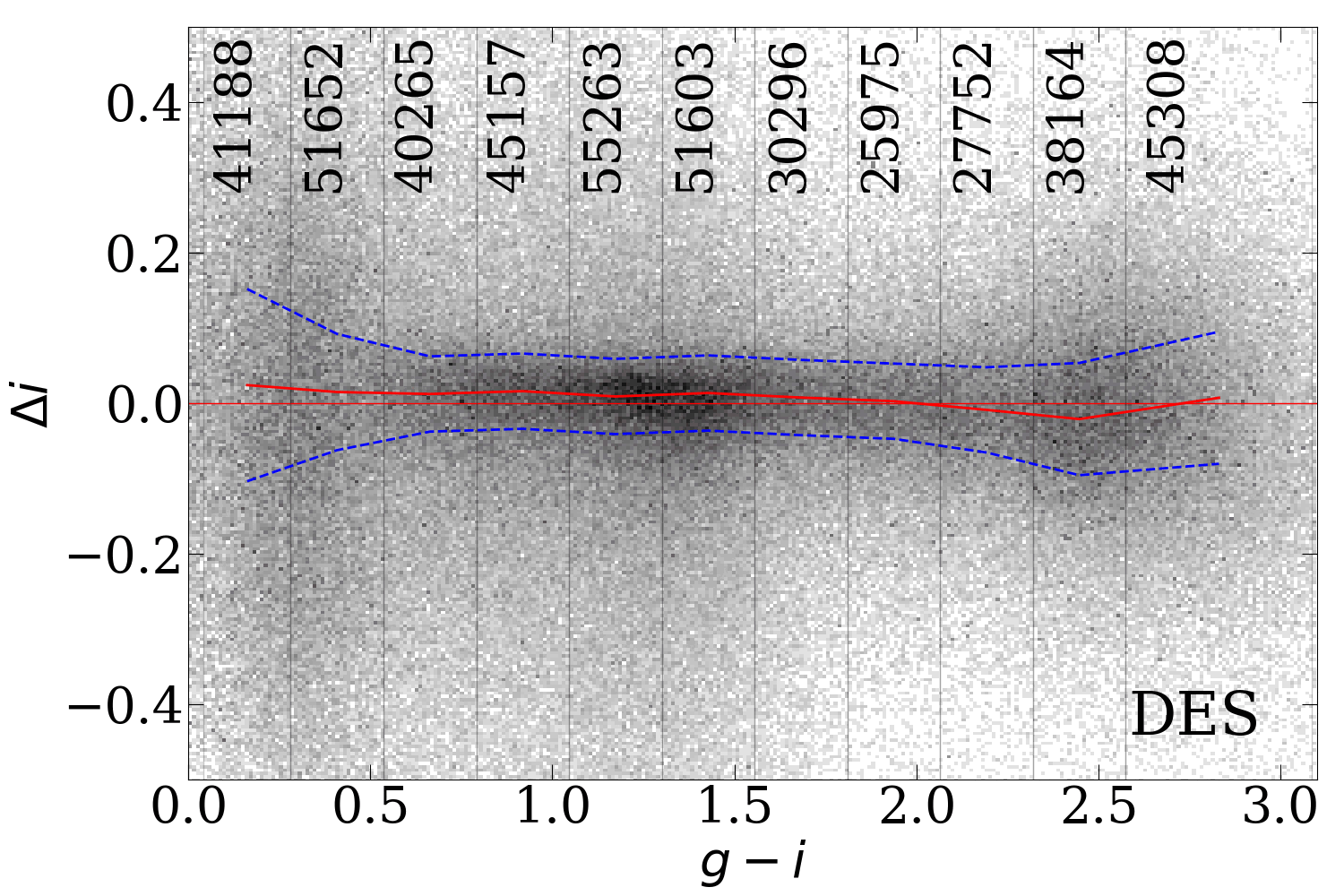}\\
\includegraphics[width=0.33\hsize]{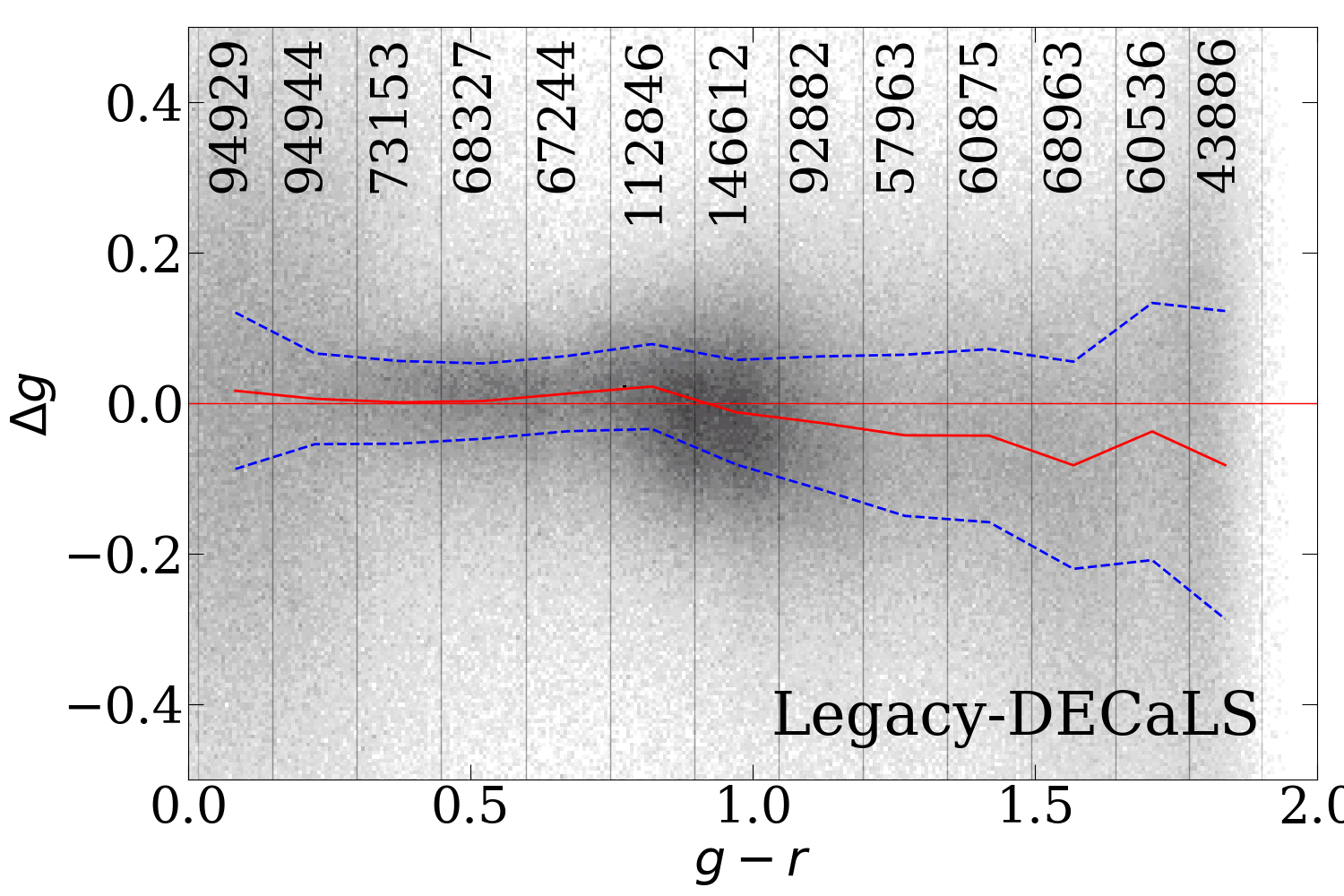}
\includegraphics[width=0.33\hsize]{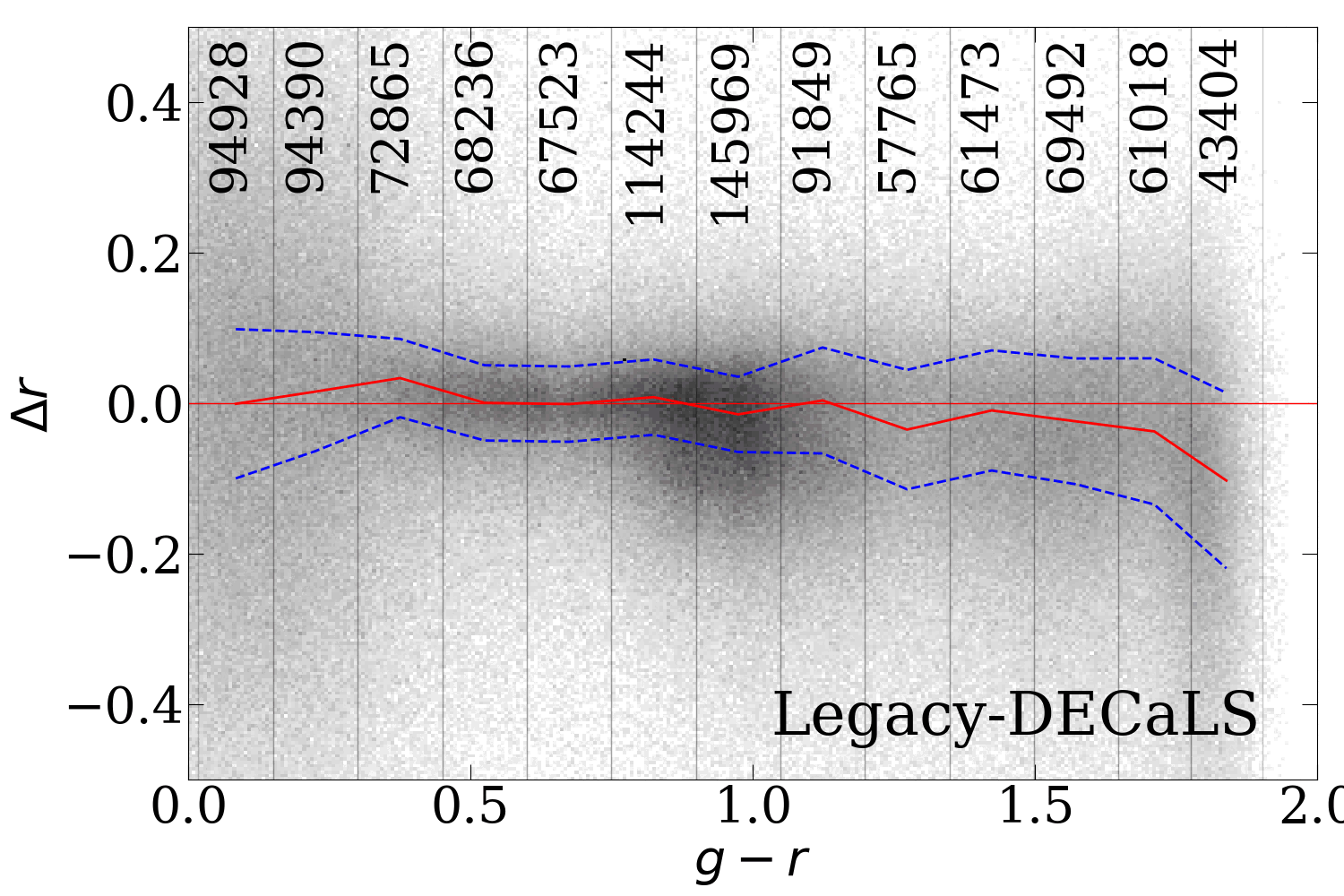}
\includegraphics[width=0.33\hsize]{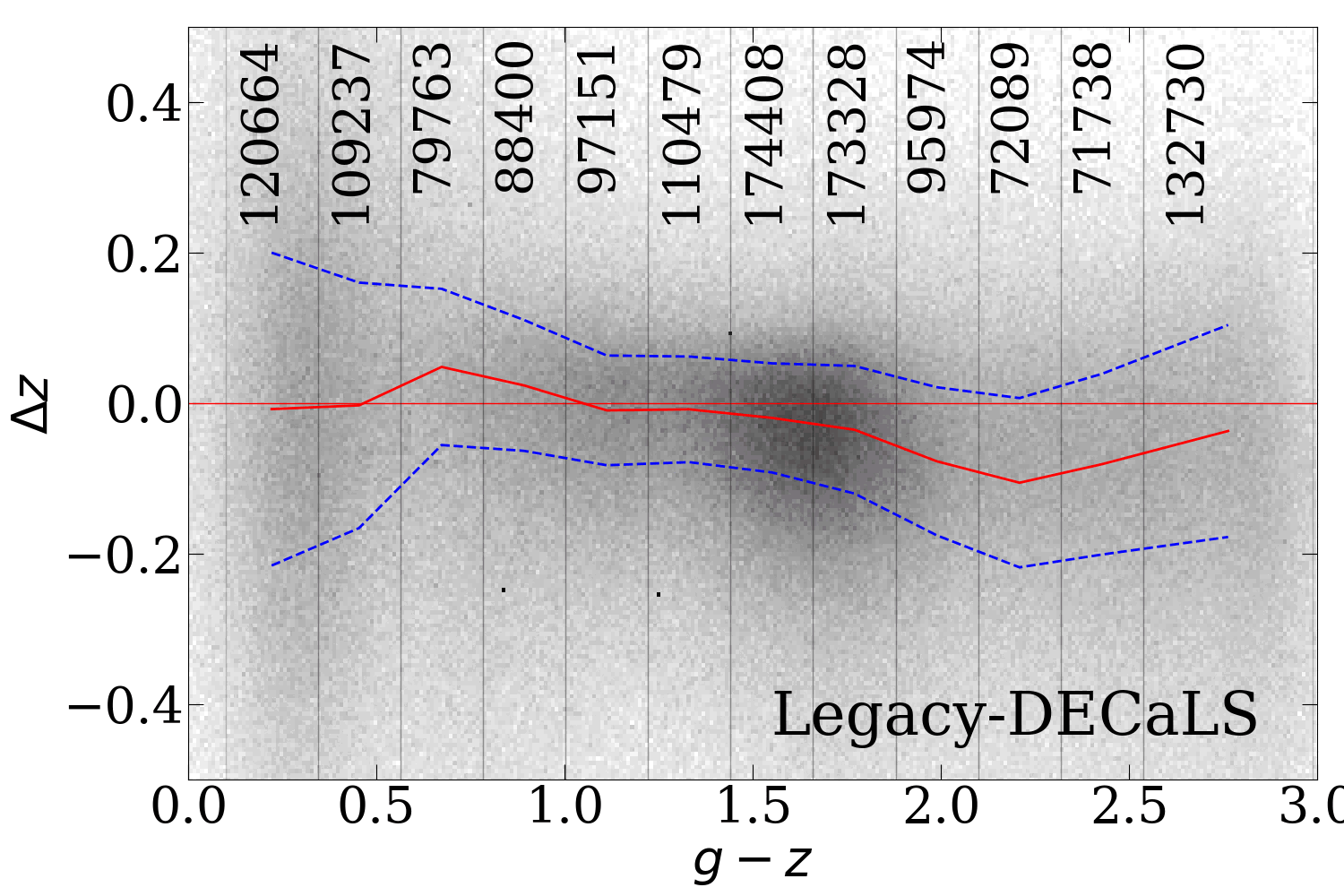}\\
\includegraphics[width=0.33\hsize]{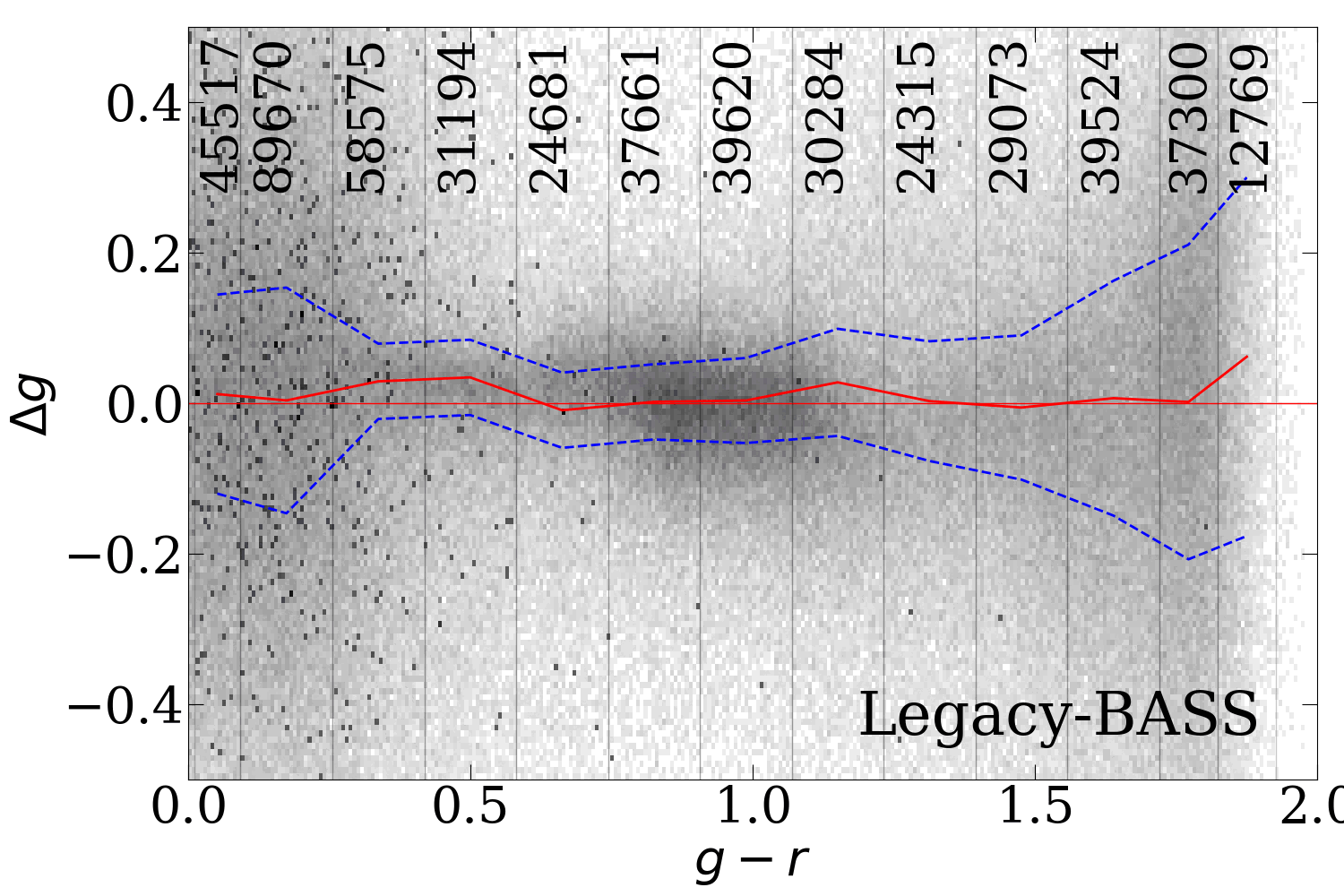}
\includegraphics[width=0.33\hsize]{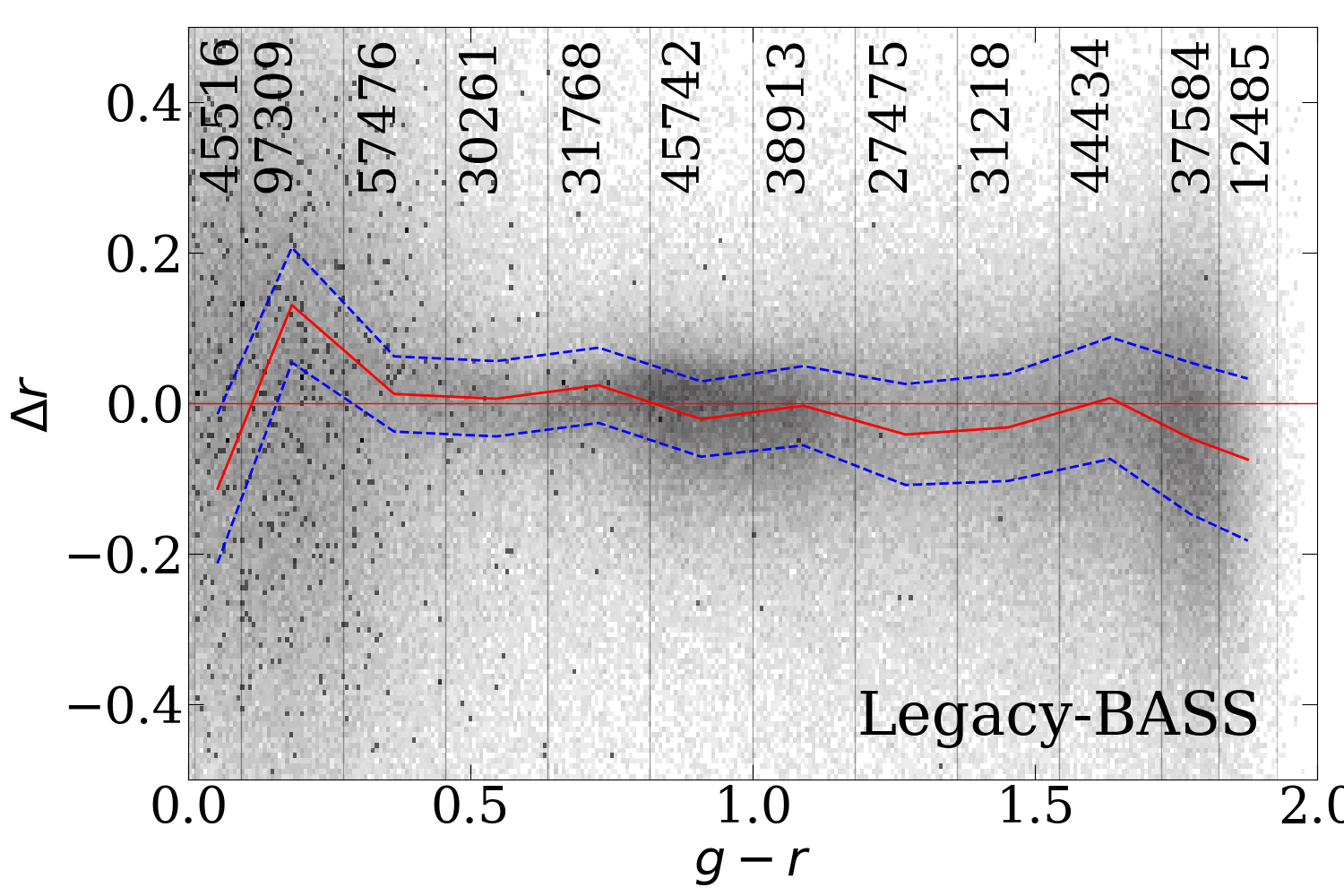}
\includegraphics[width=0.33\hsize]{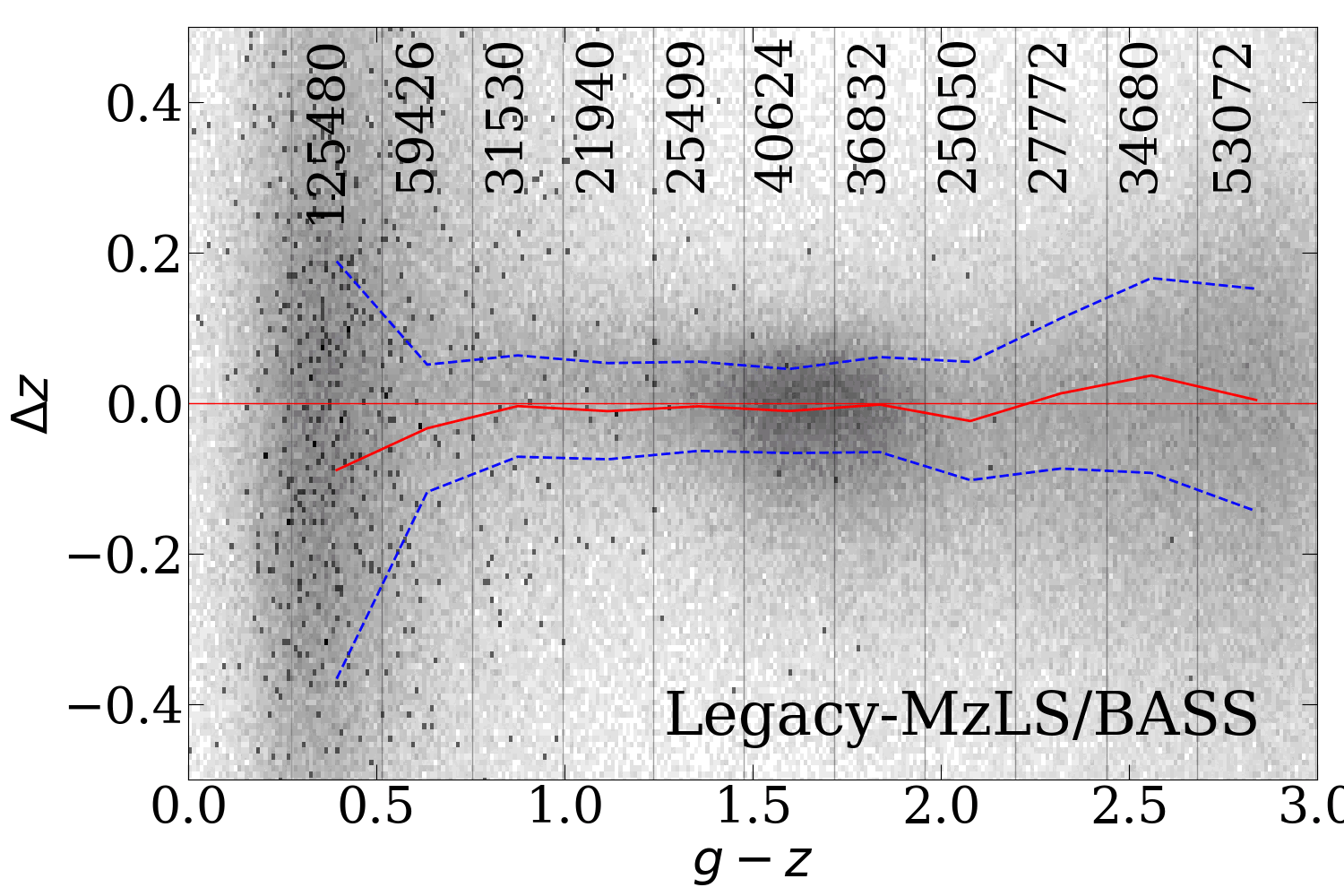}\\
\includegraphics[width=0.33\hsize]{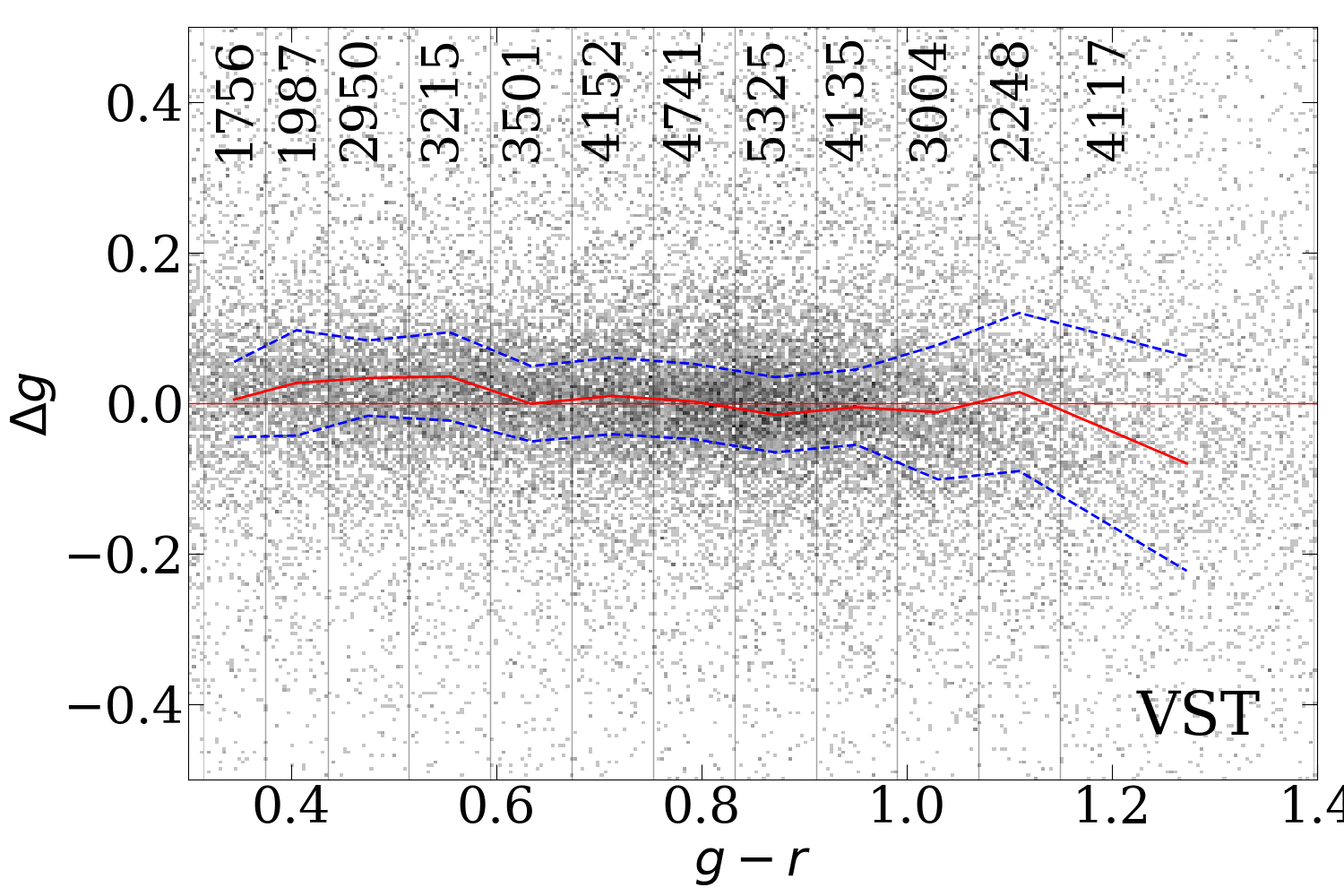}
\includegraphics[width=0.33\hsize]{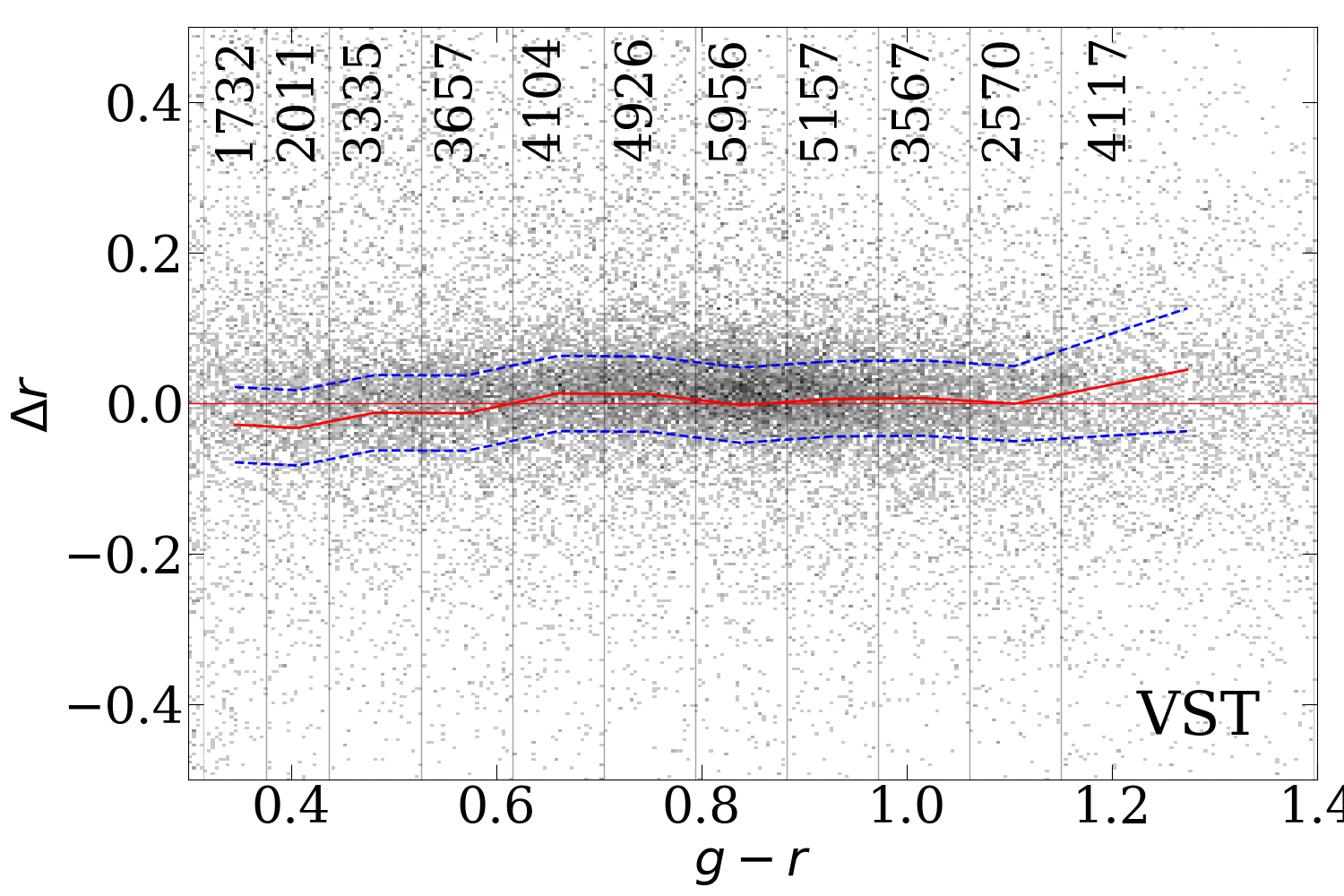}
\includegraphics[width=0.33\hsize]{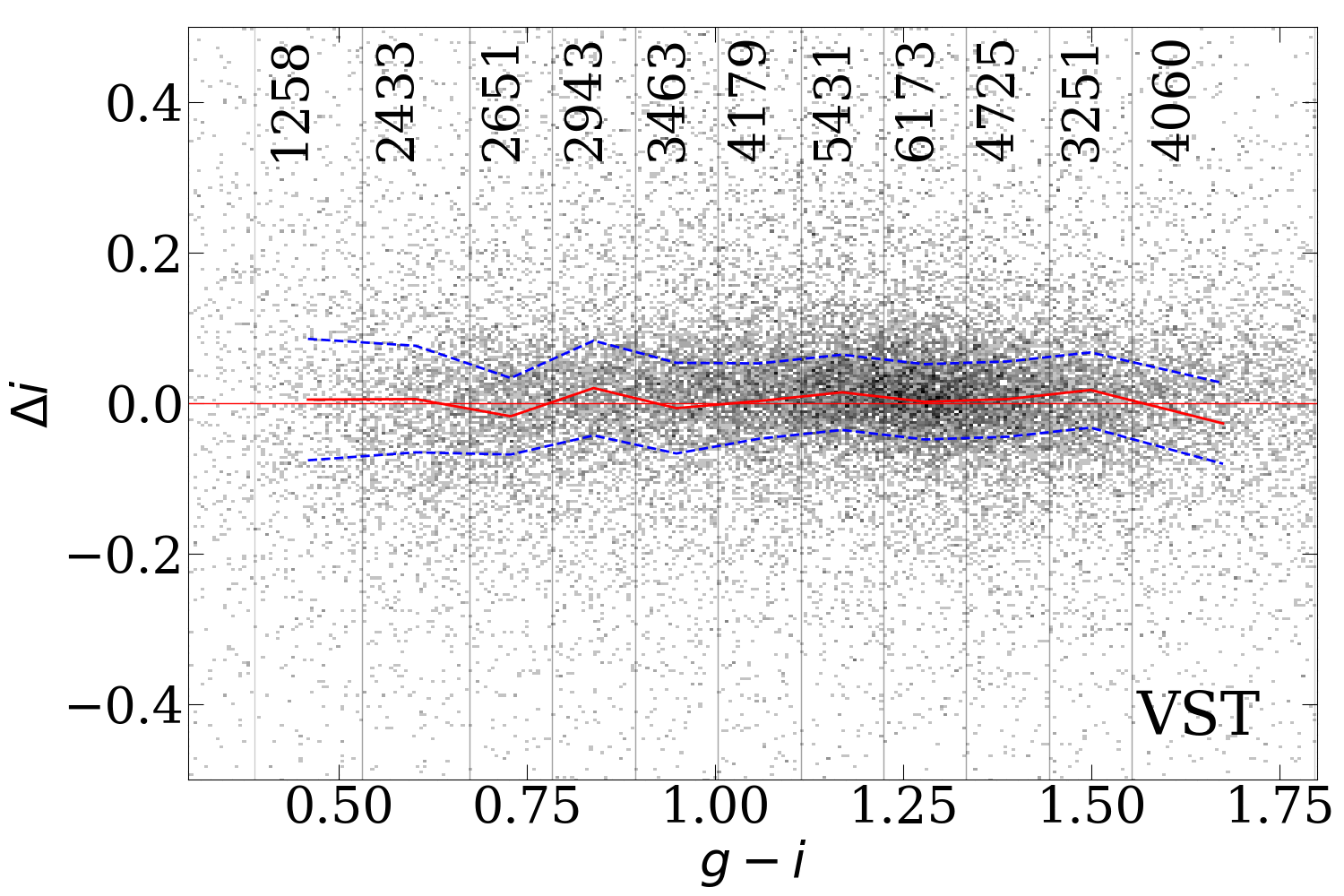}\\
\vspace{-0.4cm}
\caption{\revtwo{Fitting residuals between several surveys (converted magnitudes)} and the SDSS as a function of the color used in the \revtwo{approximation}. \revone{SDSS and DES magnitudes are Petrosian, DESI Legacy Surveys magnitudes are model-based}. The red line shows a median position in each bin, the blue line shows the median absolute deviation in each bin; their boundaries are marked with gray vertical lines. The surveys are labeled in the bottom right corner.\label{mag_resid}}
\end{figure*}

\begin{figure*}
\includegraphics[width=0.48\hsize]{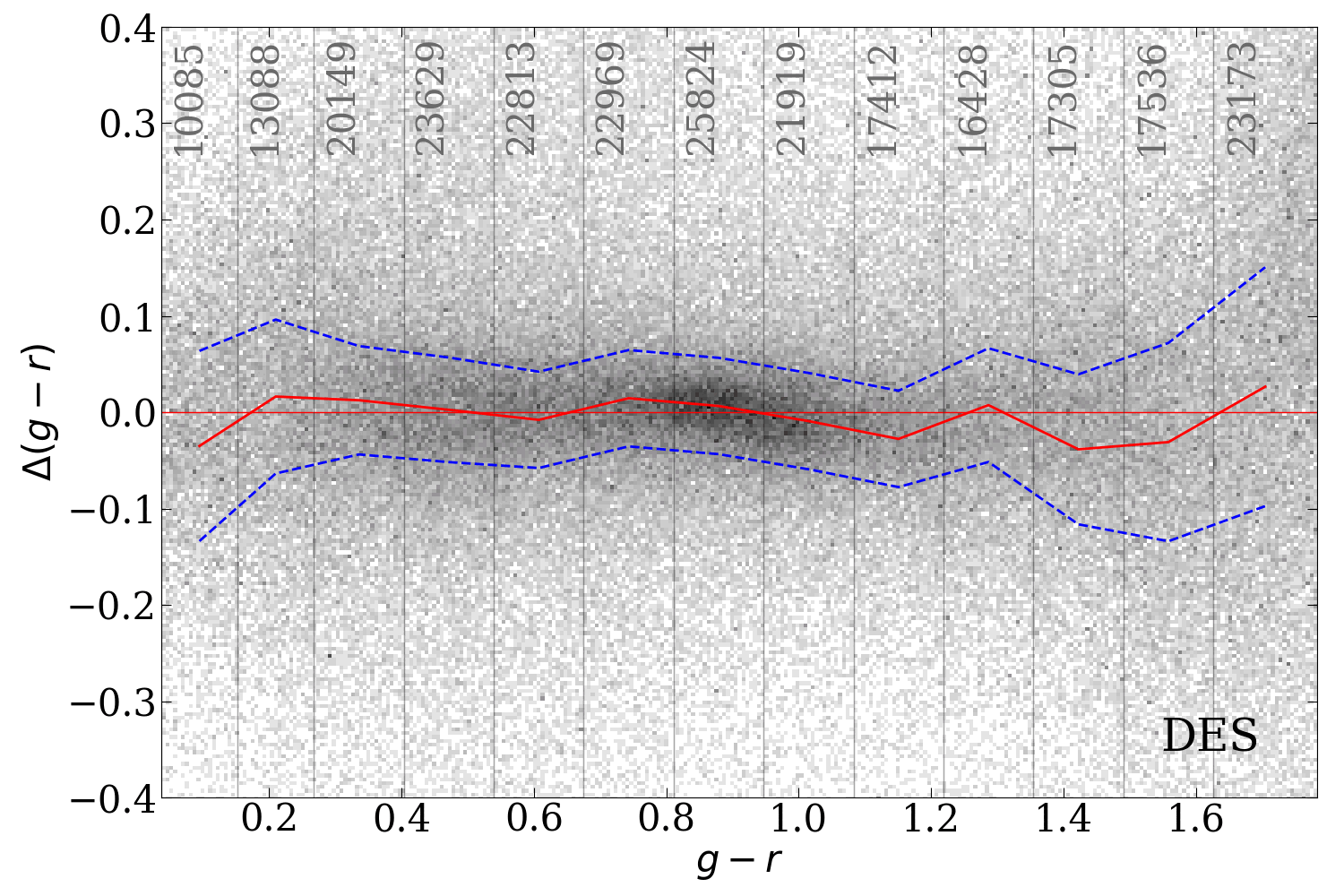}
\includegraphics[width=0.48\hsize]{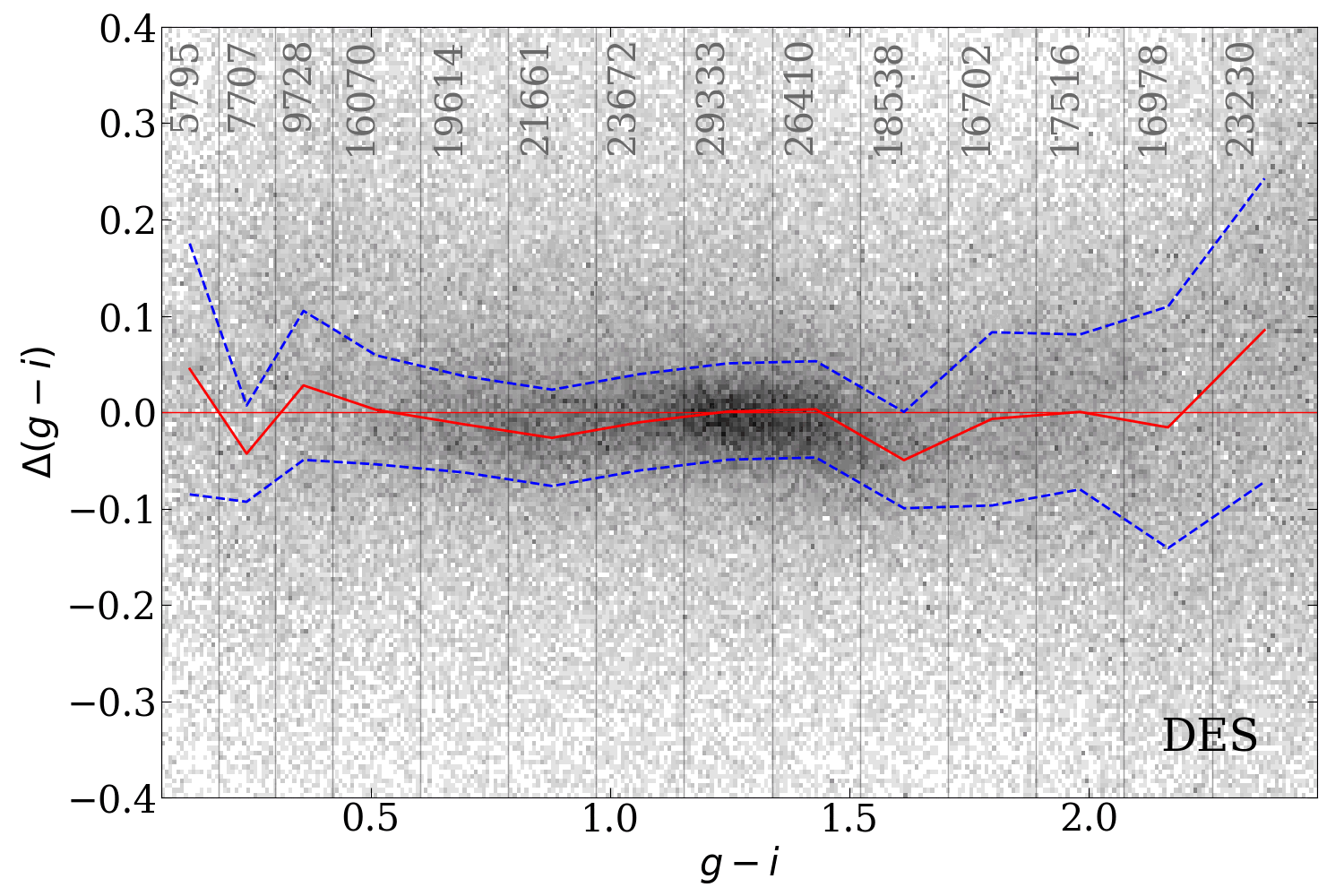}\\
\includegraphics[width=0.48\hsize]{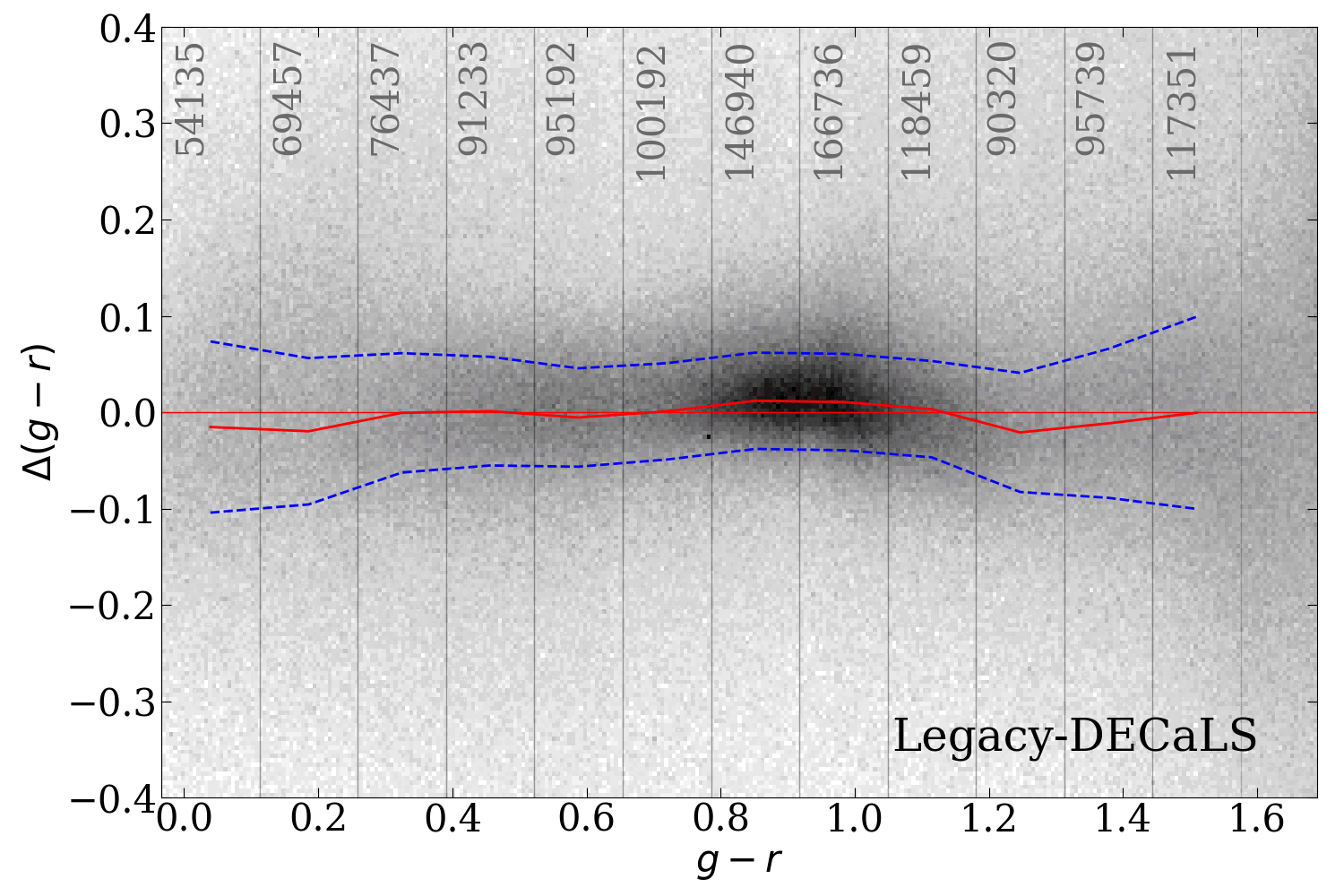}
\includegraphics[width=0.48\hsize]{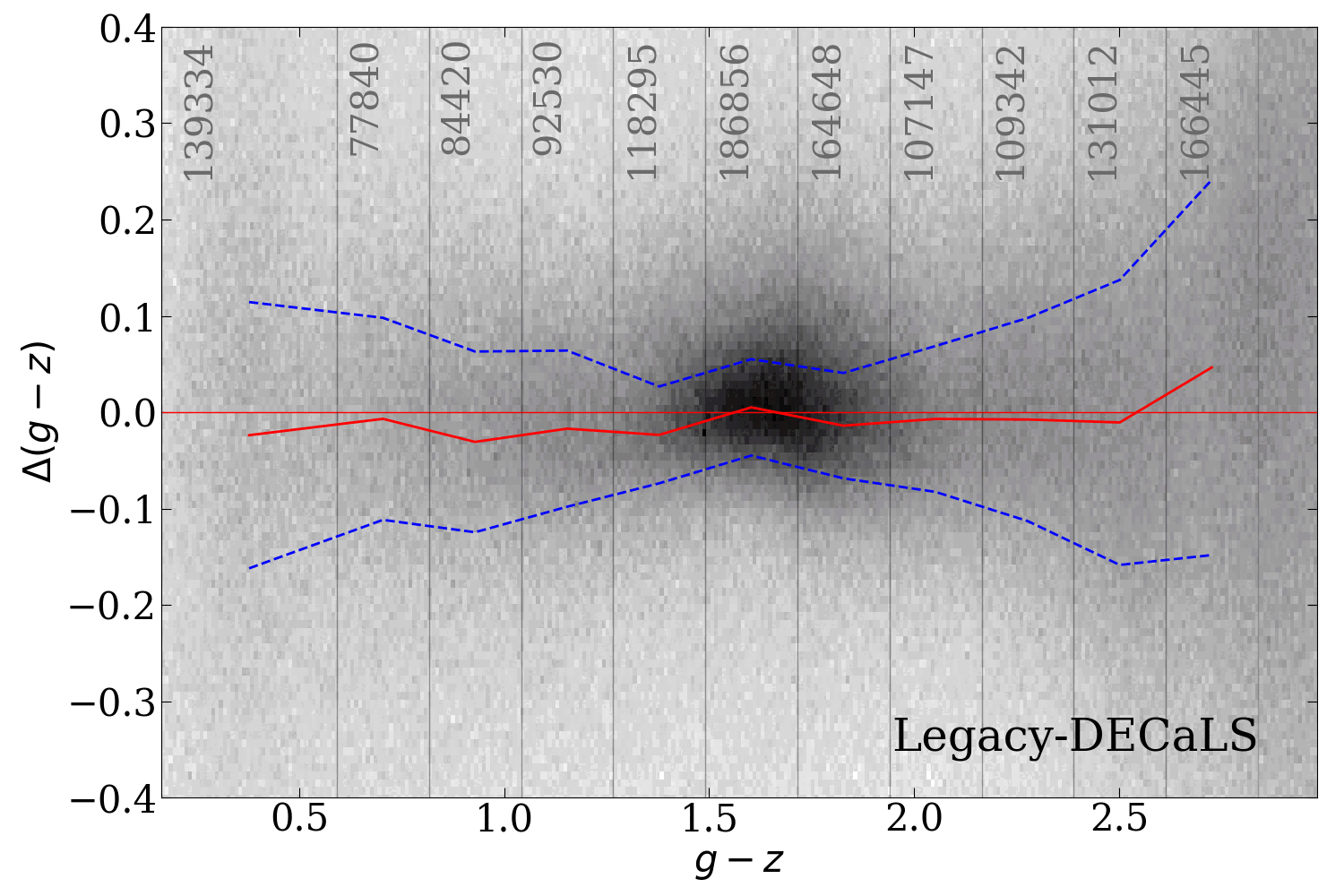}\\
\includegraphics[width=0.48\hsize]{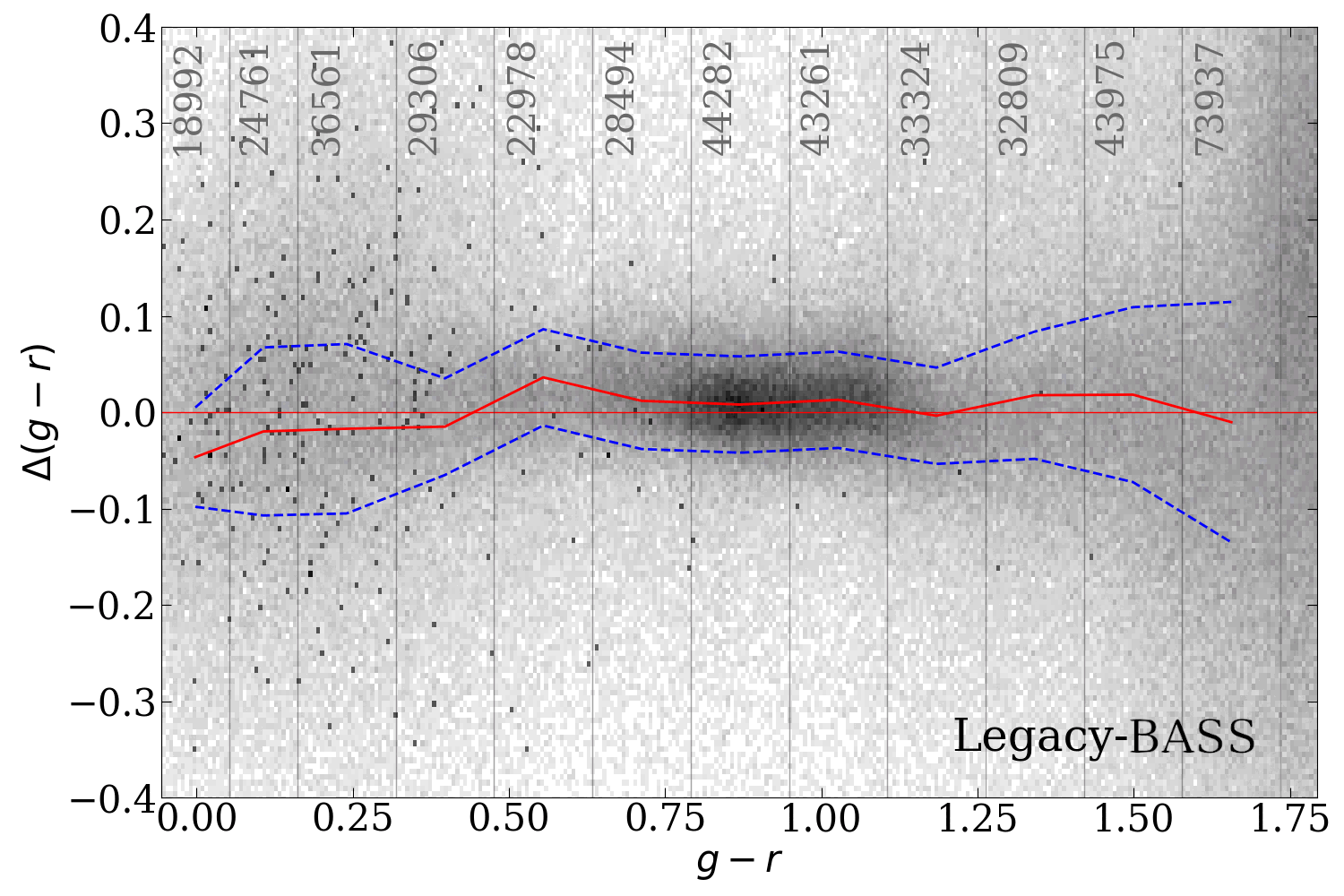}
\includegraphics[width=0.48\hsize]{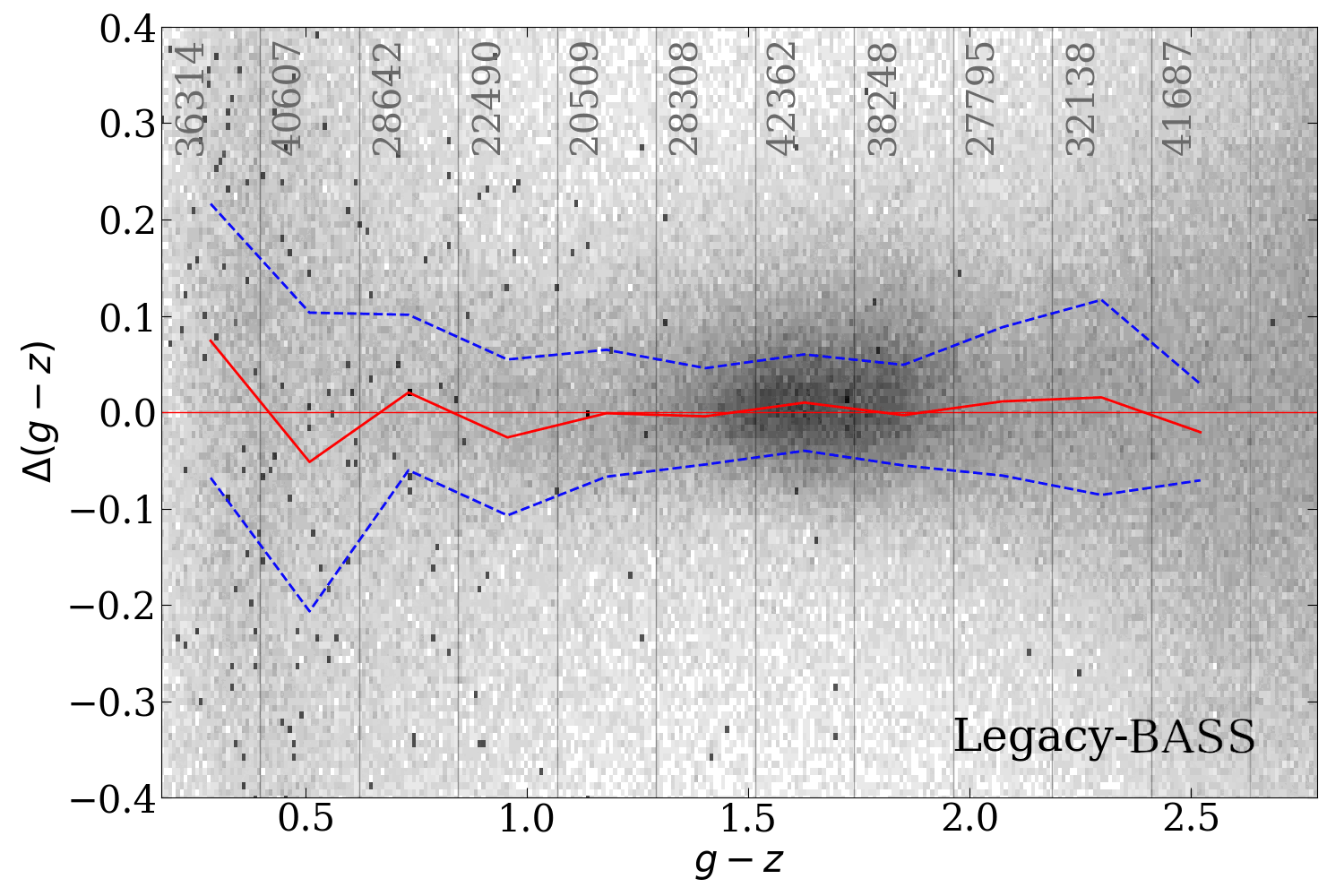}\\
\includegraphics[width=0.48\hsize]{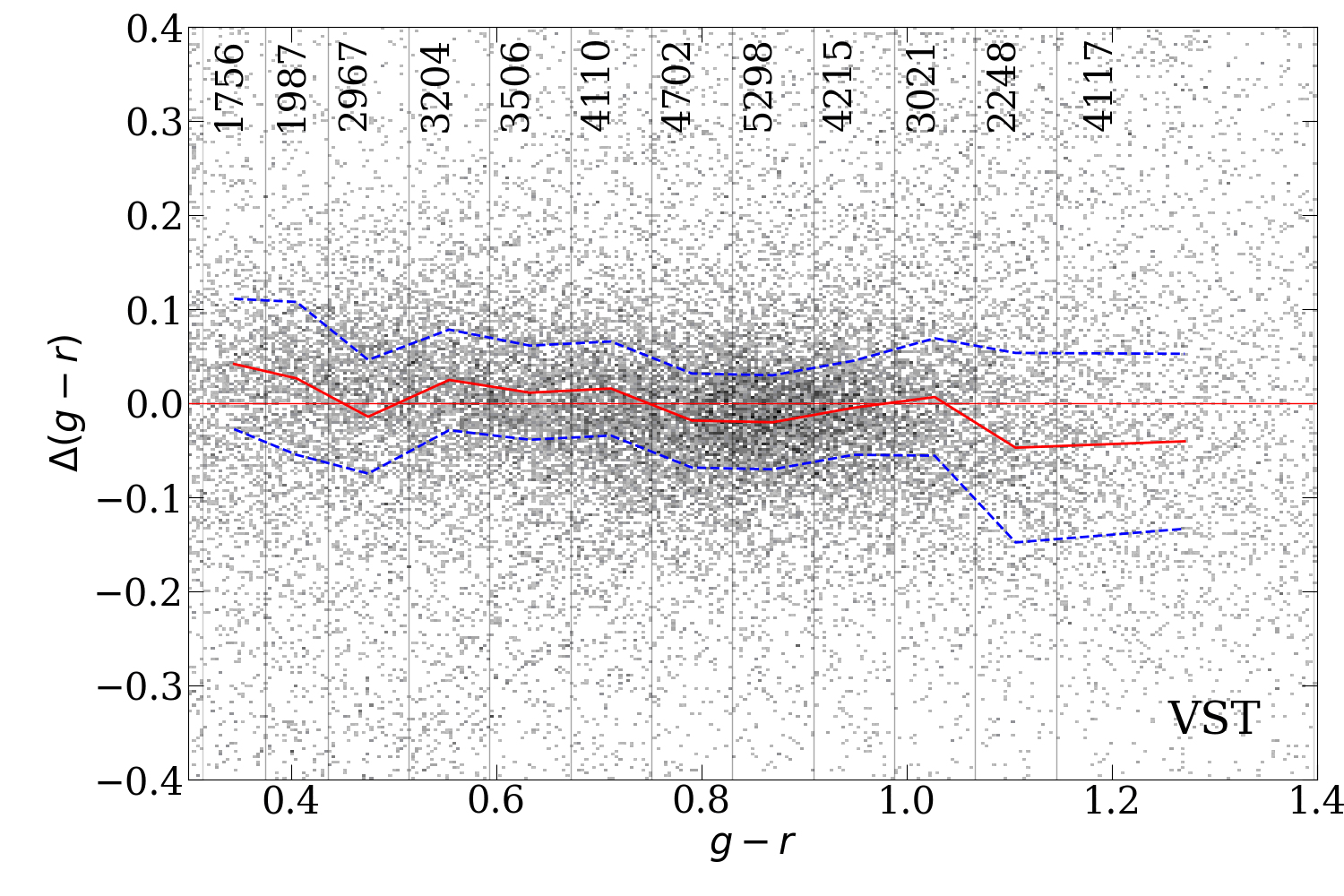}
\includegraphics[width=0.48\hsize]{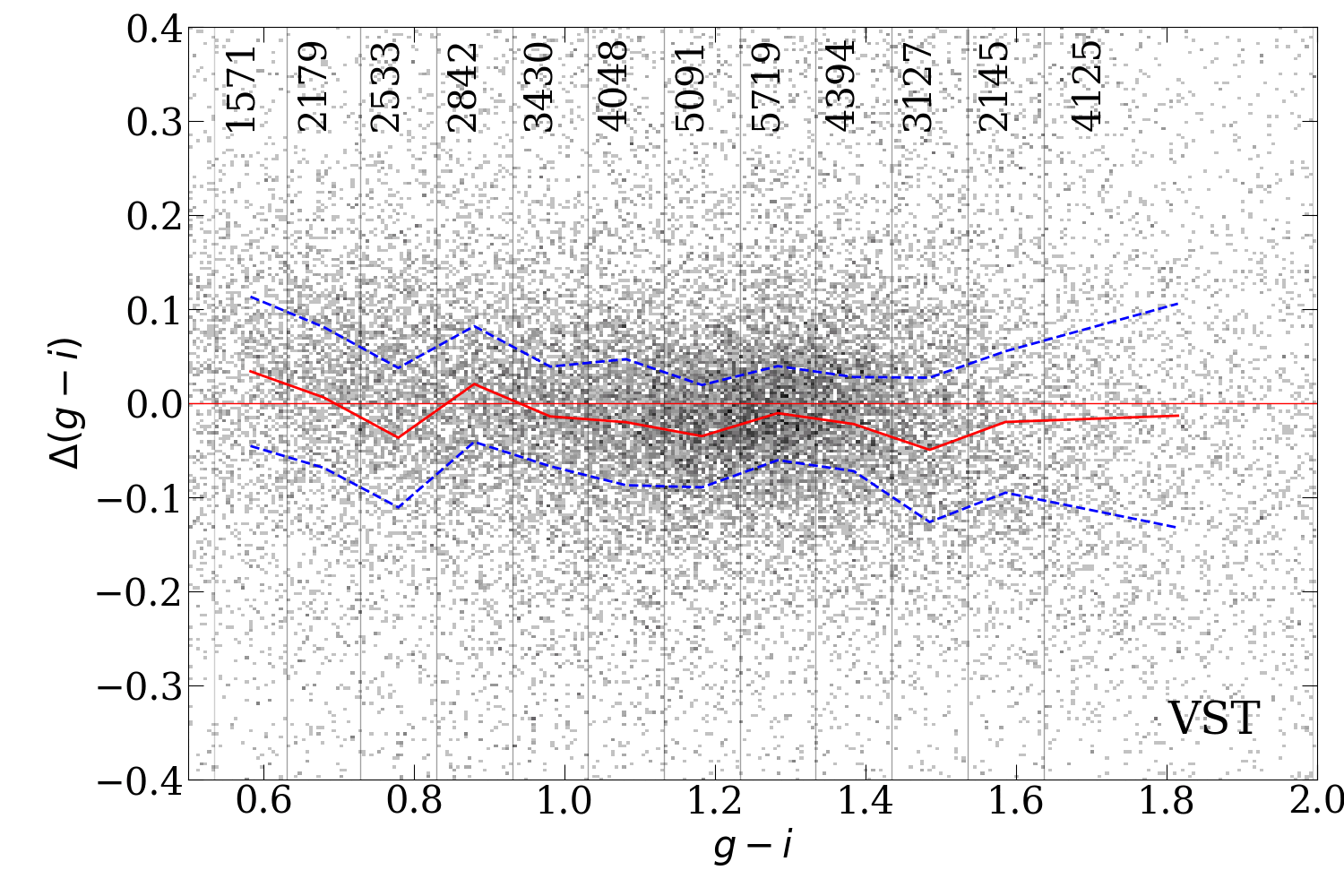}\\
\vspace{-0.6cm}
\caption{\revtwo{Same as Figure~\ref{mag_resid}, but the residuals are shown for the converted colors}.\label{color_resid}}
\end{figure*}

In this section we present color transformations of optical integrated photometric measurements of galaxies from SDSS into DES, DELVE, DESI Legacy Surveys, VST KiDS, and VST ATLAS surveys and vice versa. We discuss the applicability of our method to the PanSTARRS data where the direct computation of transformations did not \revtwo{yield} accurate results. We also provide transformations of near-infrared measurements from UKIDSS/UHS to VIKING/VHS and vice versa. Each set of transformations is provided in a form of piece-wise linear relations. To illustrate the quality of photometric transformations, we present the plots of both $\mathrm{mag_{converted}}-\mathrm{mag_{reference}}$ (Figure~\ref{mag_resid}) and color difference (Figure~\ref{color_resid}) between the converted photometry and the measurements for the same galaxies in the reference survey (e.g. $(g-r)_{\mathrm{DES}} - (g-r)_{\mathrm{SDSS}}$ for the DES$\rightarrow$SDSS conversion). 


\subsection{DES and DELVE}
\label{subsec:des_to_sdss}

We used a piece-wise linear regression of 2 or 3 segments to compute the transformations between DES and SDSS measurements. Overall, the fitting quality is very high with the residuals at a few-\%\ level for Petrosian magnitudes. The DELVE survey despite being shallower than DES was conducted with the same instrument using the same (physical) filters, and all the data reduction, processing, and analysis algorithms were identical (DES Data Management system pipeline; \citealp{2018PASP..130g4501M}). We applied the color transformations obtained for DES to DELVE data and they demonstrated full consistency (see the discussion about the reconstruction of the red sequence below).

\begingroup
\allowdisplaybreaks
\begin{align*}
\mathrm{DES \rightarrow SDSS}:\\
\scriptstyle g-r<0.656:\qquad g_{\mathrm{sdss}} = g_{\mathrm{des}} + 0.072(g_{\mathrm{des}} - r_{\mathrm{des}}) + 0.033\\
\scriptstyle 0.656<g-r<1.187:\qquad g_{\mathrm{sdss}} = \scriptstyle g_{\mathrm{des}} + 0.198(g_{\mathrm{des}} - r_{\mathrm{des}}) - 0.049\\
\scriptstyle g-r>1.187:\qquad g_{\mathrm{sdss}} = g_{\mathrm{des}} - 0.005(g_{\mathrm{des}} - r_{\mathrm{des}}) + 0.193 \\
\scriptstyle \sigma_{\mathrm{g_{\mathrm{sdss}}}}= 0.048\;mag\\
\scriptstyle g-r<0.792:\qquad r_{\mathrm{sdss}}= r_{\mathrm{des}} + \scriptstyle 0.052(g_{\mathrm{des}} - r_{\mathrm{des}}) + 0.074\\
\scriptstyle 0.792<g-r<1.148:\qquad r_{\mathrm{sdss}} = r_{\mathrm{des}} + 0.098(g_{\mathrm{des}} - r_{\mathrm{des}}) + 0.037\\
\scriptstyle g-r>1.148: \qquad r_{\mathrm{sdss}} = r_{\mathrm{des}} + 0.208(g_{\mathrm{des}} - r_{\mathrm{des}}) - 0.088\\
\scriptstyle \sigma_{\mathrm{r_{\mathrm{sdss}}}}= 0.041\;mag\\
\scriptstyle r-z<0.621:\qquad r_{\mathrm{sdss}}=r_{\mathrm{des}} + 0.034(r_{\mathrm{des}} - z_{\mathrm{des}}) + 0.089\\
\scriptstyle 0.621<r-z<1.284:\qquad r_{\mathrm{sdss}}=r_{\mathrm{des}} + 0.381(r_{\mathrm{des}} - z_{\mathrm{des}}) - 0.126\\
\scriptstyle r-z>1.284:\qquad r_{\mathrm{sdss}}=r_{\mathrm{des}} - 0.058(r_{\mathrm{des}} - z_{\mathrm{des}}) + 0.438\\
\scriptstyle \sigma_{\mathrm{r_{\mathrm{sdss}}}}= 0.042\;mag\\
\scriptstyle i-z<0.237:\qquad i_{\mathrm{sdss}} = i_{\mathrm{des}} + 0.122(i_{\mathrm{des}} - z_{\mathrm{des}}) + 0.048\\
\scriptstyle 0.237<i-z<0.400:\qquad i_{\mathrm{sdss}} = i_{\mathrm{des}} + 0.407(i_{\mathrm{des}} - z_{\mathrm{des}}) - 0.019\\
\scriptstyle i-z>0.400:\qquad i_{\mathrm{sdss}} =i_{\mathrm{des}} + 0.174(i_{\mathrm{des}} - z_{\mathrm{des}}) + 0.074\\
\scriptstyle \sigma_{\mathrm{i_{\mathrm{sdss}}}}=0.046 \;mag\\
\scriptstyle r-z<0.803:\qquad z_{\mathrm{sdss}}=z_{\mathrm{des}} + 0.060(r_{\mathrm{des}} - z_{\mathrm{des}}) + 0.031\\
\scriptstyle r-z>0.803:\qquad z_{\mathrm{sdss}}=z_{\mathrm{des}} - 0.099(r_{\mathrm{des}} - z_{\mathrm{des}}) + 0.159\\
\scriptstyle \sigma_{\mathrm{z_{\mathrm{sdss}}}}=0.061\;mag\\
\\
\mathrm{SDSS \rightarrow DES}:\\
\scriptstyle g-r<0.797:\qquad g_{\mathrm{des}} = g_{\mathrm{sdss}} - 0.132(g_{\mathrm{sdss}} - r_{\mathrm{sdss}}) - 0.001\\
\scriptstyle 0.797<g-r<1.043:\qquad g_{\mathrm{des}} = \scriptstyle g_{\mathrm{sdss}} - 0.238(g_{\mathrm{sdss}} - r_{\mathrm{sdss}}) + 0.083\\
\scriptstyle g-r>1.181:\qquad g_{\mathrm{des}} = g_{\mathrm{sdss}} - 0.165(g_{\mathrm{sdss}} - r_{\mathrm{sdss}}) + 0.007 \\
\scriptstyle \sigma_{\mathrm{g_{\mathrm{des}}}}= 0.045\;mag\\
\scriptstyle g-r<0.600:\qquad r_{\mathrm{des}}= r_{\mathrm{sdss}} + 0.027(g_{\mathrm{sdss}} - r_{\mathrm{sdss}}) - 0.117\\
\scriptstyle g-r>0.600: \qquad r_{\mathrm{des}} = r_{\mathrm{sdss}} - 0.093(g_{\mathrm{sdss}} - r_{\mathrm{sdss}})  - 0.044\\
\scriptstyle \sigma_{\mathrm{r_{\mathrm{des}}}}= 0.040\;mag\\
\scriptstyle r-z<0.659:\qquad r_{\mathrm{des}}=r_{\mathrm{sdss}} - 0.040(r_{\mathrm{sdss}} - z_{\mathrm{sdss}}) - 0.083\\
\scriptstyle 0.659<r-z<1.783:\qquad r_{\mathrm{des}}=r_{\mathrm{sdss}} - 0.276(r_{\mathrm{sdss}} - z_{\mathrm{sdss}}) + 0.072\\
\scriptstyle r-z>1.783:\qquad r_{\mathrm{des}}=r_{\mathrm{sdss}} - 0.235(r_{\mathrm{sdss}} - z_{\mathrm{sdss}}) - 0.0003\\
\scriptstyle \sigma_{\mathrm{r_{\mathrm{des}}}}= 0.036\;mag\\
\scriptstyle i-z<0.216:\qquad i_{\mathrm{des}} = i_{\mathrm{sdss}} - 0.162(i_{\mathrm{sdss}} - z_{\mathrm{sdss}}) - 0.040\\
\scriptstyle 0.216<i-z<0.583:\qquad i_{\mathrm{des}} = i_{\mathrm{sdss}} - 0.237(i_{\mathrm{sdss}} - z_{\mathrm{sdss}}) - 0.024\\
\scriptstyle i-z>0.583:\qquad i_{\mathrm{des}} =i_{\mathrm{sdss}} - 0.102(i_{\mathrm{sdss}} - z_{\mathrm{sdss}}) - 0.102\\
\scriptstyle \sigma_{\mathrm{i_{\mathrm{des}}}}=0.038 \;mag\\
\scriptstyle r-z<0.336:\qquad z_{\mathrm{des}}=z_{\mathrm{sdss}} + 0.421(r_{\mathrm{sdss}} - z_{\mathrm{sdss}}) - 0.231\\
\scriptstyle 0.336<r-z<0.868:\qquad z_{\mathrm{des}}=z_{\mathrm{sdss}}  + 0.035(r_{\mathrm{sdss}} - z_{\mathrm{sdss}}) - 0.101\\
\scriptstyle r-z>0.909:\qquad z_{\mathrm{des}}=z_{\mathrm{sdss}} + 0.123(r_{\mathrm{sdss}} - z_{\mathrm{sdss}}) - 0.177\\
\scriptstyle \sigma_{\mathrm{z_{\mathrm{des}}}}=0.063\;mag\\
\end{align*}
\endgroup

\subsection{DESI Legacy Surveys}

The DESI Legacy \revtwo{Surveys} \citep{2019AJ....157..168D} include the two parts, Northern and Southern, carried out using three different telescopes and instruments: the Northern part consists of the Beijing--Arizona Sky Survey (BASS; $g$ and $r$-band observations collected at the 2.3~m telescope using the 90Prime camera) and the Mayall $z$-band Sky Survey (MzLS; $z$-band observations collected at the 4 m Mayall telescope using the Mosaic-3 camera); the Southern part includes the data collected with the DECam instrument at the 4 m Blanco telescope (same as DES/DELVE). Therefore, one needs to derive separate sets of photometric transformations for the Northern and Southern parts of DESI Legacy Surveys data. It turns out that the DES/DELVE color transformations cannot be applied to the LS-South measurements despite being collected with the same telescope and instrument and even sometimes being derived from the same raw data, because of the differences the original observations were reduced and analyzed (see details in the next section). Therefore, we derived specific transformations for LS-South to/from SDSS photometry. The LS-South transformations yield higher residuals than those for DES/DELVE data because of higher systematic errors in the DECaLS photometry: (e.g., $\sigma_{r_{\mathrm {decals}}}=0.087\;mag$ versus $\sigma_{r_{\mathrm{des}}}= 0.044\;mag$ and $\sigma_{r_{\mathrm{vst}}}= 0.041\;mag$). Transformations for LS-South (DECaLS) and LS-North (MzLS+BASS) yield similar r.m.s. of the fitting residuals.

\begingroup
\allowdisplaybreaks
\begin{align*}
\mathrm{DECaLS \rightarrow SDSS}:\\
\scriptstyle g-r<0.712:\qquad g_{\mathrm{sdss}} = g_{\mathrm{decals}} + 0.014(g_{\mathrm{decals}} - r_{\mathrm{decals}}) + 0.055\\
\scriptstyle 0.712<g-r<0.984:\qquad g_{\mathrm{sdss}} = g_{\mathrm{decals}} + 0.404(g_{\mathrm{decals}} - r_{\mathrm{decals}}) - 0.223\\
\scriptstyle g-r>0.984:\qquad g_{\mathrm{sdss}} = g_{\mathrm{decals}} + 0.049(g_{\mathrm{decals}} - r_{\mathrm{decals}}) + 0.126\\
\scriptstyle \sigma_{\mathrm{g_{\mathrm{sdss}}}}= 0.072\;mag\\
\scriptstyle g-r<0.732:\qquad r_{\mathrm{sdss}} = r_{\mathrm{decals}} + 0.038(g_{\mathrm{decals}} - r_{\mathrm{decals}}) + 0.084\\
\scriptstyle 0.732<g-r<0.951:\qquad r_{\mathrm{sdss}} = r_{\mathrm{decals}} + 0.252(g_{\mathrm{decals}} - r_{\mathrm{decals}}) - 0.073\\
\scriptstyle g-r>0.951:\qquad r_{\mathrm{sdss}} = r_{\mathrm{decals}} + 0.082(g_{\mathrm{decals}} - r_{\mathrm{decals}}) + 0.089\\
\scriptstyle \sigma_{\mathrm{r_{\mathrm{sdss}}}}=0.087\;mag\\
\scriptstyle r-z<0.629:\qquad r_{\mathrm{sdss}} = r_{\mathrm{decals}} + 0.061(r_{\mathrm{decals}} - z_{\mathrm{decals}}) + 0.099\\
\scriptstyle 0.629<r-z<1.280:\qquad r_{\mathrm{sdss}} = r_{\mathrm{decals}} + 0.359(r_{\mathrm{decals}} - z_{\mathrm{decals}}) - 0.088\\
\scriptstyle r-z>1.280:\qquad r_{\mathrm{sdss}} = r_{\mathrm{decals}} + 0.054(r_{\mathrm{decals}} - z_{\mathrm{decals}}) + 0.302\\
\scriptstyle \sigma_{\mathrm{r_{\mathrm{sdss}}}}=0.073\;mag\\
\scriptstyle r-z>0.779:\qquad z_{\mathrm{sdss}} = z_{\mathrm{decals}} + 0.191(r_{\mathrm{decals}} - z_{\mathrm{decals}}) - 0.023\\
\scriptstyle r-z<0.779:\qquad z_{\mathrm{sdss}} = z_{\mathrm{decals}} - 0.066(r_{\mathrm{decals}} - z_{\mathrm{decals}}) + 0.177\\
\scriptstyle \sigma_{\mathrm{z_{\mathrm{sdss}}}}=0.079\;mag\\
\\
\mathrm{SDSS \rightarrow DECaLS}:\\
\scriptstyle g-r<0.657:\qquad g_{\mathrm{decals}} = g_{\mathrm{sdss}} - 0.034(g_{\mathrm{sdss}} - r_{\mathrm{sdss}}) - 0.035\\
\scriptstyle 0.657<g-r<0.973:\qquad g_{\mathrm{decals}} = g_{\mathrm{sdss}} - 0.356(g_{\mathrm{sdss}} - r_{\mathrm{sdss}}) + 0.176\\
\scriptstyle g-r>0.973:\qquad g_{\mathrm{decals}} = g_{\mathrm{sdss}} - 0.113(g_{\mathrm{sdss}} - r_{\mathrm{sdss}}) - 0.060\\
\scriptstyle \sigma_{\mathrm{g_{\mathrm{decals}}}}= 0.071\;mag\\
\scriptstyle g-r<0.662:\qquad r_{\mathrm{decals}} = r_{\mathrm{sdss}} + 0.007(g_{\mathrm{sdss}} - r_{\mathrm{sdss}}) - 0.112\\
\scriptstyle 0.662<g-r<0.968:\qquad r_{\mathrm{decals}} = r_{\mathrm{sdss}} - 0.214(g_{\mathrm{sdss}} - r_{\mathrm{sdss}}) + 0.034\\
\scriptstyle g-r>0.968:\qquad r_{\mathrm{decals}} = r_{\mathrm{sdss}} - 0.057(g_{\mathrm{sdss}} - r_{\mathrm{sdss}}) - 0.117\\
\scriptstyle \sigma_{\mathrm{r_{\mathrm{decals}}}}=0.065\;mag\\
\scriptstyle r-z<0.619:\qquad r_{\mathrm{decals}} = r_{\mathrm{sdss}} - 0.036(r_{\mathrm{sdss}} - z_{\mathrm{sdss}}) - 0.107\\
\scriptstyle 0.619<r-z<1.787:\qquad r_{\mathrm{decals}} = r_{\mathrm{sdss}} - 0.247(r_{\mathrm{sdss}} - z_{\mathrm{sdss}}) + 0.022\\
\scriptstyle r-z>1.787:\qquad r_{\mathrm{decals}} = r_{\mathrm{sdss}} - 0.202(r_{\mathrm{sdss}} - z_{\mathrm{sdss}}) - 0.056\\
\scriptstyle \sigma_{\mathrm{r_{\mathrm{decals}}}}=0.068\;mag\\
\scriptstyle r-z>0.787:\qquad z_{\mathrm{decals}} = z_{\mathrm{sdss}} - 0.093(r_{\mathrm{sdss}} - z_{\mathrm{sdss}}) - 0.041\\
\scriptstyle 0.787>r-z>0.893:\qquad z_{\mathrm{decals}} = z_{\mathrm{sdss}} + 0.050(r_{\mathrm{sdss}} - z_{\mathrm{sdss}}) - 0.154\\
\scriptstyle r-z<0.893:\qquad z_{\mathrm{decals}} = z_{\mathrm{sdss}} + 0.2356(r_{\mathrm{sdss}} - z_{\mathrm{sdss}}) - 0.320\\
\scriptstyle \sigma_{\mathrm{z_{\mathrm{decals}}}}=0.144\;mag\\
\\
\mathrm{MzLS+BASS \rightarrow SDSS}:\\
\scriptstyle g-r<0.664:\qquad g_{\mathrm{sdss}} = g_{\mathrm{bass}} - 0.072(g_{\mathrm{bass}} - r_{\mathrm{bass}}) + 0.071\\
\scriptstyle 0.664<g-r<1.102:\qquad g_{\mathrm{sdss}} = g_{\mathrm{bass}} + 0.241(g_{\mathrm{bass}} - r_{\mathrm{bass}}) - 0.137\\
\scriptstyle g-r>1.102:\qquad g_{\mathrm{sdss}} = g_{\mathrm{bass}} + 0.017(g_{\mathrm{bass}} - r_{\mathrm{bass}}) + 0.110\\
\scriptstyle \sigma_{\mathrm{g_{\mathrm{sdss}}}}= 0.066\;mag\\
\scriptstyle g-r<0.732:\qquad r_{\mathrm{sdss}} = r_{\mathrm{bass}} - 0.053(g_{\mathrm{bass}} - r_{\mathrm{bass}}) + 0.120\\
\scriptstyle 0.732<g-r<0.951:\qquad r_{\mathrm{sdss}} = r_{\mathrm{bass}} + 0.265(g_{\mathrm{bass}} - r_{\mathrm{bass}}) - 0.099\\
\scriptstyle g-r>0.951:\qquad r_{\mathrm{sdss}} = r_{\mathrm{bass}} + 0.048(g_{\mathrm{bass}} - r_{\mathrm{bass}}) + 0.108\\
\scriptstyle \sigma_{\mathrm{r_{\mathrm{sdss}}}}=0.061\;mag\\
\scriptstyle r-z<0.619:\qquad r_{\mathrm{sdss}} = r_{\mathrm{bass}} + 0.040(r_{\mathrm{bass}} - z_{\mathrm{mzls}}) + 0.102\\
\scriptstyle 0.619<r-z<1.302:\qquad r_{\mathrm{sdss}} = r_{\mathrm{bass}} + 0.262(r_{\mathrm{bass}} - z_{\mathrm{mzls}}) - 0.035\\
\scriptstyle r-z>1.302:\qquad r_{\mathrm{sdss}} = r_{\mathrm{bass}} + 0.081(r_{\mathrm{bass}} - z_{\mathrm{mzls}}) + 0.201\\
\scriptstyle \sigma_{\mathrm{r_{\mathrm{sdss}}}}=0.072\;mag\\
\scriptstyle r-z<0.692:\qquad z_{\mathrm{sdss}} = z_{\mathrm{mzls}} + 0.235(r_{\mathrm{bass}} - z_{\mathrm{mzls}}) - 0.053\\
\scriptstyle 0.692<r-z<0.797:\qquad z_{\mathrm{sdss}} = z_{\mathrm{mzls}} + 0.047(r_{\mathrm{bass}} - z_{\mathrm{mzls}}) + 0.078\\
\scriptstyle r-z<0.797:\qquad z_{\mathrm{sdss}} = z_{\mathrm{mzls}} - 0.076(r_{\mathrm{bass}} - z_{\mathrm{mzls}}) + 0.175\\
\scriptstyle \sigma_{\mathrm{z_{\mathrm{sdss}}}}=0.081\;mag\\
\\
\mathrm{SDSS \rightarrow MzLS+BASS:}\\
\scriptstyle g-r<0.689:\qquad g_{\mathrm{bass}} = g_{\mathrm{sdss}} + 0.007(g_{\mathrm{sdss}} - r_{\mathrm{sdss}}) - 0.025\\
\scriptstyle 0.689<g-r<0.845:\qquad g_{\mathrm{bass}} = g_{\mathrm{sdss}} - 0.455(g_{\mathrm{sdss}} - r_{\mathrm{sdss}}) + 0.293\\
\scriptstyle g-r>0.845:\qquad g_{\mathrm{bass}} = g_{\mathrm{sdss}} - 0.107(g_{\mathrm{sdss}} - r_{\mathrm{sdss}}) - 0.001\\
\scriptstyle \sigma_{\mathrm{g_{\mathrm{bass}}}}= 0.067\;mag\\
\scriptstyle g-r<0.603:\qquad r_{\mathrm{bass}} = r_{\mathrm{sdss}} + 0.007(g_{\mathrm{sdss}} - r_{\mathrm{sdss}}) - 0.106\\
\scriptstyle 0.603<g-r<0.930:\qquad r_{\mathrm{bass}} = r_{\mathrm{sdss}} - 0.188(g_{\mathrm{sdss}} - r_{\mathrm{sdss}}) + 0.011\\
\scriptstyle g-r>0.930:\qquad r_{\mathrm{bass}} = r_{\mathrm{sdss}} - 0.030(g_{\mathrm{sdss}} - r_{\mathrm{sdss}}) - 0.135\\
\scriptstyle \sigma_{\mathrm{r_{\mathrm{bass}}}}=0.060\;mag\\
\scriptstyle r-z<0.592:\qquad r_{\mathrm{bass}} = r_{\mathrm{sdss}} - 0.045(r_{\mathrm{sdss}} - z_{\mathrm{sdss}}) - 0.091\\
\scriptstyle 0.592<r-z<1.240:\qquad r_{\mathrm{bass}} = r_{\mathrm{sdss}} - 0.198(r_{\mathrm{sdss}} - z_{\mathrm{sdss}}) - 0.001\\
\scriptstyle r-z>1.240:\qquad r_{\mathrm{bass}} = r_{\mathrm{sdss}} - 0.218(r_{\mathrm{sdss}} - z_{\mathrm{sdss}}) + 0.023\\
\scriptstyle \sigma_{\mathrm{r_{\mathrm{bass}}}}=0.068\;mag\\
\scriptstyle r-z<0.309:\qquad z_{\mathrm{mzls}} = z_{\mathrm{mzls}} - 0.349(r_{\mathrm{sdss}} - z_{\mathrm{mzls}}) + 0.036\\
\scriptstyle 0.309<r-z<1.015:\qquad z_{\mathrm{mzls}} = z_{\mathrm{mzls}} - 0.077(r_{\mathrm{sdss}} - z_{\mathrm{mzls}}) - 0.047\\
\scriptstyle r-z<1.015:\qquad z_{\mathrm{mzls}} = z_{\mathrm{mzls}} + 0.247(r_{\mathrm{sdss}} - z_{\mathrm{mzls}}) - 0.377\\
\scriptstyle \sigma_{\mathrm{z_{\mathrm{mzls}}}}=0.078\;mag\\
\end{align*}
\endgroup

\subsection{VST ATLAS and KiDS}

VST ATLAS is a wide-field survey of the Southern hemisphere of the sky; KiDS covers a smaller area, but at a higher depth. Both surveys were conducted with the 2.6 m ESO VST telescope and include the $u$ band in addition to $griz$ even though the $u$ band images are very shallow, especially in VST ATLAS. Both surveys have a small overlap with SDSS, what makes the analysis more difficult \revtwo{because} of a much smaller number of measurements, which we can use to fit color transformations. Nevertheless, it was still possible to calculate the transformations to/from SDSS. \revthree{We use VST ATLAS DR3 data because it is accessible via Virtual Observatory protocols through CDS Vizier. However, we compared DR3 versus DR4 photometric measurements and found no statistically significant differences between them, therefore the color transformations presented here can be used for VST ATLAS DR4.} The VST ATLAS catalog contains Petrosian magnitudes, while only Kron magnitudes are provided by KiDS, which introduces additional systematic uncertainties to the transformations of KiDS to/from SDSS. The fitting residuals are rather small for the $griz$ bands (0.039 to 0.065~mag; similar to DES) and much larger for the $u$ band in the KiDS survey (0.13~mag). \revone{We do not provide $u$-band transformations between VST ATLAS and SDSS because of the limited depth of both surveys in the u band for galaxies as compared to stars. The VST ATLAS color transformations published in \citet{2015MNRAS.451.4238S} provide a very good quality of the converted color. We find a very small difference about 0.02~mag in the mean $r$-band magnitudes. Nevertheless, here we provide color transformations obtained using our approach for consistency with other surveys.}

\begingroup
\allowdisplaybreaks
\begin{align*}
\mathrm{VST\;ATLAS \rightarrow SDSS:}\\
\scriptstyle g-r<0.309:\qquad g_{\mathrm{sdss}} = g_{\mathrm{vst}} - 0.051(g_{\mathrm{vst}} - r_{\mathrm{vst}}) + 0.040\\
\scriptstyle 0.309<g-r<1.059:\qquad g_{\mathrm{sdss}} =  g_{\mathrm{vst}} + 0.010(g_{\mathrm{vst}} - r_{\mathrm{vst}}) + 0.021\\
\scriptstyle g-r>1.059:\qquad g_{\mathrm{sdss}} = g_{\mathrm{vst}} - 0.128(g_{\mathrm{vst}} - r_{\mathrm{vst}}) + 0.169\\
\scriptstyle \sigma_{\mathrm{g_{\mathrm{sdss}}}}= 0.052\;mag\\
\scriptstyle g-r<0.248:\qquad r_{\mathrm{sdss}} = r_{\mathrm{vst}} + 1.206(g_{\mathrm{vst}} - r_{\mathrm{vst}}) - 0.284\\
\scriptstyle [0.248<g-r<0.821:\qquad r_{\mathrm{sdss}} =r_{\mathrm{vst}} + 0.080(g_{\mathrm{vst}} - r_{\mathrm{vst}}) - 0.005\\
\scriptstyle g-r>0.821:\qquad r_{\mathrm{sdss}} =r_{\mathrm{vst}} + 0.021(g_{\mathrm{vst}} - r_{\mathrm{vst}}) + 0.042\\
\scriptstyle \sigma_{\mathrm{r_{\mathrm{sdss}}}}= 0.041\;mag\\
\scriptstyle r-i<0.024:\qquad i_{\mathrm{sdss}} =i_{\mathrm{vst}} + 0.636(r_{\mathrm{vst}} - i_{\mathrm{vst}}) - 0.138\\
\scriptstyle 0.024<r-i<0.310:\qquad i_{\mathrm{sdss}} =i_{\mathrm{vst}} + 0.547(r_{\mathrm{vst}} - i_{\mathrm{vst}}) - 0.136\\
\scriptstyle r-i>0.310:\qquad i_{\mathrm{sdss}} =i_{\mathrm{vst}} - 0.004(r_{\mathrm{vst}} - i_{\mathrm{vst}}) + 0.035\\
\scriptstyle \sigma_{\mathrm{i_{\mathrm{sdss}}}}= 0.047\;mag\\
\scriptstyle i-z<-0.297:\qquad z_{\mathrm{sdss}} =z_{\mathrm{vst}} + 1.011(i_{\mathrm{vst}} - z_{\mathrm{vst}}) + 0.071\\
\scriptstyle -0.297<i-z<0.356:\qquad z_{\mathrm{sdss}} =z_{\mathrm{vst}} + 0.504(i_{\mathrm{vst}} - z_{\mathrm{vst}}) - 0.079\\
\scriptstyle i-z>0.356:\qquad z_{\mathrm{sdss}} =z_{\mathrm{vst}} + 0.079(i_{\mathrm{vst}} - z_{\mathrm{vst}}) + 0.072\\
\scriptstyle \sigma_{\mathrm{z_{\mathrm{sdss}}}}= 0.067\;mag\\
\\
\mathrm{SDSS \rightarrow VST\;ATLAS:}\\
\scriptstyle g-r<1.056:\qquad g_{\mathrm{vst}} = g_{\mathrm{sdss}} - 0.079(g_{\mathrm{sdss}} - r_{\mathrm{sdss}}) + 0.038\\
\scriptstyle g-r>1.056:\qquad g_{\mathrm{vst}} = g_{\mathrm{sdss}} + 0.051(g_{\mathrm{sdss}} - r_{\mathrm{sdss}}) - 0.099\\
\scriptstyle \sigma_{\mathrm{g_{\mathrm{vst}}}}= 0.041\;mag\\
\scriptstyle g-r<0.544:\qquad r_{\mathrm{vst}} = r_{\mathrm{sdss}} + 0.044(g_{\mathrm{sdss}} - r_{\mathrm{sdss}}) - 0.058\\
\scriptstyle g-r>0.544:\qquad r_{\mathrm{vst}} = r_{\mathrm{sdss}} - 0.043(g_{\mathrm{sdss}} - r_{\mathrm{sdss}}) - 0.010\\
\scriptstyle \sigma_{\mathrm{r_{\mathrm{vst}}}}=0.036\;mag\\
\scriptstyle i_{\mathrm{vst}} = i_{\mathrm{sdss}}  + 0.128(r_{\mathrm{sdss}} - i_{\mathrm{sdss}}) - 0.071\\
\scriptstyle \sigma_{\mathrm{i_{\mathrm{vst}}}}=0.041\;mag\\
\scriptstyle i-z<0.089:\qquad z_{\mathrm{vst}}  =z_{\mathrm{sdss}} + 0.492(i_{\mathrm{sdss}} - z_{\mathrm{sdss}}) - 0.194\\
\scriptstyle 0.089<i-z<0.399:\qquad z_{\mathrm{vst}} = z_{\mathrm{sdss}} + 0.413(i_{\mathrm{sdss}} - z_{\mathrm{sdss}}) - 0.187\\
\scriptstyle i-z>0.399:\qquad z_{\mathrm{vst}} =z_{\mathrm{sdss}} + 0.561(i_{\mathrm{sdss}} - z_{\mathrm{sdss}}) - 0.246\\
\scriptstyle \sigma_{\mathrm{z_{\mathrm{vst}}}}=0.061\;mag\\
\\
\\
\mathrm{KiDS \rightarrow SDSS:}\\
\scriptstyle u-g<1.809:\qquad u_{\mathrm{sdss}} =u_{\mathrm{KiDS}} + 0.037(u_{\mathrm{KiDS}} - g_{\mathrm{KiDS}}) - 0.105\\
\scriptstyle u-g>1.809:\qquad u_{\mathrm{sdss}}= u_{\mathrm{KiDS}} - 0.460(u_{\mathrm{KiDS}} - g_{\mathrm{KiDS}}) + 0.795\\
\scriptstyle \sigma_{\mathrm{u_{\mathrm{sdss}}}}= 0.130\;mag\\
\scriptstyle g_{\mathrm{sdss}} = g_{\mathrm{KiDS}} - 0.005(g_{\mathrm{KiDS}} - r_{\mathrm{KiDS}}) + 0.030\\
\scriptstyle \sigma_{\mathrm{g_{\mathrm{sdss}}}}= 0.052\;mag\\
\scriptstyle r_{\mathrm{sdss}} = r_{\mathrm{KiDS}} - 0.002(g_{\mathrm{KiDS}} - r_{\mathrm{KiDS}}) + 0.035\\
\scriptstyle \sigma_{\mathrm{r_{\mathrm{sdss}}}}= 0.039\;mag\\
\scriptstyle r-i<0.477:\qquad i_{\mathrm{sdss}} = i_{\mathrm{KiDS}} + 0.129(r_{\mathrm{KiDS}} - i_{\mathrm{KiDS}}) - 0.020\\
\scriptstyle r-i>0.477:\qquad i_{\mathrm{sdss}} = i_{\mathrm{KiDS}} - 0.231(r_{\mathrm{KiDS}} - i_{\mathrm{KiDS}}) + 0.152\\
\scriptstyle \sigma_{\mathrm{i_{\mathrm{sdss}}}}= 0.046\;mag\\
\\
\mathrm{SDSS \rightarrow KiDS:}\\
\scriptstyle u-g<1.253:\qquad u_{\mathrm{KiDS}} =u_{\mathrm{sdss}} - 0.549(u_{\mathrm{sdss}} - g_{\mathrm{sdss}}) + 0.679\\
\scriptstyle 1.253<u-g<1.579:\qquad u_{\mathrm{KiDS}}= u_{\mathrm{sdss}} + 0.444(u_{\mathrm{sdss}} - g_{\mathrm{sdss}}) - 0.565\\
\scriptstyle u-g>1.579:\qquad u_{\mathrm{KiDS}}= u_{\mathrm{sdss}} - 0.622(u_{\mathrm{sdss}} - g_{\mathrm{sdss}}) + 1.118)\\
\scriptstyle \sigma_{\mathrm{u_{\mathrm{KiDS}}}}=0.131 \;mag\\
\scriptstyle g-r<0.872:\qquad g_{\mathrm{KiDS}} =g_{\mathrm{sdss}} - 0.142(g_{\mathrm{sdss}} - r_{\mathrm{sdss}}) + 0.083\\
\scriptstyle 0.872<g-r<1.452:\qquad g_{\mathrm{KiDS}}= g_{\mathrm{sdss}} - 0.019(g_{\mathrm{sdss}} - r_{\mathrm{sdss}}) - 0.023\\
\scriptstyle g-r>1.452:\qquad g_{\mathrm{KiDS}}=  g_{\mathrm{sdss}} - 0.175(g_{\mathrm{sdss}} - r_{\mathrm{sdss}}) + 0.201)\\
\scriptstyle \sigma_{\mathrm{g_{\mathrm{KiDS}}}}=0.049 \;mag\\
\scriptstyle g-r<1.153:\qquad r_{\mathrm{KiDS}} =r_{\mathrm{sdss}} - 0.005(g_{\mathrm{sdss}} - r_{\mathrm{sdss}}) - 0.026\\
\scriptstyle g-r>1.153:\qquad r_{\mathrm{KiDS}}= r_{\mathrm{sdss}} + 0.022(g_{\mathrm{sdss}} - r_{\mathrm{sdss}}) - 0.058\\
\scriptstyle \sigma_{\mathrm{r_{\mathrm{KiDS}}}}=0.037 \;mag\\
\scriptstyle g-r<0.335:\qquad i_{\mathrm{KiDS}} =i_{\mathrm{sdss}} + 0.147(r_{\mathrm{sdss}} - i_{\mathrm{sdss}}) - 0.096\\
\scriptstyle g-r>0.335:\qquad i_{\mathrm{KiDS}}= i_{\mathrm{sdss}} + 0.193(r_{\mathrm{sdss}} - i_{\mathrm{sdss}}) - 0.111\\
\scriptstyle \sigma_{\mathrm{i_{\mathrm{KiDS}}}}=0.045 \;mag\\
\end{align*}
\endgroup


\subsection{PanSTARRS}

\begin{figure*}
\includegraphics[width=0.48\hsize]{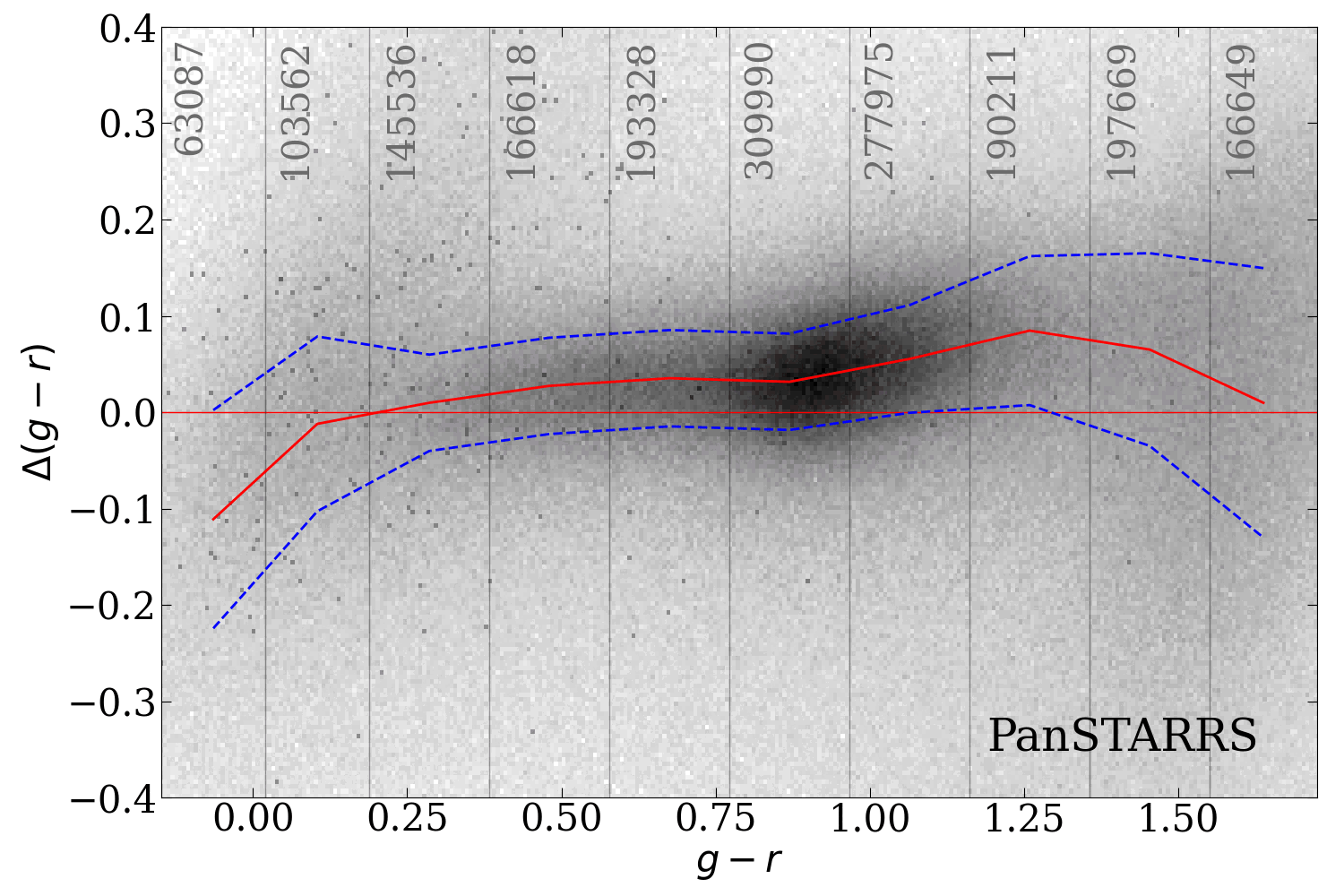}
\includegraphics[width=0.48\hsize]{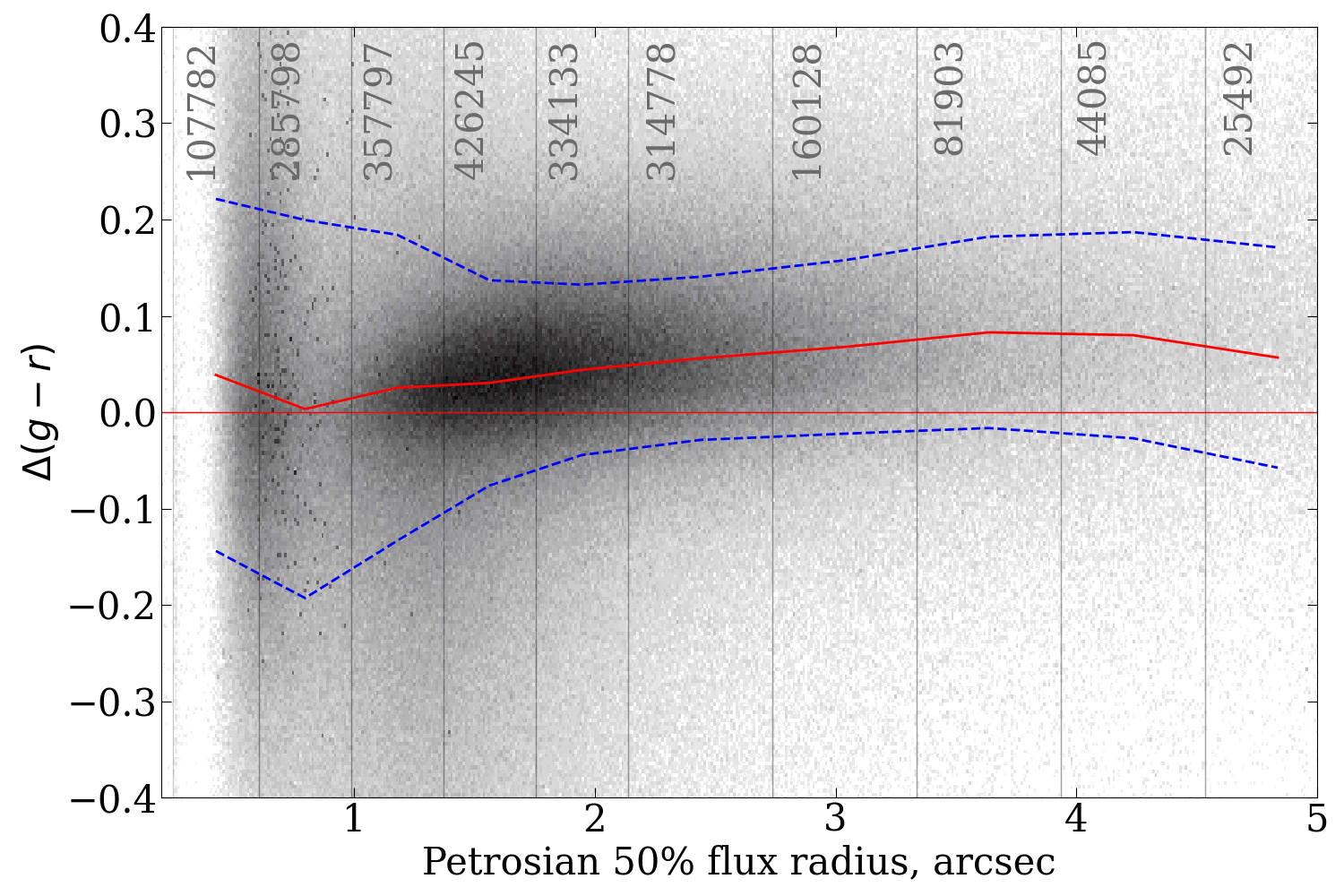}\\
\vspace{-0.4cm}
\caption{\revone{Color residuals for PanSTARRS and the SDSS as a function of the color and Petrosian radius after the conversions from \citet{2016ApJ...822...66F}.} The symbols are the same as in Figure~\ref{mag_resid}. \revone{SDSS magnitudes are Petrosian, PanSTARRS magnitudes are Kron.} \label{ps}}
\end{figure*}

PanSTARRS data demonstrate a highly nonlinear structure of magnitude differences from SDSS magnitudes as a function of color. Therefore, the method of piecewise linear regression described above could not be applied to the PanSTARRS data with a reasonable number of segments, and we used color transformations based on PSF measurements from the literature~\citet{2016ApJ...822...66F}. These transformations yield high systematic differences for the corrected values of Kron magnitudes compared to SDSS Petrosian magnitudes (see Figure~\ref{ps}), much higher than those we observed in the case of VST KiDS. We notice that the differences with SDSS increase with the size of galaxy. \revone{PanSTARRS also contains Petrosian magnitudes, which can be better for comparison with SDSS data, but we do not consider them because they are not available through Virtual Observatory services \revtwo{which we used to access the data}.}

\subsection{VHS and VIKING}

The wide-field infrared surveys VHS and VIKING conducted at the 4 m ESO VISTA telescope were translated to the UKIRT system using the same method as for the optical surveys. Both surveys use identical prescriptions for data reduction, processing and analysis and, hence, were treated as a single dataset in our study. For the approximation, we used a piecewise linear regression with 2 segments over the entire color range. The fitting residuals are somewhat higher than for optical surveys (0.07--0.1~mag) but the residuals do not show any trends with the color.

\begingroup
\allowdisplaybreaks
\begin{align*}
\mathrm{VISTA \rightarrow UKIRT:}\\
\scriptstyle Y_{\mathrm{ukirt}} = Y_{\mathrm{vista}} - 0.342(Y_{\mathrm{vista}} - J_{\mathrm{vista}}) - 0.226\\
\scriptstyle \sigma_{\mathrm{Y_{\mathrm{ukirt}}}}= 0.077\;mag\\
\scriptstyle J_{\mathrm{ukirt}} = J_{\mathrm{vista}} - 0.288(J_{\mathrm{vista}} - Y_{\mathrm{vista}}) - 0.372\\
\scriptstyle \sigma_{\mathrm{J_{\mathrm{ukirt}}}}= 0.102\;mag\\
\scriptstyle H_{\mathrm{ukirt}} = H_{\mathrm{vista}} - 0.239(H_{\mathrm{vista}} - Ks_{\mathrm{vista}}) - 0.358\\
\scriptstyle \sigma_{\mathrm{H_{\mathrm{ukirt}}}}= 0.099\;mag\\
\scriptstyle K_{\mathrm{ukirt}} = Ks_{\mathrm{vista}} - 0.091(Ks_{\mathrm{vista}} - H_{\mathrm{vista}}) - 0.320\\
\scriptstyle \sigma_{\mathrm{Ks_{\mathrm{ukirt}}}}= 0.103\;mag\\
\\
\mathrm{UKIRT \rightarrow VISTA:}\\
\scriptstyle Y-J<0.216:\qquad Y_{\mathrm{vista}} = Y_{\mathrm{ukirt}} + 0.078(Y_{\mathrm{ukirt}} - J_{\mathrm{ukirt}}) + 0.260\\
\scriptstyle Y-J>0.216:\qquad Y_{\mathrm{vista}} = Y_{\mathrm{ukirt}} - 0.109(Y_{\mathrm{ukirt}} - J_{\mathrm{ukirt}}) + 0.300\\
\scriptstyle \sigma_{\mathrm{Y_{\mathrm{vista}}}}= 0.074\;mag\\
\scriptstyle J_{\mathrm{vista}} = J_{\mathrm{ukirt}} + 0.713(J_{\mathrm{ukirt}} - Y_{\mathrm{ukirt}}) + 0.122\\
\scriptstyle \sigma_{\mathrm{J_{\mathrm{vista}}}}= 0.086\;mag\\
\scriptstyle H-K<0.200:\qquad H_{\mathrm{vista}} = H_{\mathrm{ukirt}} - 0.069(H_{\mathrm{ukirt}} - Ks_{\mathrm{ukirt}}) + 0.354\\
\scriptstyle H-K>0.200:\qquad H_{\mathrm{vista}} = H_{\mathrm{ukirt}} - 0.332(H_{\mathrm{ukirt}} - Ks_{\mathrm{ukirt}}) + 0.407\\
\scriptstyle \sigma_{\mathrm{H_{\mathrm{vista}}}}= 0.098\;mag\\
\scriptstyle H-K<-0.218:\qquad  K_{\mathrm{vista}} = Ks_{\mathrm{ukirt}} + 0.490(Ks_{\mathrm{ukirt}} - H_{\mathrm{ukirt}}) + 0.375\\
\scriptstyle H-K>-0.218:\qquad  K_{\mathrm{vista}} = Ks_{\mathrm{ukirt}} + 0.210(Ks_{\mathrm{ukirt}} - H_{\mathrm{ukirt}}) + 0.314\\
\scriptstyle \sigma_{\mathrm{Ks_{\mathrm{vista}}}}= 0.101\;mag\\
\end{align*}
\endgroup

\begin{figure*}
\includegraphics[width=0.48\hsize]{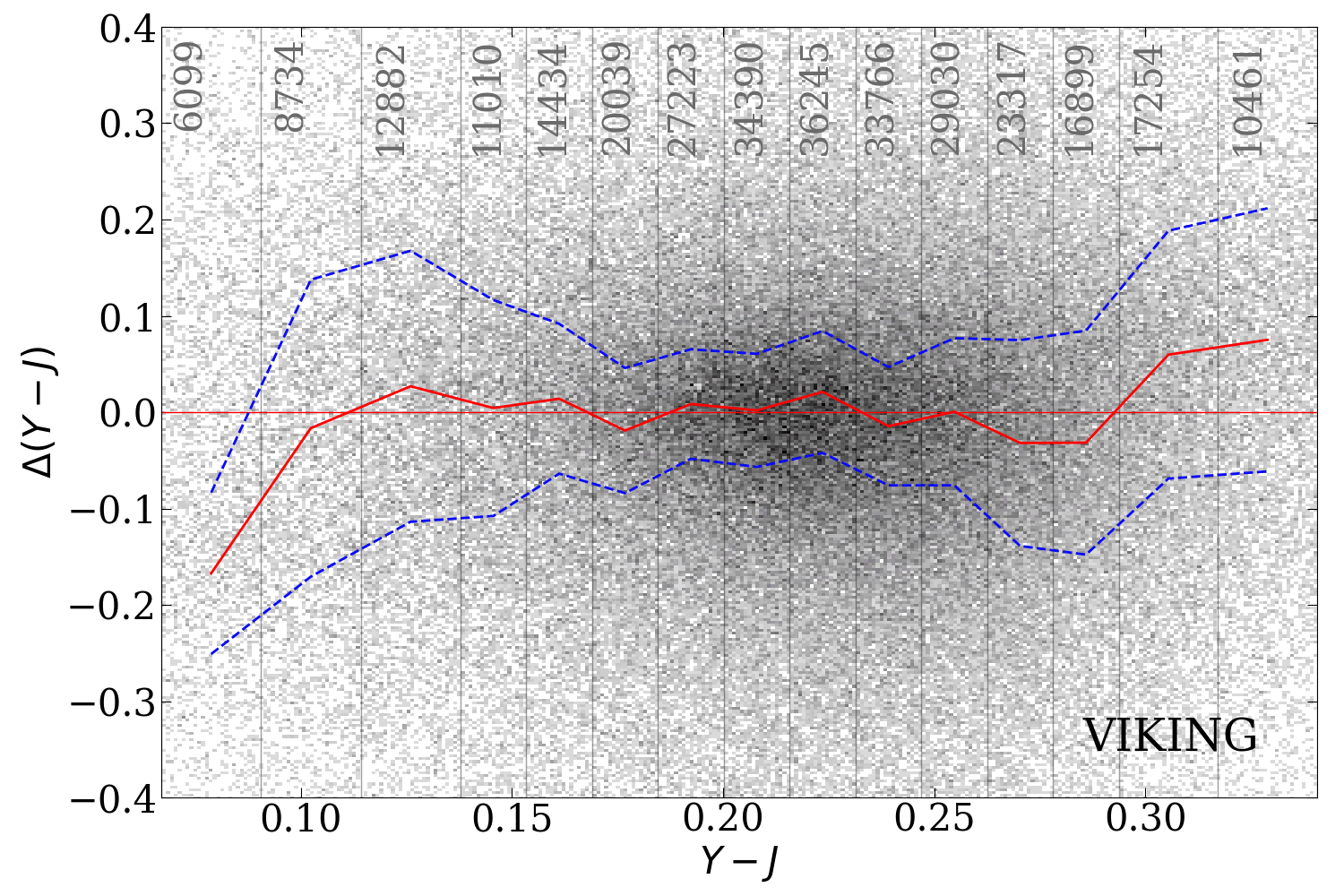}
\includegraphics[width=0.48\hsize]{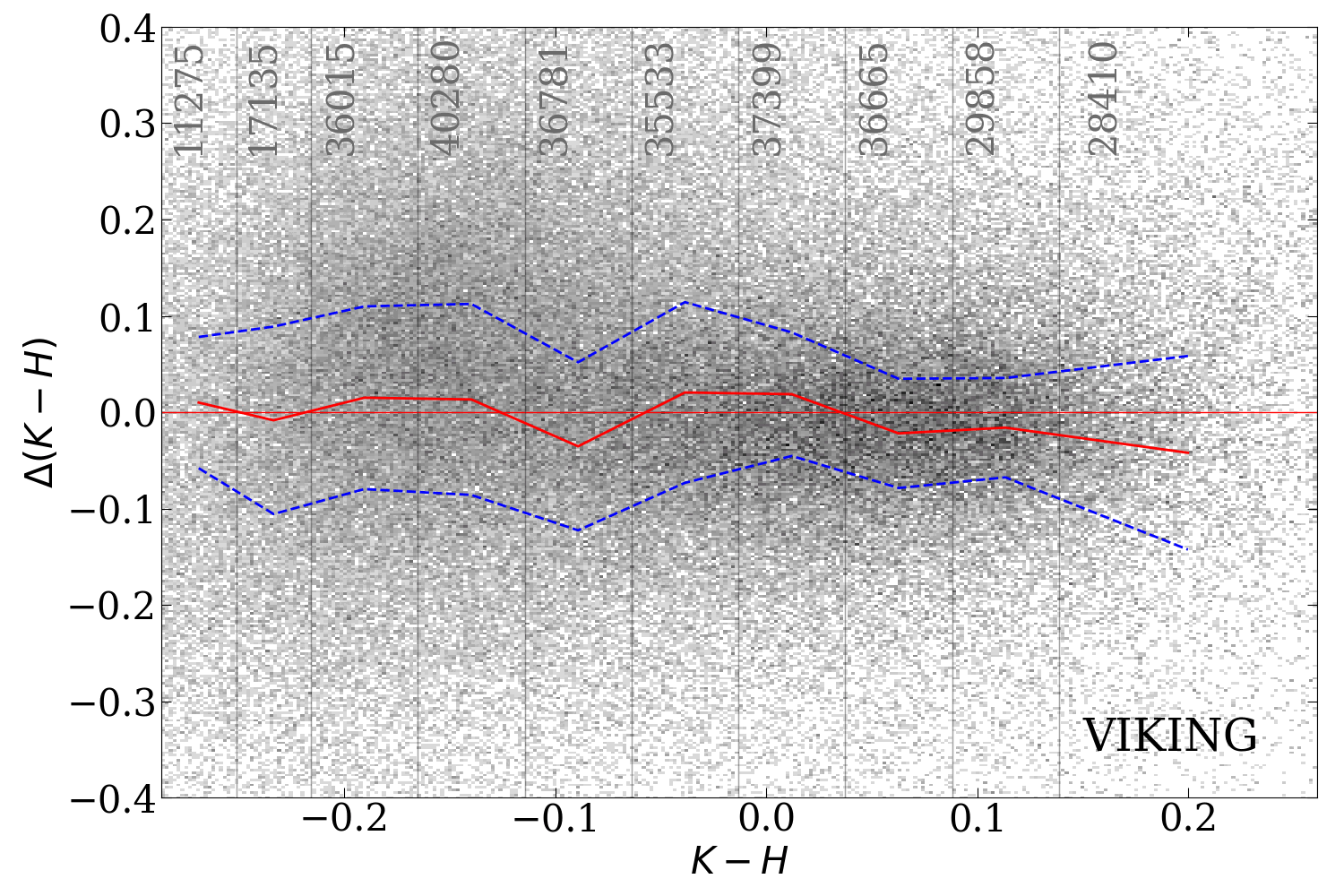}\\
\vspace{-0.4cm}
\caption{Same as Figure~\ref{color_resid} for the difference between VIKING and UKIDSS. \revone{All the magnitudes are Petrosian.}\label{ir}}
\end{figure*}

\subsection{Web-based Service and Code Availability}

We implemented the derived transformations on the project web-site \url{https://colors.voxastro.org/} in the form of a simple calculator that can convert individual measurements. We also provide the conversion formulae in the form of code snippets in {\sc idl} and {\sc python} available from the same website. Both implementations yield identical results provided the same input data. The input parameters for the function {\sc calc\_color\_transf} are: (i) a name of a filter added to the name of a survey for the input data, e.g. for magnitudes in DES g band this argument is `g\_des'; (ii) the same as (i) for an output values; (iii) a list of input total magnitude measurements; (iv) a list of input color values. The size of an output array equals to the size of input lists in (i) -- (ii), which should also be equal to each other.

\section{Discussion}

In the previous section we derived photometric transformations for galaxies in several wide-field surveys in the optical and infrared domains to the SDSS and UKIRT photometric systems. Here we discuss the application of the derived transformations to aperture magnitudes, test them on colors of early-type galaxies, explain the shortcomings of the DESI Legacy Surveys photometric measurements and propose a solution on how to improve them, and apply the derived photometric transformations to spectroscopic samples of nearby and intermediate-redshift galaxies.


\subsection{Applying the transformations to aperture magnitudes}

While our main goal was to correct integrated photometric measurements to build galaxy SEDs consistent among different surveys, it is also important to be able to correct aperture photometric measurements, e.g. to compute the photometry in the apertures matching fiber sizes in spectroscopic surveys.

\begin{figure}
\centering
\includegraphics[width=1\hsize]{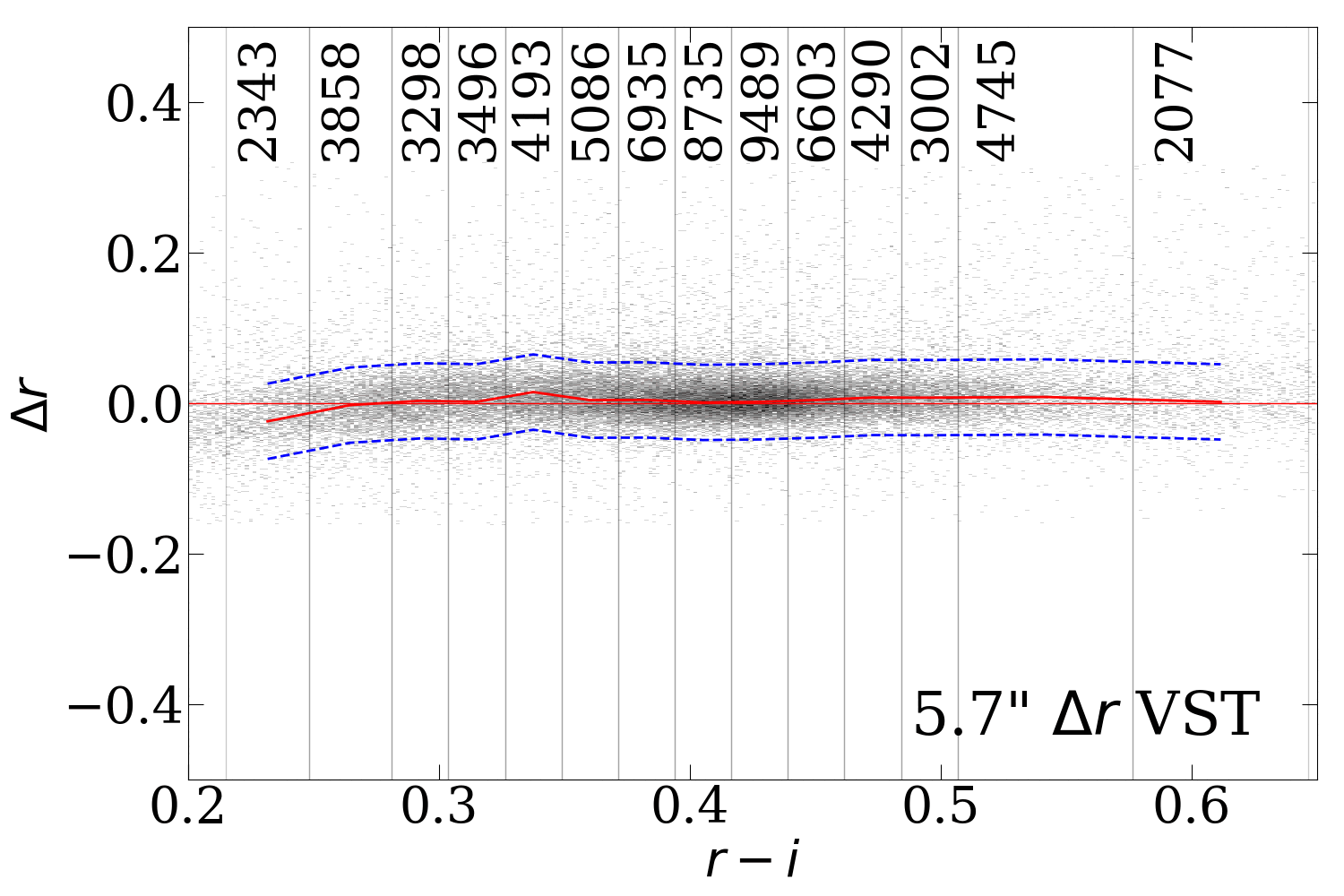}
\caption{Residuals in the $r$-band between the magnitudes from VST ATLAS and DES in 5.7$''$ apertures after color transformations into the SDSS system.
\label{aper5}}
\end{figure}

We applied the transformations \revtwo{derived for total magnitudes in Section \ref{sec:results}} to magnitudes extracted from circular apertures with the diameters $1.''5$ (Hectospec fibers), $2''$, $3''$ (SDSS fibers) and $5.''7$, the largest aperture in VST ATLAS and WFCAM/VIRCAM infrared surveys. For photometric surveys that do not include measurements in these apertures, the magnitudes were calculated using quadratic interpolation from the available aperture magnitudes. In Table~\ref{tab_sample} we list the available aperture diameters for each of the surveys. Because the extrapolation beyond the maximum available aperture is subject to biases, we use the converted optical photometry in the $1.''5$, $2''$ and $3''$ apertures (estimating $1.''5$ aperture magnitudes also requires an extrapolation ``inwards,'' however it is done into a smaller aperture than the original measurements and therefore is only affected the shape of the inner part of a galaxy light profile convolved with the atmospheric seeing). Then, we compare the converted aperture measurements from each survey to those in either SDSS or UKIDSS/UHS. Because we do not have SDSS magnitudes in the $5.''7$ aperture, we compare the VST and DES surveys where such measurements are available with each other (see Figure~\ref{aper5}).

The aperture values \revtwo{calculated using transformations for total magnitudes} demonstrate different offsets from the reference survey depending on the size of the aperture without noticeable color trends in most color/survey/aperture combinations. At the same time, the offset values vary a lot for different surveys even for the same aperture size and reference photometric band, e.g., 3$''$ DES photometry in the $z$-band is offset with respect to the SDSS by 0.187~mag, and the 3$''$ VST magnitudes are offset by 0.312~mag. The offset values \revtwo{were} taken into account to correctly translate the aperture photometry \revtwo{and} are presented in Table~\ref{tab_shifts}. These offsets are caused primarily by the difference in the mean seeing quality of the imaging surveys we consider, and also by the difference in depth between the surveys, which are absorbed by our color transformations of integrated magnitudes and have to be taken back from the transformations when dealing with aperture fluxes, which should not be seriously affected by the depth.

\revthree{The VST Atlas survey includes two types of an aperture magnitude, ``Default point source aperture corrected magnitude'' and ``Default extended source aperture magnitude.''}
\revfive{We provide offsets only for ``Default extended source aperture magnitude'' since this is more correct for extended objects.}

\begin{figure*}
\centering
\includegraphics[trim=0 4.5cm 0cm 0cm, clip,width=0.24\hsize]{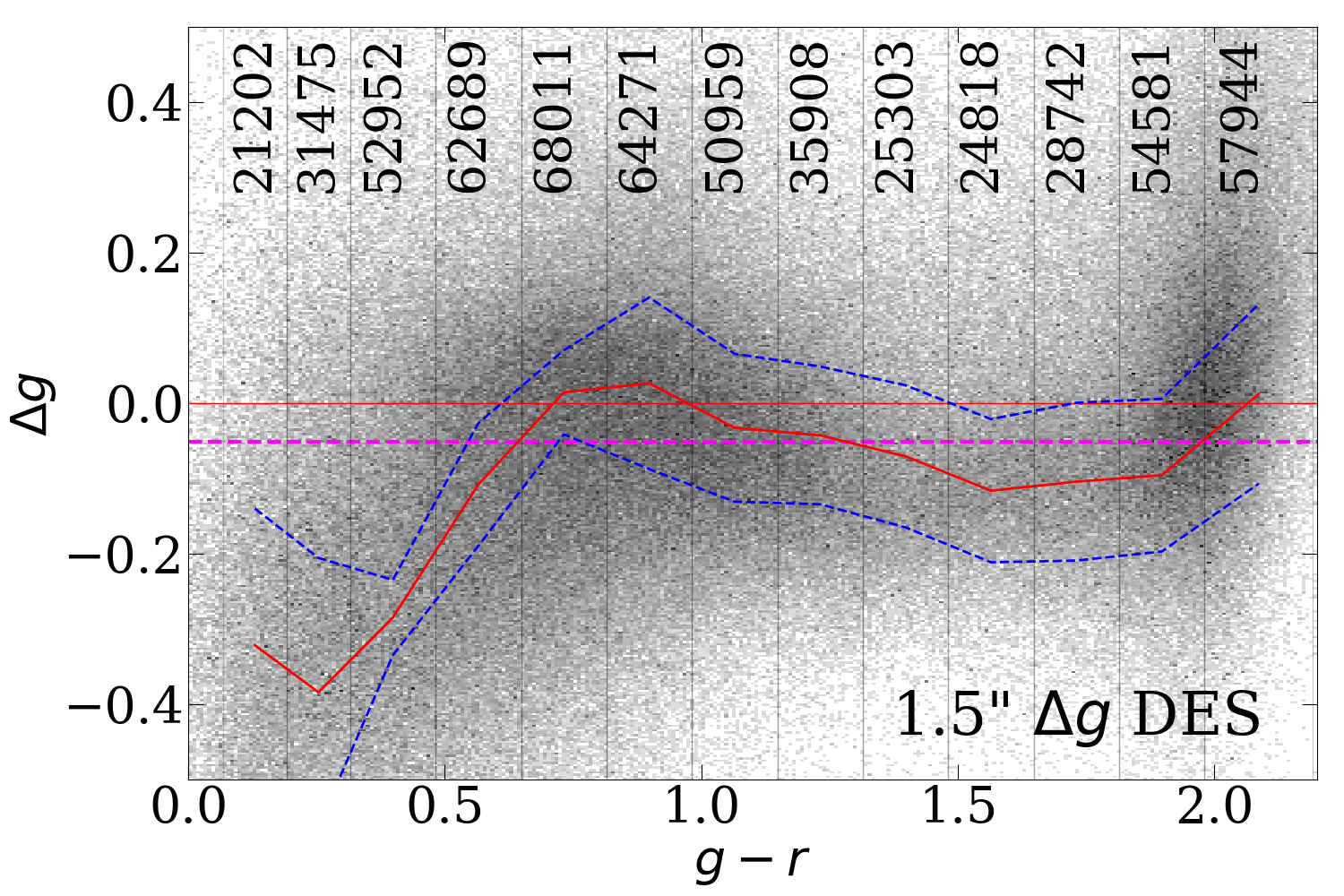}
\includegraphics[trim=0 4.5cm 0cm 0cm, clip,width=0.24\hsize]{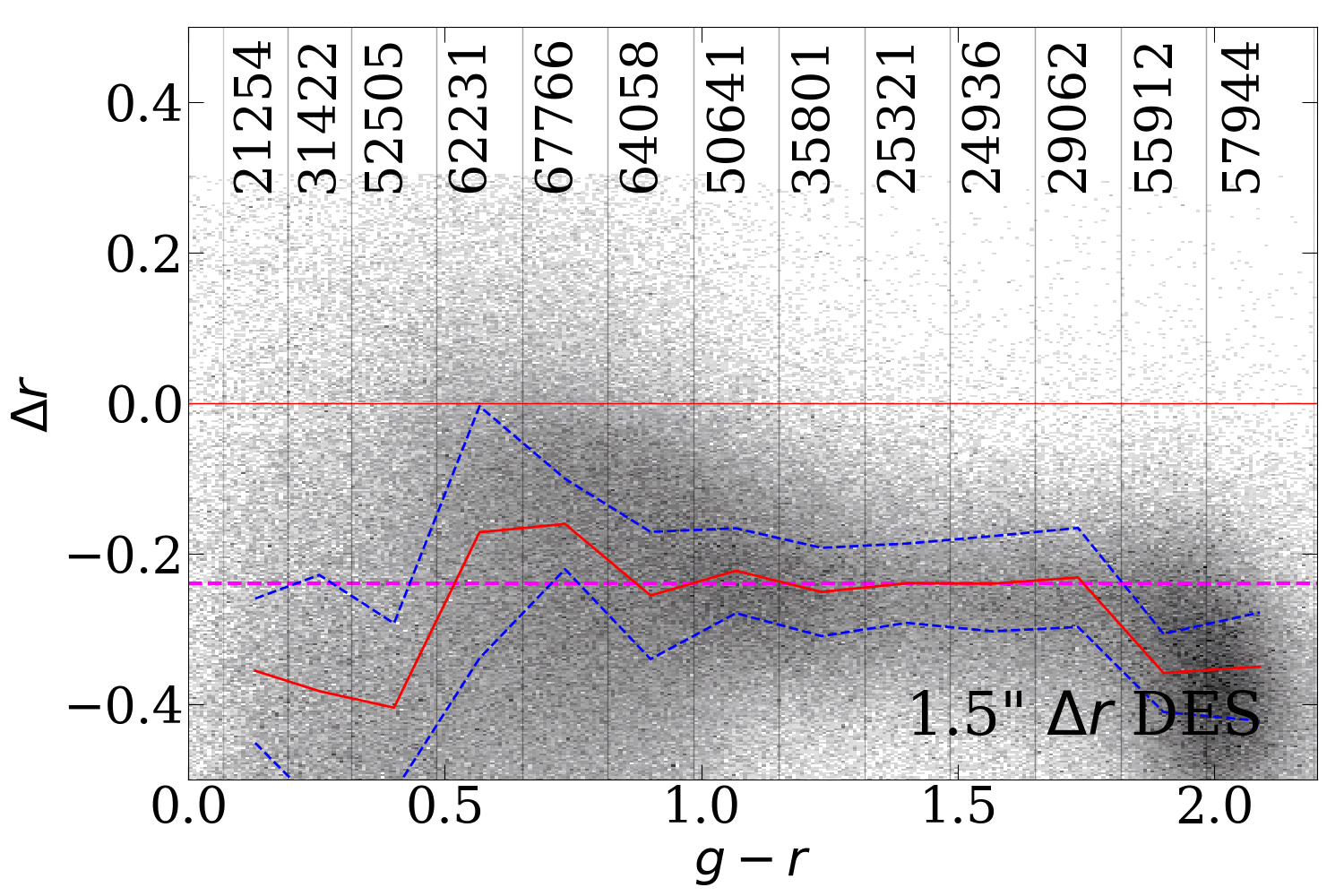}
\includegraphics[trim=0 4.5cm 0cm 0cm, clip,width=0.24\hsize]{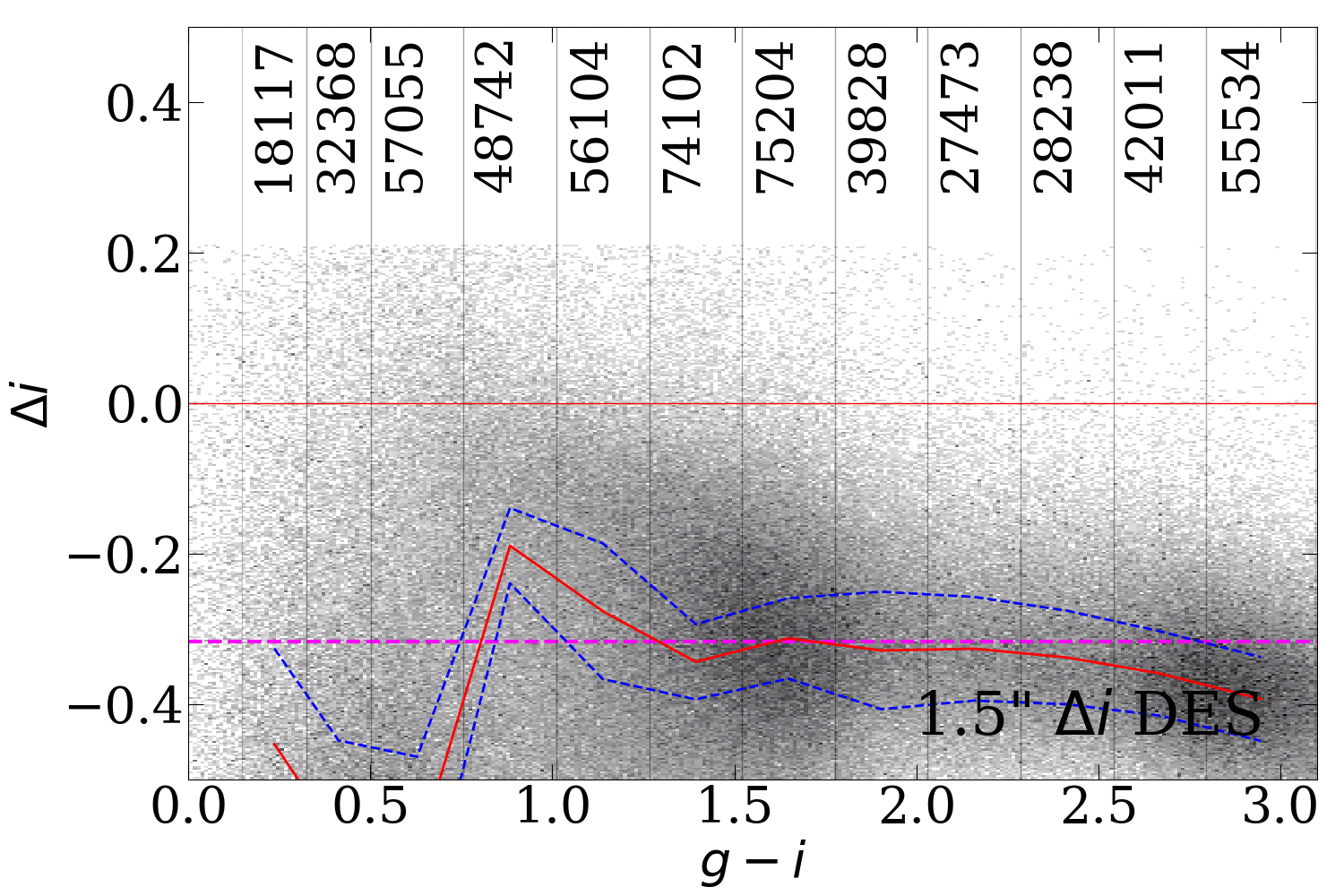}
\includegraphics[trim=0 4.5cm 0cm 0cm, clip,width=0.24\hsize]{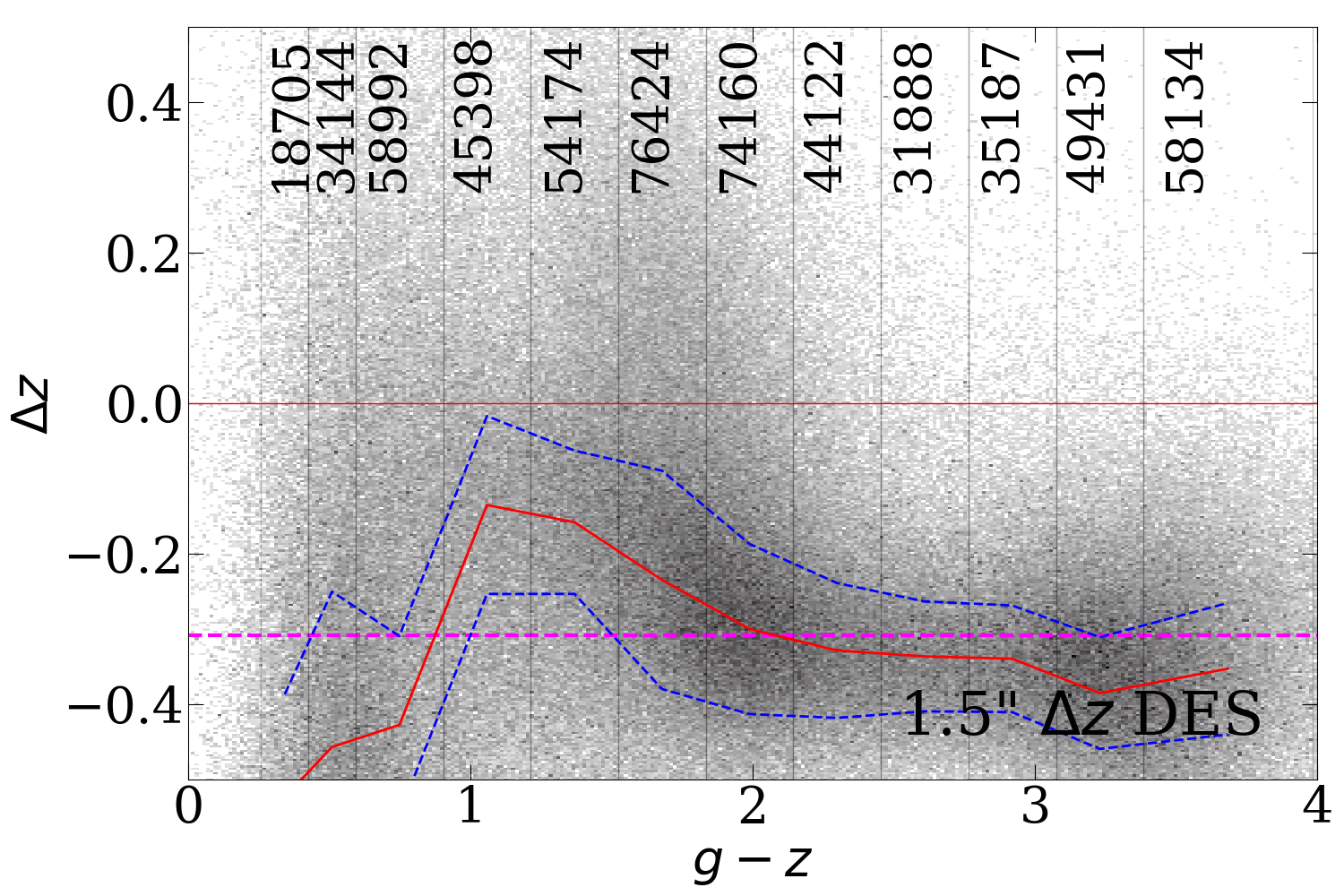}\\
\includegraphics[trim=0 4.5cm 0cm 0cm, clip,width=0.24\hsize]{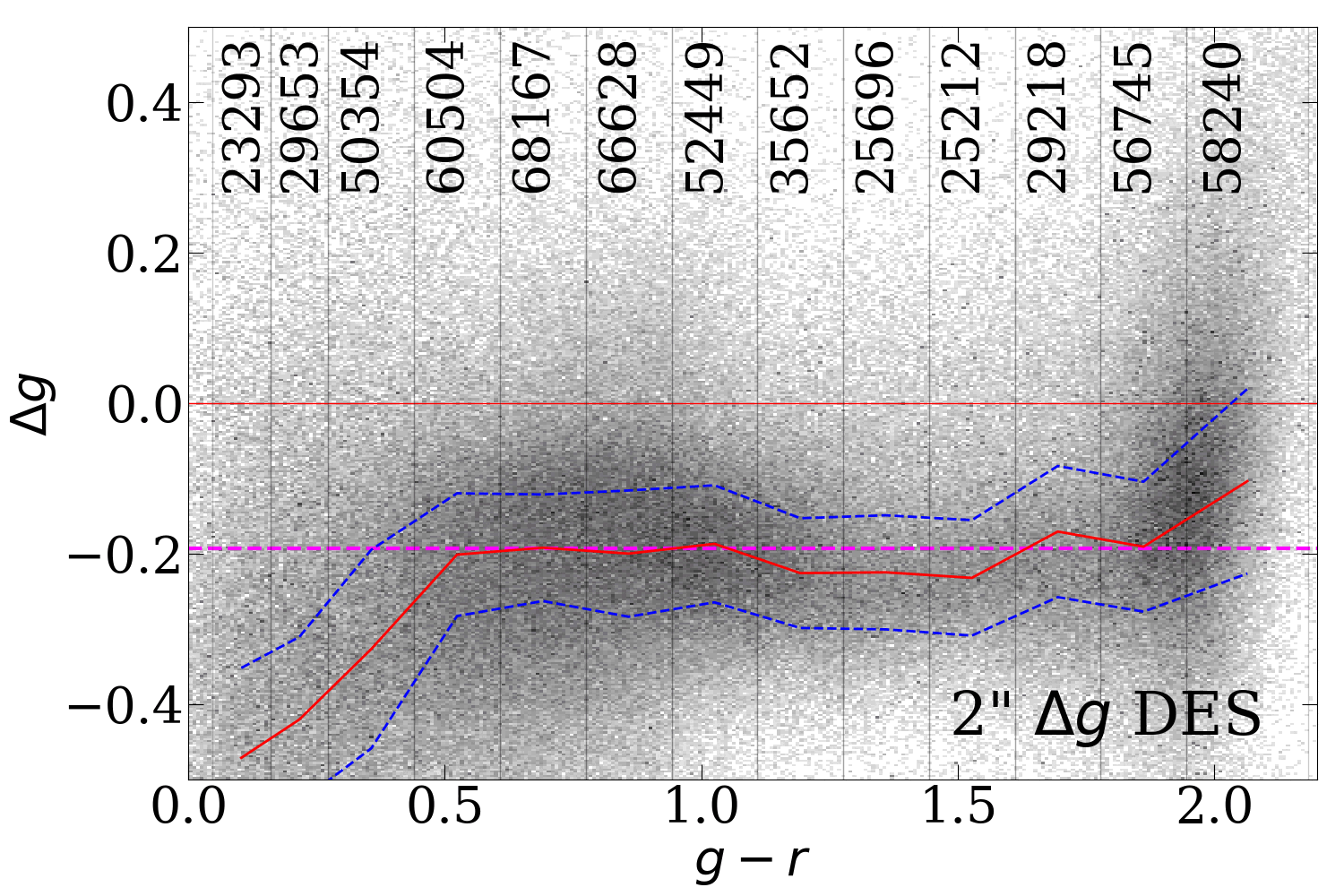}
\includegraphics[trim=0 4.5cm 0cm 0cm, clip,width=0.24\hsize]{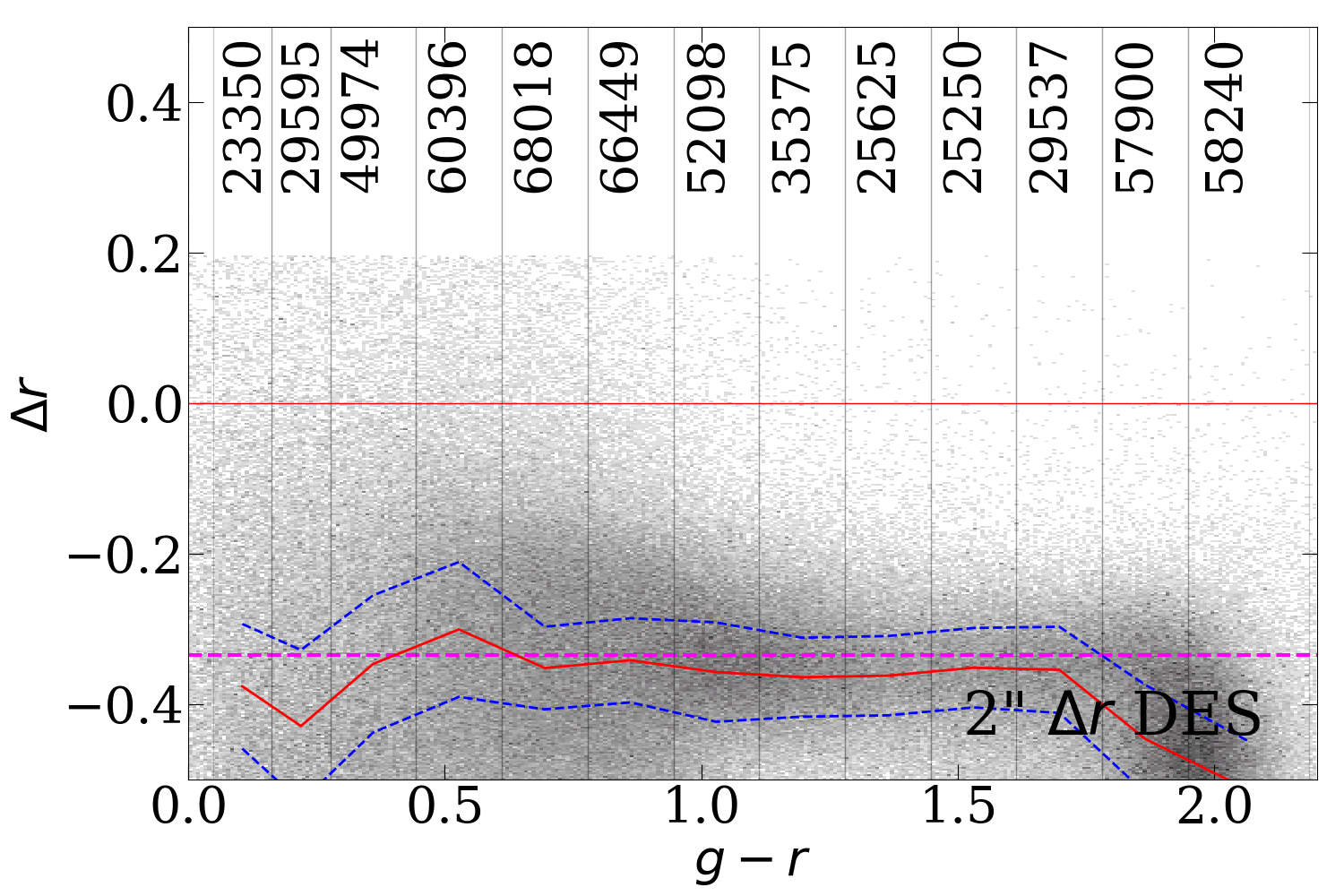}
\includegraphics[trim=0 4.5cm 0cm 0cm, clip,width=0.24\hsize]{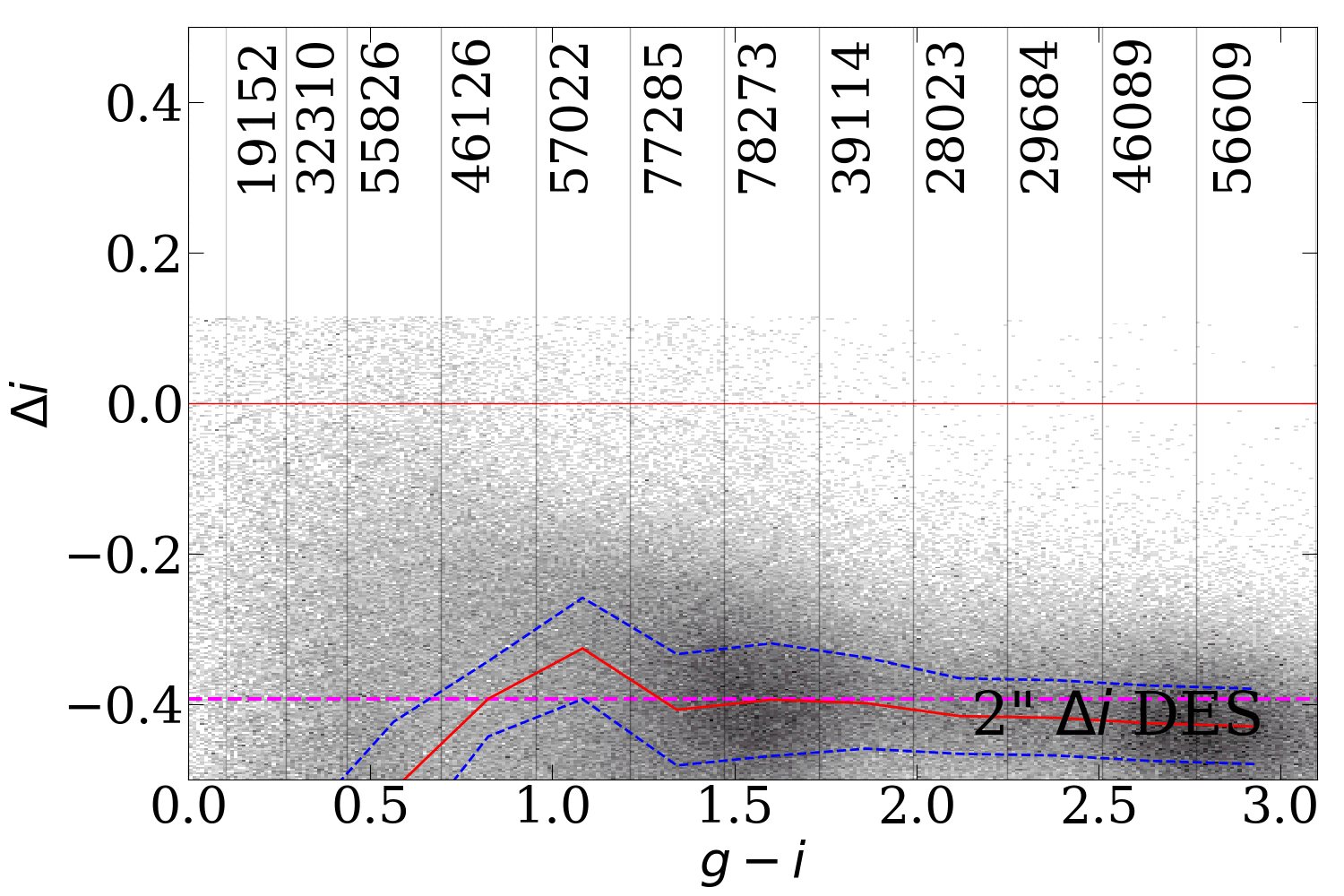}
\includegraphics[trim=0 4.5cm 0cm 0cm, clip,width=0.24\hsize]{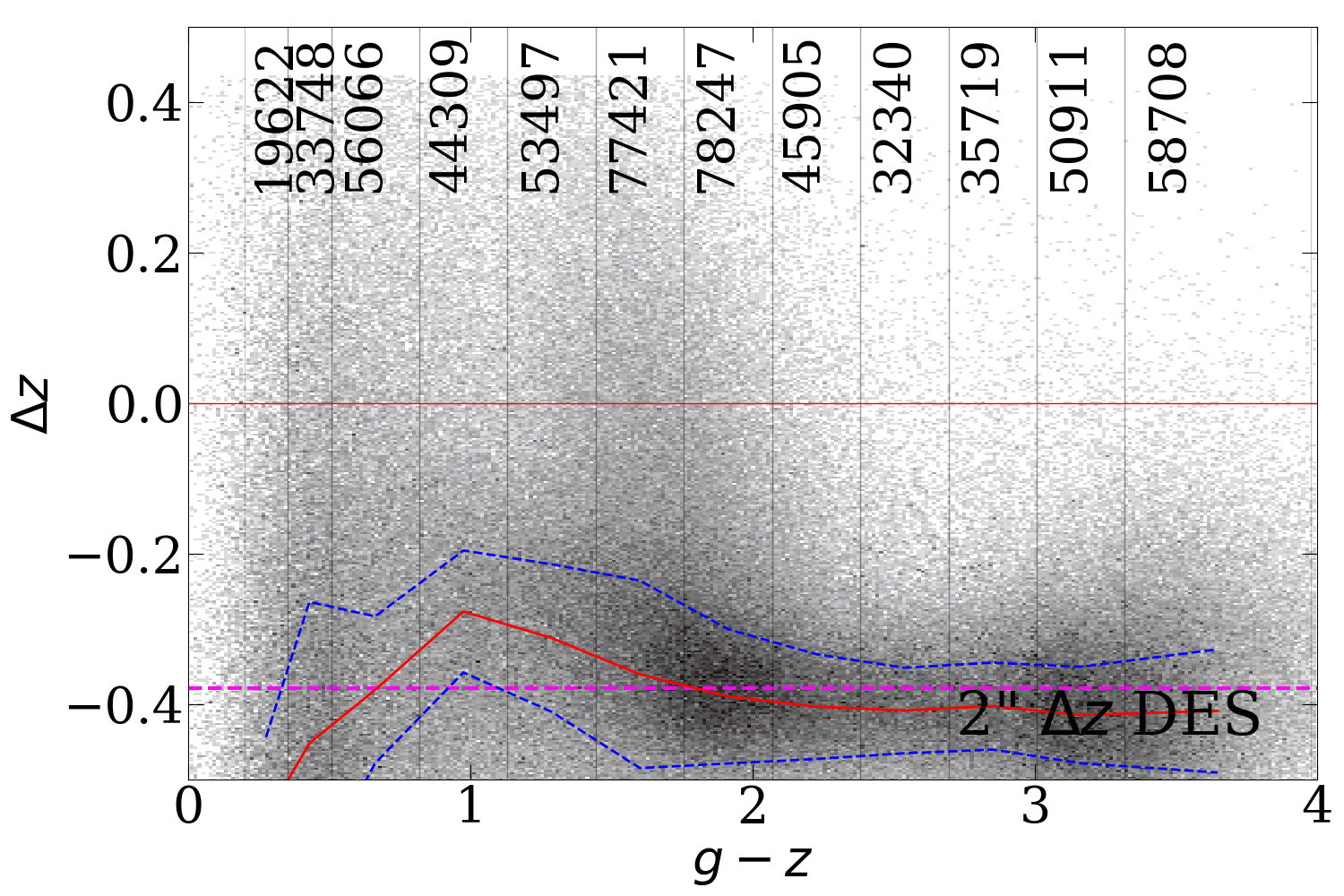}\\
\includegraphics[width=0.24\hsize]{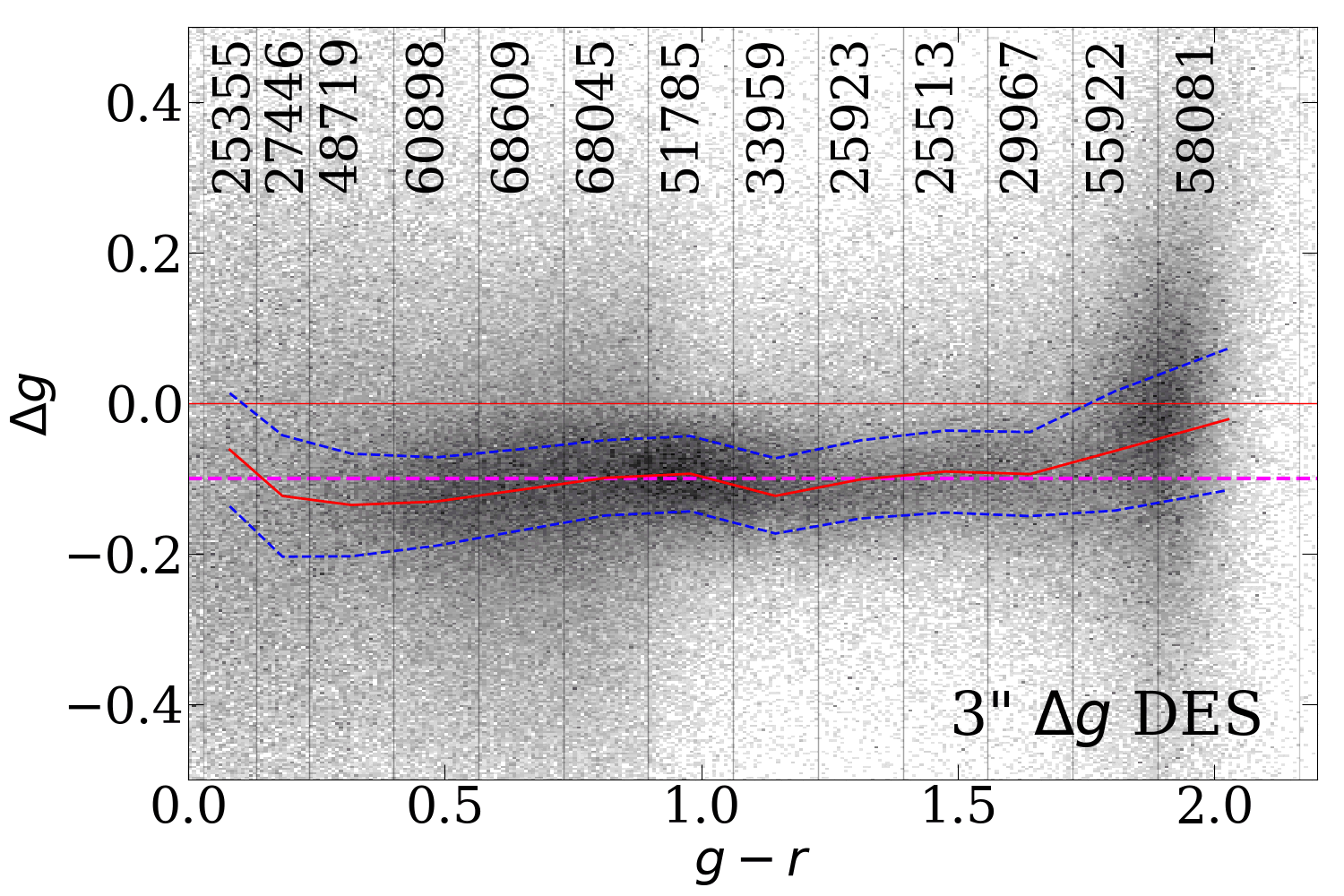}
\includegraphics[width=0.24\hsize]{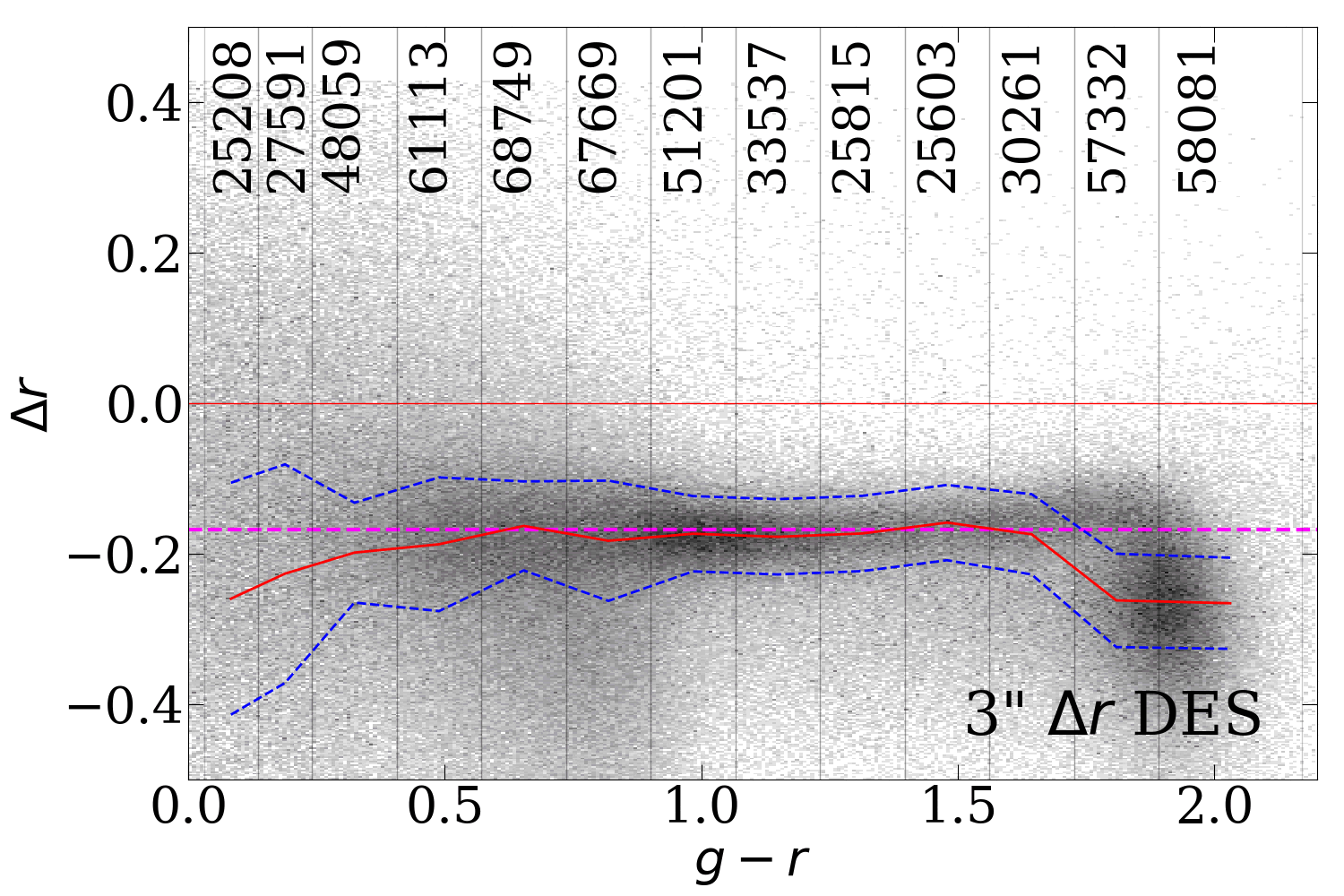}
\includegraphics[width=0.24\hsize]{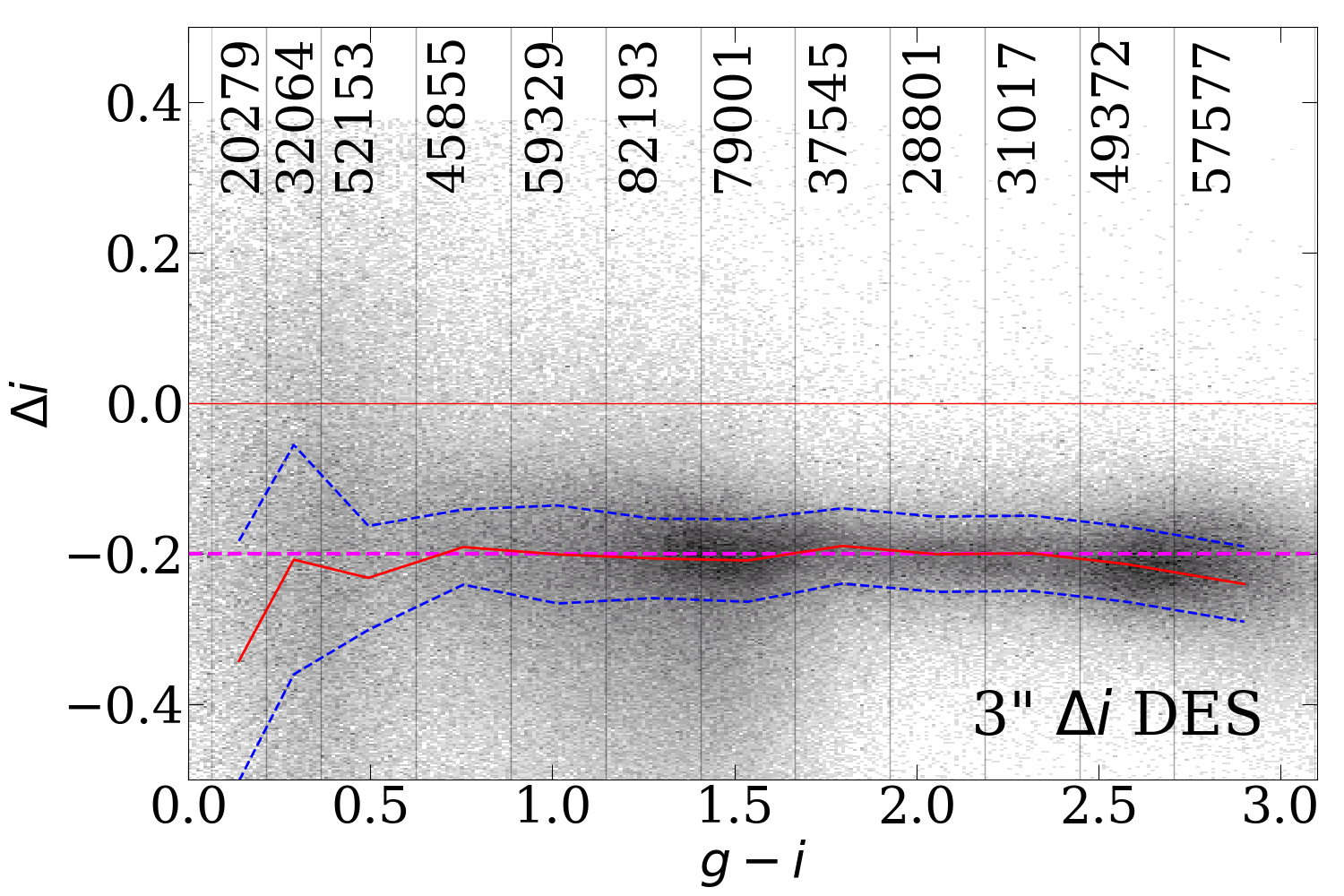}
\includegraphics[width=0.24\hsize]{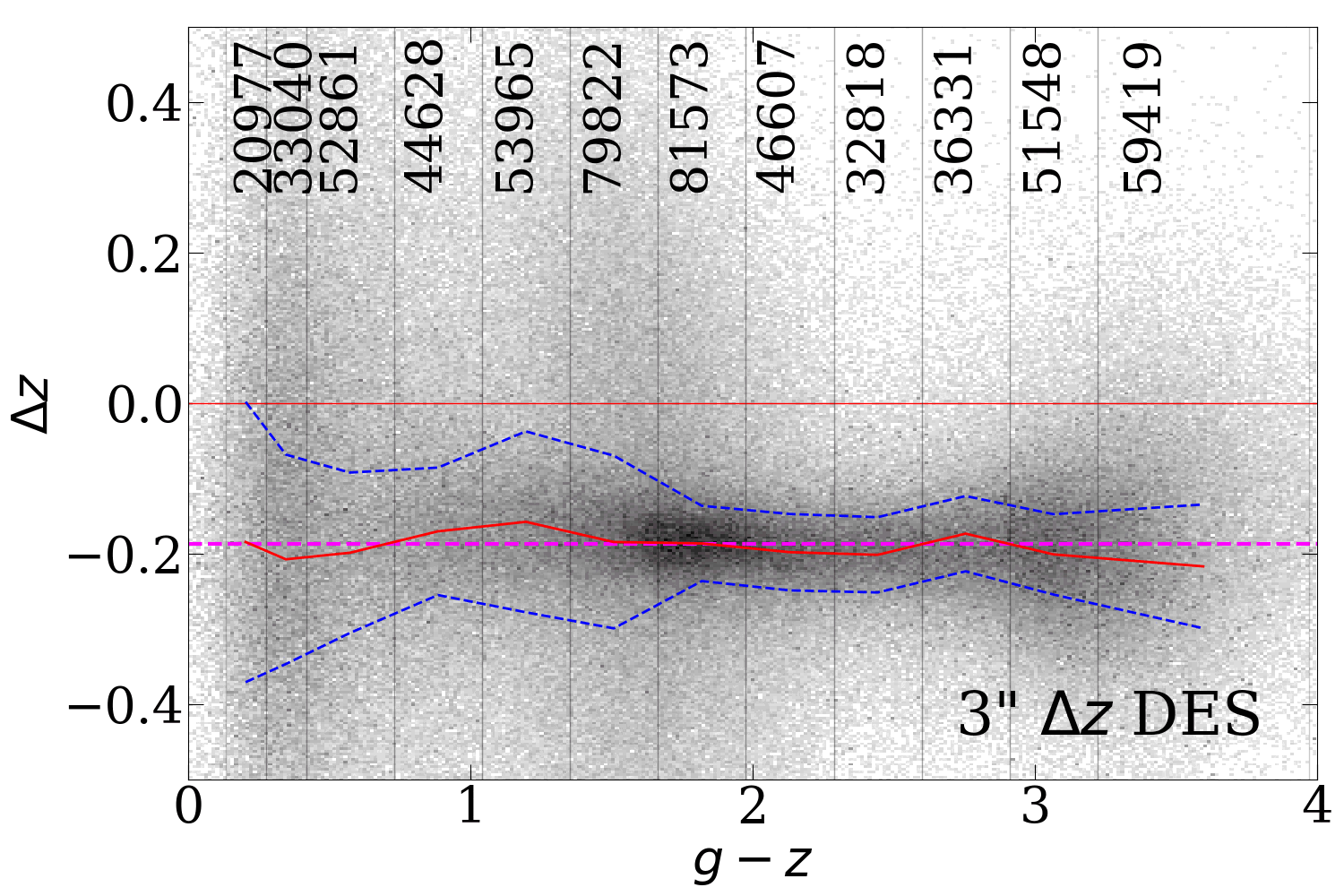}\\
\caption{Magnitude difference \revtwo{between DES corrected using our transformations for total magnitudes and SDSS system as a function of} color for 1.5$''$, 2$''$ and 3$''$ apertures. \revone{All colors are computed in the same apertures. The zero level corresponding to the offsets from Table~\ref{tab_shifts} is shown by a magenta line.}
\label{apers}}
\end{figure*}

\begin{figure*}
\centering
\includegraphics[width=0.32\hsize]{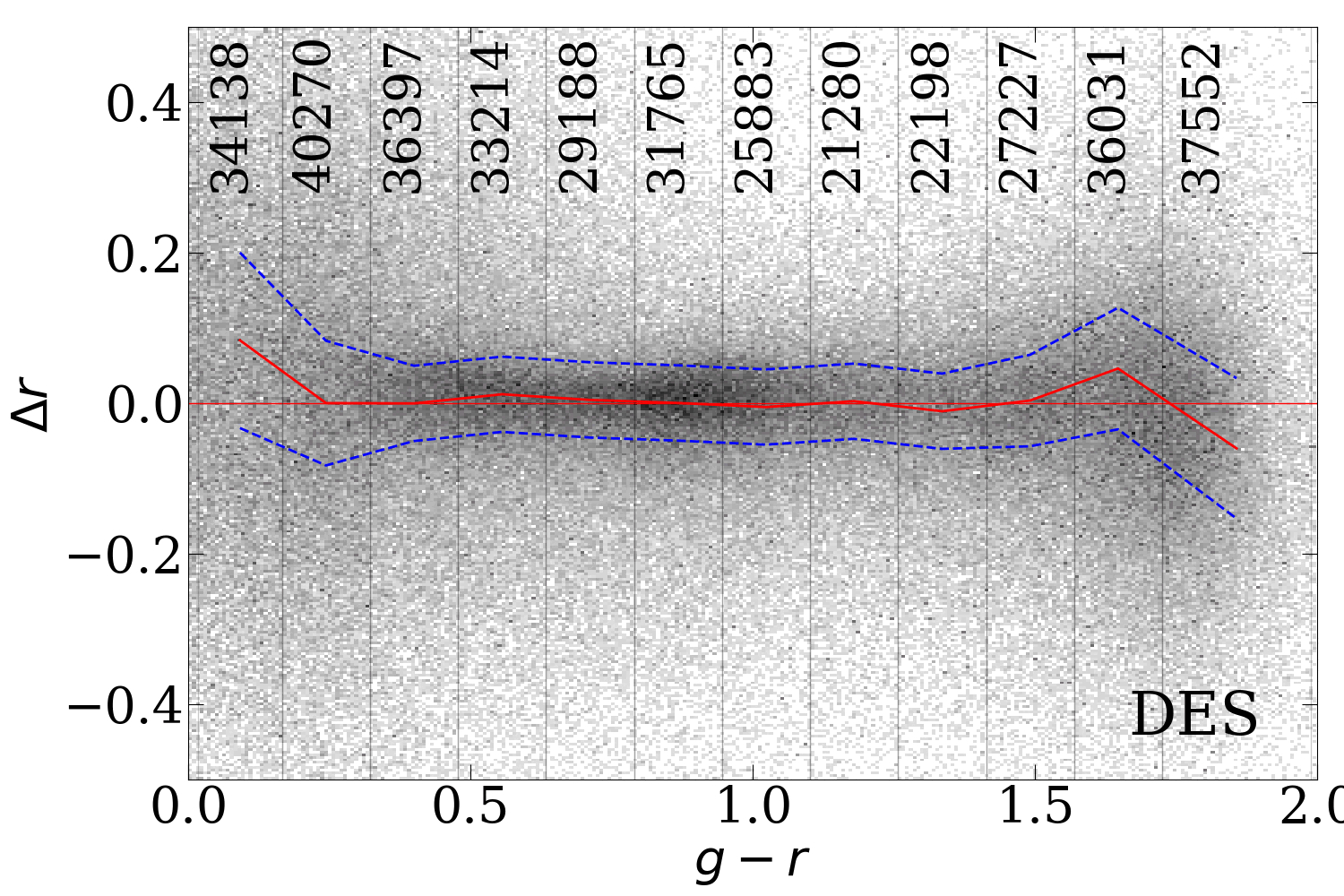}
\includegraphics[width=0.32\hsize]{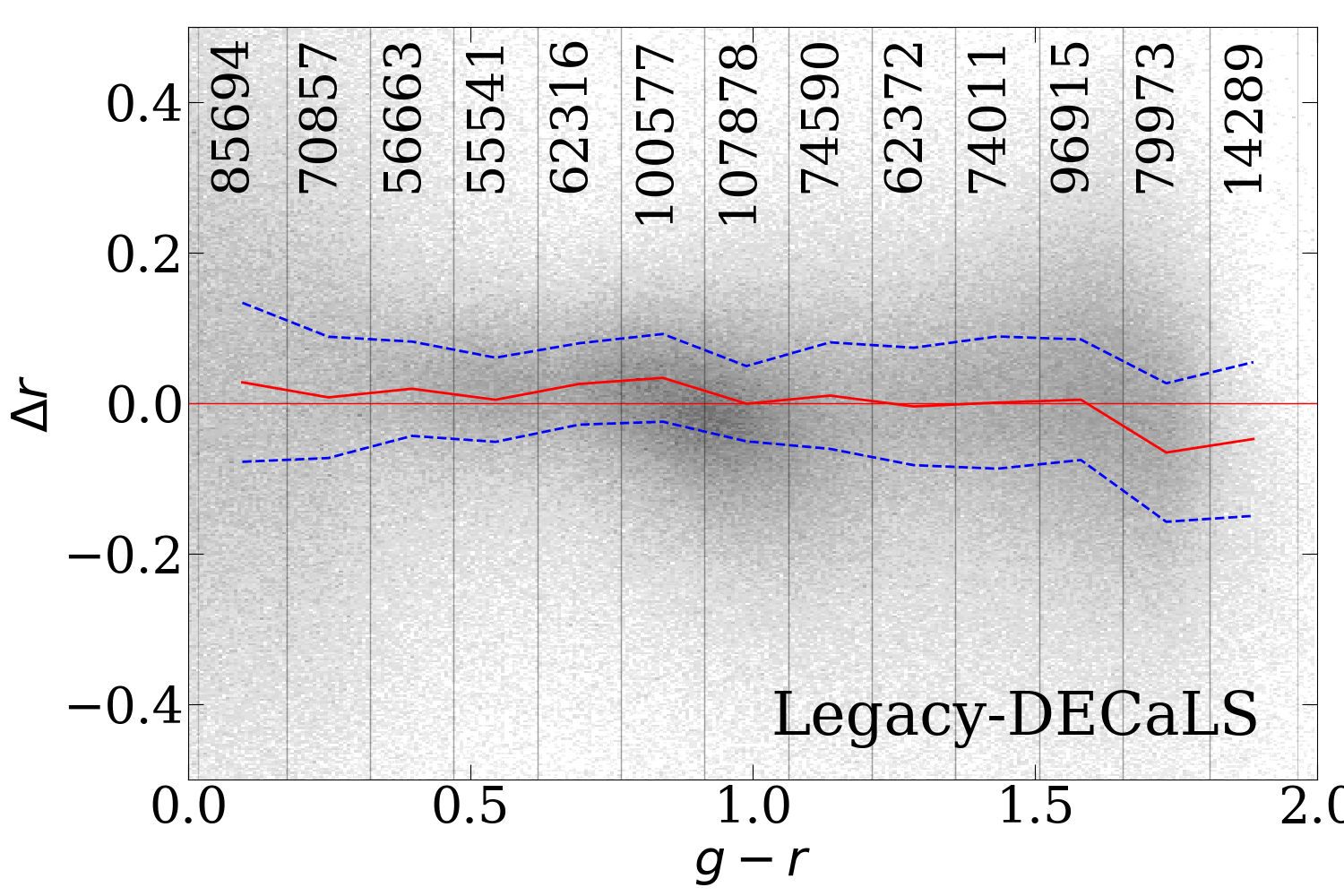}
\includegraphics[width=0.32\hsize]{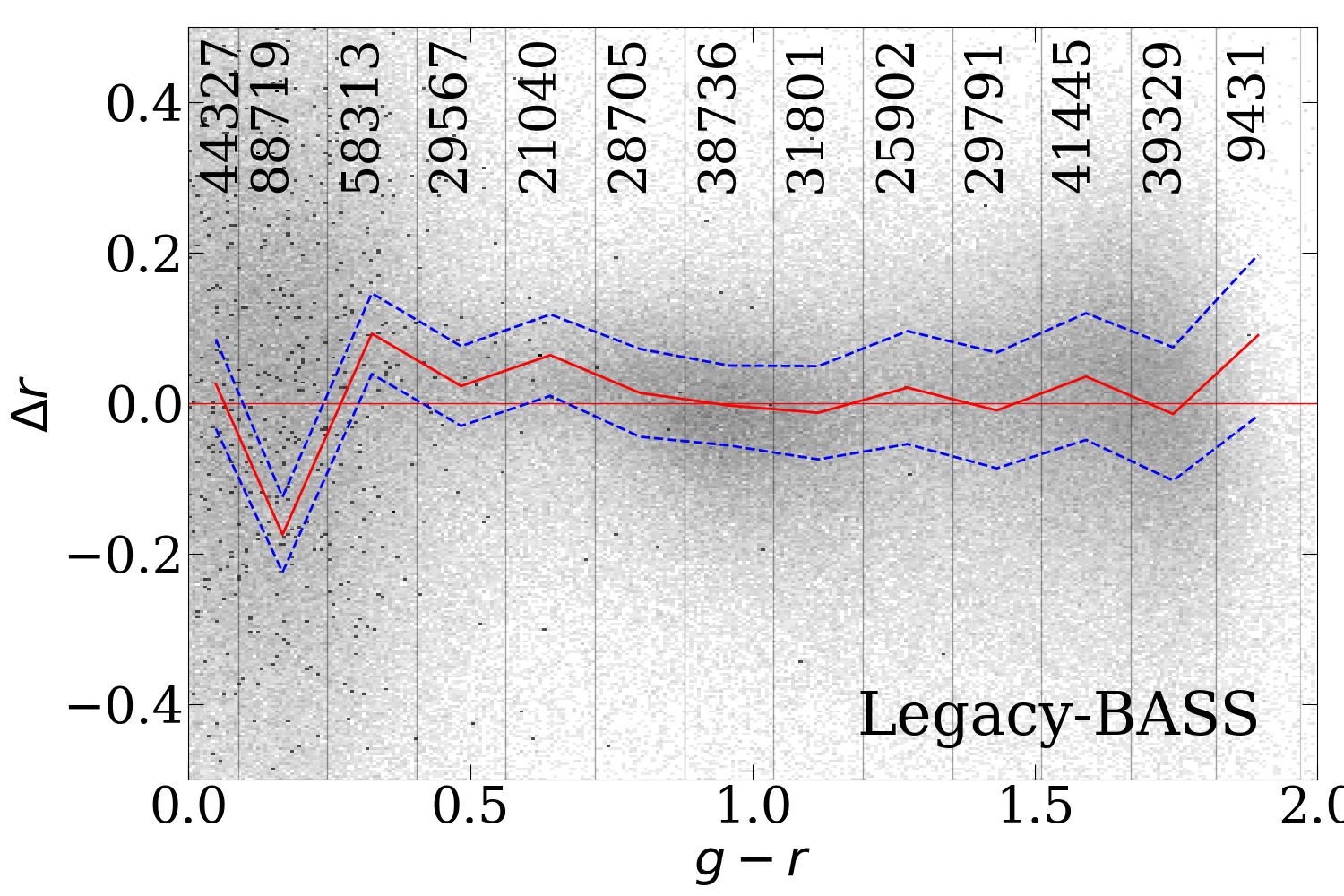}\\
\includegraphics[width=0.32\hsize]{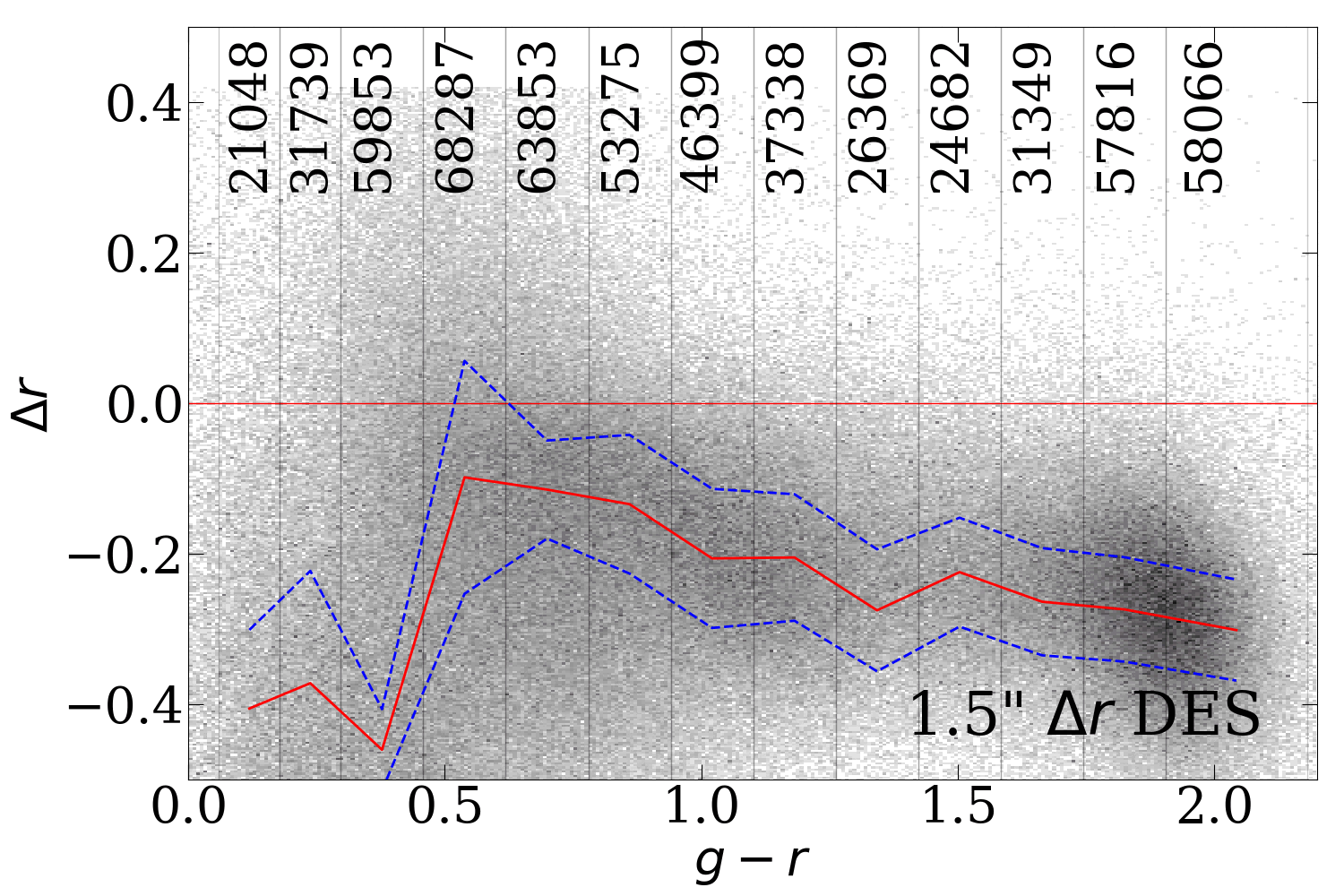}
\includegraphics[width=0.32\hsize]{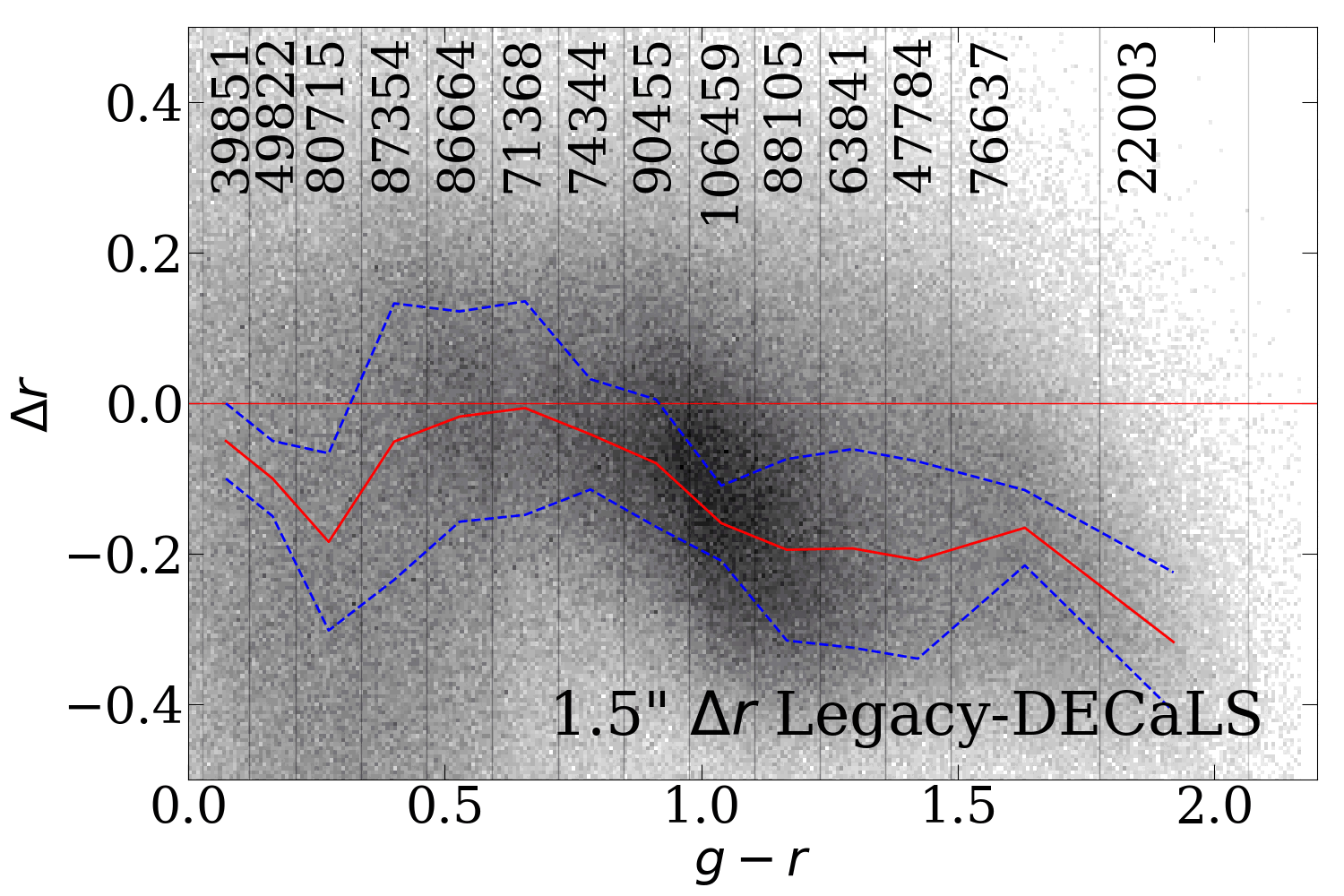}
\includegraphics[width=0.32\hsize]{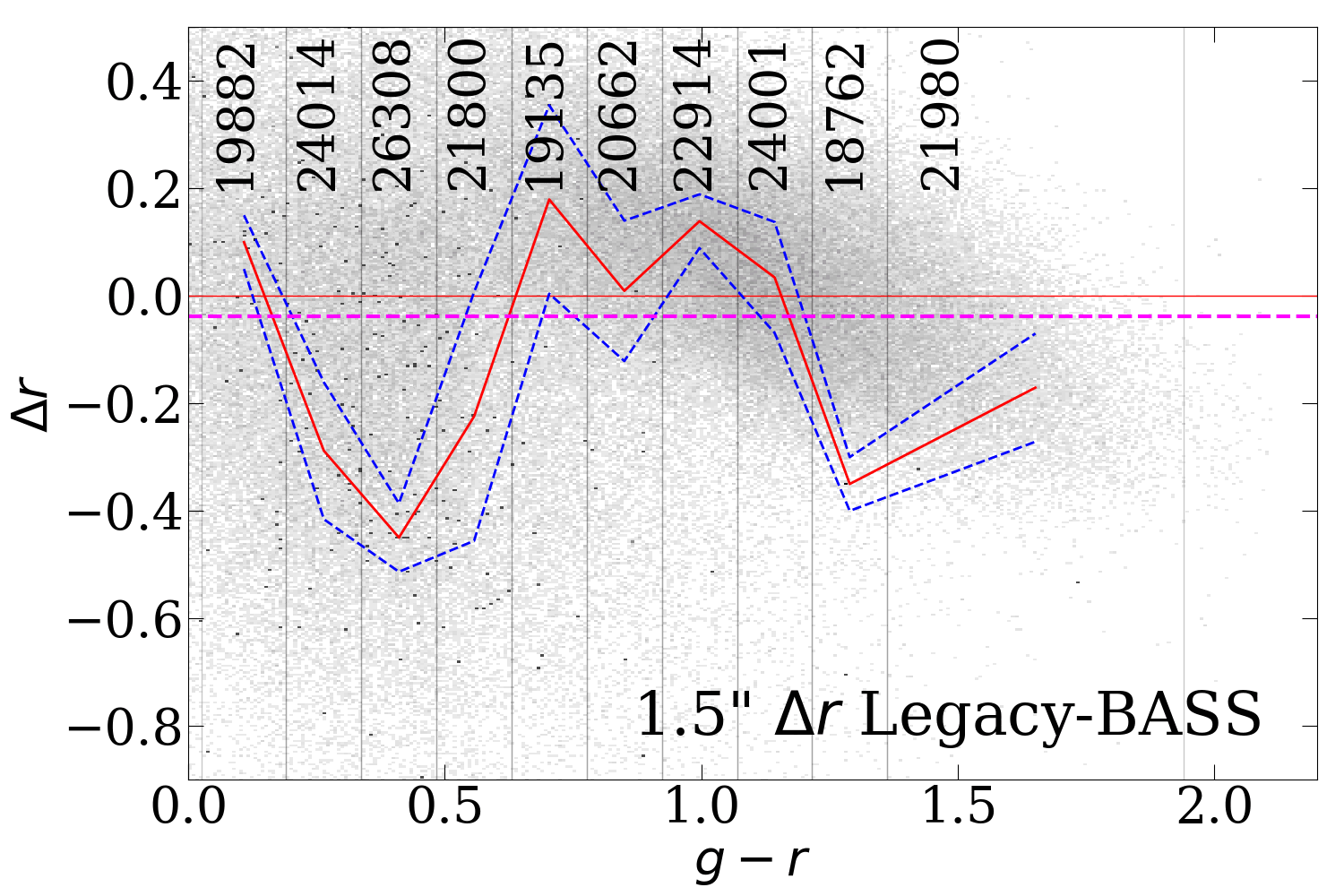}\\
\includegraphics[width=0.32\hsize]{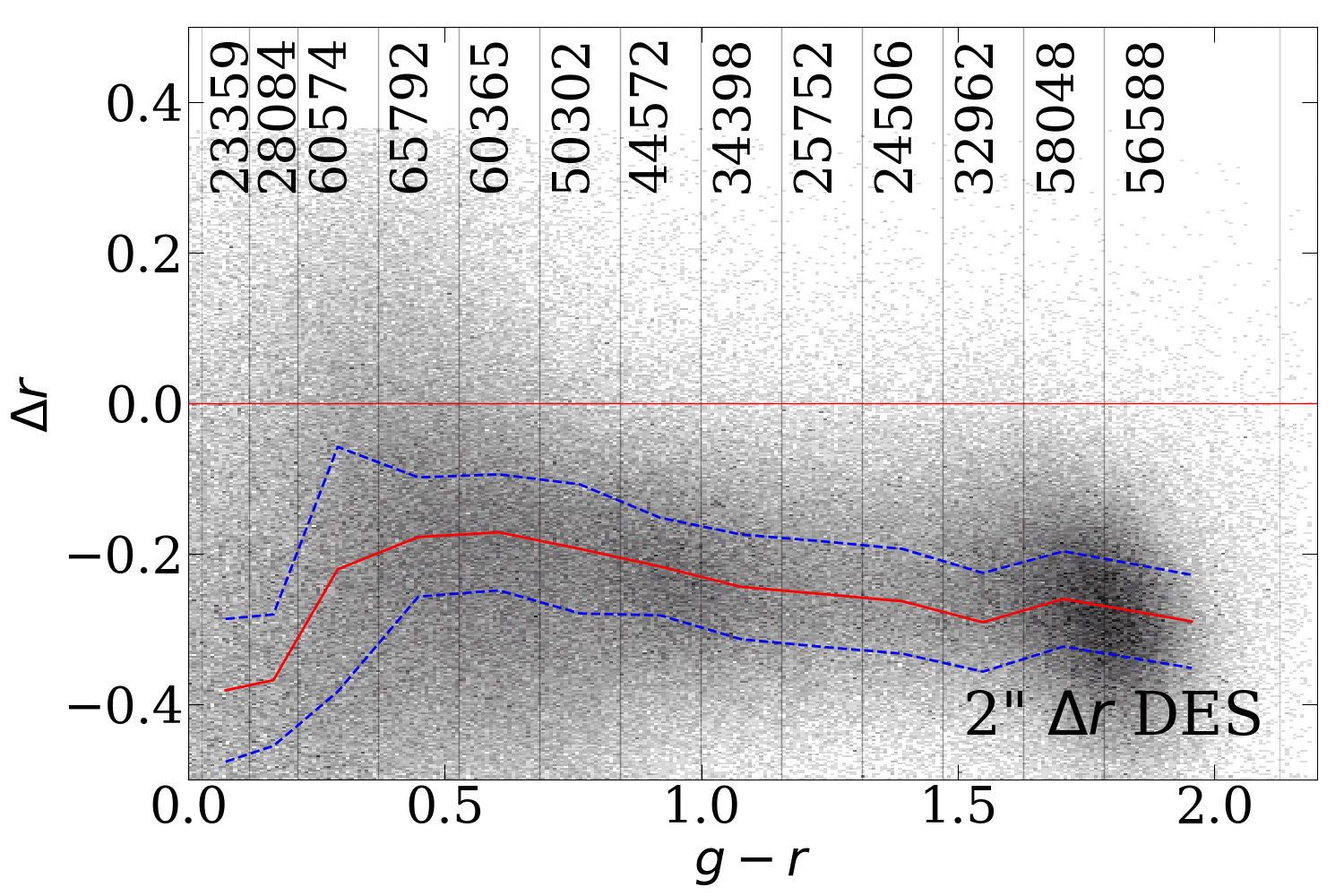}
\includegraphics[width=0.32\hsize]{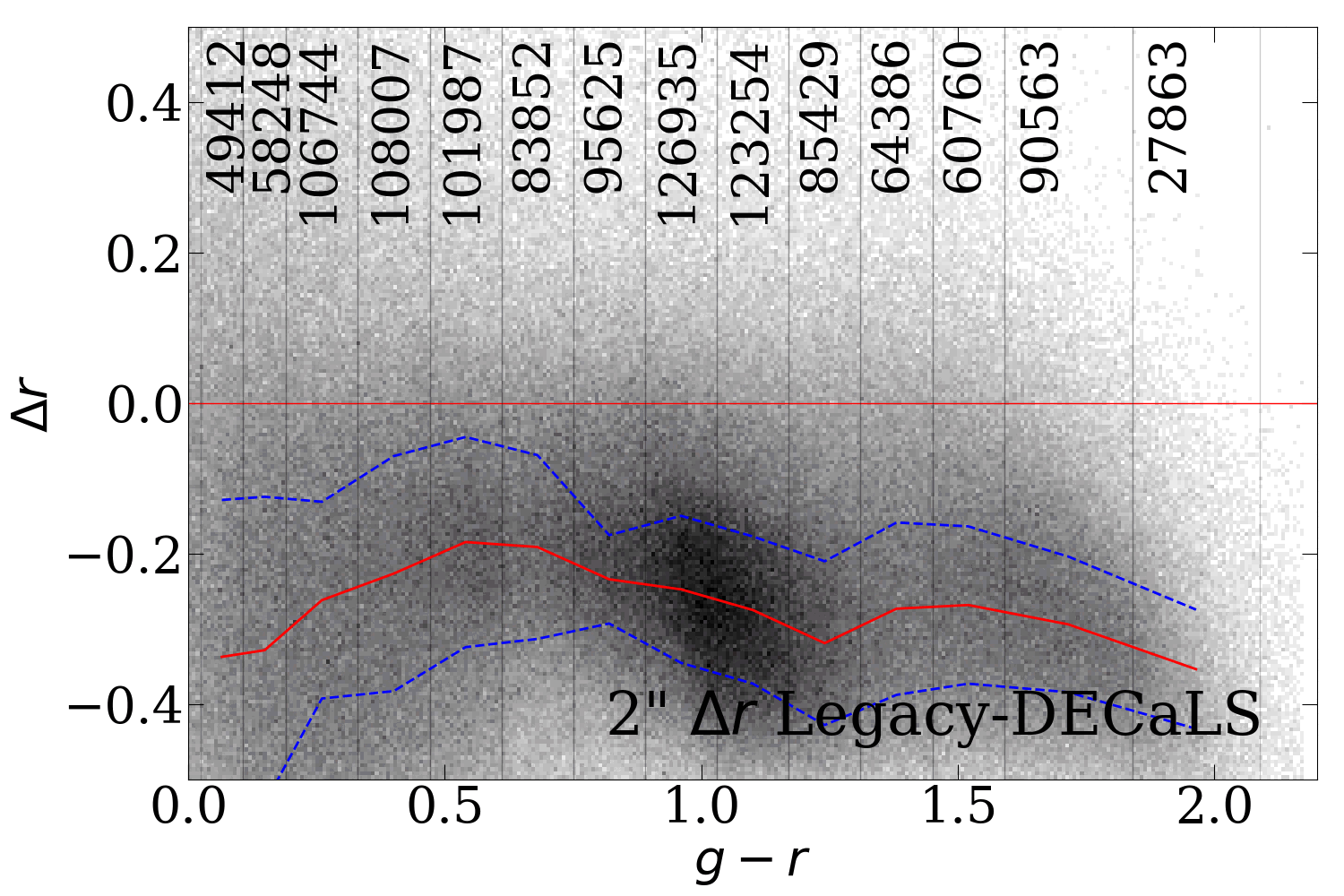}
\includegraphics[width=0.32\hsize]{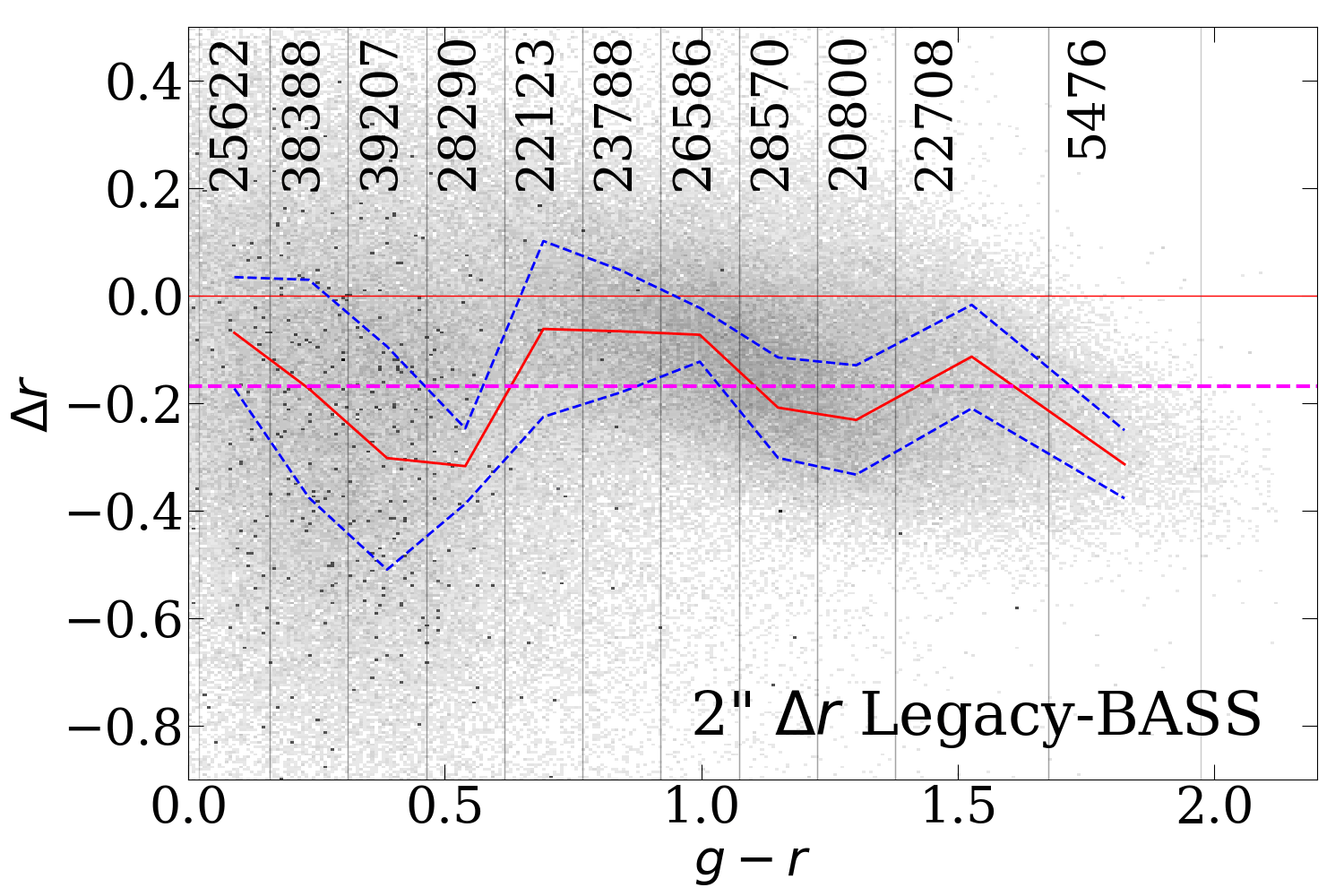}\\
\includegraphics[width=0.32\hsize]{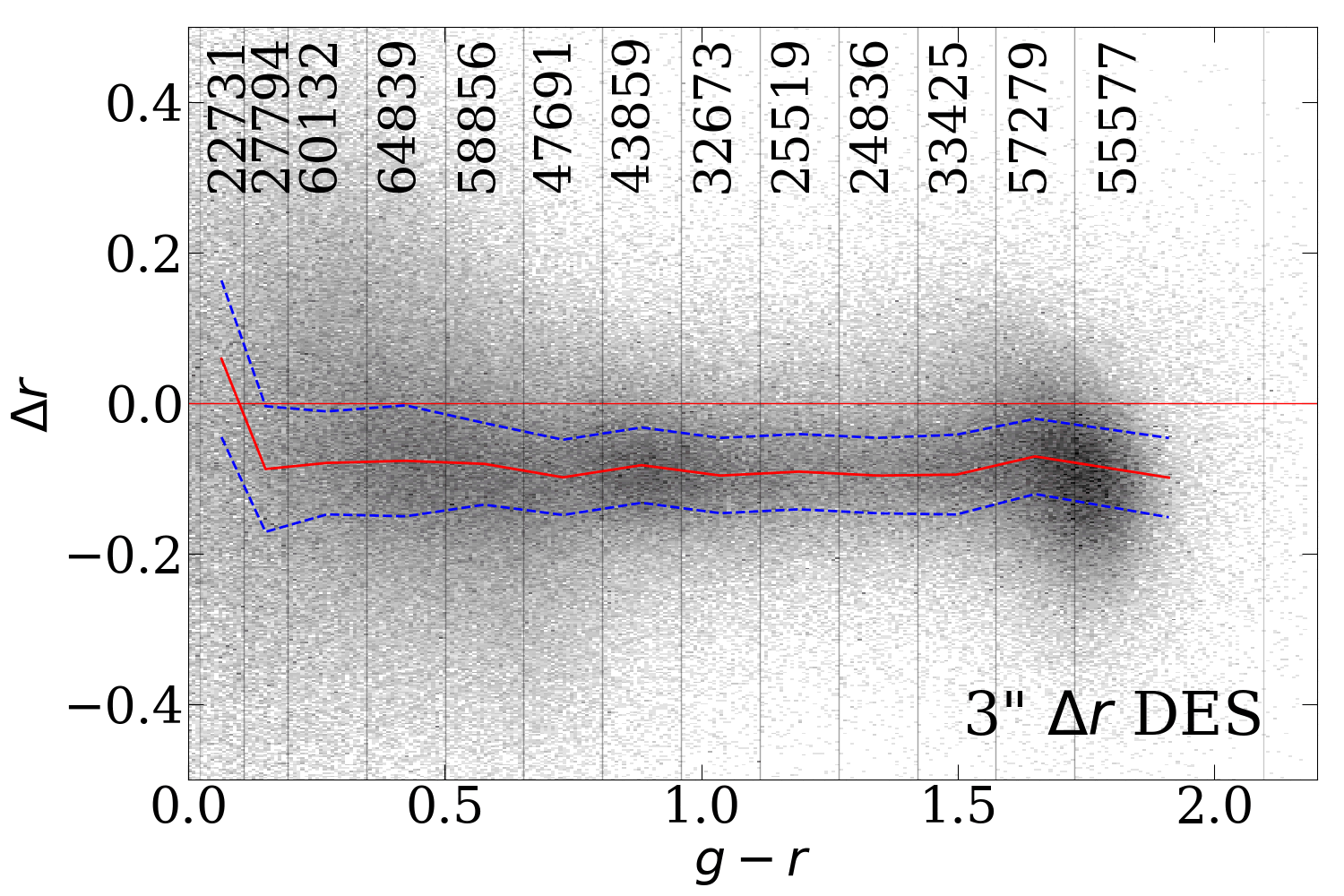}
\includegraphics[width=0.32\hsize]{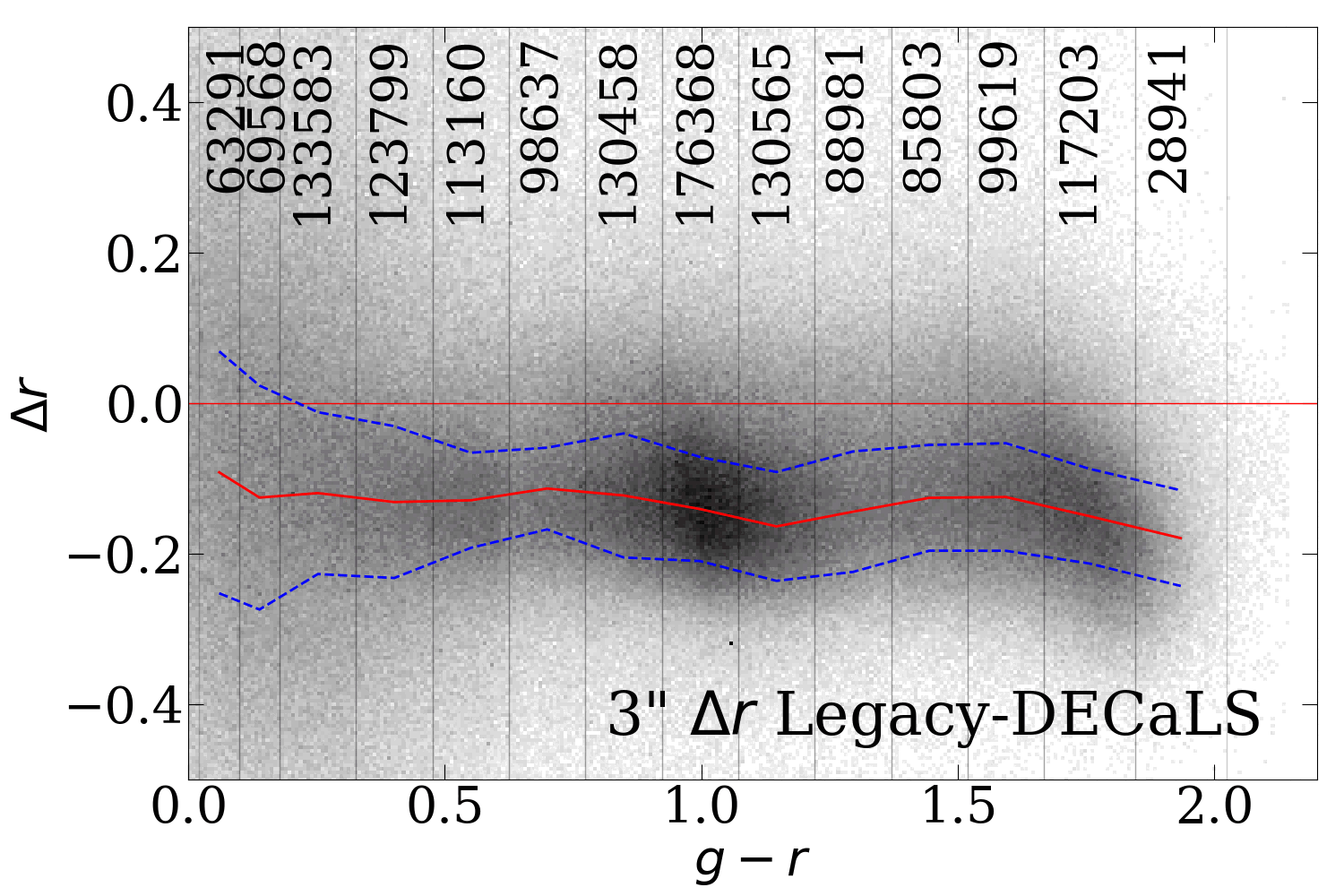}
\includegraphics[width=0.32\hsize]{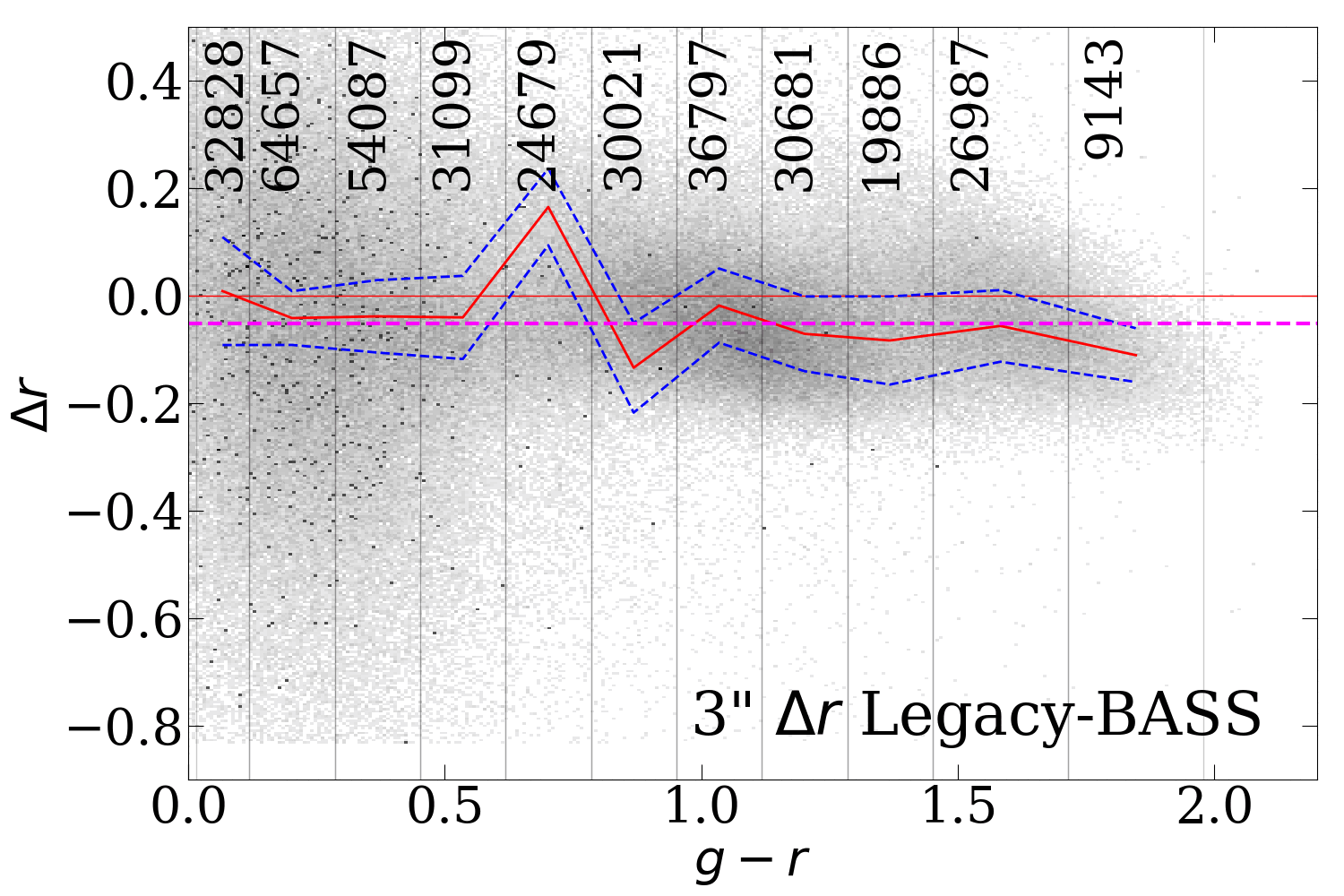}\\
\caption{
The magnitude difference \revtwo{between DES and DESI Legacy Surveys corrected using our transformations for total magnitudes and SDSS system as a function of color} for total, 1.5$''$, 2$''$ and 3$''$ apertures for $r$-band corrected with $r-z$ color. \revone{All colors are computed in the same apertures. DES and SDSS total magnitudes are Petrosian, DESI Legacy Surveys total magnitudes are model based.}
\label{rz_fit}}
\end{figure*}

\revone{As one can see in Figure~\ref{apers},~\ref{rz_fit},~\ref{apers_full1},~\ref{apers_full2}, the aperture magnitude differences for small-sized apertures ($1.''5$ and $2''$) have substantial dispersion about a mean value, it is higher for $1.''5$ than for $2''$, and the dispersion drops significantly for $3''$ values getting close to that of total magnitudes. This effect is trivially explained by the mean image quality difference among the considered surveys when measuring aperture fluxes of extended sources having a great variety of intrinsic light profiles. For SDSS the average seeing FWHM is $1.''7$ \citep{2009ApJS..182543A}, while it is close to $1''$ for DECaLS \citep{2019AJ....157..168D} and is subarcsecond for VST ATLAS \citep{2015MNRAS.451.4238S}. Therefore, the smaller the aperture size, the larger fraction of light will be lost for a compact object in a survey with a worse delivered image quality. On the other hand, for extended galaxies without much central light concentration, the seeing quality will have a much lesser effect, dropping to zero in the limiting case of a constant surface brightness object. If we look at non-variable stars, which are constant point sources, they will form very tight sequences without color dependence having the mean value, which can be calculated by integrating two-dimensional Gaussians corresponding to the typical values of seeing quality in the two surveys inside the apertures and then converting the flux ratio into a magnitude difference. Larger aperture sizes, which significantly exceed the seeing FWHM (e.g. $3''$), enclose a higher fraction of the central light concentration of a galaxy, therefore the mean offset becomes smaller and the dispersion of points around the mean values also decreases.} \revtwo{This explanation is consistent with the results shown in Figure \ref{aper5}, which demonstrates a zero difference between the VST and DES in $5.''7$ apertures computed using the transformations for total magnitudes. This is because (i) $5.''7$ is a relatively large aperture to being affected, and (ii) the VST and DES surveys have similar seeing FWHM (sub-arcsecond and arcsecond), while the SDSS has a FWHM of $1.''7$. }

\begin{table}
\caption{Mean offset values in different apertures \revtwo{used for correction of aperture magnitudes into SDSS bands (Survey value - SDSS value)}. \label{tab_shifts}} 
\begin{center}
\begin{tabular}{lcccc} 
 \hline
Aperture & \textit{g} & \textit{r} & \textit{i} & \textit{z} \\
  & mag & mag & mag  & mag \\
 \hline
 DES $1.''5\:^*$ & -0.051 & -0.239 & -0.317 & -0.308 \\
 DES $2''\:^*$ & -0.193 &-0.334 & -0.393 & -0.379 \\
 DES $3''\:^*$ & -0.100 & -0.168 & -0.200 & -0.187 \\
 \revfive{VST $1.''5\:^*$ }& -0.187 & -0.233 & -0.330 & -0.359 \\
 \revfive{VST $2''$ }& -0.209 & -0.239 & -0.312 & -0.330 \\
 \revfive{VST $3''\:^*$ }& -0.115 & -0.123 & -0.158 & -0.159 \\
DECaLS $1.''5$ & -0.055 & -0.124 & & -0.197 \\
DECaLS $2''$ & -0.190 & -0.241 & &-0.284 \\
DECaLS $3''$ & -0.078 & -0.110 & &-0.112 \\
MzLS+BASS $1.''5$ & 0.133 & -0.038 & & -0.261 \\
MzLS+BASS $2''$ & -0.040 & -0.168 & & -0.325 \\
MzLS+BASS $3''$ & 0.023 & -0.052 & & -0.127 \\
\hline 
Aperture & \textit{Y} & \textit{J} & \textit{H} & \textit{K} \\
  & mag & mag & mag  & mag \\
 \hline
VISTA $1.''5\:^*$ & 0.015 & 0.048 & 0.033 & 0.059 \\
VISTA $2''$ & 0.012 & 0.044 & 0.023 & 0.050 \\
VISTA $3''\:^*$ & 0.011 & 0.040 & 0.009 & 0.035 \\
 \hline
 \end{tabular} 
 \end{center} 
\footnotesize{\revone{$^*\:$Magnitudes were calculated using quadratic
interpolation from the available aperture magnitudes (see Table  \ref{tab_sample})}}
 \end{table}

In Figure~\ref{apers} we see a clump that consists of a large number of data points in the $g$ and $r$ bands residuals between DES and SDSS in the 3$''$ aperture at the mean color $g-r = 1.8$~mag, which lies substantially off the principal sequence formed by the rest of the galaxies in the sample. This clump includes exclusively intermediate-redshift ($z > 0.4$) luminous red galaxies (LRGs) and it disappears in the residuals in the $i$ and $z$ bands. The most prominent difference is seen in the $r$ band and we attribute it to difference among the filter throughput curves: the DES $r$ extends sufficiently further into the red ($\lambda_{r,\mathrm{cutoff,DES}}=720$~nm versus $\lambda_{r,\mathrm{cutoff,SDSS}}=670$~nm). The difference in the $g$ band has an opposite sign and a similar mean absolute offset but larger spread because the SDSS $g$-band photometry has very poor quality for LRGs because they are \revtwo{red and therefore faint in $g$ band}. As a result, the color difference $\Delta(g-r)$ in the converted colors remains very close to zero. The clump originates from the sharp increase of the number of objects in the sample at $z>0.4$ because of the LRG selection in eBOSS. They present at colors $(g-r)<1.8$~mag but in relatively small quantities, however the LRG selection kicks in at around $z=0.4$. LRGs have a very strong redshift dependence of $g-r$ and it is very sensitive to the difference between the exact filter throughput curves in SDSS and DES, and one needs a redder color, e.g. $r-z$ to properly describe the trend. For luminous red galaxies at redshifts $z>0.4$, the $g-r$ color saturates at $\sim$1.7~mag and therefore it cannot be used for parameterization of the color transformation. Therefore, for very red galaxies we recommend using the $r-z$-based transformations (see color differences for the $r-z$ parameterization in Figure~\ref{rz_fit}.)


\begin{figure*}
\centering
\includegraphics[width=1\hsize]{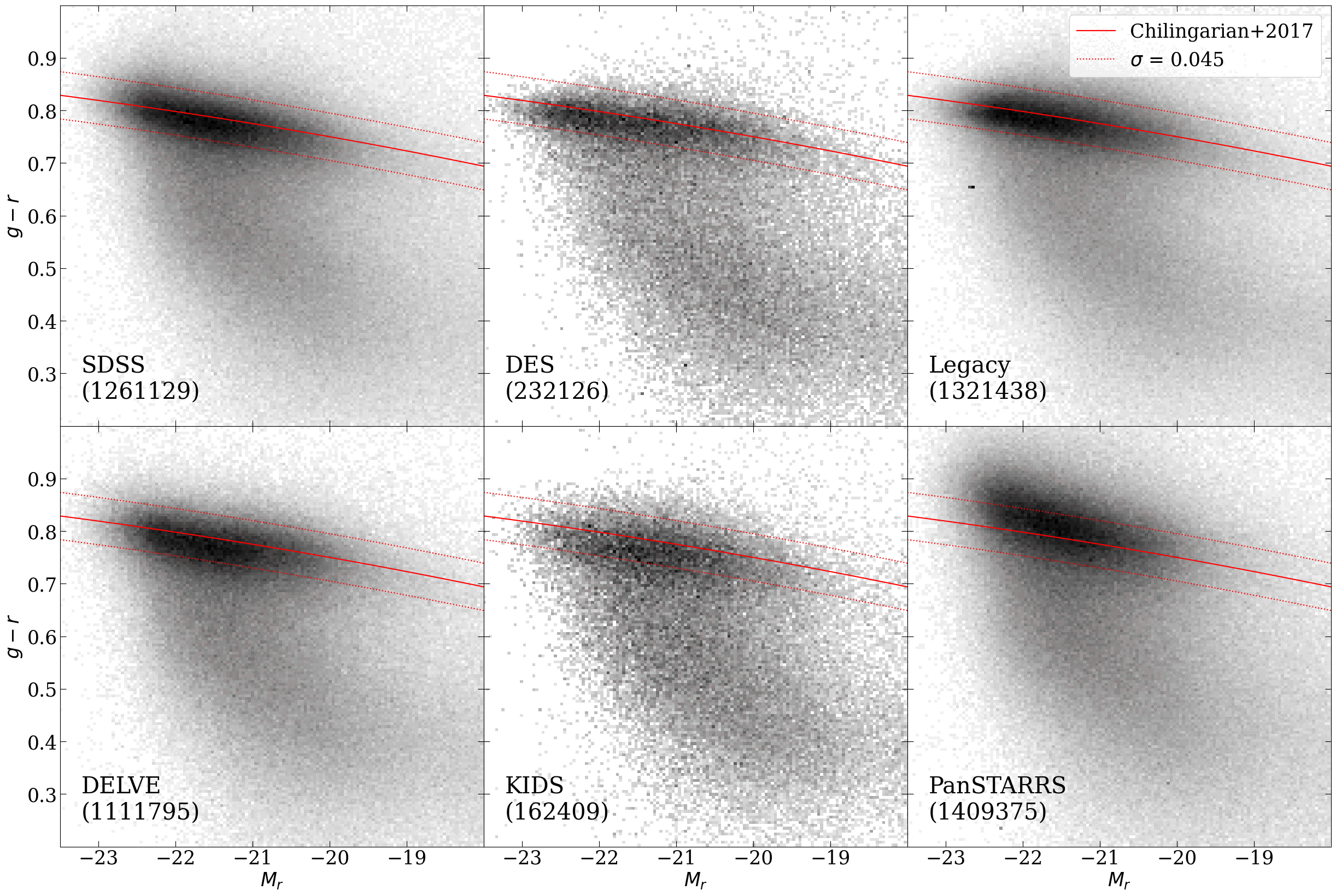}
\caption{Optical color–magnitude diagram for \revtwo{extinction and k-} corrected magnitudes from different surveys. The red line shows \revtwo{the} red sequence \revtwo{parametrization} from \citet{RCSED}. A number of objects used for each histogram is shown in the left corner of each plot.
\label{rs}}
\end{figure*}

\subsection{Red sequence in different catalogs}
\label{subsection:rs}

As we mentioned in the Introduction, the optical color--magnitude diagram constructed from integrated \revtwo{extinction and} $k$-corrected photometric measurements represents the ultimate test of the quality of the obtained photometric relations: early-type galaxies form a very narrow `red sequence' -- its position and tightness reflects the quality of the input photometric data. It is also important to know the red sequence shape in different colors for practical astrophysical purposes, such as selecting early-type candidate members in galaxy clusters using photometric data. \citet{RCSED} published the red sequence shape derived for different filter combinations in the SDSS DR7. 

In Figure~\ref{rs} we present the $(g-r)$~versus~$M_r$ color--magnitude plots the 5 surveys we used in our combined catalog transformed into the SDSS photometric system using our results and the PanSTARRS survey converted using the literature transformations. We restricted the redshift range to $0.02 < z < 0.2$. We applied $k$-corrections to the converted measurements according to \citet{2010MNRAS.405.1409C,2012MNRAS.419.1727C} \revtwo{and corrections for the Galactic extinction \citep{2011ApJ...737..103S}}. We overplot the polynomial approximation for the red sequence presented in \citet{RCSED}. It turns out that the red sequence has the same slope and location in all surveys except PanSTARRS, which validates the photometric transformations we derived in this work. The red sequence in DES, DELVE, and KiDS has a similar or smaller scatter compared to SDSS, and also a very small number of very red outliers ($>+2\sigma$ from the mean red sequence position). The distributions of galaxies in the $(g-z)$~versus~$M_z$ and $(g-i)$~versus~$M_i$ color spaces behave similarly: DES, DELVE, and KiDS show fully consistent results with the SDSS as presented in the RCSED catalog paper \citep{RCSED}.

\begin{figure*}
\centering
\includegraphics[width=0.9\hsize]{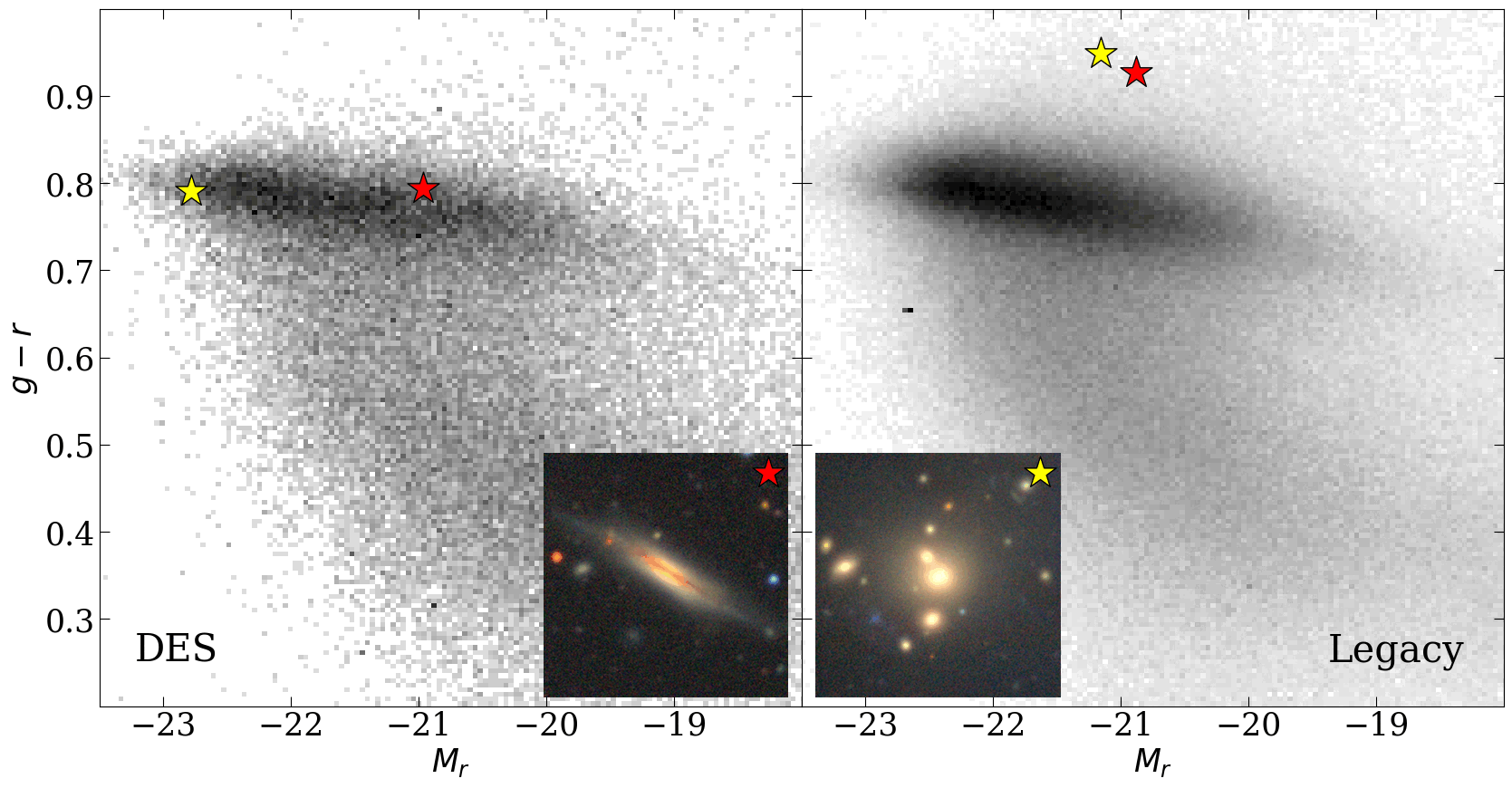}
\caption{Positions of an edge-on galaxy with a dust lane (J010252.85+050631.7, red star) and a BCG (J232653.85-524149.5, yellow star) in the center of a massive galaxy cluster. The stars symbols show their positions on the diagrams for DES (left) and DESI Legacy Surveys (right). The cutouts were \revtwo{made using data} from DESI Legacy Surveys Data Release 9. \label{rs_comp}}
\end{figure*}

The red sequence based on the DESI Legacy Surveys data, both from DECaLS and MzLS demonstrates a larger scatter than those from DES and DELVE despite being based on the data from the same facility, however, for the most luminous late-type galaxies, the scatter value is similar or even smaller compared to DES likely because of the better number statistics. There is also a substantial number of very red objects above $+2\sigma$ off the red sequence in the intermediate-luminosity range -- we see a similar behavior in the VST ATLAS data. We have inspected some of the objects off from the red sequence by $>4 \sigma$ and they all settle within 1$\sigma$ of the DES red sequence (see Figure \ref{rs_comp}). The first object \revtwo{(marked by red star)} probably has a segmentation problem related to the presence of the dust lane. The second galaxy \revtwo{(marked by yellow star)} is the central galaxy in a cluster and it has issues with deblending caused by the presence of numerous satellites in its vicinity. These examples illustrate that compared to other surveys, total magnitudes in the DESI Legacy Surveys have a higher intrinsic scatter, which we investigate in the next subsection.

The red sequence based on PanSTARRS total magnitudes converted into SDSS using the transformations from the literature is wider than that in SDSS and has the same tilt as SDSS only for small and spatially unresolved objects. It `bends up' for spatially extended galaxies, which is probably caused by imperfect sky subtraction for extended targets. The red sequence is also substantially thicker and there is a large number of red outliers.

\subsection{DESI Legacy Surveys integrated photometry}

\begin{figure}
\centering
\includegraphics[width=1\hsize]{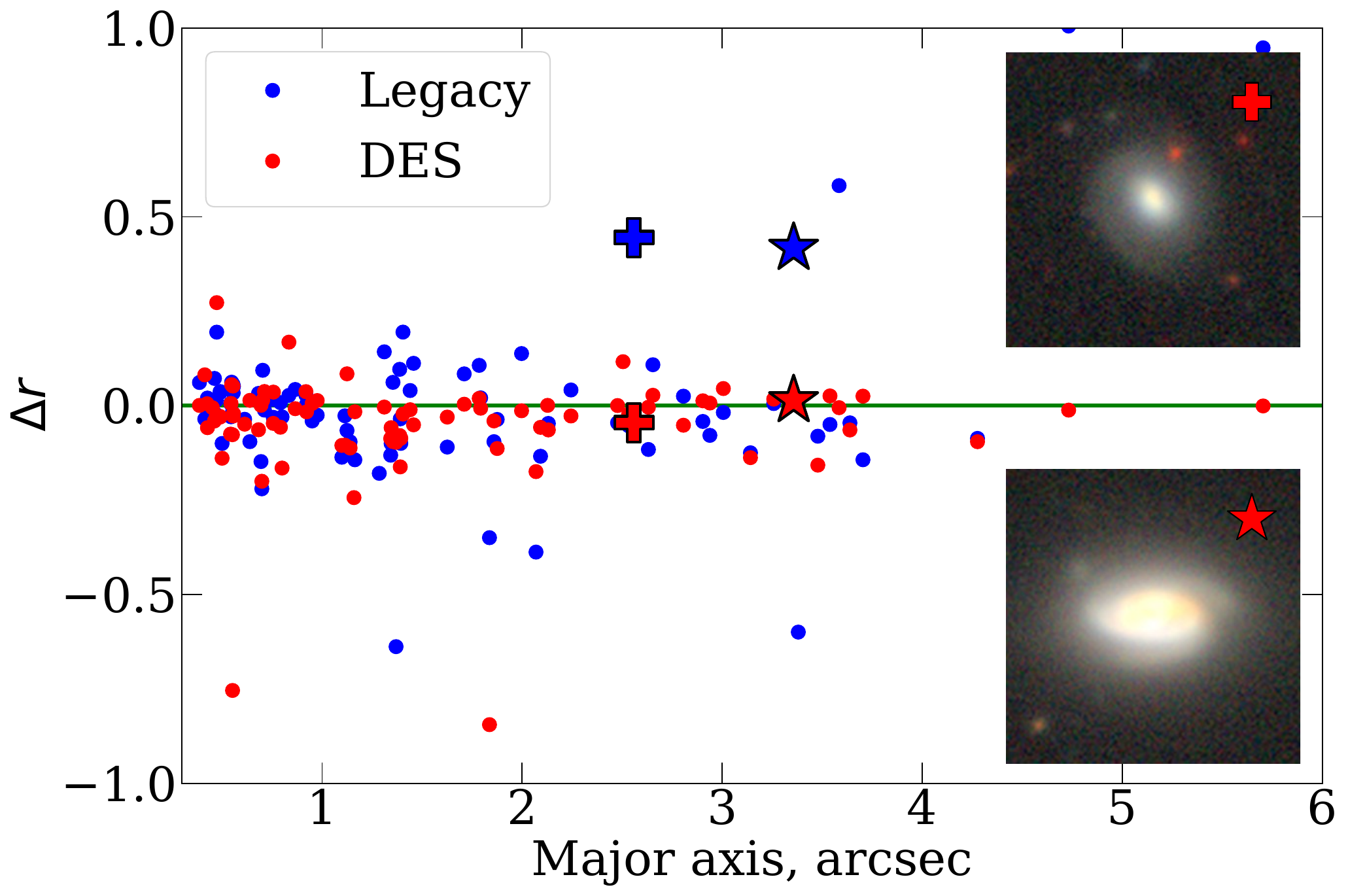}
\caption{Comparison of the total magnitude measurements from DES (red) and DECaLS (blue) catalogs with those from {\sc Source Extractor} output for the same region of the sky. We \revtwo{plot} the difference between total magnitudes in the catalog and in the {\sc Source Extractor} output as a function of Petrosian radius. 
\label{abell}}
\end{figure}

While computing photometric transformation for different imaging surveys carried out with the DECam instrument we noted that while DES-based transformations yielded excellent results for DELVE, they \revone{do not work properly} for DESI Legacy Surveys: the residuals displayed strong systematics with a nonlinear color dependence. \revone{This was the reason why we used other transformations for DESI Legacy Surveys, which work well for it as shown at Figures \ref{mag_resid} and \ref{color_resid}.} DELVE and DES used the same data processing and photometric calibration tools while DESI Legacy Surveys used totally different solutions. 


There are two possible reasons for this behavior: (i) difference in the pipeline processing of raw images, e.g. a sky subtraction algorithm clipping outer parts of extended targets in one of the pipelines; (ii) difference in the source extraction and deblending algorithms. To identify the primary source of inconsistencies, we decided first to test the source extraction. DES and DELVE used \textsc{Source Extractor} \citep{1996A&AS..117..393B} while DECaLS used \textsc{tractor} \citep{2016ascl.soft04008L} implemented in {\sc python}. To perform the test, we downloaded a $16'\times16'$ region of the sky in the $r$-band centered on the galaxy cluster A~168 from DECaLS using Astro Data Lab. Galaxy clusters are convenient for such a comparison because they include a large number of galaxies at the same redshift, \revtwo{many} of which \revtwo{reside on} the red sequence and have regular elliptical morphology, which simplifies source extraction and measurement. Then we ran \textsc{Source Extractor} with a default configuration on this image and compared the results with those presented in DES for about 100 sources. In Figure~\ref{abell} we present the comparsion of {\sc Source Extractor}-based DECaLS photometry and the published values derived by {\sc tractor} against DES. It turns out that using {\sc Source Extractor} on DECaLS images yields fully consistent results with DES with a zero mean deviation and an r.m.s of 0.044~mag compared to 0.075~mag for DECaLS photometry. We also see fewer `catastrophic outliers' especially among extended galaxies. In the insets of Figure~\ref{abell} we present two examples of such outliers. Both seem to have problems with deblending: one galaxy has a red star projected on its outer part, which was not separated from it by {\sc tractor} and caused the unphysically red color of this object in DECaLS; the second galaxy was not separated from a satellite in DECaLS. The increased r.m.s. is probably due to a less stable algorithm used in {\sc tractor} to define a surface brightness threshold for extended sources.

This test provides an explanation why the color--magnitude plot from DESI Legacy Surveys shows a puffed-up red sequence and a high number of red outliers (see the discussion in \ref{subsection:rs}). We also see a way to improve extended object photometry in DECaLS: the images from DECaLS should be analyzed using {\sc Source Extractor} to bring the photometry to the consistency with DES and DELVE.

%

\subsection{Photometry for CfA public spectroscopic datasets}

As an example use case for our color transformation, we compiled integrated photometric measurements in the optical and NIR bands for the galaxy samples from the two CfA public archives for FAST (73,849 spectra for 42,694 objects) and Hectospec (211,075 spectra for 164,439 objects). Both datasets originate from data archives and cover a very wide area that extends beyond the SDSS footprint, so that a combination of SDSS, DELVE, DES, and BASS/MzLS was needed to compile an optical photometric catalog.

The CfA FAST archive \citep{2021AJ....161....3M} contains 141,531 spectra for 72,247 objects and include observations of nearby galaxies, supernovae, various types of stars, and planetary nebulae collected between 1994 and 2021 using the FAST spectrograph operated at the 1.5 m Tillinghast telescope at the Fred Lawrence Whipple Observatory. The spectra were automatically processed with the FAST data reduction pipeline. We kept 42,694 objects which we identified as extragalactic targets using either their redshifts or a cross-identification in Hyperleda and for which we could find photometric measurements in wide-field surveys.

The CfA Hectospec archive includes the results of spectroscopic observations from the multi-fiber spectrograph Hectospec operated at the 6.5 m MMT \citep{2005PASP..117.1411F}. These are the results of publicly released programs typically older than 5~yr processed with the Hectospec pipeline. The sample includes observations of low-to-intermediate redshift galaxies, extragalactic star clusters, bright stars, H~{\sc ii} regions, and planetary nebulae in nearby galaxies as well as some distant quasars. We kept 164,439 distinct objects, which we identified as galaxies with available photometric measurements and excluded extragalactic star clusters and planetary nebulae as well as Galactic and Local group stars.

\begin{figure*}
\centering
\includegraphics[width=1\hsize]{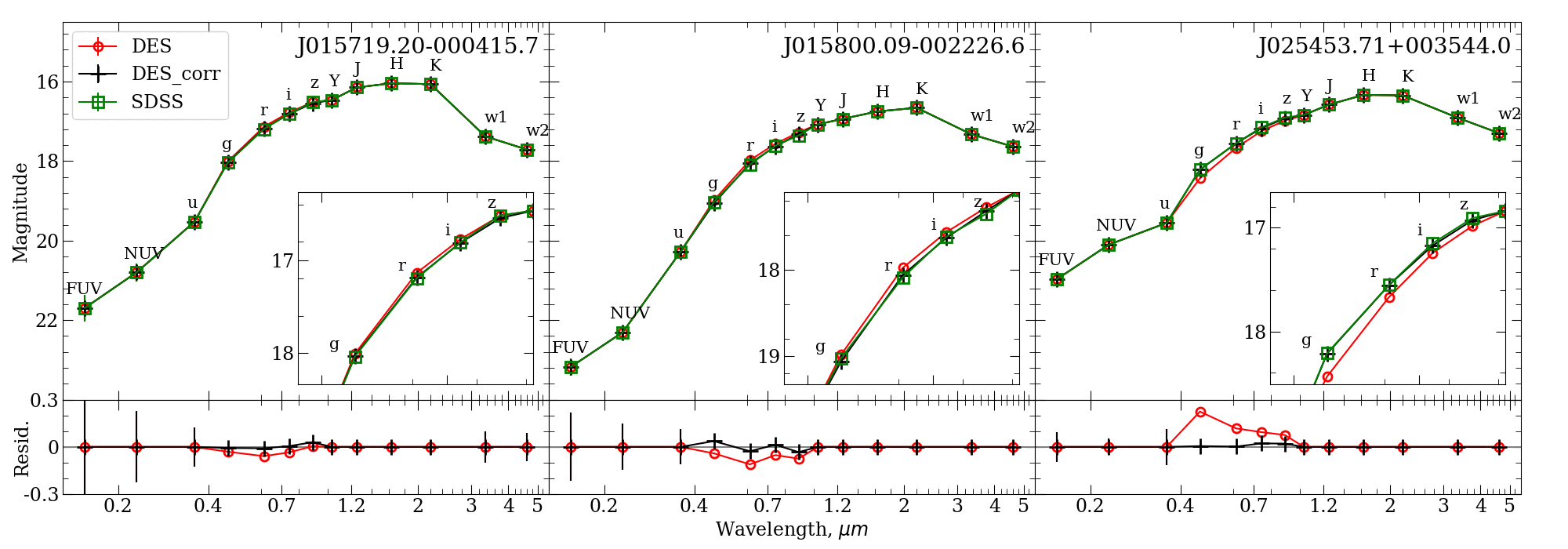}
\caption{Spectral energy distributions for three galaxies from the CfA Hectospec archive using SDSS, uncorrected and converted DES optical photometry. The data in the FUV and NUV bands are from GALEX DR7, \textit{YJHK} are from UKIDSS, \textit{w1} and \textit{w2} are from unWISE. Lower panels: difference between SDSS and original (red) or converted (red) DES.\label{sed}}
\end{figure*}

To compile a photometric catalog, we converted the Petrosian magnitudes from DES, DELVE, and BASS/MzLS, VHS, and VIKING into the SDSS and UKIDSS photometric systems, and corrected them for the foreground extinction. We complemented these measurements with extinction-corrected GALEX and unWISE photometry. Some examples of multi-wavelength SEDs for Hectospec galaxies are shown in Figure~\ref{sed}: we demonstrate the differences of the magnitudes from the original SDSS measurements before (red) and after (black) applying our color transformations. The converted DES measurements show excellent agreement with SDSS. Our photometric catalog will be included into the CfA data archive in the form of additional properties for each extragalactic object.

\begin{figure*}
\centering
\includegraphics[width=1\hsize]{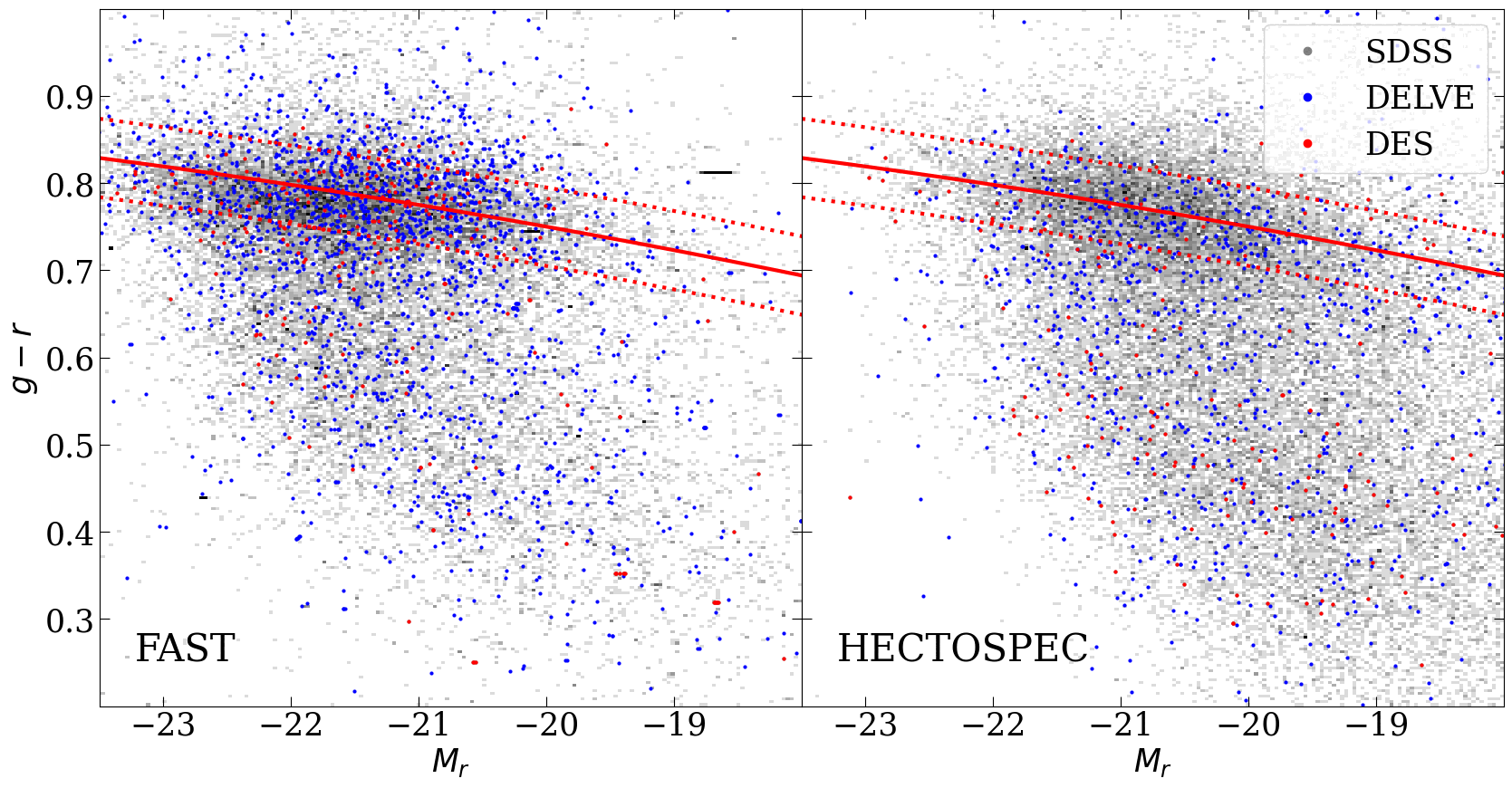}
\caption{\revtwo{Extinction and} $k$-corrected color–-magnitude diagrams for FAST (left) and Hectospec (right) galaxies. Photometric measurements taken from SDSS are shown by a grey 2D-histogram, DELVE and DES photometry are shown by blue and red dots respectively. \label{rs_comp_fast}}
\end{figure*}

As an additional quality check, we computed $k$-corrections for the converted optical magnitudes using prescriptions from \citet{2012MNRAS.419.1727C} and built fully corrected color-magnitude diagrams for FAST and Hectospec galaxies, which we show in Figure~\ref{rs_comp_fast}. The red sequence position and tightness based on DES and DELVE data perfectly match those from SDSS published by the RCSED project \citep{RCSED}.


\subsection{On the quality of integrated photometry of galaxies in optical surveys}

By applying photometric transformations derived in this work to the data originating from various surveys and comparing them against SDSS and each other, and also quantifying the properties of the color-magnitude diagrams (e.g. red sequence location and tightness), we can make the following assessment of the quality of integrated photometric measurements of galaxies:
\begin{itemize}
    \item DES provides the highest quality of optical photometry of galaxies compared to all other wide-field surveys and can be used as a `gold standard' for many scientific applications and future projects, e.g. to check the quality of photometric calibration from the forthcoming Rubin Legacy Survey of Space and Time. It also allows for refined photometric tasks, e.g. color-based search for rare galaxies \citep{CZ15,2021NatAs...5.1308G} or simultaneous fitting of spectral and photometric information \citep{Nburstsphot} to infer detailed star formation histories of galaxies using complex parametric stellar population models \citep[see e.g.][]{2019arXiv190913460G}.
    \item DELVE is a shallower equivalent of DES with similar photometric quality and low systematic errors even though higher statistical uncertainties because of the lower depth. Its main advantage is a very wide sky coverage of 17,000~deg$^2$ vs. 5000~deg$^2$ in DES.
    \item VST ATLAS and KiDS conducted at the 2.6 m ESO VST also have minimal systematic photometric errors and are complementary to DES and DELVE. In addition, they supply $u$-band photometry not available in the surveys done with DECam providing a Southern hemisphere to SDSS.
    \item Both parts of the DESI Legacy Surveys, DECaLS and BASS/MzLS exhibit significant systematics in the extended source photometry, which exceeds statistical uncertainties by a factor of a few. Nevertheless, the photometric measurements can be adequately transformed into the SDSS photometric system and can be used for certain applications of the photometric/SED-based selection even though the photometric uncertainties in any type of an SED fitting using DESI Legacy Surveys data should be artificially increased to account for the systematics. We attribute it mostly to the source extraction algorithm, so that a complete re-processing of DECaLS data using a different software package (e.g. {\sc Source Extractor}) should drastically improve the quality of photometry \revtwo{of galaxies}.
    \item PanSTARRS photometry deviates from SDSS nonlinearly with color, and the deviations in all bands show significant trends with the galaxy size. This is noticeable not only by the dependence between residuals and sizes, but also by the red sequence -- it has a correct slope and is tight only for spatially unresolved objects. Further investigations are needed to fully understand the sources of this systematics, but it will likely be improved when using a different source extraction software \citep[see e.g.][for an application of {\sc Source Extractor} to PanSTARRS data]{2022MNRAS.511.3063M}.
\end{itemize}

\section*{Acknowledgements}

V.T., I.C. and K.G. acknowledge the RScF grant No. 19-12-00281 for supporting the analysis of the photometric data and the Interdisciplinary Scientific and Educational School of Moscow University ``Fundamental and Applied Space Research.'' I.K. acknowledges the RScF grant No. 21-72-00036 for supporting the development of the web-site for online transformations of photometric measurements. IC's research is supported by the Telescope Data Center at Smithsonian Astrophysical Observatory. This research has made use of TOPCAT, developed by Mark Taylor at the University of Bristol; Aladin, developed by the Centre de Données Astronomiques de Strasbourg (CDS).

Funding for the SDSS and SDSS-II has been provided by the Alfred P. Sloan Foundation, the Participating Institutions, the National Science Foundation, the U.S. Department of Energy, the National Aeronautics and Space Administration, the Japanese Monbukagakusho, the Max Planck Society, and the Higher Education Funding Council for England. The SDSS Web Site is \href{http://www.sdss.org/}.

The DESI Legacy Surveys consist of three individual and complementary projects: the Dark Energy Camera Legacy Survey (DECaLS; Proposal ID \#2014B-0404; PIs: David Schlegel and Arjun Dey), the Beijing-Arizona Sky Survey (BASS; NOAO Prop. ID \#2015A-0801; PIs: Zhou Xu and Xiaohui Fan), and the Mayall z-band Legacy Survey (MzLS; Prop. ID \#2016A-0453; PI: Arjun Dey). DECaLS, BASS and MzLS together include data obtained, respectively, at the Blanco telescope, Cerro Tololo Inter-American Observatory, NSF’s NOIRLab; the Bok telescope, Steward Observatory, University of Arizona; and the Mayall telescope, Kitt Peak National Observatory, NOIRLab. Pipeline processing and analyses of the data were supported by NOIRLab and the Lawrence Berkeley National Laboratory (LBNL). The DESI Legacy Surveys project is honored to be permitted to conduct astronomical research on Iolkam Du’ag (Kitt Peak), a mountain with particular significance to the Tohono O’odham Nation.

NOIRLab is operated by the Association of Universities for Research in Astronomy (AURA) under a cooperative agreement with the National Science Foundation. LBNL is managed by the Regents of the University of California under contract to the U.S. Department of Energy.

This project used data obtained with the Dark Energy Camera (DECam), which was constructed by the Dark Energy Survey (DES) collaboration. Funding for the DES Projects has been provided by the U.S. Department of Energy, the U.S. National Science Foundation, the Ministry of Science and Education of Spain, the Science and Technology Facilities Council of the United Kingdom, the Higher Education Funding Council for England, the National Center for Supercomputing Applications at the University of Illinois at Urbana-Champaign, the Kavli Institute of Cosmological Physics at the University of Chicago, Center for Cosmology and Astro-Particle Physics at the Ohio State University, the Mitchell Institute for Fundamental Physics and Astronomy at Texas A\&M University, Financiadora de Estudos e Projetos, Fundacao Carlos Chagas Filho de Amparo, Financiadora de Estudos e Projetos, Fundacao Carlos Chagas Filho de Amparo a Pesquisa do Estado do Rio de Janeiro, Conselho Nacional de Desenvolvimento Cientifico e Tecnologico and the Ministerio da Ciencia, Tecnologia e Inovacao, the Deutsche Forschungsgemeinschaft and the Collaborating Institutions in the Dark Energy Survey. The Collaborating Institutions are Argonne National Laboratory, the University of California at Santa Cruz, the University of Cambridge, Centro de Investigaciones Energeticas, Medioambientales y Tecnologicas-Madrid, the University of Chicago, University College London, the DES-Brazil Consortium, the University of Edinburgh, the Eidgenossische Technische Hochschule (ETH) Zurich, Fermi National Accelerator Laboratory, the University of Illinois at Urbana-Champaign, the Institut de Ciencies de l\'Espai (IEEC/CSIC), the Institut de Fisica d\'Altes Energies, Lawrence Berkeley National Laboratory, the Ludwig Maximilians Universitat Munchen and the associated Excellence Cluster Universe, the University of Michigan, NSF\'s NOIRLab, the University of Nottingham, the Ohio State University, the University of Pennsylvania, the University of Portsmouth, SLAC National Accelerator Laboratory, Stanford University, the University of Sussex, and Texas A\&M University.

BASS is a key project of the Telescope Access Program (TAP), which has been funded by the National Astronomical Observatories of China, the Chinese Academy of Sciences (the Strategic Priority Research Program “The Emergence of Cosmological Structures” Grant \# XDB09000000), and the Special Fund for Astronomy from the Ministry of Finance. The BASS is also supported by the External Cooperation Program of Chinese Academy of Sciences (Grant \# 114A11KYSB20160057), and Chinese National Natural Science Foundation (Grant \# 12120101003, \# 11433005).

The DESI Legacy Surveys team makes use of data products from the Near-Earth Object Wide-field Infrared Survey Explorer (NEOWISE), which is a project of the Jet Propulsion Laboratory/California Institute of Technology. NEOWISE is funded by the National Aeronautics and Space Administration.

The Legacy Surveys imaging of the DESI footprint is supported by the Director, Office of Science, Office of High Energy Physics of the U.S. Department of Energy under Contract No. DE-AC02-05CH1123, by the National Energy Research Scientific Computing Center, a DOE Office of Science User Facility under the same contract; and by the U.S. National Science Foundation, Division of Astronomical Sciences under Contract No. AST-0950945 to NOAO.

Based on data products from observations made with ESO Telescopes at the La Silla Paranal Observatory under program IDs 177.A-3016, 177.A-3017 and 177.A-3018, and on data products produced by Target/OmegaCEN, INAF-OACN, INAF-OAPD and the KiDS production team, on behalf of the KiDS consortium. OmegaCEN and the KiDS production team acknowledge support by NOVA and NWO-M grants. Members of INAF-OAPD and INAF-OACN also acknowledge the support from the Department of Physics \& Astronomy of the University of Padova, and of the Department of Physics of Univ. Federico II (Naples). 

\bibliography{main}{}
\bibliographystyle{aasjournal}

\appendix

\begin{figure*}[h]
\centering
\includegraphics[trim=0 2.5cm 0cm 0cm, clip,width=0.245\hsize]{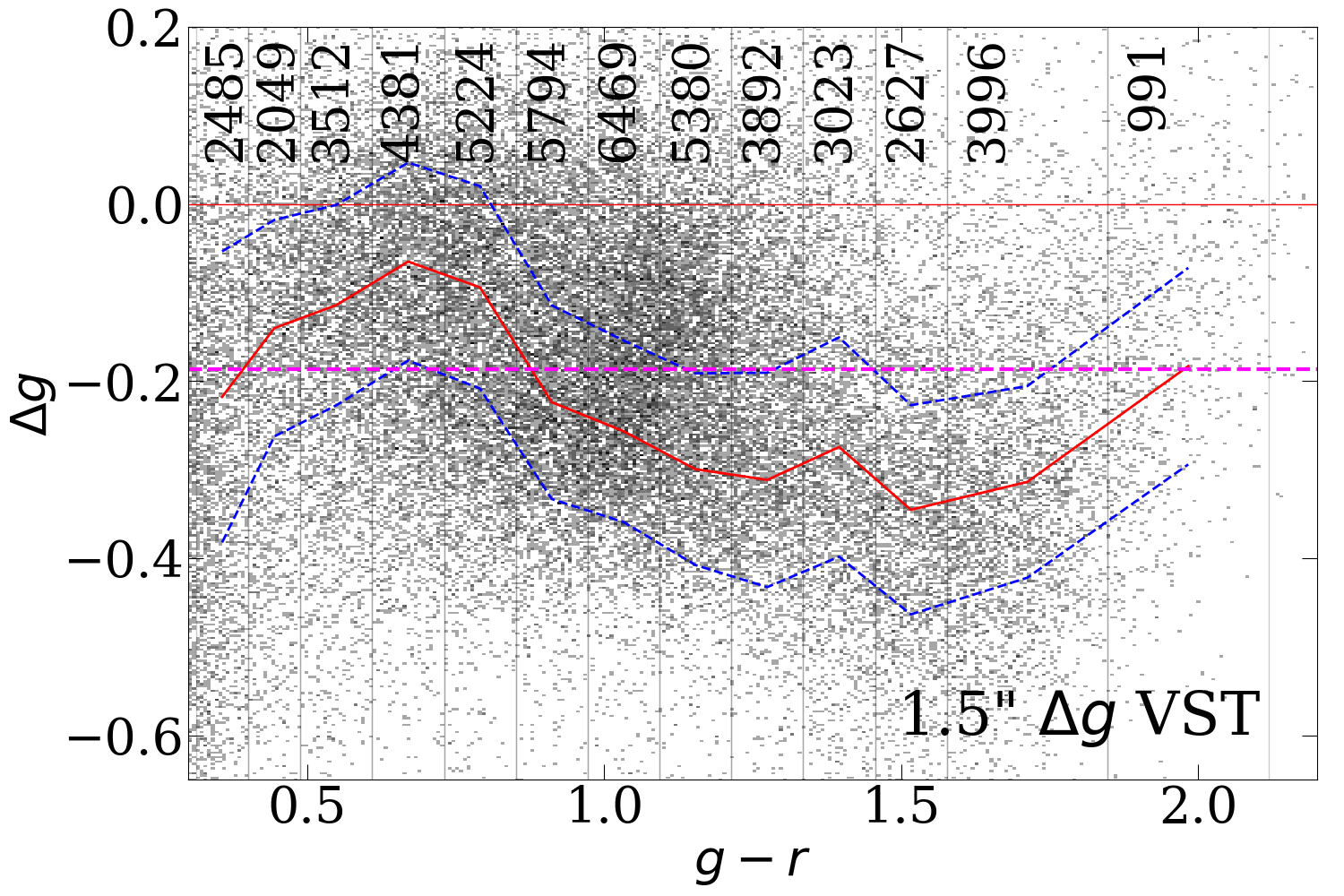}
\includegraphics[trim=0 2.5cm 0cm 0cm, clip,width=0.245\hsize]{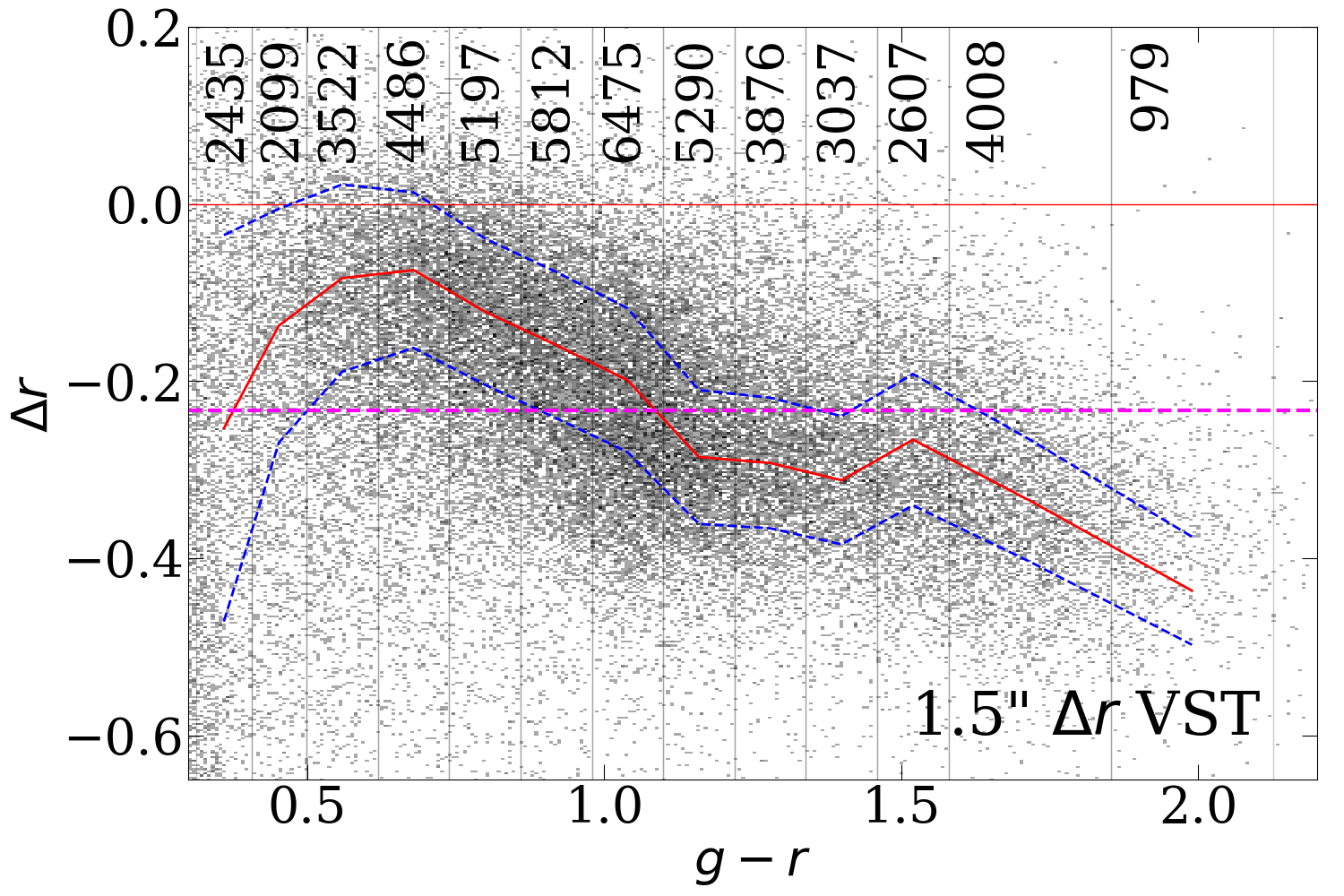}
\includegraphics[trim=0 2.5cm 0cm 0cm, clip,width=0.245\hsize]{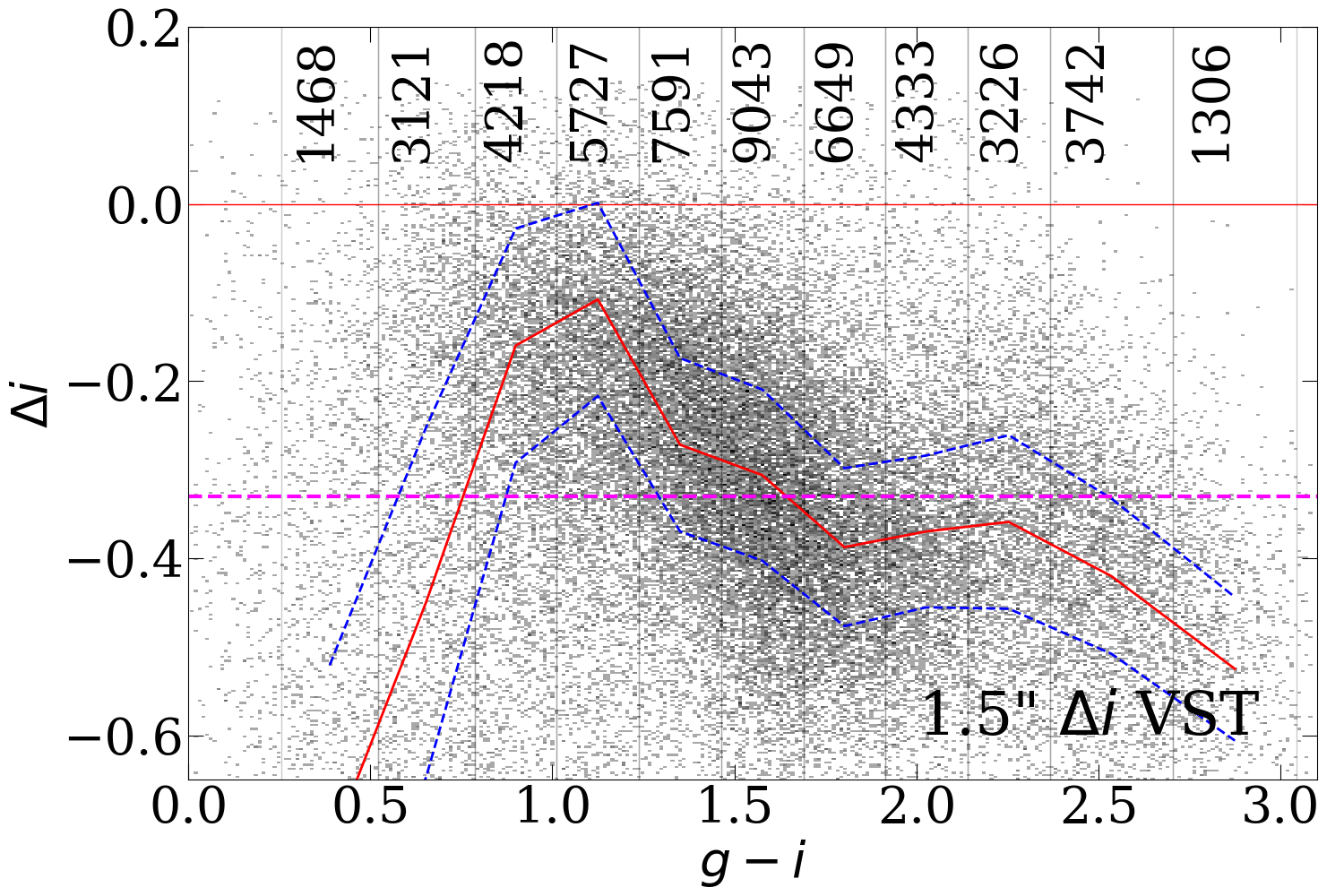}
\includegraphics[trim=0 2.5cm 0cm 0cm, clip,width=0.245\hsize]{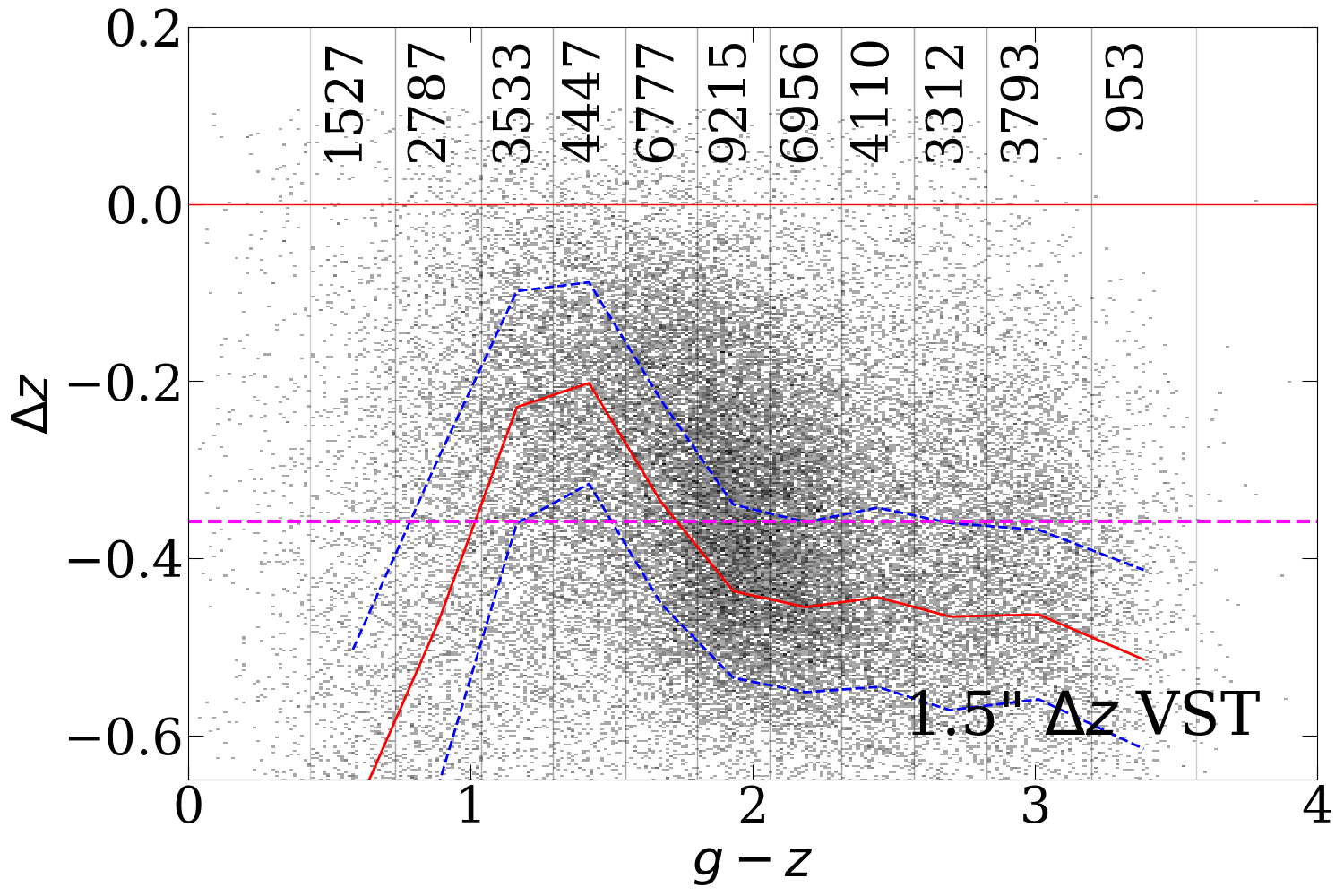}\\
\includegraphics[trim=0 2.5cm 0cm 0cm, clip,width=0.245\hsize]{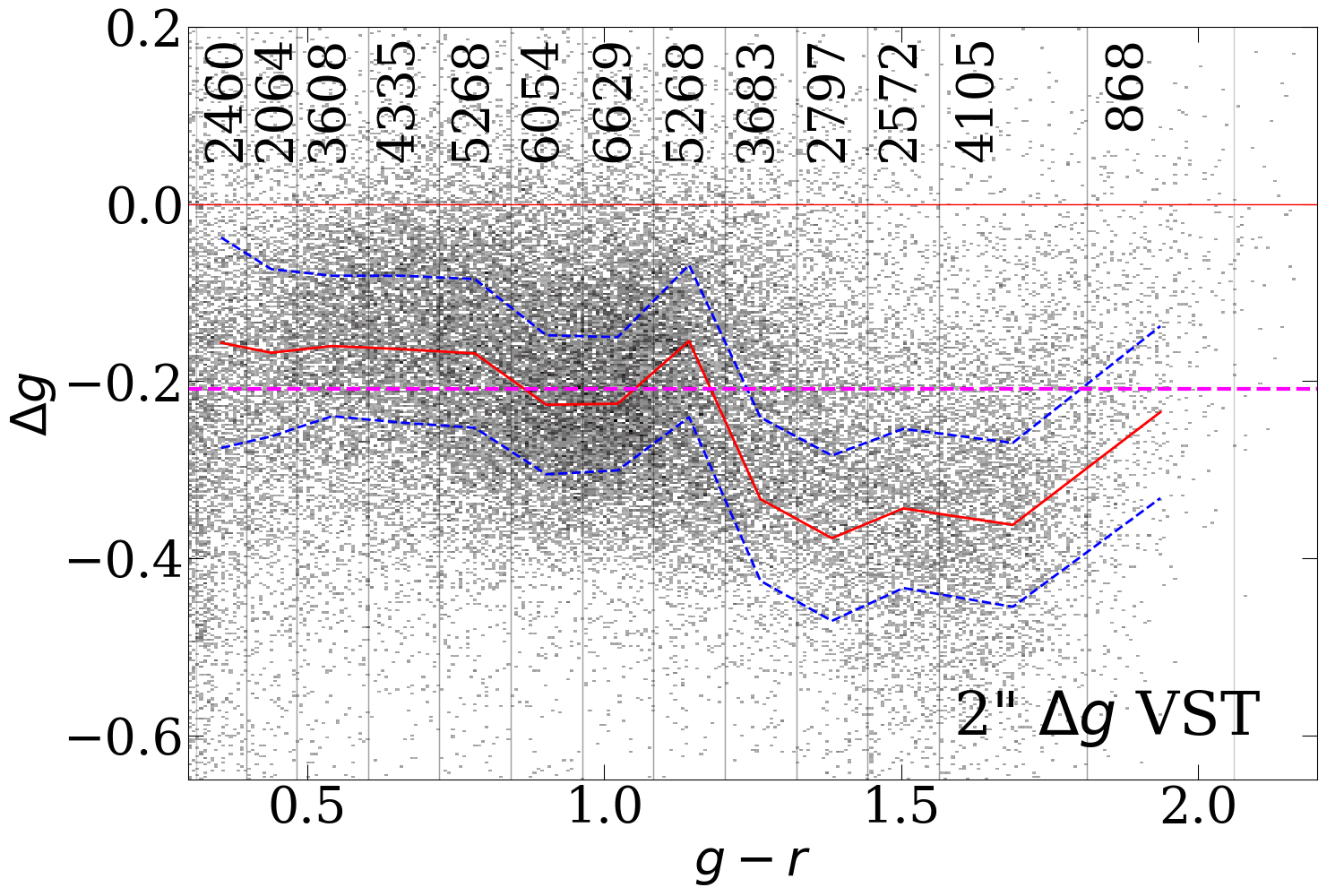}
\includegraphics[trim=0 2.5cm 0cm 0cm, clip,width=0.245\hsize]{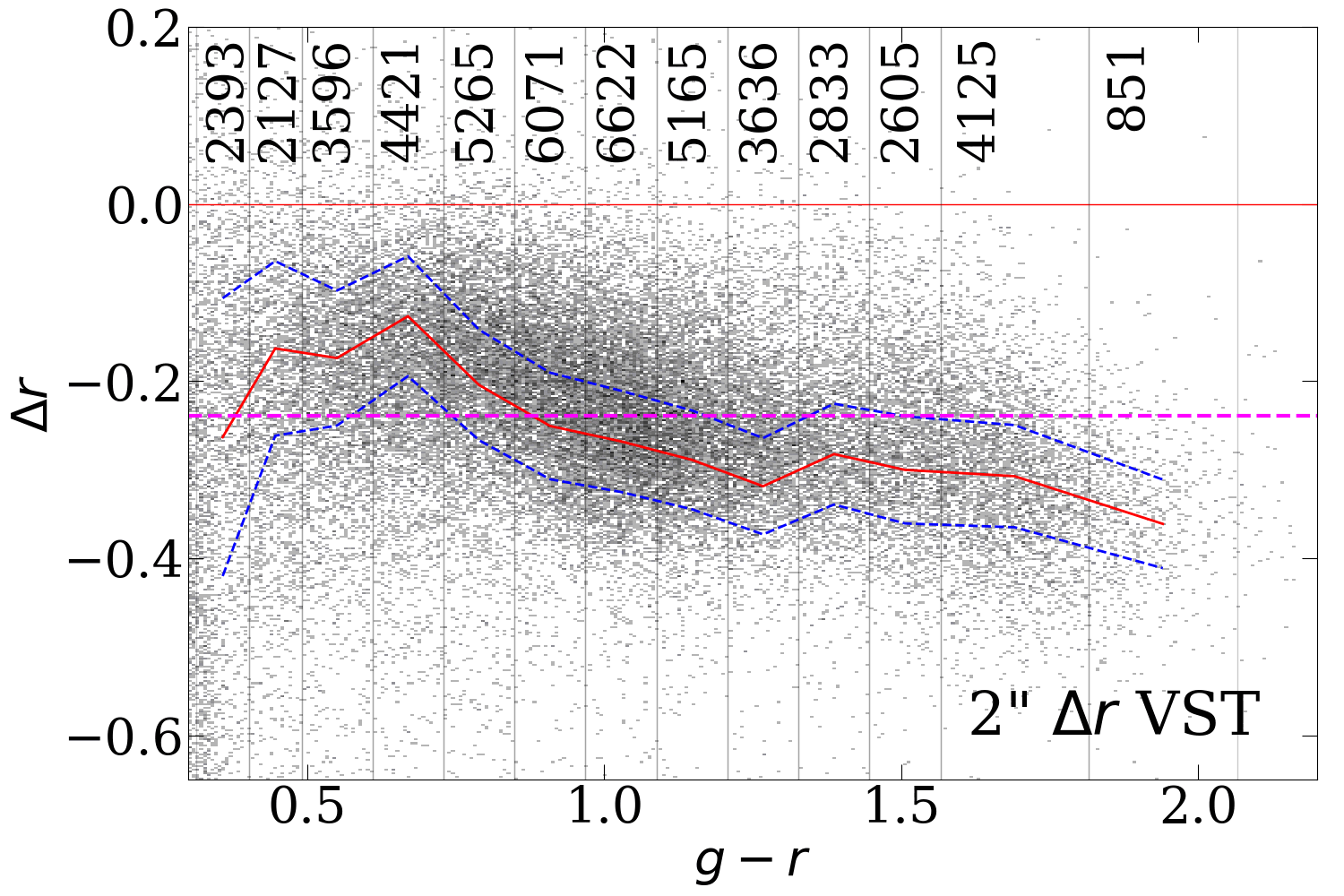}
\includegraphics[trim=0 2.5cm 0cm 0cm, clip,width=0.245\hsize]{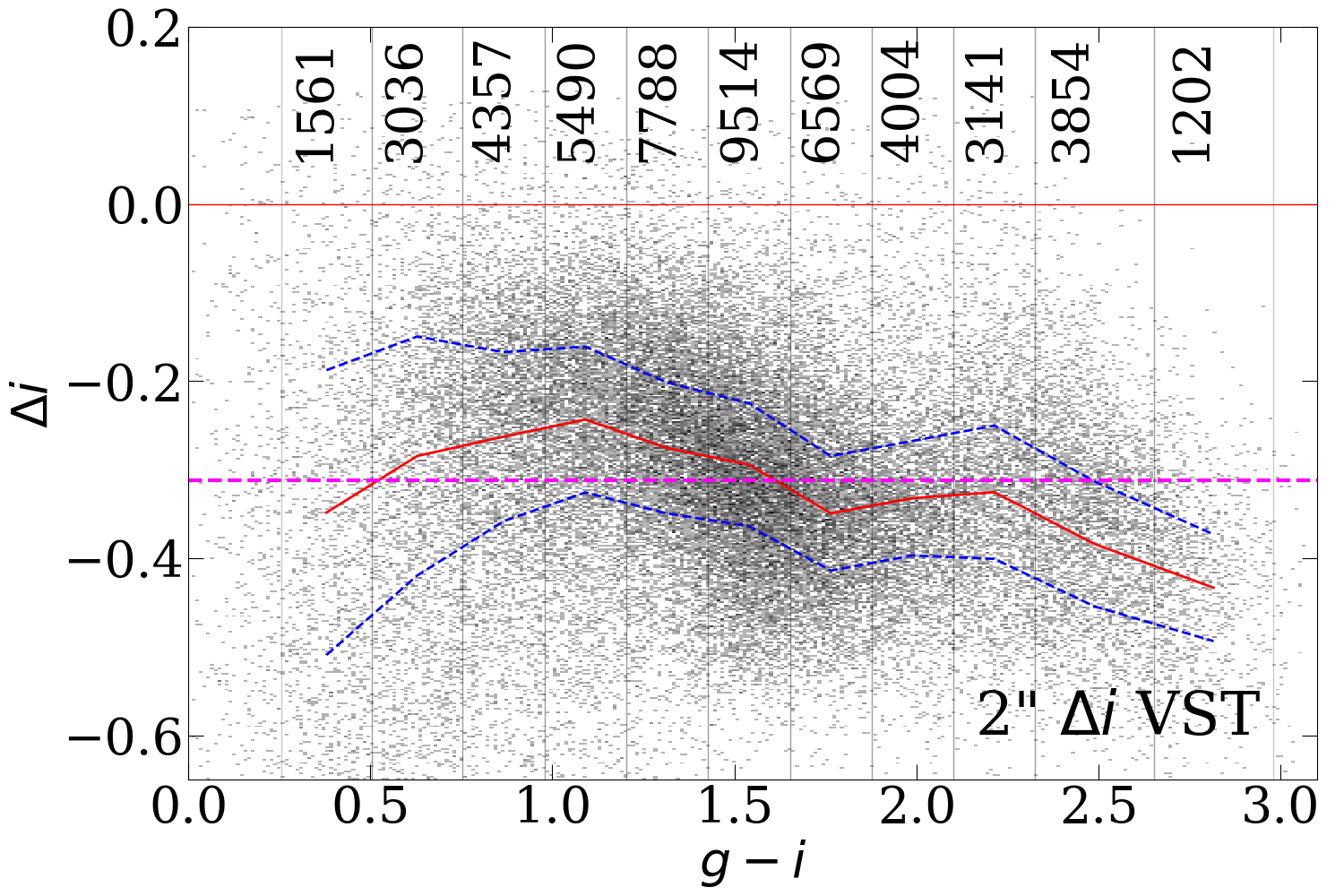}
\includegraphics[trim=0 2.5cm 0cm 0cm, clip,width=0.245\hsize]{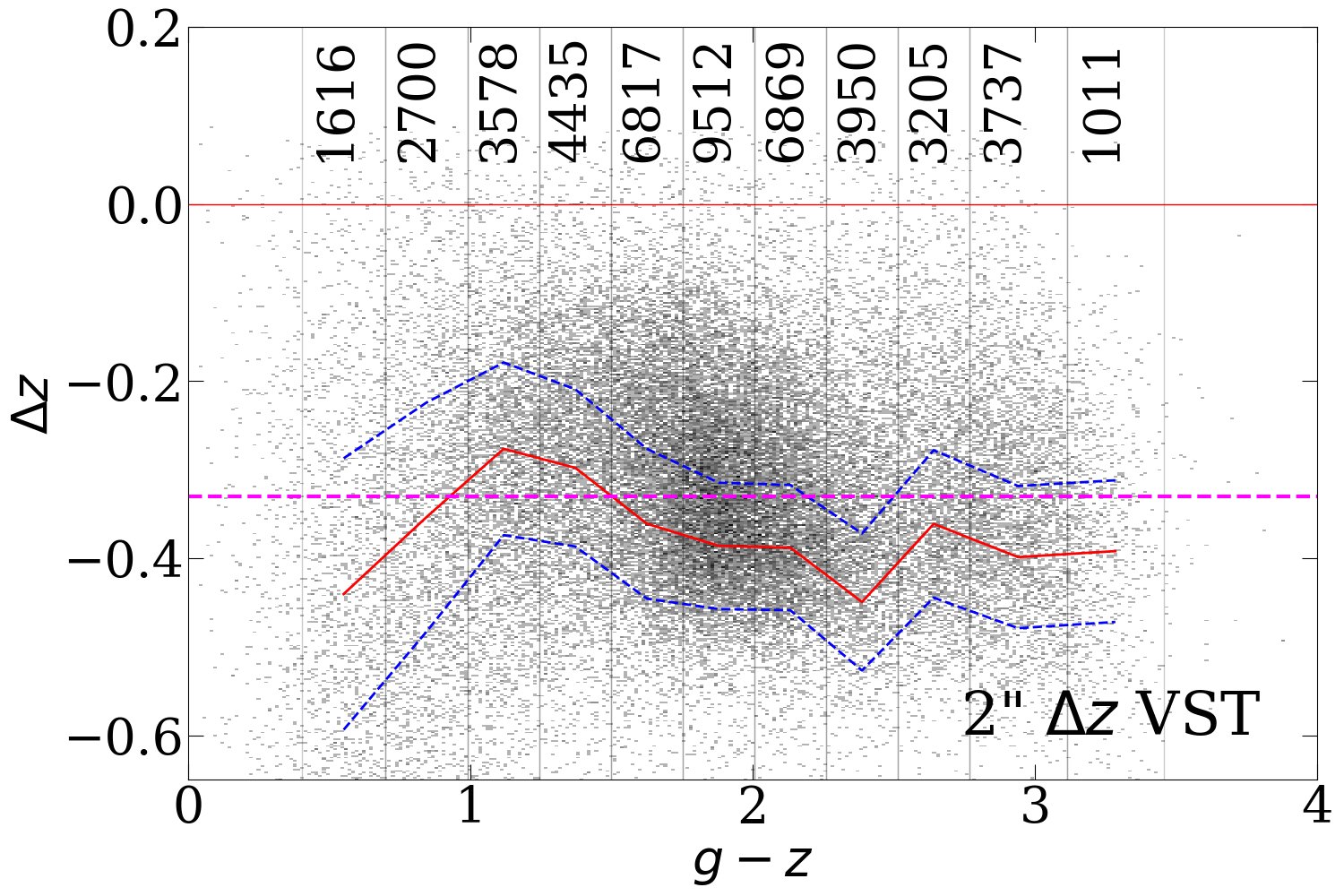}\\
\includegraphics[trim=0 0cm 0cm 0cm, clip,width=0.245\hsize]{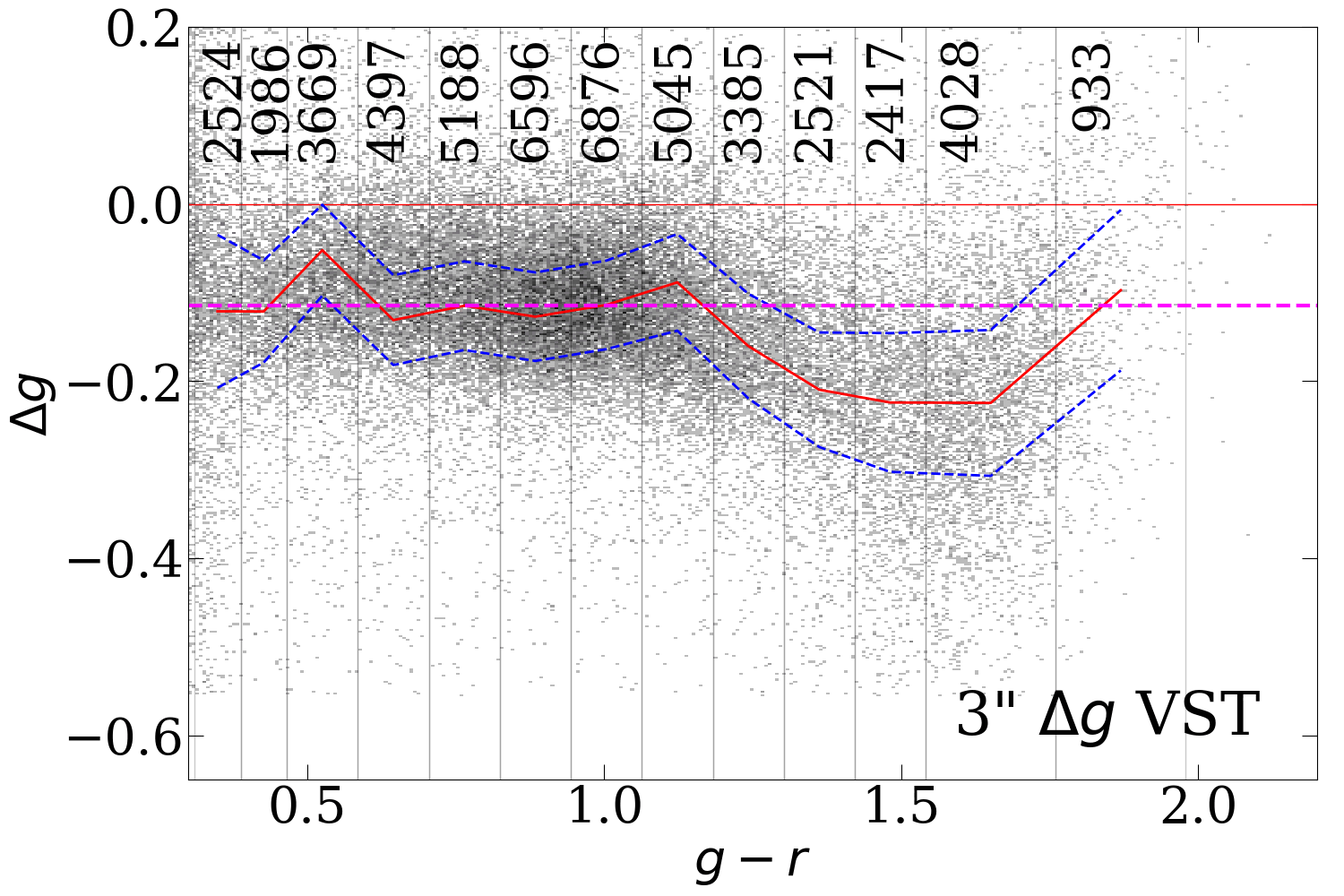}
\includegraphics[trim=0 0cm 0cm 0cm, clip,width=0.245\hsize]{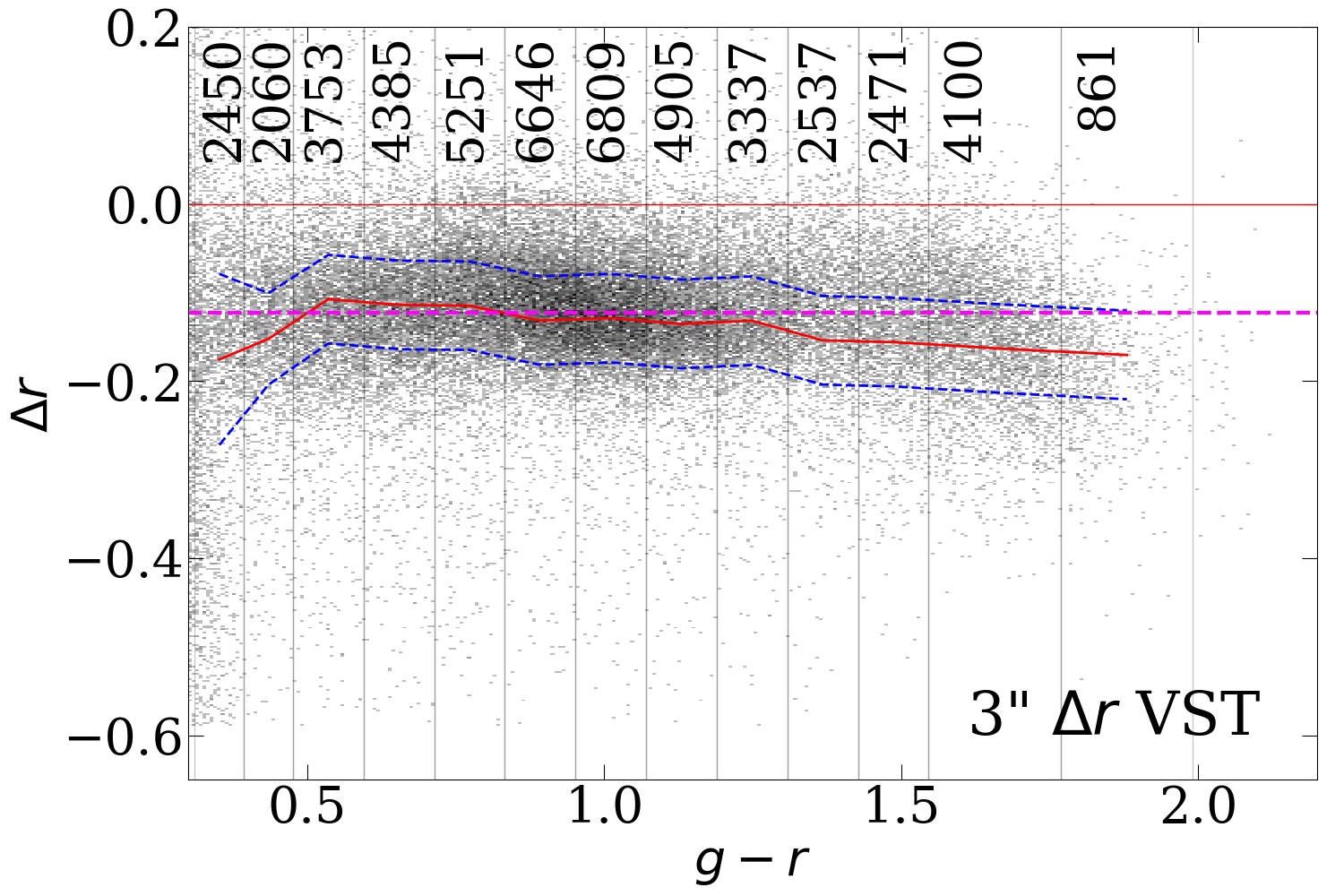}
\includegraphics[trim=0 0cm 0cm 0cm, clip,width=0.245\hsize]{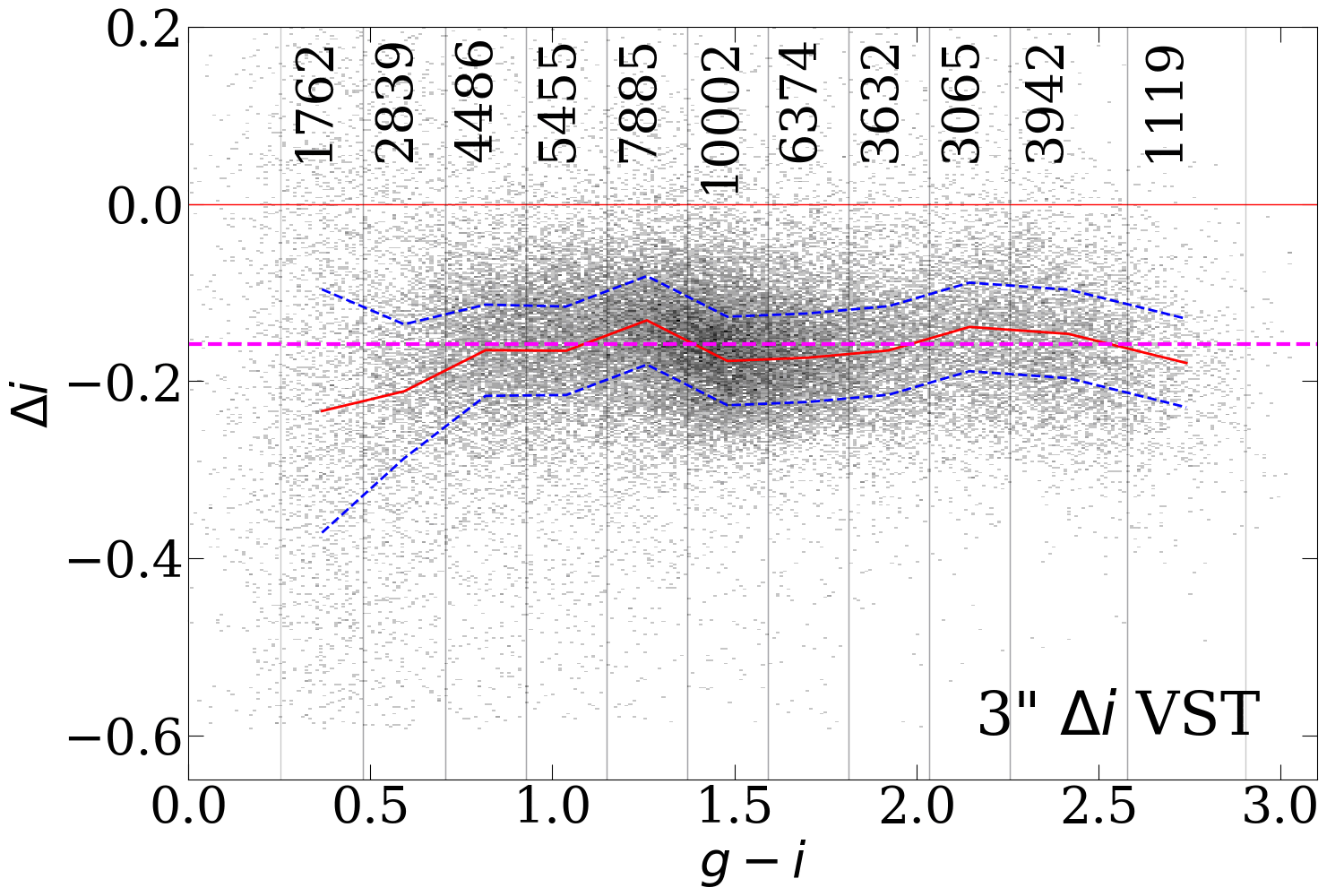}
\includegraphics[trim=0 0cm 0cm 0cm, clip,width=0.245\hsize]{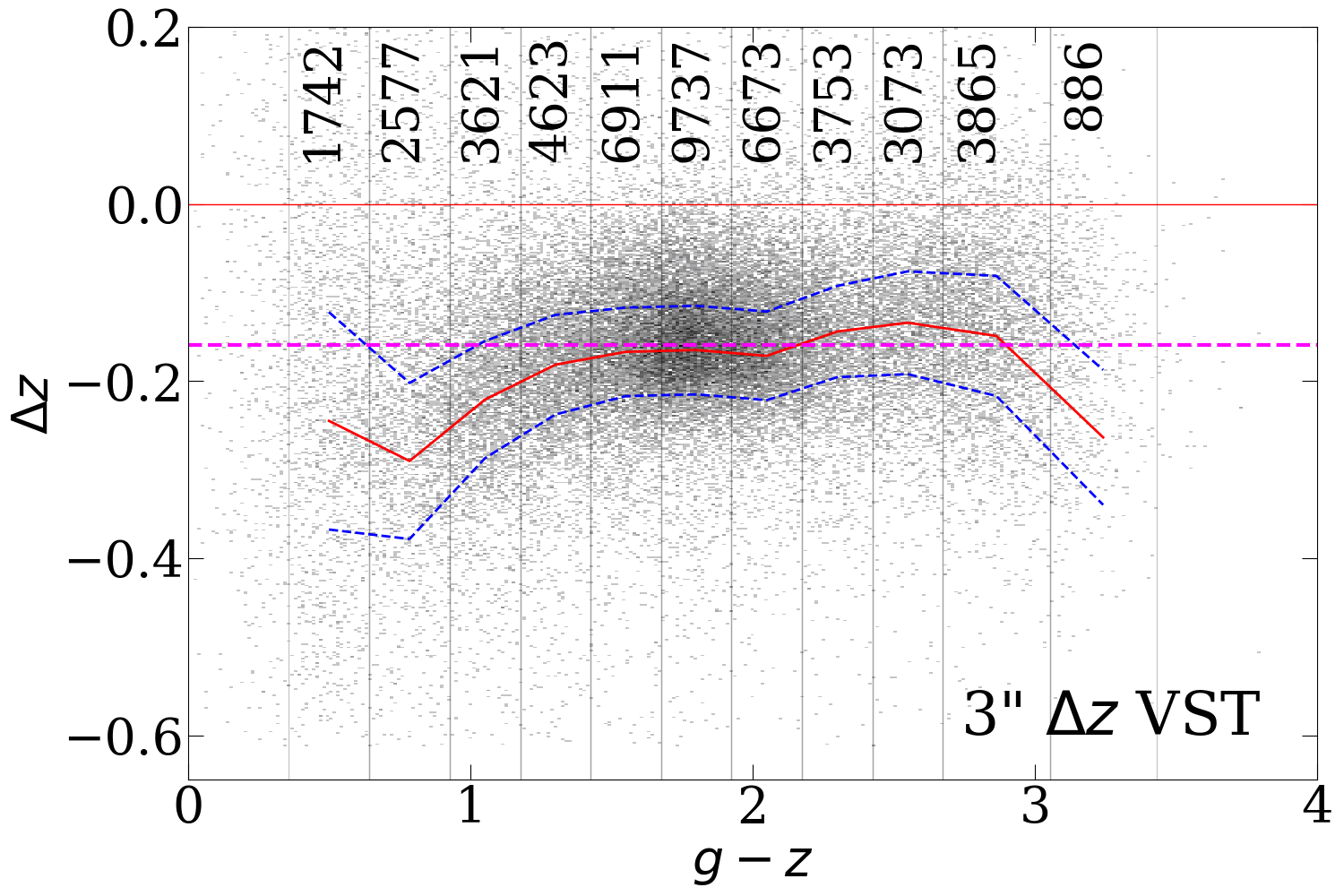}
\caption{Magnitude difference \revtwo{between VST Atlas corrected by transformations for total magnitudes and SDSS} for 1.5$''$, 2$''$ and 3$''$ apertures. \revone{All colors are computed in the same apertures. The zero corresponding to the offsets from Table~\ref{tab_shifts} is shown by a magenta line.}
\label{apers_full1}}
\end{figure*}

\begin{figure*}
\centering
\includegraphics[trim=0 2.5cm 0cm 0cm, clip,width=0.245\hsize]{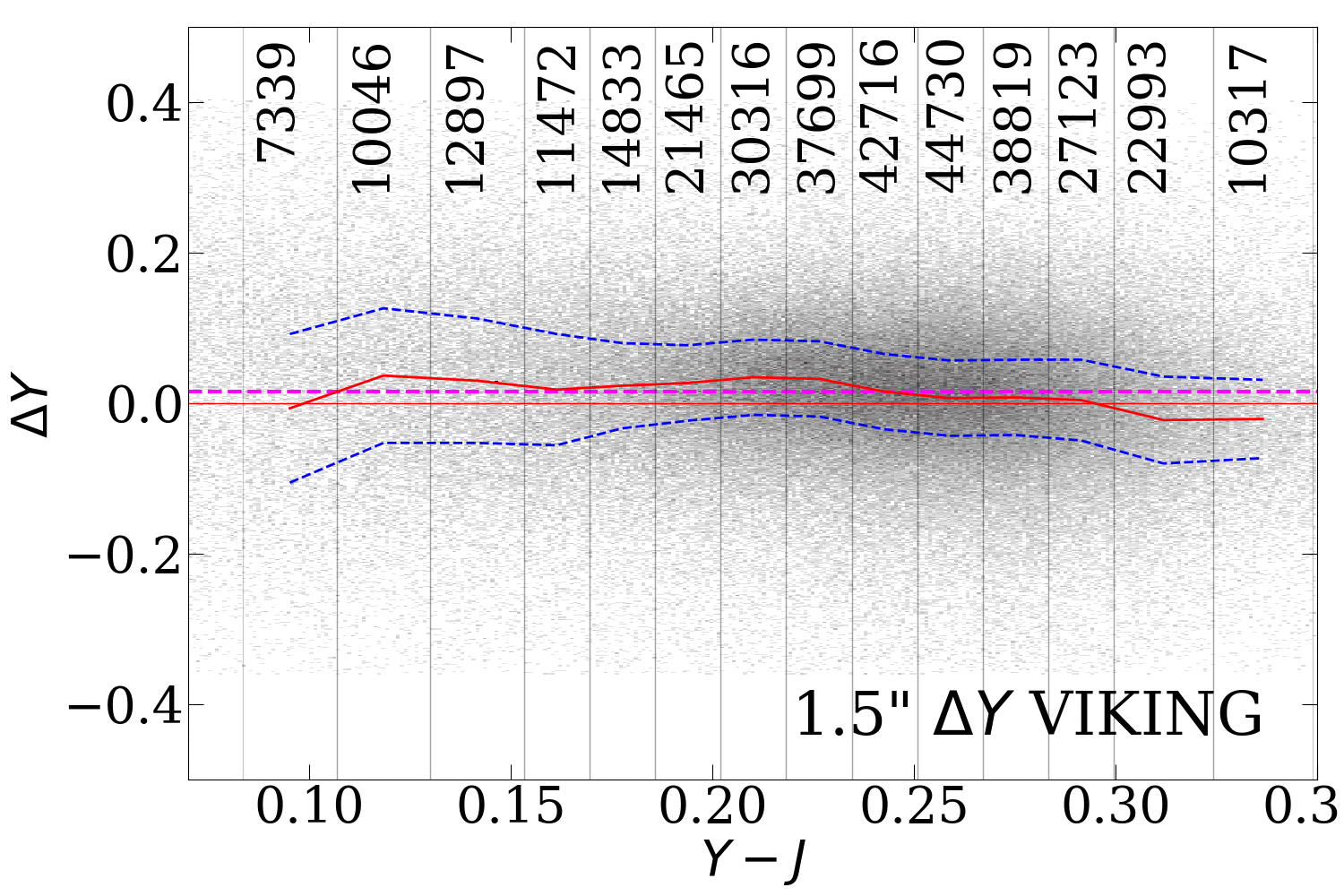}
\includegraphics[trim=0 2.5cm 0cm 0cm, clip,width=0.245\hsize]{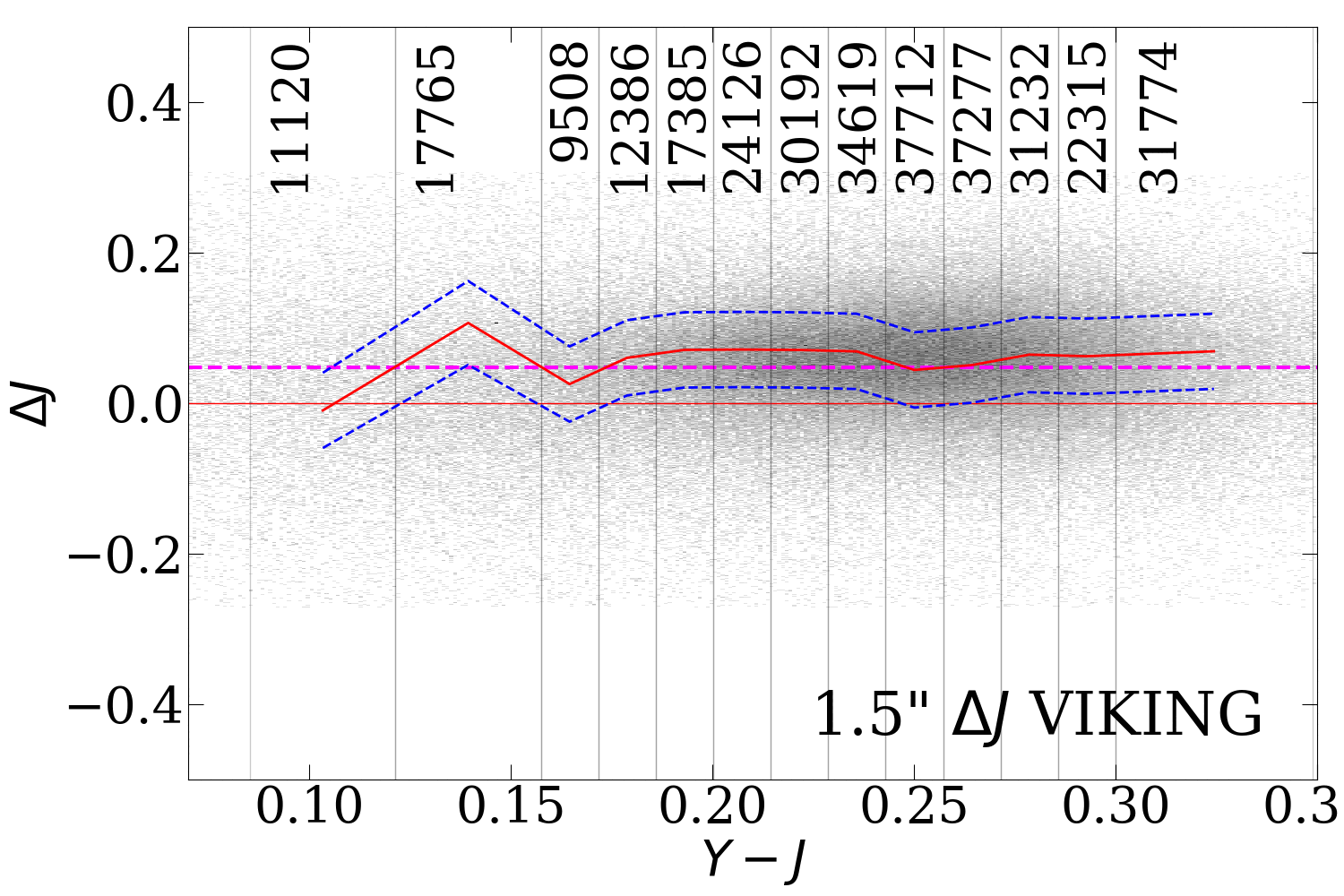}
\includegraphics[trim=0 2.5cm 0cm 0cm, clip,width=0.245\hsize]{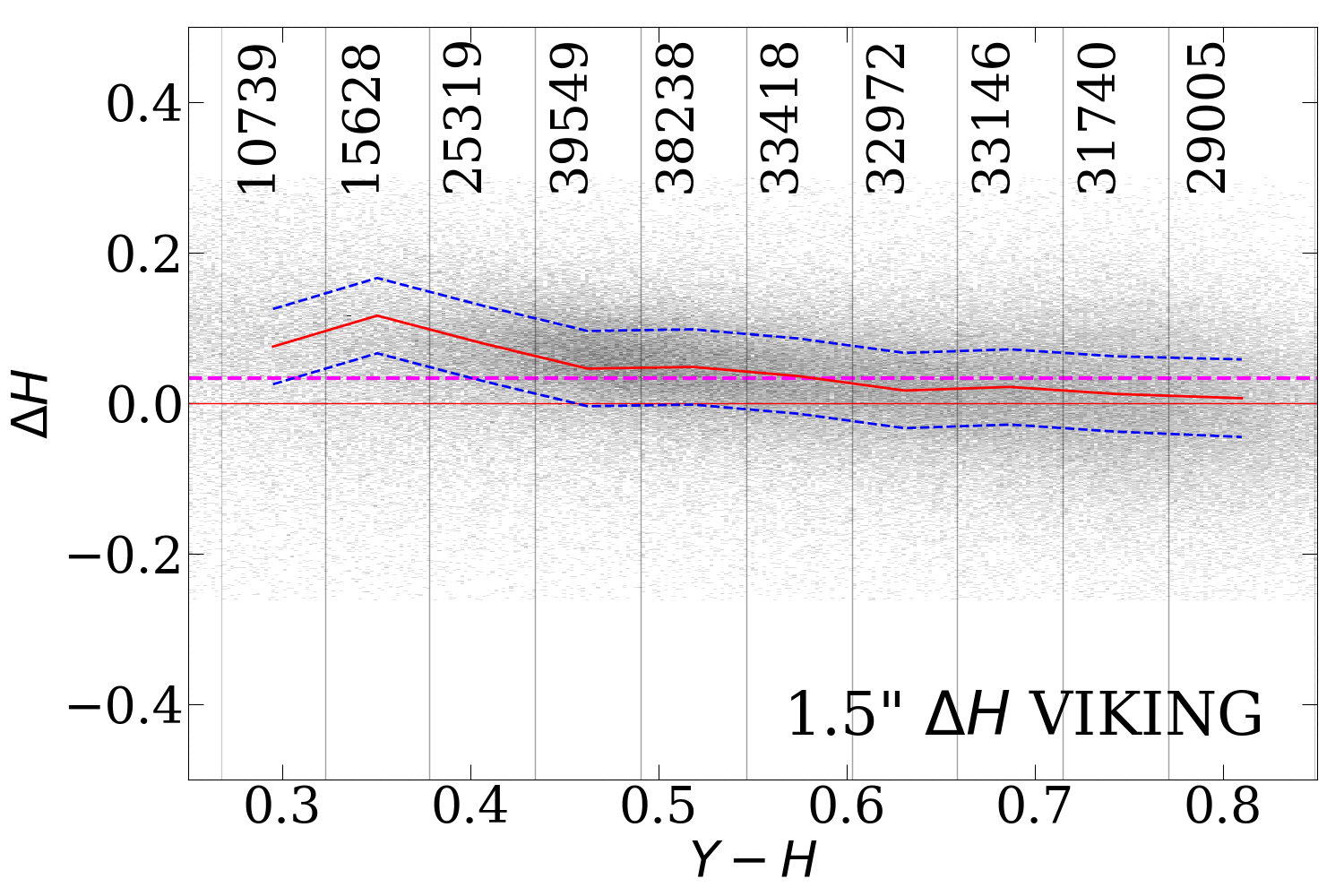}
\includegraphics[trim=0 2.5cm 0cm 0cm, clip,width=0.245\hsize]{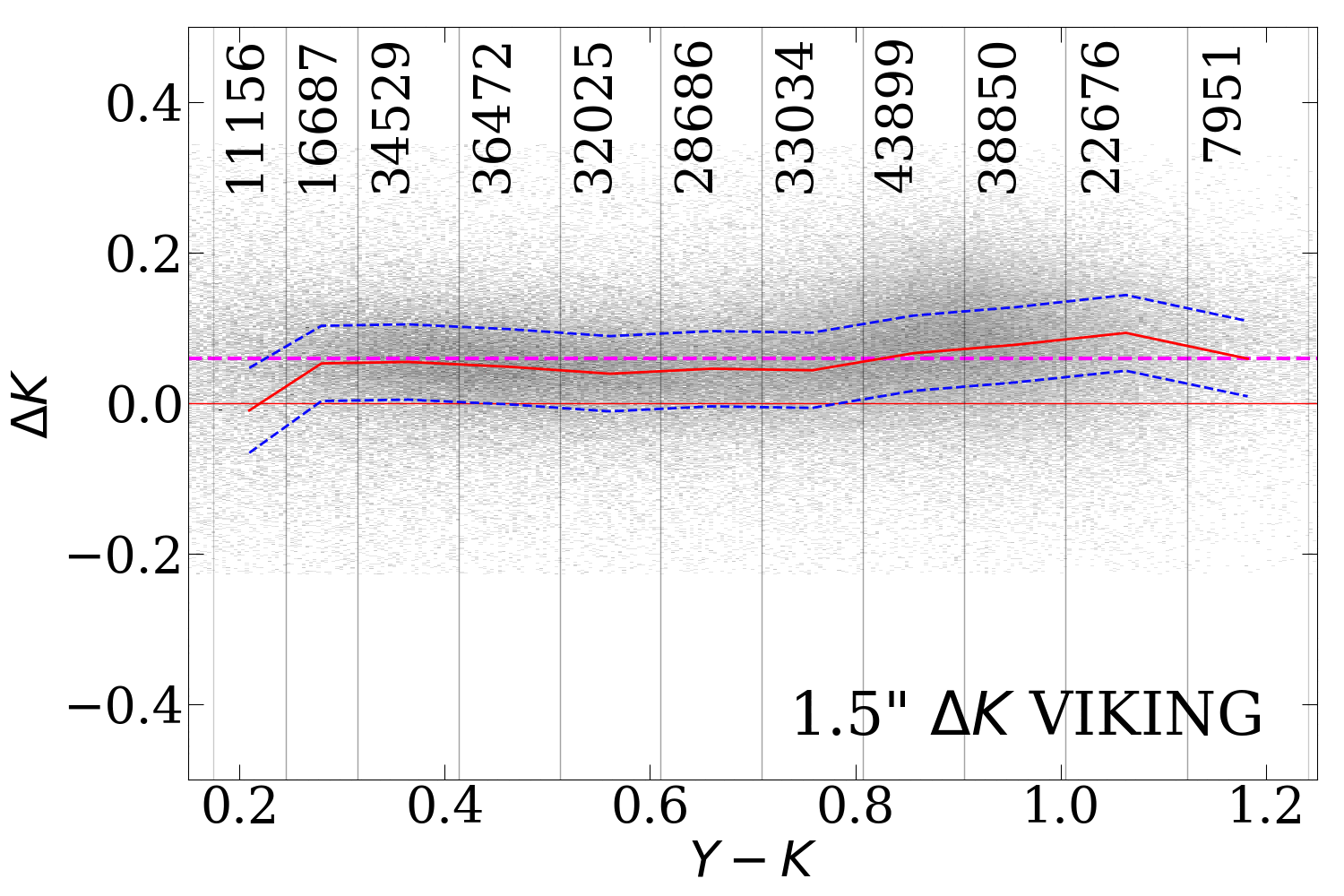}\\
\includegraphics[trim=0 2.5cm 0cm 0cm, clip,width=0.245\hsize]{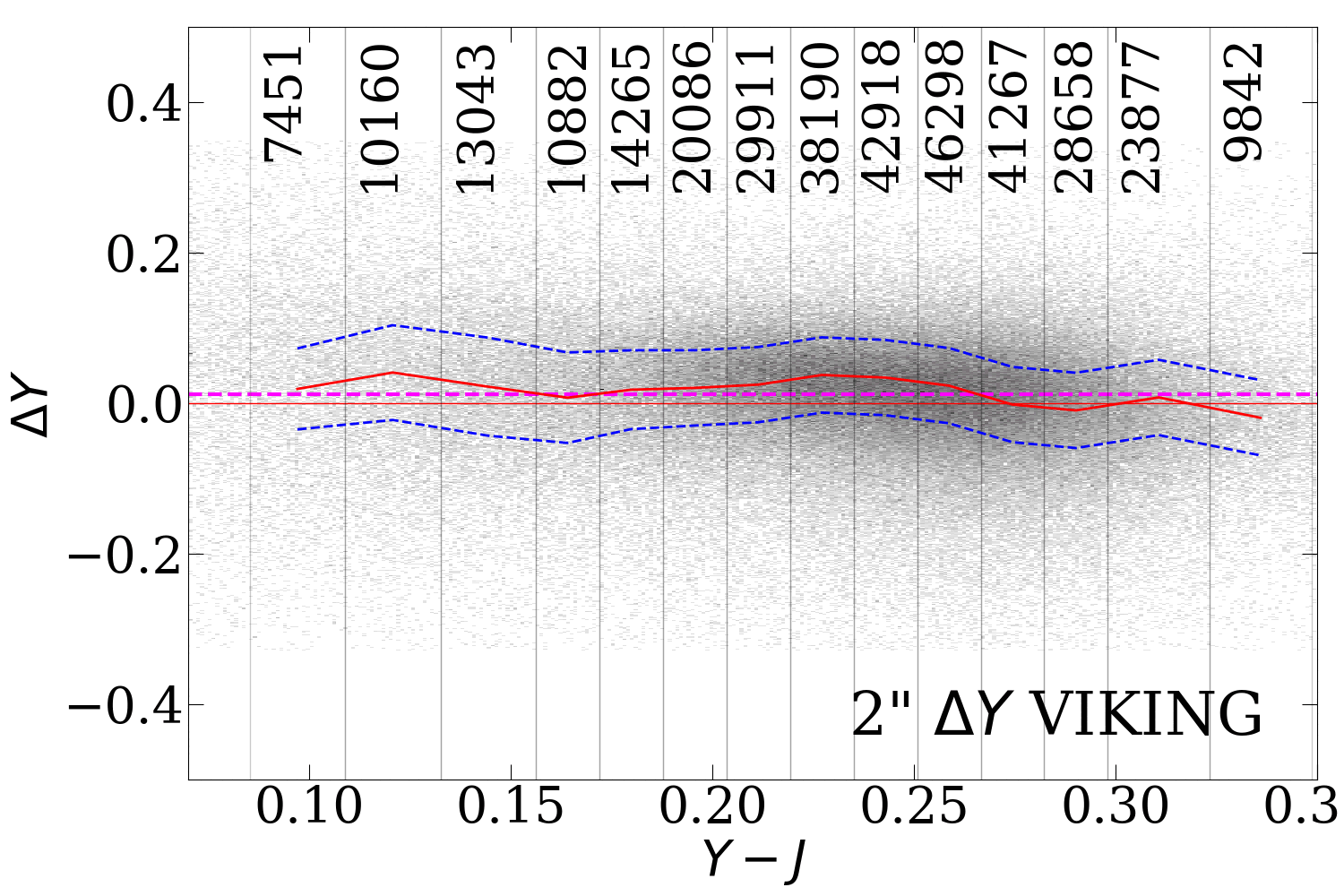}
\includegraphics[trim=0 2.5cm 0cm 0cm, clip,width=0.245\hsize]{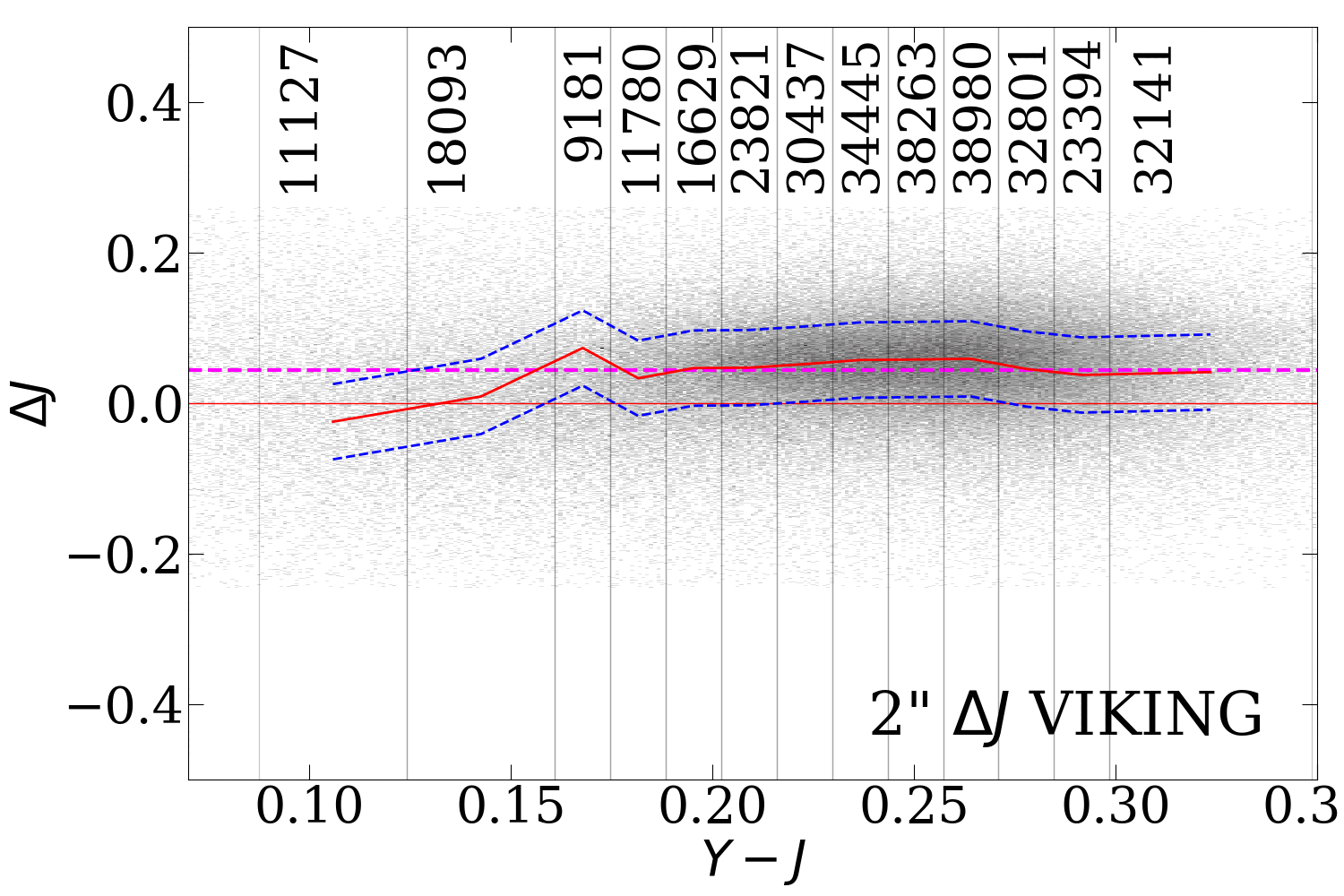}
\includegraphics[trim=0 2.5cm 0cm 0cm, clip,width=0.245\hsize]{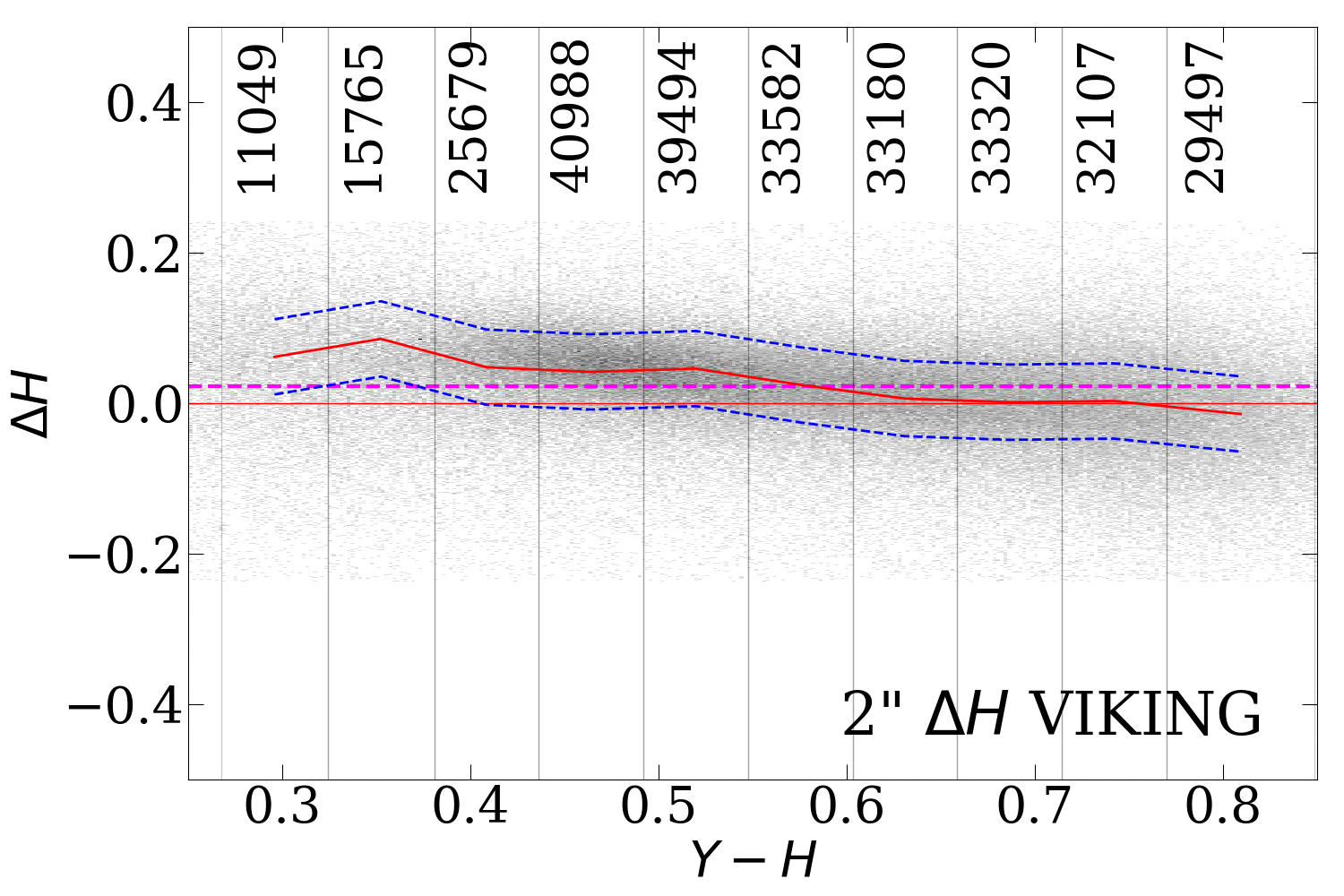}
\includegraphics[trim=0 2.5cm 0cm 0cm, clip,width=0.245\hsize]{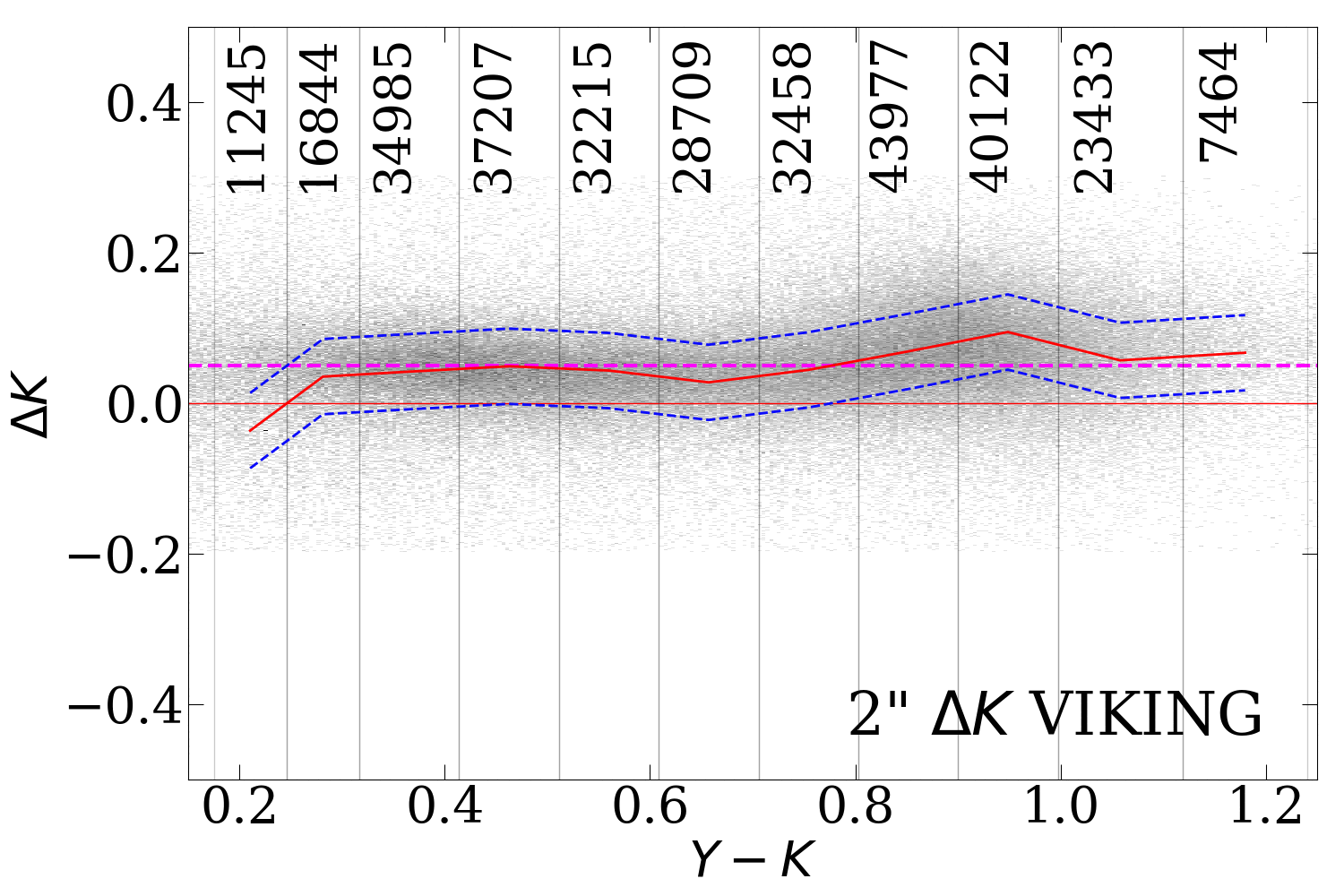}\\
\includegraphics[width=0.245\hsize]{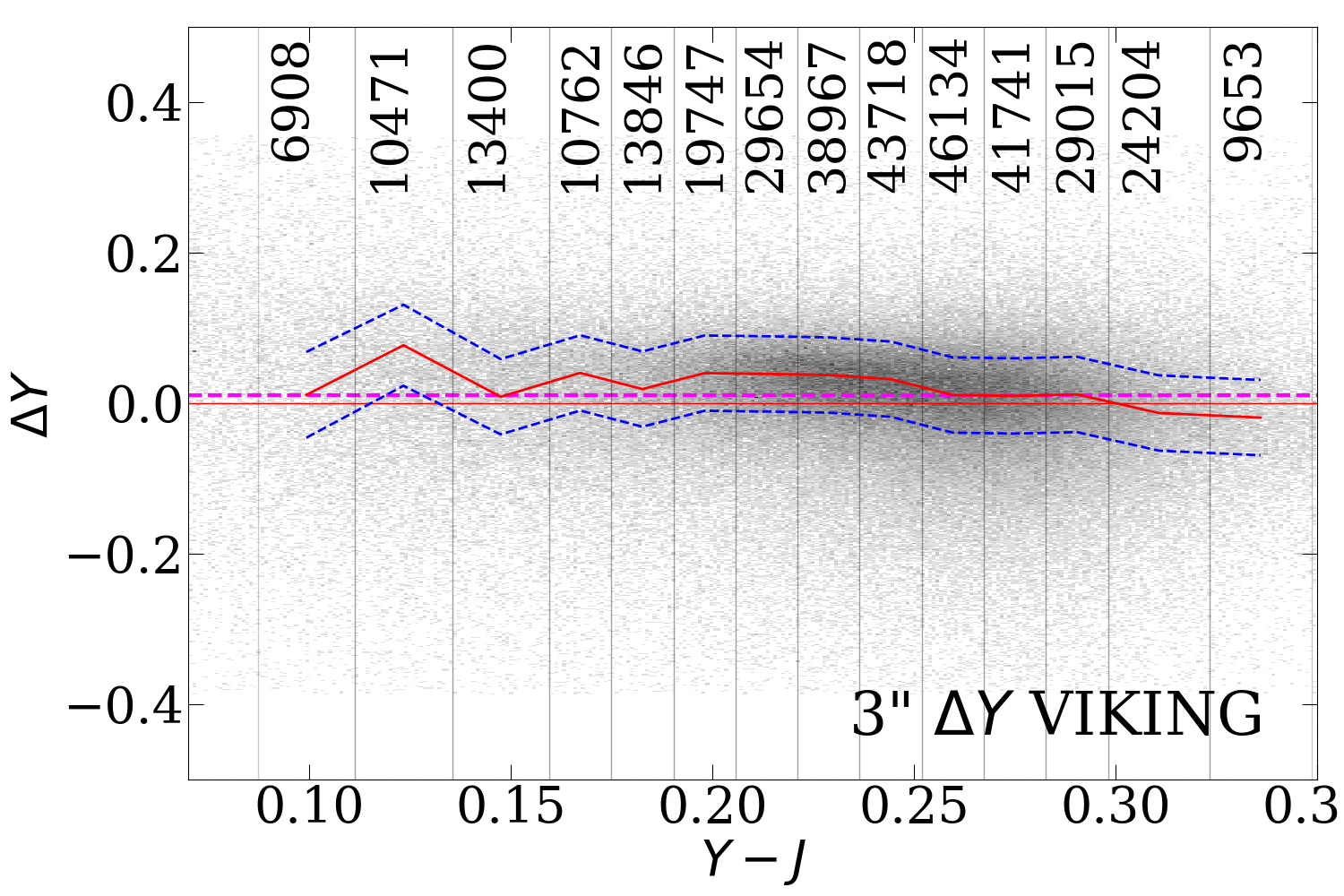}
\includegraphics[width=0.245\hsize]{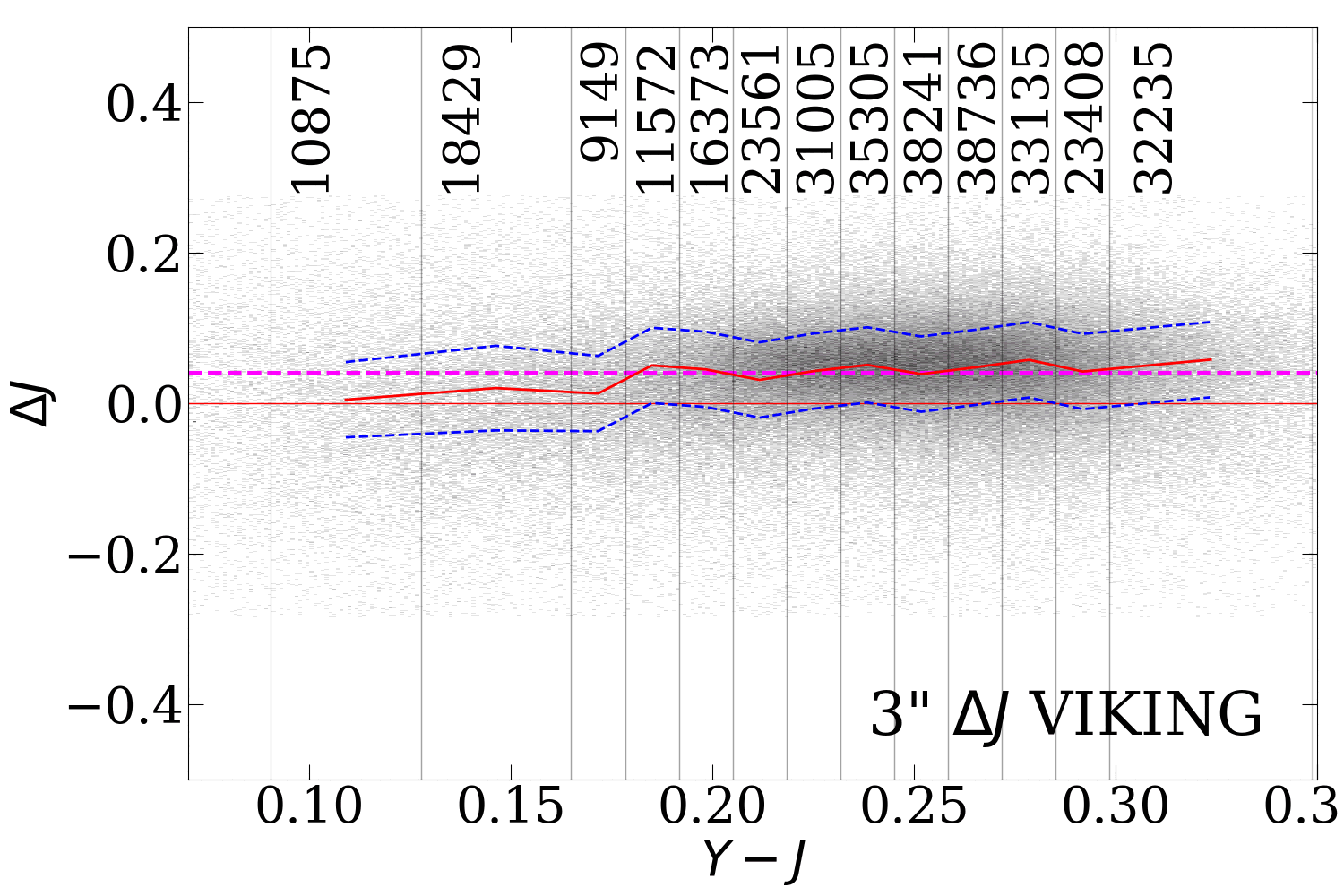}
\includegraphics[width=0.245\hsize]{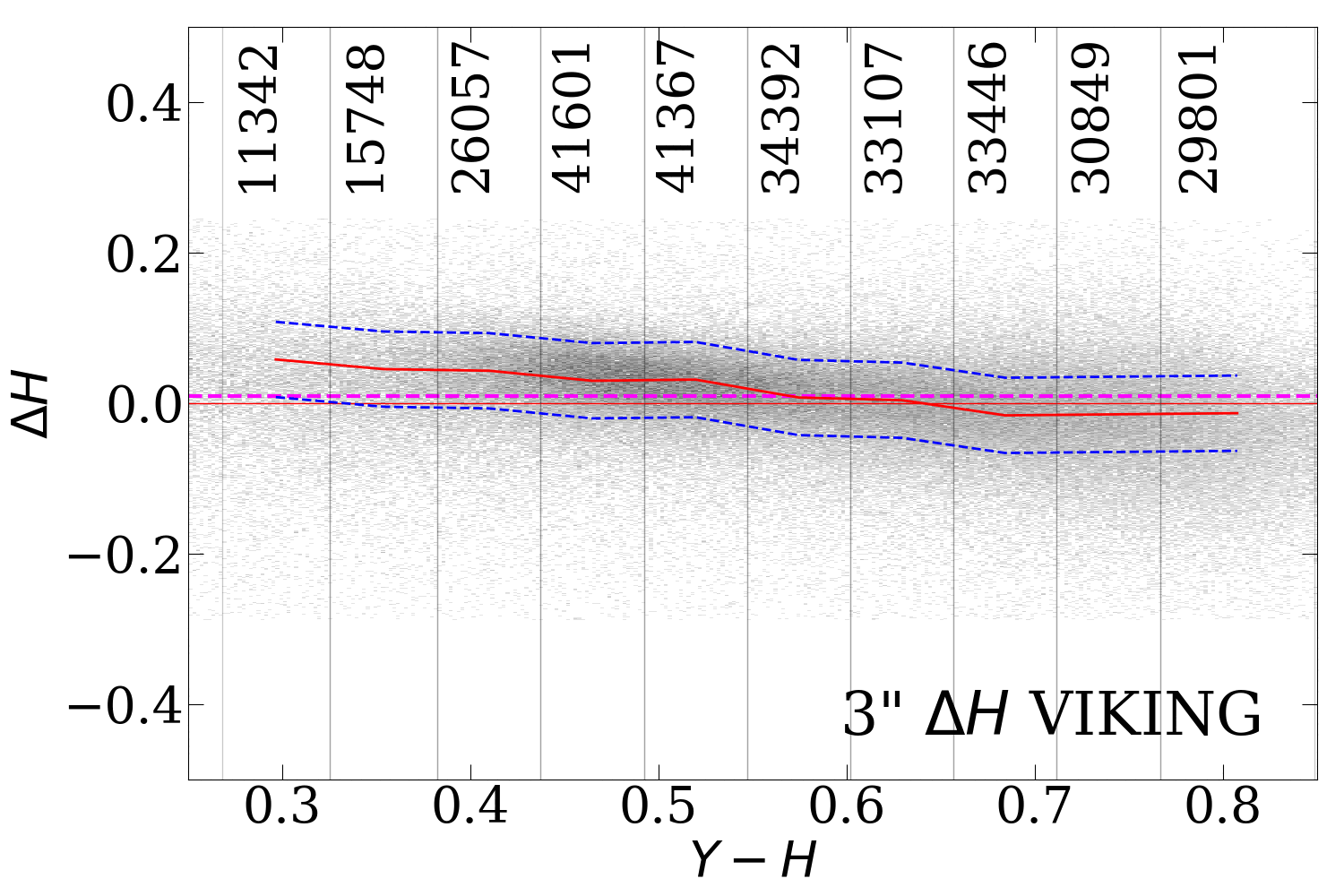}
\includegraphics[width=0.245\hsize]{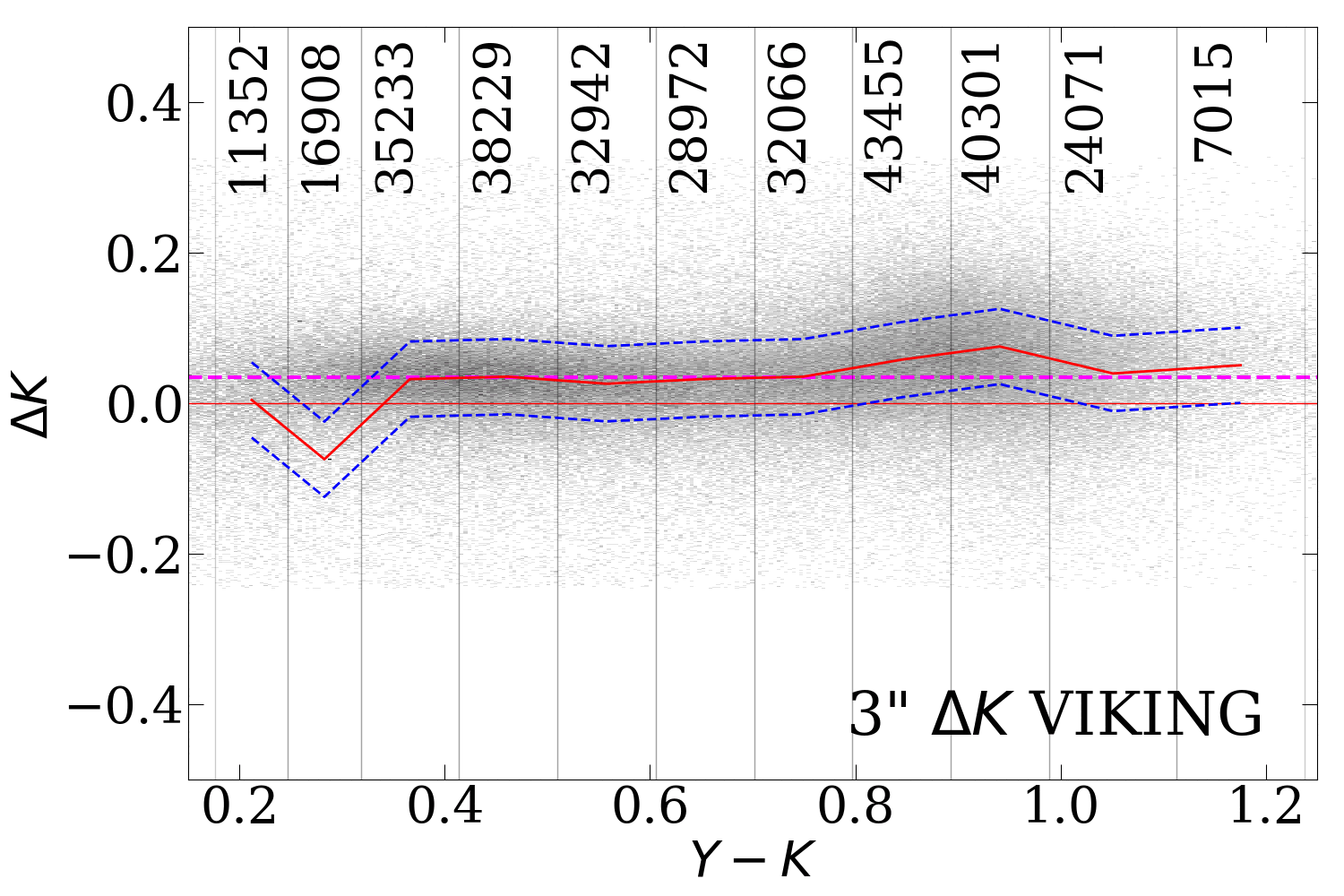}\\
\caption{
\revone{Same as Figure \ref{apers_full1} for magnitude difference in the \revtwo{VIKING} system and \revtwo{UKIDSS}.}
\label{apers_ir}}
\end{figure*}

\begin{figure*}
\centering
\includegraphics[trim=0.6cm 2.5cm 0cm 0cm, clip,width=0.2942\hsize]{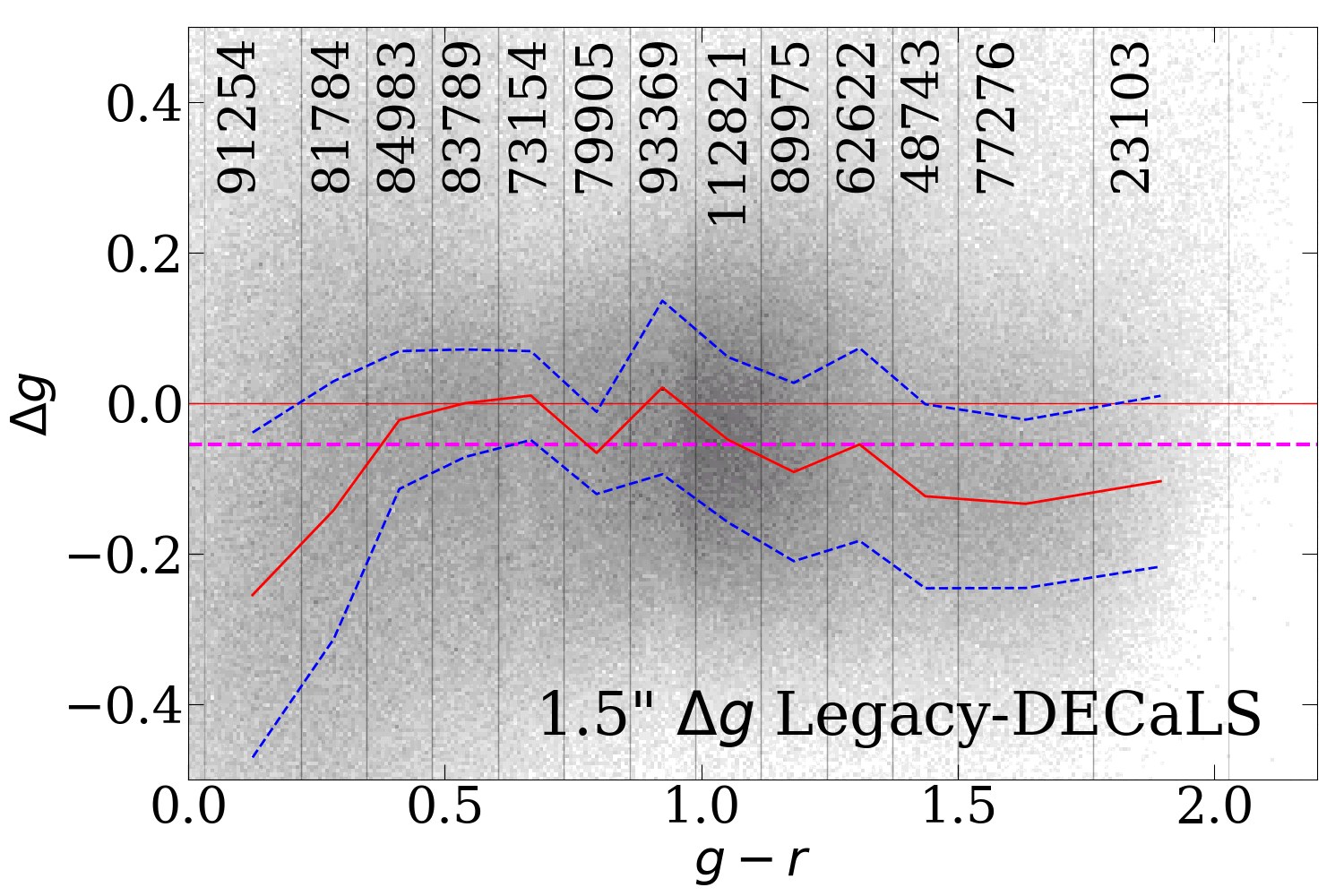}
\includegraphics[trim=3.0cm 2.5cm 0cm 0cm, clip,width=0.28\hsize]{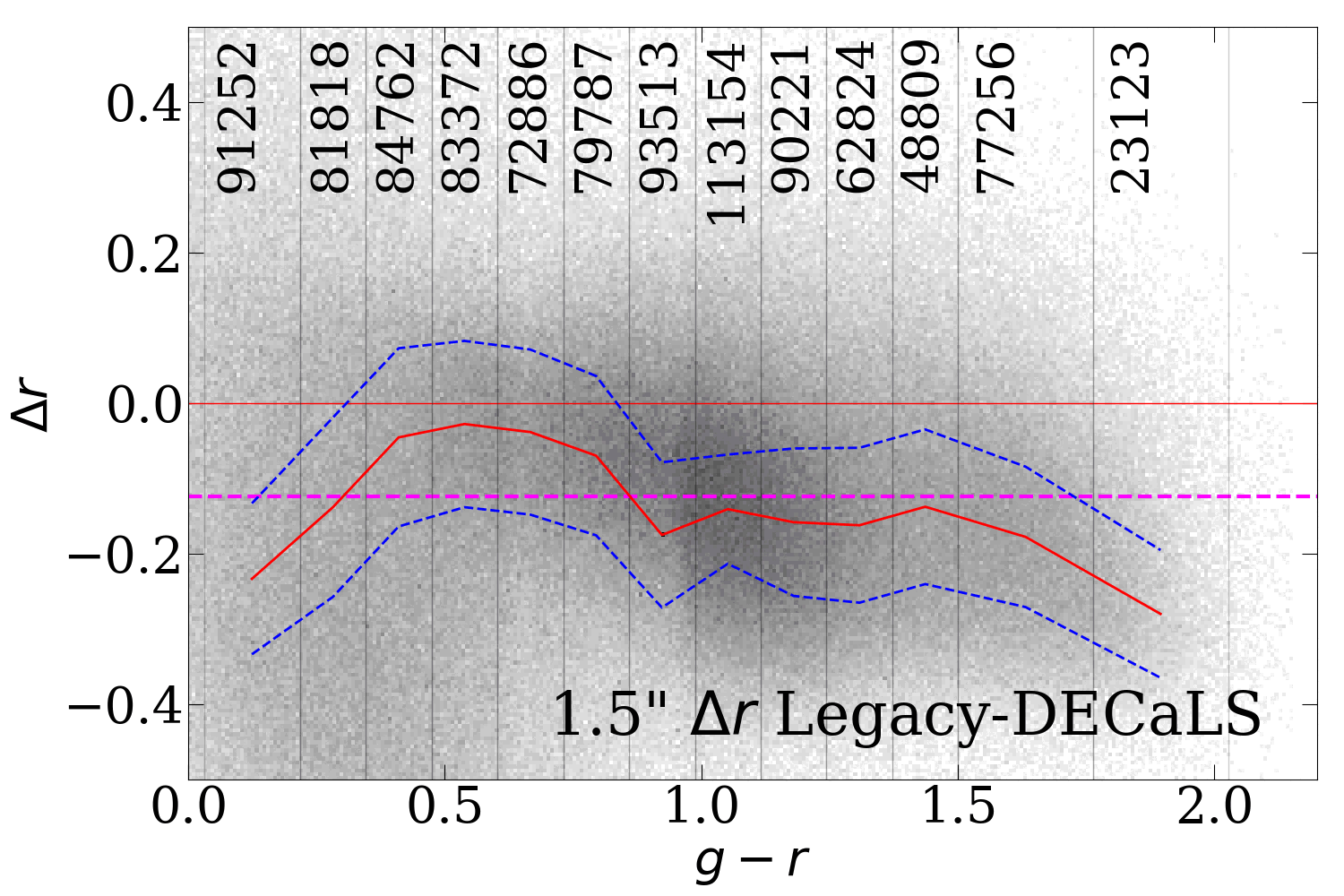}
\includegraphics[trim=3.0cm 2.5cm 0cm 0cm, clip,width=0.28\hsize]{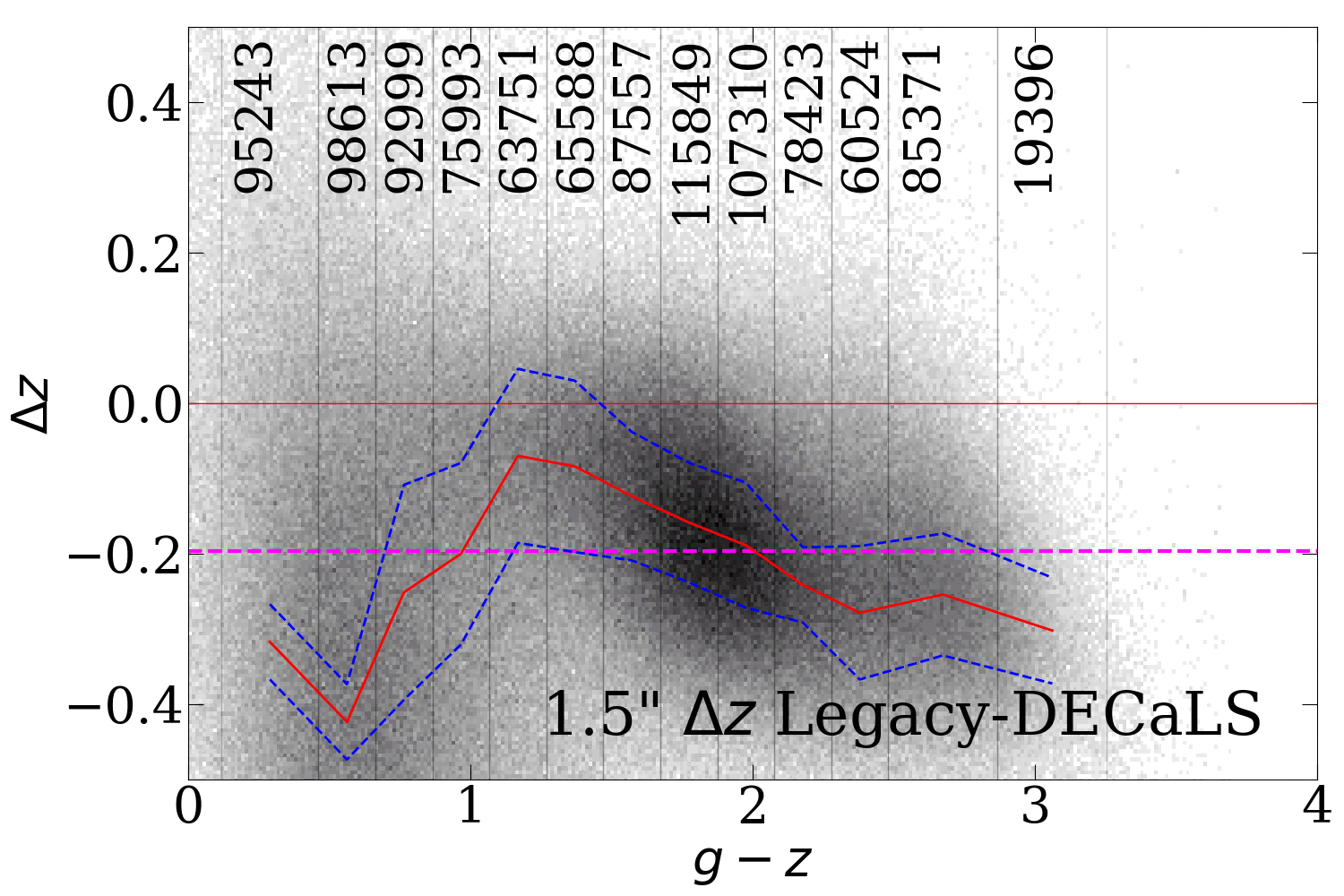}
\includegraphics[trim=0.6cm 2.5cm 0cm 0cm, clip,width=0.2942\hsize]{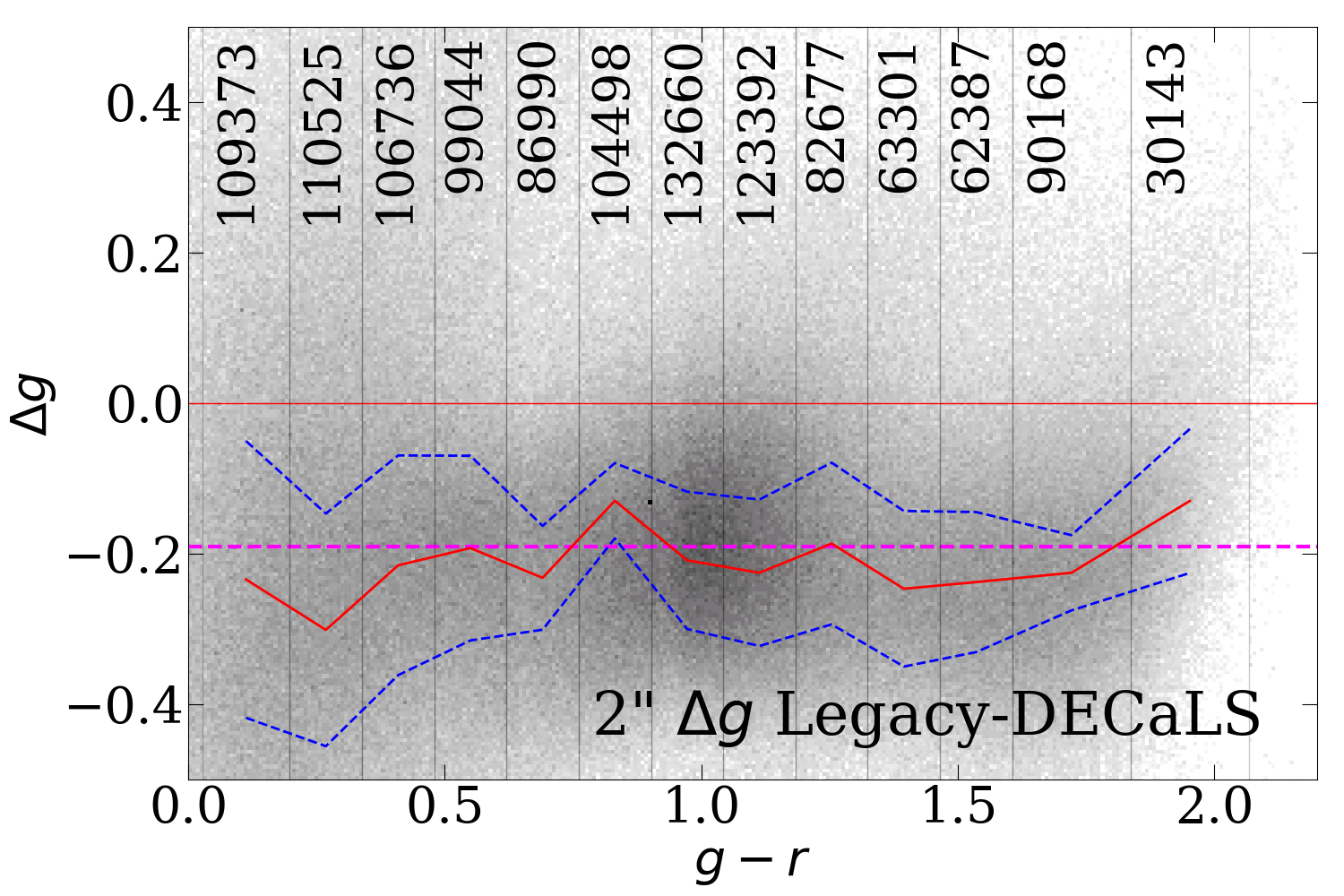}
\includegraphics[trim=3.0cm 2.5cm 0cm 0cm, clip,width=0.28\hsize]{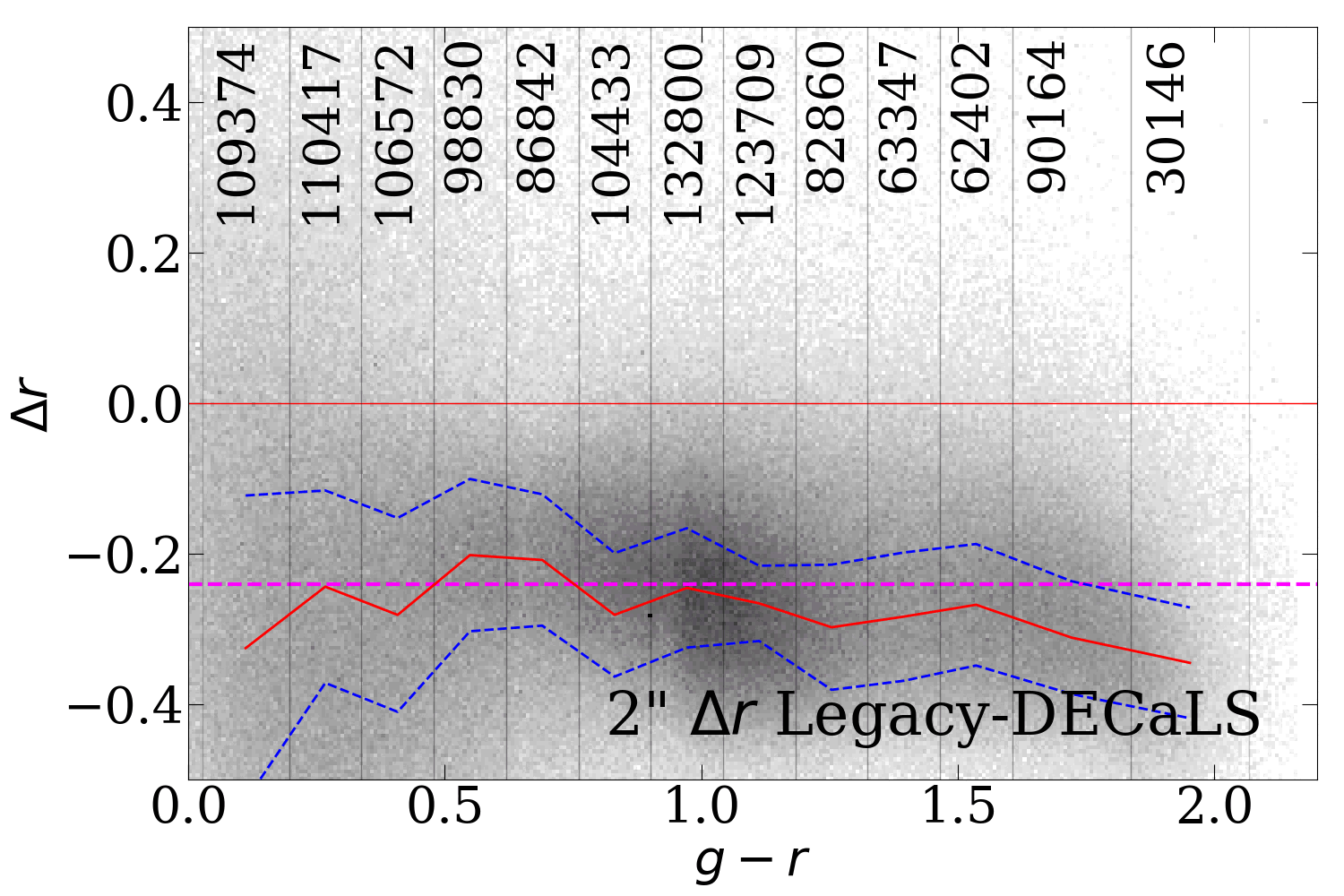}
\includegraphics[trim=3.0cm 2.5cm 0cm 0cm, clip,width=0.28\hsize]{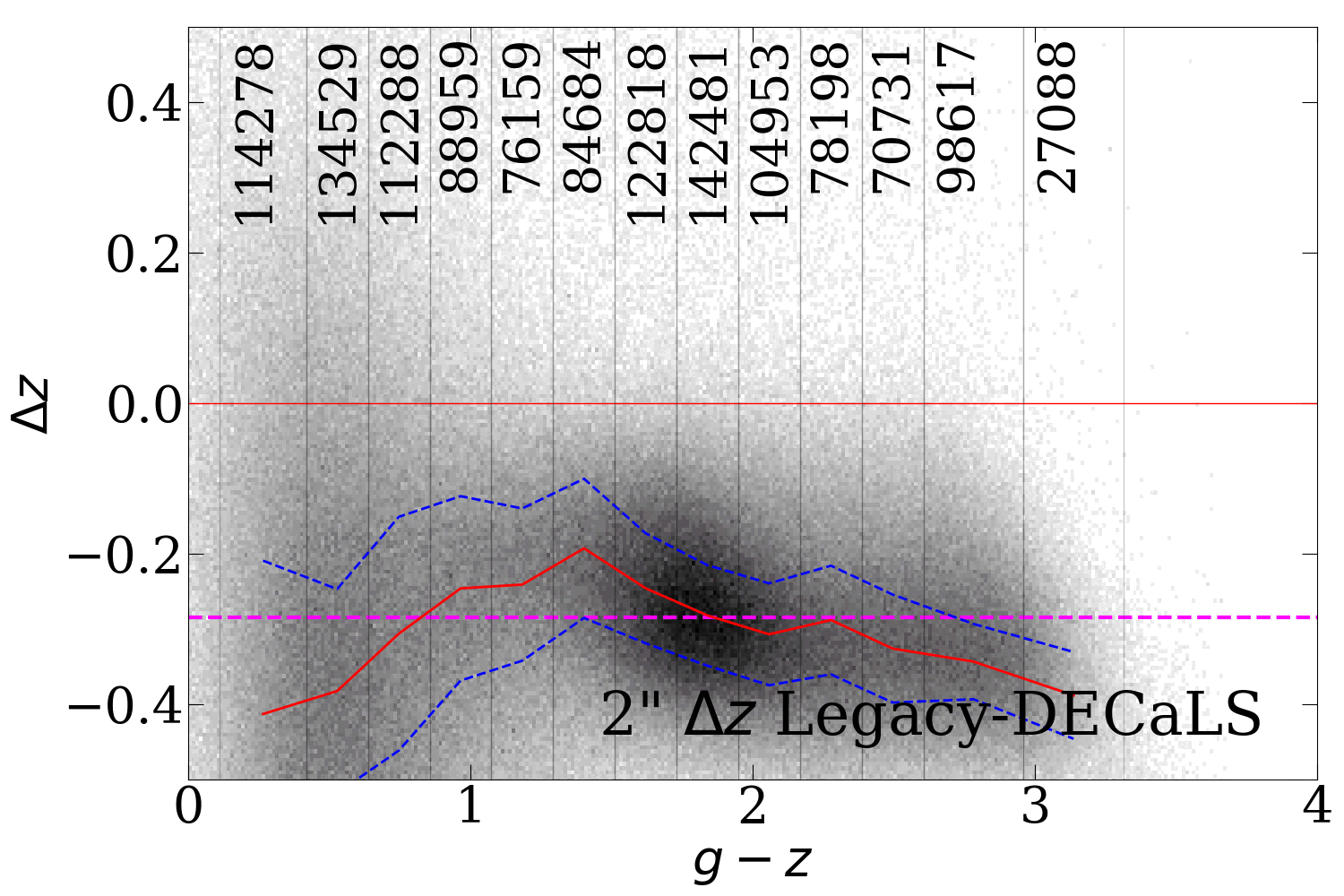}
\includegraphics[trim=0.6cm 2.5cm 0cm 0cm, clip,width=0.2942\hsize]{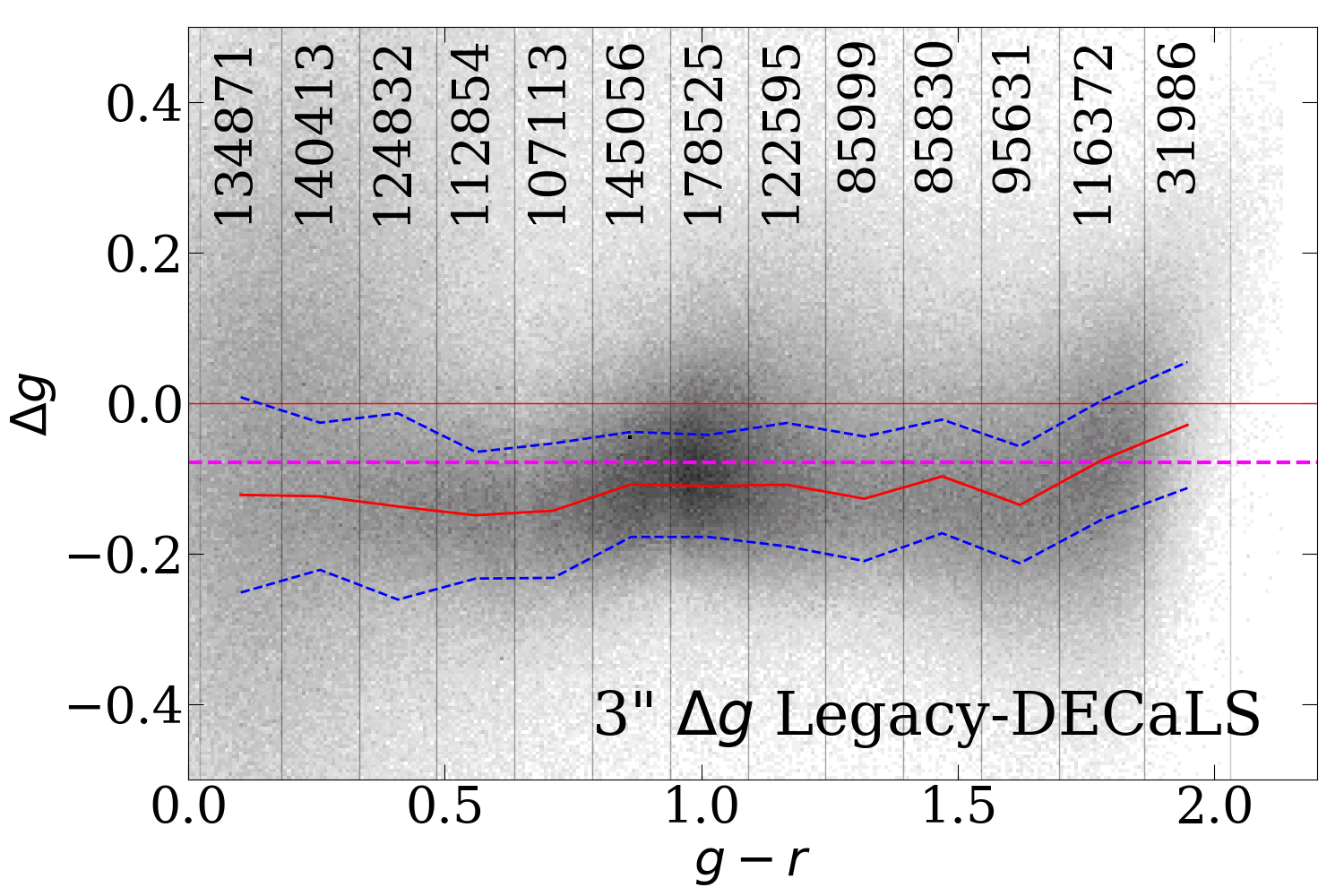}
\includegraphics[trim=3.cm 2.5cm 0cm 0cm, clip,width=0.28\hsize]{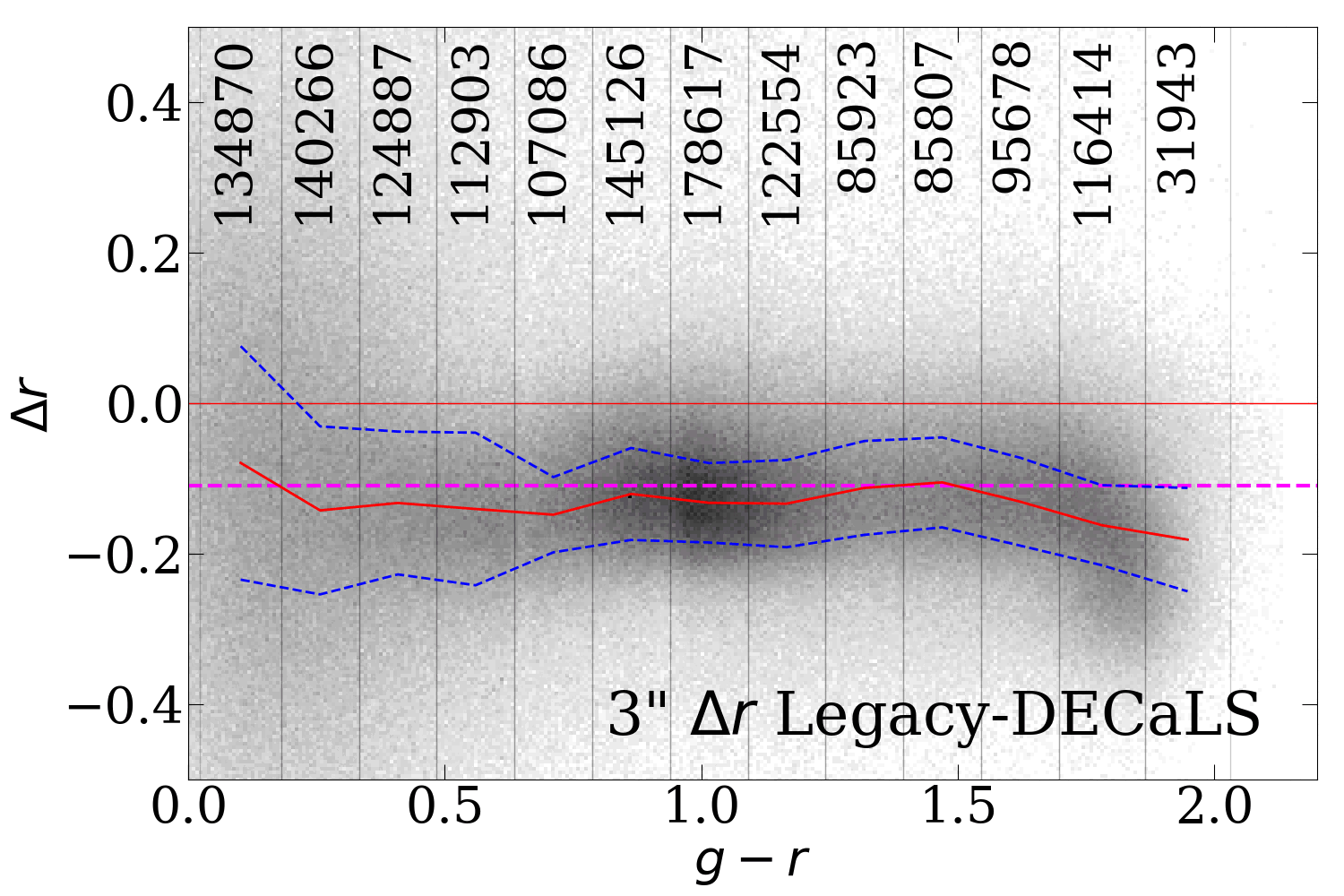}
\includegraphics[trim=3.0cm 2.cm 0cm 0cm, clip,width=0.28\hsize]{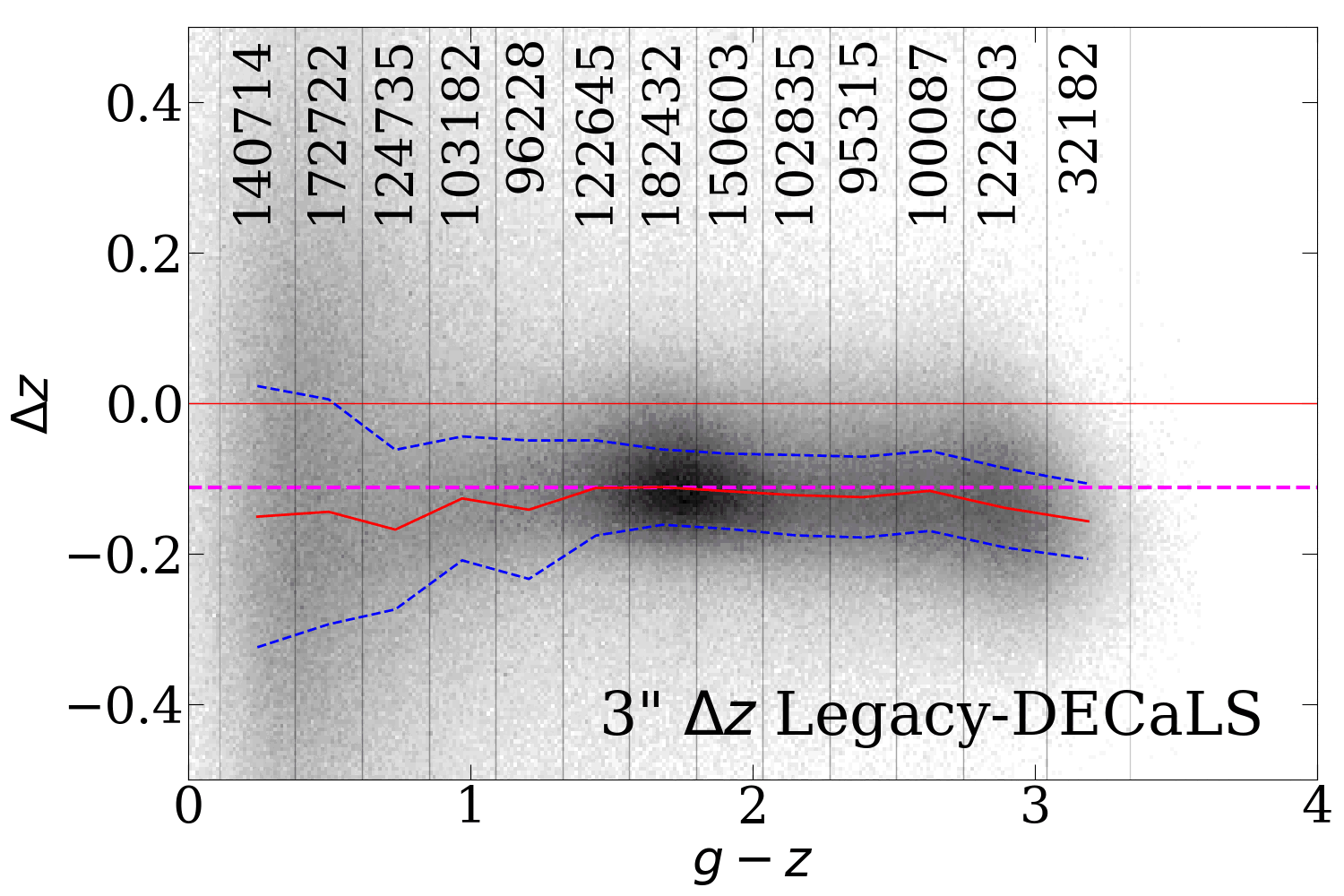}
\includegraphics[trim=0.6cm 2.cm 0cm 0cm, clip,width=0.2942\hsize]{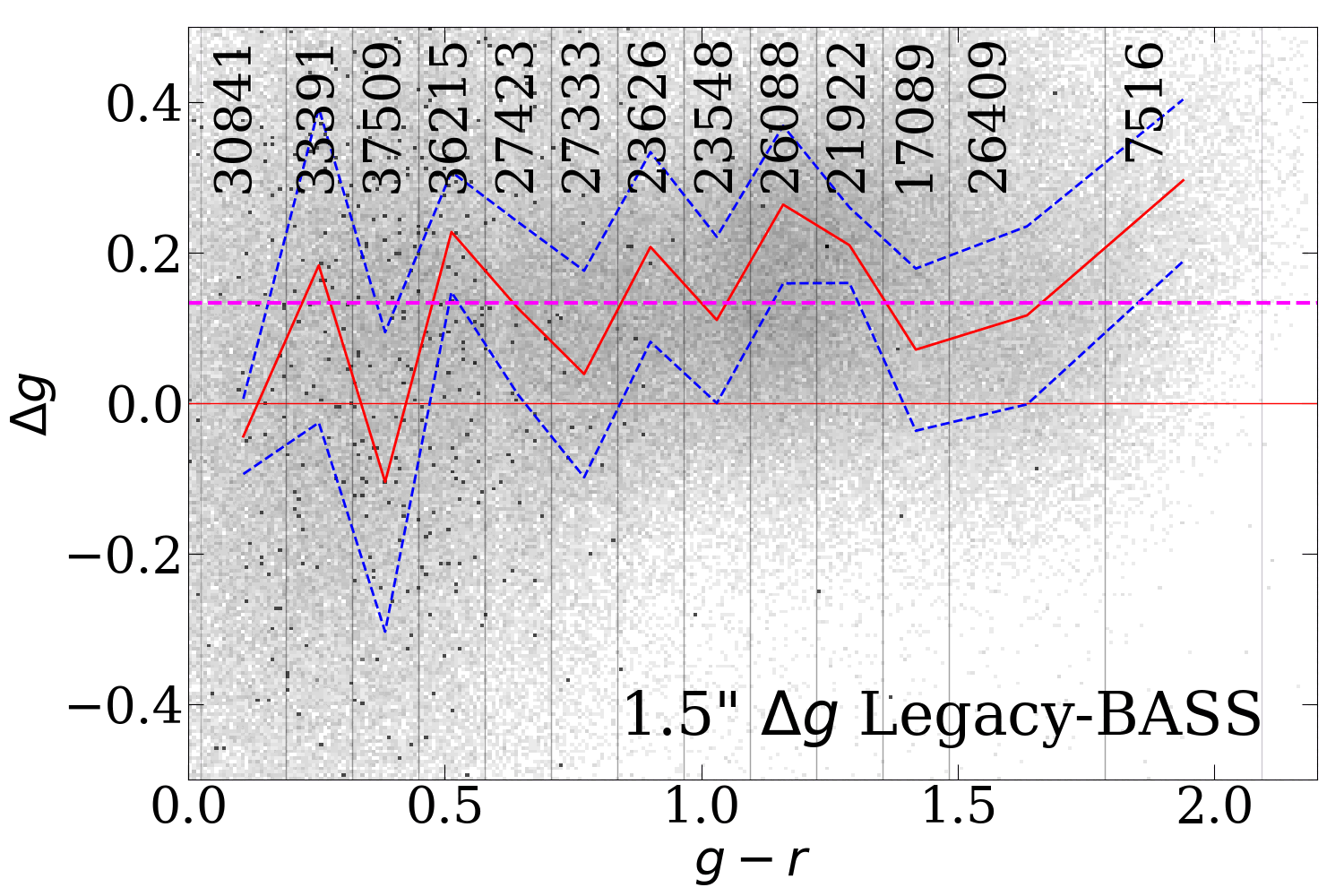}
\includegraphics[trim=3.0cm 2.cm 0cm 0cm, clip,width=0.275\hsize]{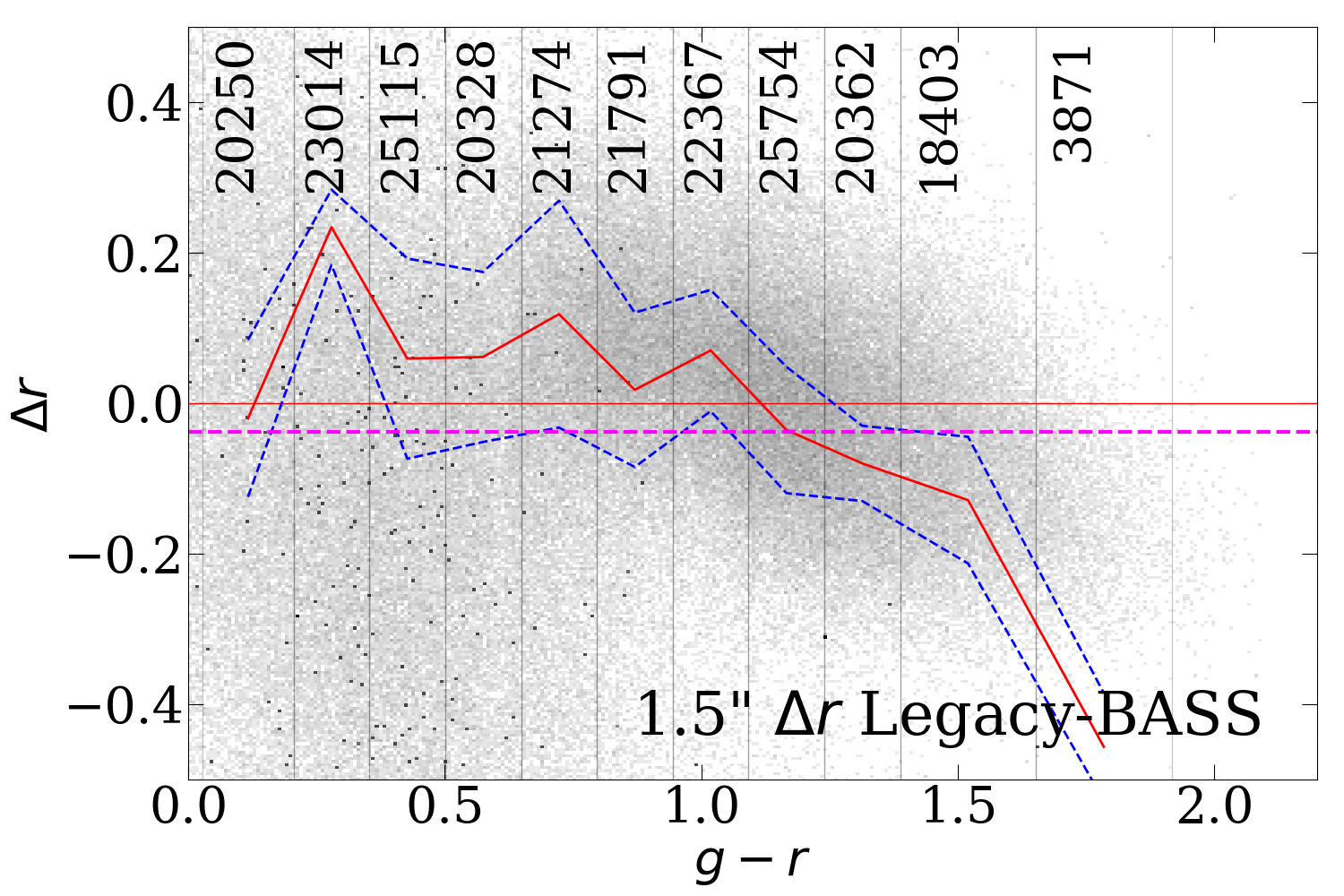}
\includegraphics[trim=3.0cm 2.cm 0cm 0cm, clip,width=0.28\hsize]{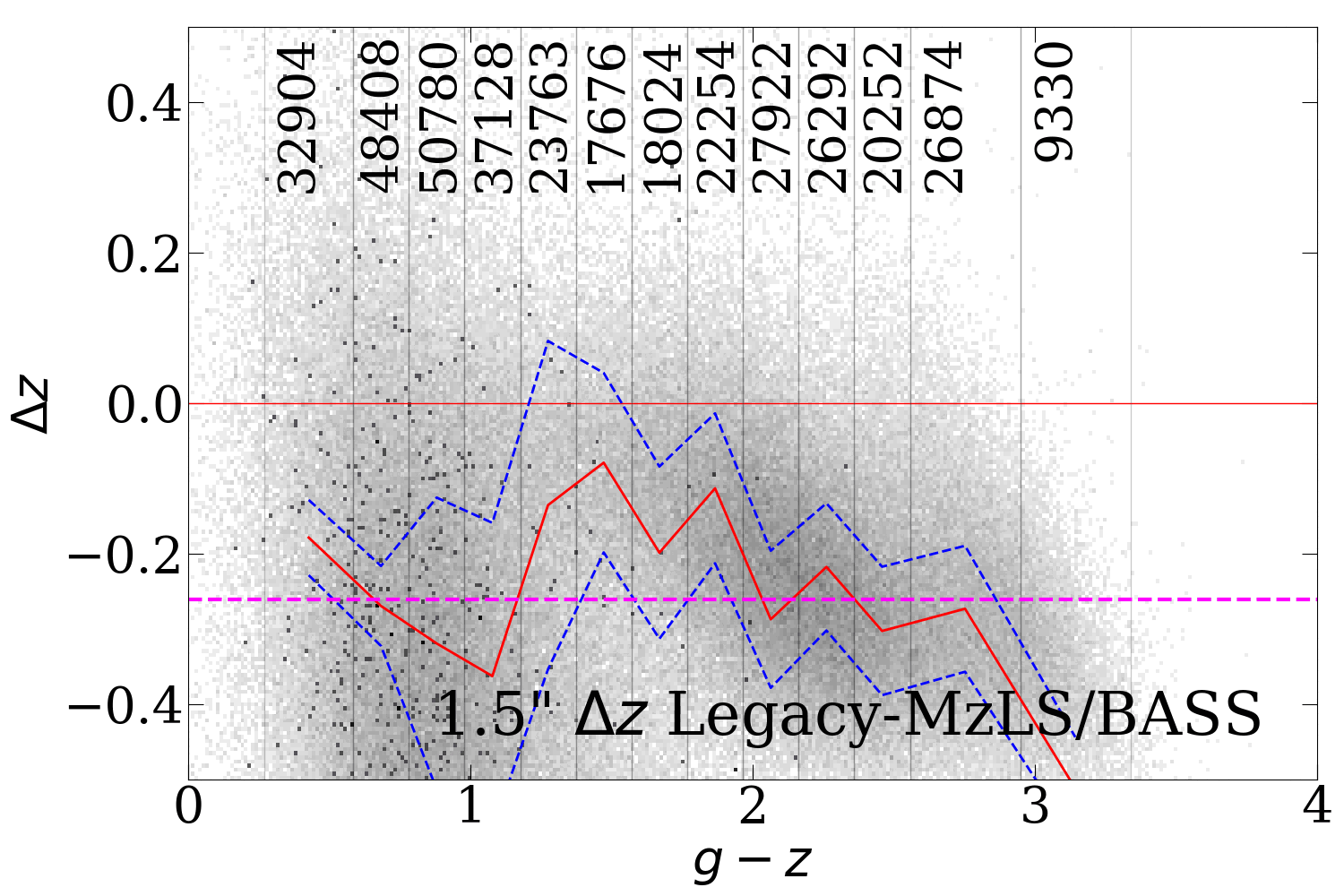}
\includegraphics[trim=0.6cm 2.cm 0cm 0cm, clip,width=0.2942\hsize]{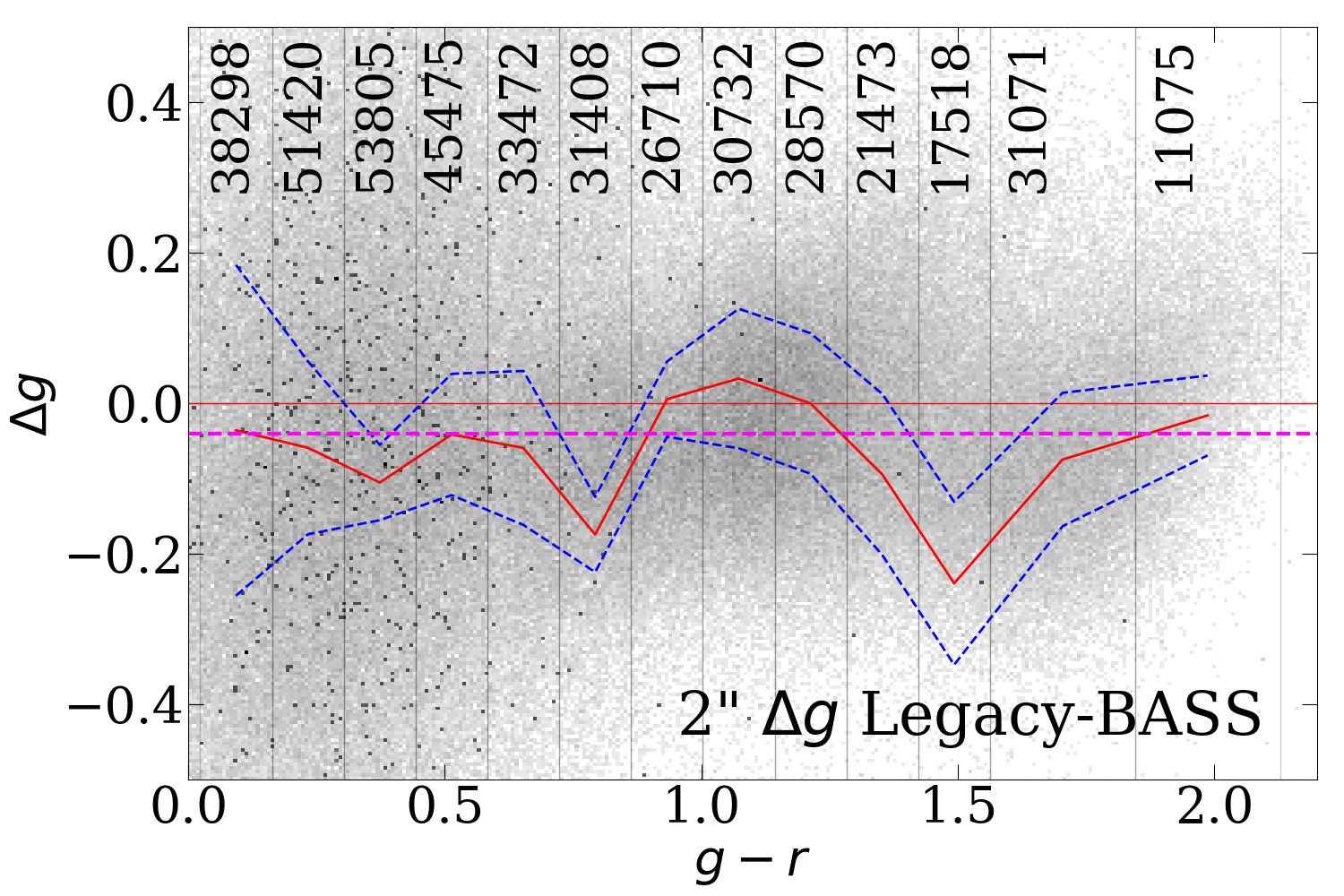}
\includegraphics[trim=3.0cm 2.cm 0cm 0cm, clip,width=0.275\hsize]{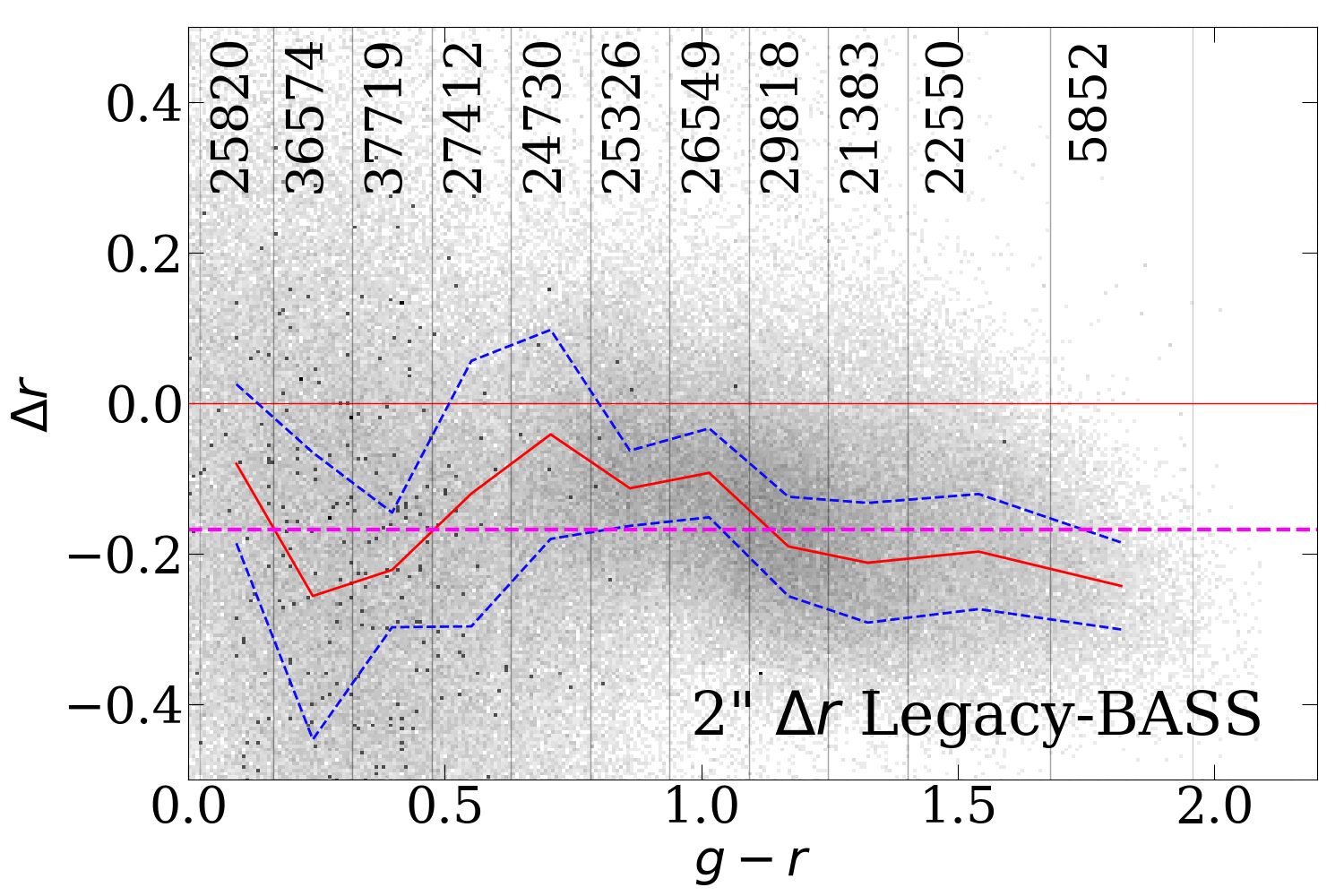}
\includegraphics[trim=3.0cm 2.cm  0cm 0cm, clip,width=0.28\hsize]{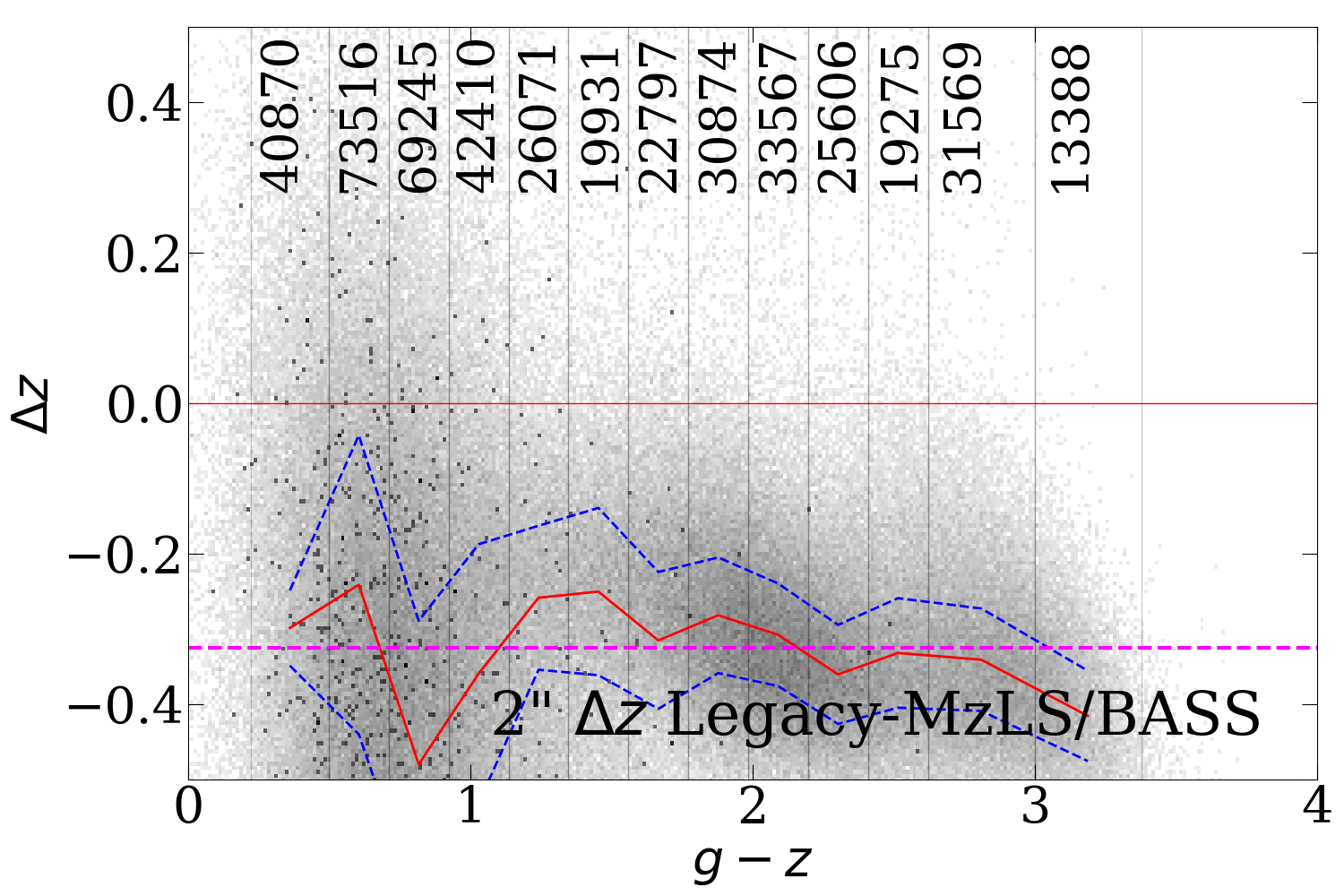}
\includegraphics[trim=0.6cm 0cm 0cm 0cm, clip,width=0.2942\hsize]{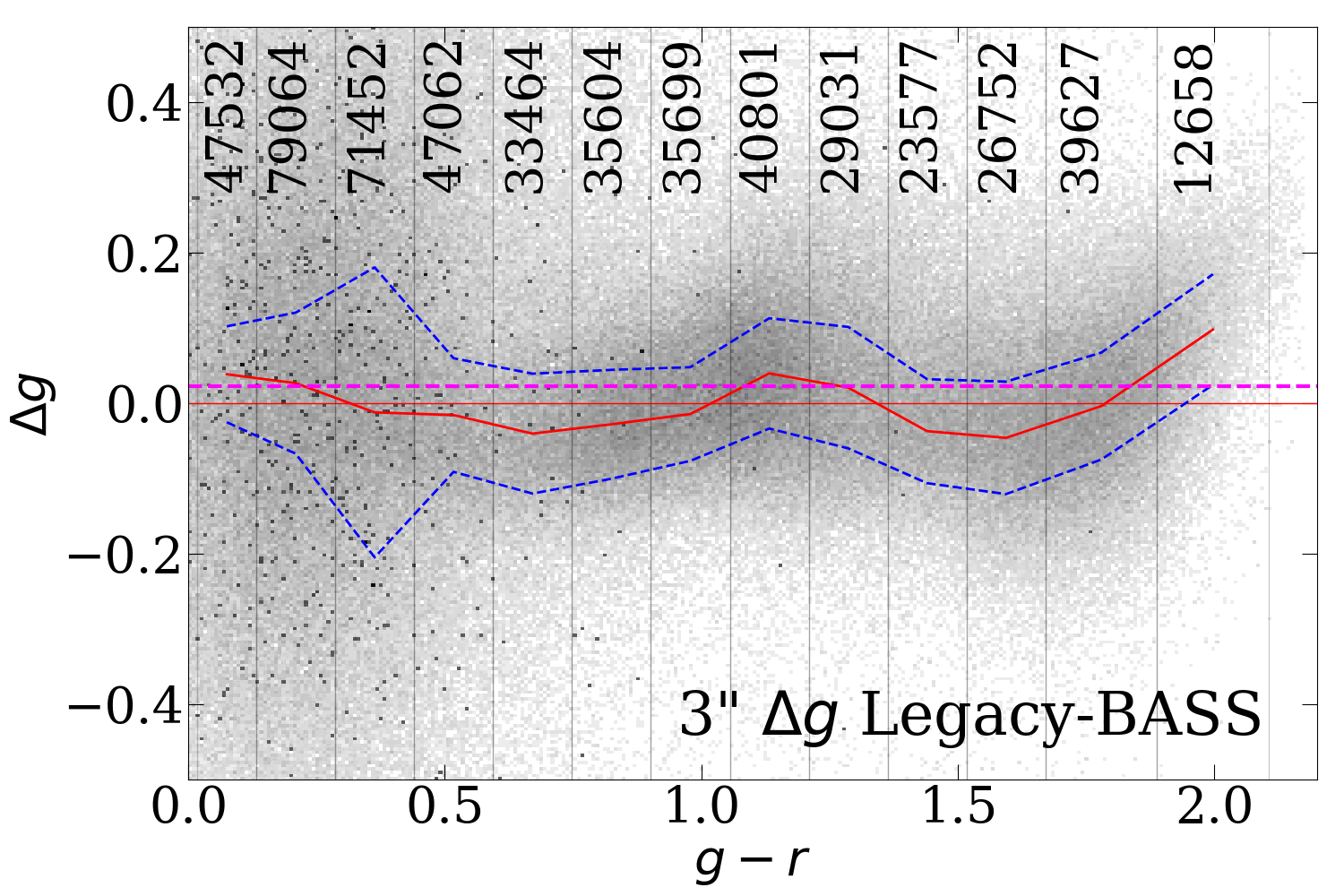}
\includegraphics[trim=3.0cm 0cm 0cm 0cm, clip,width=0.275\hsize]{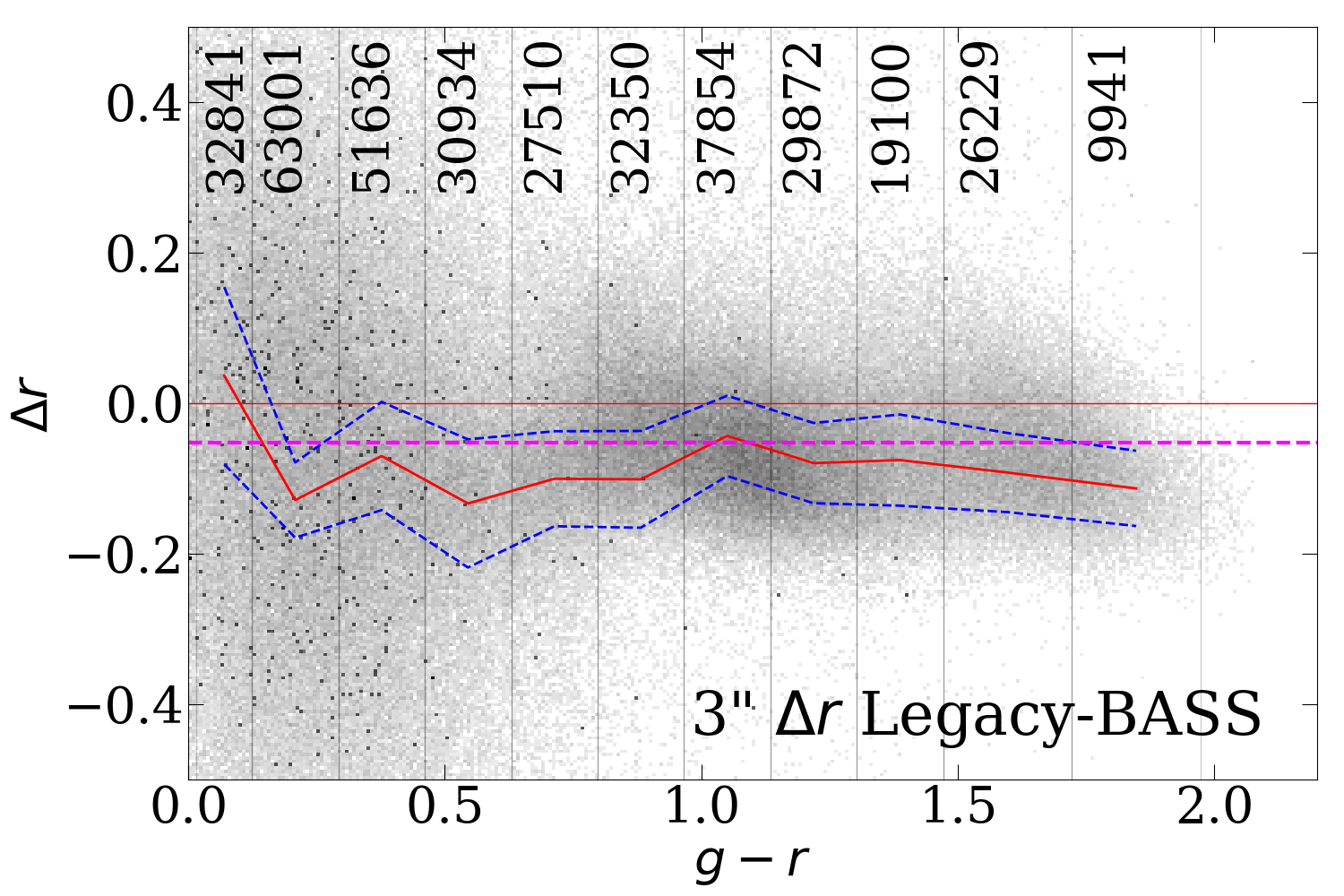}
\includegraphics[trim=3.0cm 0cm 0cm 0cm, clip,width=0.28\hsize]{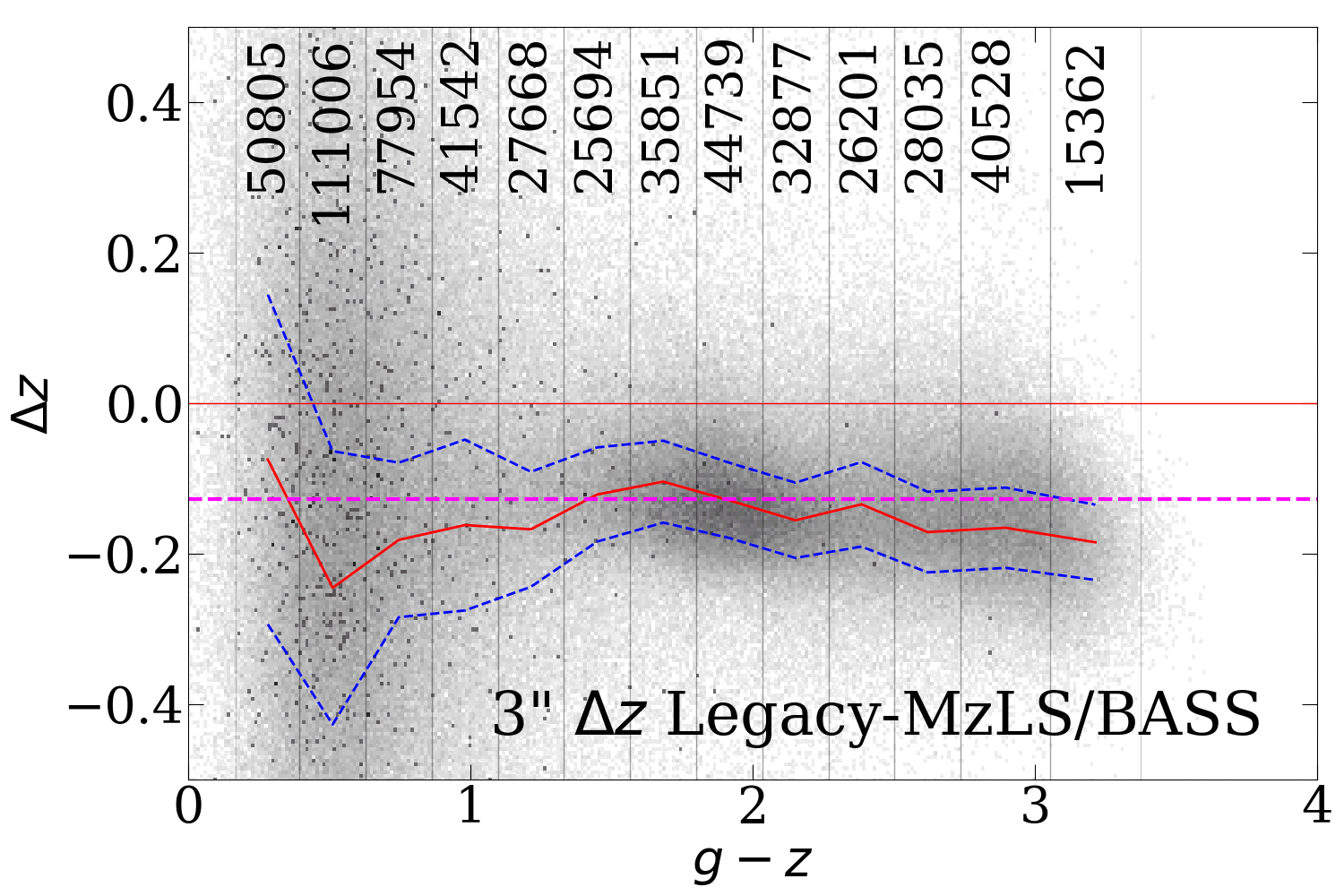}
\caption{Magnitude difference \revtwo{between DESI Legacy Surveys corrected by transformations for total magnitudes and SDSS system} for 1.5$''$, 2$''$ and 3$''$ apertures. \revone{All colors are computed in the same apertures. The zero corresponding to the offsets from Table~\ref{tab_shifts} is shown by a magenta line.}
\label{apers_full2}}
\end{figure*}



\end{document}